\documentclass[ twoside,openright,titlepage,numbers=noenddot,headinclude,%
                footinclude=true,cleardoublepage=empty,abstractoff, %
                BCOR=5mm,paper=a4,fontsize=11pt,%
                ngerman,american,%
                dottedtoc,%
                longbibliography,
                ]{scrreprt}

\PassOptionsToPackage{eulerchapternumbers,listings,%
				 pdfspacing,%
				 subfig,beramono,parts}{classicthesis}										

\usepackage{ifthen}
\newboolean{enable-backrefs} %
\setboolean{enable-backrefs}{false} %

\usepackage[svgnames]{xcolor}
\usepackage{setspace}

\newcommand{\myTitle}{Strong gravity physics of dynamical horizons in black hole mergers and its imprint in their gravitational radiation}

\newcommand{\myName}{Vaishak Prasad\xspace}

\newcommand{\myFaculty}{Put data here\xspace}

\newcommand{\myUni}{Put data here\xspace}

\newcommand{\myVersion}{version 1\xspace}

\newcommand{\be}{\begin{equation}}
\newcommand{\ee}{\end{equation}}
\newcommand{\bea}{\begin{eqnarray}}
\newcommand{\eea}{\end{eqnarray}}

\def\P[#1]#2{\Pi^{#1}_{\phantom{a}#2}}

\newcounter{dummy} %
\providecommand{\mLyX}{L\kern-.1667em\lower.25em\hbox{Y}\kern-.125emX\@}

\PassOptionsToPackage{latin9}{inputenc}	%
 \usepackage{inputenc}				

 \usepackage{babel}					

\PassOptionsToPackage{square,numbers}{natbib}
\usepackage{natbib}				

\PassOptionsToPackage{fleqn}{amsmath}		%
 \usepackage{amsmath}

\PassOptionsToPackage{T1}{fontenc} %
	\usepackage{fontenc}     
\usepackage{textcomp} %
\usepackage{scrhack} %
\usepackage{xspace} %
\usepackage{mparhack} %
\usepackage{fixltx2e} %
\usepackage{setspace} %
\PassOptionsToPackage{printonlyused,smaller}{acronym}
	\usepackage{acronym} %
\usepackage[parfill]{parskip}    %
\usepackage{graphicx}
\usepackage{amssymb}
\usepackage{epstopdf}
\usepackage{amsmath}
\usepackage{hyperref}
\usepackage{mathrsfs}
\usepackage{paralist}
\usepackage{varioref}  %
\usepackage{multirow}

\usepackage{lscape}
\usepackage{booktabs}
\usepackage{longtable}
\usepackage{supertabular,booktabs}
\usepackage{adjustbox}

\usepackage[bottom]{footmisc}
\usepackage[caption=false]{subfig}
\labelformat{part}{Part #1}
\labelformat{chapter}{Chapter #1}
\labelformat{section}{Section #1} 
\labelformat{subsection}{Section #1} 
\labelformat{subsubsection}{Section #1}
\labelformat{equation}{Equation~#1} 
\labelformat{figure}{Figure~#1} 
\labelformat{subfigure}{Figure~\thefigure#1} 
\labelformat{table}{Table~#1}
\labelformat{appendix}{Appendix~#1}

\usepackage{textcomp}
\usepackage{enumerate}

\usepackage{titlesec}

\usepackage{amsfonts}
\usepackage{mathrsfs}

\usepackage{color}
\usepackage[normalem]{ulem}

\expandafter\ifx\csname package@font\endcsname\relax\else
\expandafter\expandafter
\expandafter\usepackage
\expandafter\expandafter
\expandafter{\csname package@font\endcsname}%
\fi
\usepackage{graphics}
\usepackage{comment}
\includecomment{comment}
\titleclass{\subsubsubsection}{straight}[\subsection]

\newcounter{subsubsubsection}[subsubsection]
\renewcommand\thesubsubsubsection{\thesubsubsection.\arabic{subsubsubsection}}
\renewcommand\theparagraph{\thesubsubsubsection.\arabic{paragraph}} %

\titleformat{\subsubsubsection}
  {\normalfont\normalsize\bfseries}{\thesubsubsubsection}{1em}{}
\titlespacing*{\subsubsubsection}
{0pt}{3.25ex plus 1ex minus .2ex}{1.5ex plus .2ex}

\makeatletter
\renewcommand\paragraph{\@startsection{paragraph}{5}{\z@}%
  {3.25ex \@plus1ex \@minus.2ex}%
  {-1em}%
  {\normalfont\normalsize\bfseries}}
\renewcommand\subparagraph{\@startsection{subparagraph}{6}{\parindent}%
  {3.25ex \@plus1ex \@minus .2ex}%
  {-1em}%
  {\normalfont\normalsize\bfseries}}
\def\toclevel@subsubsubsection{4}
\def\toclevel@paragraph{5}
\def\toclevel@paragraph{6}
\def\l@subsubsubsection{\@dottedtocline{4}{7em}{4em}}
\def\l@paragraph{\@dottedtocline{5}{10em}{5em}}
\def\l@subparagraph{\@dottedtocline{6}{14em}{6em}}
\makeatother

\usepackage{tabularx} %
\usepackage{caption}
\usepackage{subfig}

\usepackage{listings} 
\PassOptionsToPackage{pdftex,hyperfootnotes=false,pdfpagelabels}{hyperref}
	\usepackage{hyperref}  %
\PassOptionsToPackage{pdftex}{graphicx}
	\usepackage{graphicx} 

\newcommand{\backrefnotcitedstring}{\relax}%
\newcommand{\backrefcitedsinglestring}[1]{(Cited on page~#1.)}
\newcommand{\backrefcitedmultistring}[1]{(Cited on pages~#1.)}
\ifthenelse{\boolean{enable-backrefs}}%
{%
		\PassOptionsToPackage{hyperpageref}{backref}
		\usepackage{backref} %
		   \renewcommand*{\backref}[1]{}  %
		   \renewcommand*{\backrefalt}[4]{%
		      \ifcase #1 %
		         \backrefnotcitedstring%
		      \or%
		         \backrefcitedsinglestring{#2}%
		      \else%
		         \backrefcitedmultistring{#2}%
		      \fi}%
}{\relax}    

\hypersetup{%
    colorlinks=true, linktocpage=true, pdfstartpage=3, pdfstartview=FitV,%
    breaklinks=true, pdfpagemode=UseNone, pageanchor=true, pdfpagemode=UseOutlines,%
    plainpages=false, bookmarksnumbered, bookmarksopen=true, bookmarksopenlevel=1,%
    hypertexnames=true, pdfhighlight=/O,%
    urlcolor=webbrown, linkcolor=RoyalBlue, citecolor=webgreen, %
    pdftitle={\myTitle},%
    pdfauthor={\textcopyright\ \myName, \myUni, \myFaculty},%
    pdfsubject={},%
    pdfkeywords={},%
    pdfcreator={pdfLaTeX},%
    pdfproducer={LaTeX with hyperref and classicthesis}%
}   

\makeatletter
\@ifpackageloaded{babel}%
    {%
       \addto\extrasamerican{%
				}%
       \addto\extrasngerman{%
				}%
    }{\relax}
\makeatother

\usepackage{classicthesis} 
\usepackage{lmodern}
\usepackage{tikz}
\usepackage{threeparttable}
\usetikzlibrary{calc}

\ifthenelse{\boolean{@linedheaders}}%
    {%
    \titleformat{\chapter}[display]%
        {\relax}
        {\raggedleft{\color{halfgray}\chapterNumber\thechapter} \\ }
        {0pt}%
        {\color{Blue!70!black}\huge\titlerule\vspace*{.9\baselineskip}\raggedright\spacedallcaps}%
        [\normalcolor\normalsize\vspace*{.8\baselineskip}\titlerule]%
    }
    {%
    \titleformat{\chapter}[display]%
        {\relax}
        {\mbox{}\oldmarginpar{\vspace*{-3\baselineskip}\color{halfgray}\chapterNumber\thechapter}}
        {0pt}%
        {\color{Blue!70!black}\huge\raggedright\spacedallcaps}%
        [\normalcolor\normalsize\vspace*{.8\baselineskip}\titlerule]%
    }

\def\be{\begin{equation}}
	\def\ee{\end{equation}}
\def\bea{\begin{eqnarray}}
	\def\eea{\end{eqnarray}}

\usepackage[tikz]{bclogo}
\usepackage{listings}

\definecolor{codegreen}{rgb}{0,0.6,0}
\definecolor{codegray}{rgb}{0.5,0.5,0.5}
\definecolor{codepurple}{rgb}{0.58,0,0.82}
\definecolor{backcolour}{rgb}{0.0,0.0,0}
\definecolor{codewhite}{rgb}{1,1,1}

\lstdefinestyle{mystyle}{
	backgroundcolor=\color{backcolour},   
	commentstyle=\color{codegreen},
	keywordstyle=\color{magenta},
	numberstyle=\tiny\color{codegray},
	stringstyle=\color{codepurple},
	basicstyle=\ttfamily\footnotesize\color{codewhite},
	breakatwhitespace=false,         
	breaklines=true,                 
	captionpos=b,                    
	keepspaces=true,                 
	numbers=left,                    
	numbersep=5pt,                  
	showspaces=false,                
	showstringspaces=false,
	showtabs=false,                  
	tabsize=2
}

\numberwithin{equation}{chapter}

\usepackage[toc, titletoc, title, page]{appendix}

\KOMAoptions{toc=indentunnumbered,
	chapterentrydots=true%
}

\begin{document}

\newcommand{\sukanta}{\textcolor{blue}}
\newcommand{\vaishak}{\textcolor{red}}

\newcommand{\myh}{2.8cm}
\newcommand{\myw}{7.6cm}

\newcommand{\rhat}{\hat{r}}
\newcommand{\iotahat}{\hat{\iota}}
\newcommand{\phihat}{\hat{\phi}}
\newcommand{\h}{\mathfrak{h}}
\newcommand{\vek}[1]{\boldsymbol{#1}}

\renewcommand\eqref{\ref}   

\frenchspacing %

\raggedbottom %

\selectlanguage{american} %

\pagenumbering{roman} %

\pagestyle{plain} %

\begin{titlepage}
	\begin{addmargin}[-3cm]{-4.5cm}
	
	\begin{tikzpicture}[remember picture, overlay]
  \draw[line width = 2.4pt] ($(current page.north west) + (1in,-1in)$) rectangle ($(current page.south east) + (-1in,1in)$);
        \end{tikzpicture} 
        \begin{tikzpicture}[remember picture, overlay]
  \draw[line width = 1.2pt] ($(current page.north west) + (1.05in,-1.05in)$) rectangle ($(current page.south east) + (-1.05in,1.05in)$);
    \end{tikzpicture}
    \begin{center}
        \large  
        \hfill
        \vfill
        \begingroup
        \begin{center}
            \color{midnightblue}\spacedallcaps{\Large Strong gravity physics\xspace} 
        \end{center} 
        \endgroup
        \begingroup
        \begin{center}
            \color{midnightblue}\spacedallcaps{\Large of dynamical horizons\xspace} 
         \end{center}    
        \endgroup  
        \begingroup
        \begin{center}
            \color{midnightblue}\spacedallcaps{\Large in black hole mergers\xspace} 
         \end{center}    
        \endgroup  
        
        \begingroup
        \begin{center}
            \color{midnightblue}\spacedallcaps{\Large and its imprint in their\xspace} 
         \end{center}    
        \endgroup  
        \begingroup
        \begin{center}
        	\color{midnightblue}\spacedallcaps{\Large gravitational radiation\xspace} 
        \end{center}    
        \endgroup  
        \medskip
        \bigskip
        \bigskip
        \textls[90]{\scshape{\textbf{Vaishak Prasad}}}
        \medskip
        \bigskip
        \bigskip
        \begingroup
        \begin{center}
            {\textbf{Thesis}\xspace} 
        \end{center} 
        \endgroup
        \begingroup
        \begin{center}
            {submitted for the degree of\xspace} 
        \end{center}
        \endgroup  
        \begingroup
        \begin{center}
            {\textit{\textbf{Doctor of Philosophy} }(in Physics)\xspace} 
         \end{center}    
        \endgroup  
        \begin{center}
         to
        \end{center}
        \begin{center}
         {\Large Jawaharlal Nehru University }
         
        \end{center}

        \bigskip
        \includegraphics[height=2.4cm]{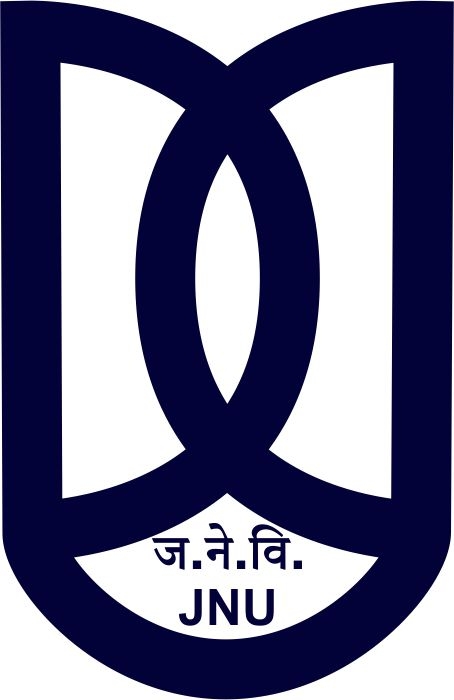} \quad
        \bigskip
        \bigskip

        \vfill
    \bigskip
      \includegraphics[width=2.1cm]{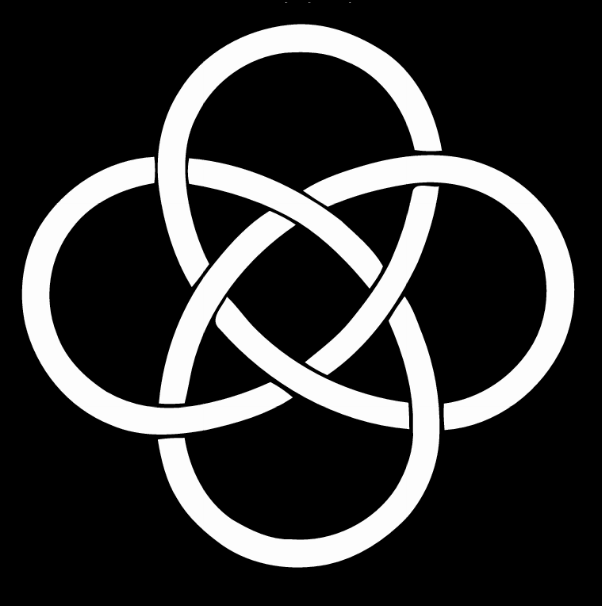}\quad \quad

        \vfill                      

        \begin{center}
        Inter-University Centre for Astronomy and Astrophysics 
        \end{center}
        \begin{center}
        Pune, India-411007
        \end{center}
        \medskip
        \bigskip
        \bigskip
        \bigskip
        \bigskip
        \bigskip
        \bigskip
        \bigskip

    \end{center}  
  \end{addmargin}       
\end{titlepage}   

\setstretch{1.5}
\cleardoublepage\begin{titlepage}
    
    \begin{flushright}
        \large  
        \hfill
        \vfill
       
        \medskip

     \begin{center}
    	{\Large \color{midnightblue} \textit{This thesis is dedicated to my fellow Indians\\ and \\ my lovely family for their unconditional love and support.}}
\end{center}
        \medskip
        \bigskip
        \bigskip
        \bigskip
        \bigskip

    \end{flushright}  
\end{titlepage}   

\cleardoublepage\pdfbookmark[1]{Certificate}{Certificate}

\bigskip

\begingroup
\let\clearpage\relax
\let\cleardoublepage\relax
\let\cleardoublepage\relax

\,\;$\;$
\bigskip
\bigskip
\bigskip
\bigskip
\bigskip
\bigskip
\bigskip

\chapter*{Certificate}

This is to certify that the thesis entitled `\textbf{Strong gravity physics of dynamical horizons in binary black hole mergers and its imprint in their gravitational radiation}' submitted by Mr. Vaishak Prasad for the award of the degree of Doctor of Philosophy of Jawaharlal Nehru University, New Delhi is his
original work. This has not been published in or submitted to any other University for any
other Degree or Diploma.

\bigskip
\bigskip
\bigskip 
\bigskip
\bigskip

\bigskip

\def\s#1#2{\vbox{\hsize=4.5cm
    \kern2cm
    \hrule\kern2ex
    \hbox to \hsize{\strut\hfil #1 \hfil}
    \hbox to \hsize{\strut\hfil #2 \hfil}}}

\hbox to \hsize{\s{Prof. Somak Raychaudhury}{(Director, IUCAA)}\hfil
    \s{Prof. Sukanta Bose}{(Thesis Supervisor)}}

\endgroup

\cleardoublepage\pdfbookmark[1]{Declaration}{Declaration}

\bigskip

\begingroup
\let\clearpage\relax
\let\cleardoublepage\relax
\let\cleardoublepage\relax

\,\;$\;$
\bigskip
\bigskip
\bigskip
\bigskip
\bigskip
\bigskip
\bigskip

\chapter*{Declaration}

I hereby declare that the work reported in this thesis is entirely original. This thesis is
composed independently by me at the Inter University Centre for Astronomy and Astrophysics,
Pune under the supervision of Prof. Sukanta Bose. I further declare
that the subject matter presented in the thesis has not previously formed the basis for the
award of any degree, diploma, membership, associateship, fellowship, or any other similar
title of any University or Institution.

\bigskip
\bigskip
\bigskip 
\bigskip
\bigskip

\bigskip

\def\s#1#2{\vbox{\hsize=4.5cm
    \kern2cm
    \hrule\kern1ex
    \hbox to \hsize{\strut\hfil #1 \hfil}
    \hbox to \hsize{\strut\hfil #2 \hfil}}}

\hbox to \hsize{\s{Prof. Sukanta Bose}{(Thesis Supervisor)}\hfil
    \s{Vaishak Prasad}{(Ph.D. Candidate)}}

\endgroup

\tableofcontents
\listoftables
\listoffigures
\cleardoublepage\pdfbookmark[1]{Acknowledgements}{Acknowledgements}

\bigskip

\begingroup
\let\clearpage\relax
\let\cleardoublepage\relax
\let\cleardoublepage\relax

\chapter*{Acknowledgements}

First and foremost, I would like to thank my Supervisor for the unstinted support and patience throughout my PhD years. Prof. Sukanta Bose always provided me with great independence, given the exploratory nature of this work. I would also thank Dr. Anshu Gupta for her discussions and support. She has always been there for discussions. I thank Badri Krishnan for timely guidance and discussions and support; I especially thank my supervisor, Badri Krishnan, for patiently listening to my ideas and my endeavors, and providing feedback. Prof. Erik Schnetter for encouragement and support. All of these people patiently persisted, listened to, and encouraged me even when the going got hard.

I thank Prof. Sanjit Mitra and Prof. Aseem Paranjape, who were a part of my research advisory committee team for encouragement, support, and constructive criticism.

Importantly, I thank my family who have continually supported me, patiently all through these years.

I would like to express my gratitude to the Director Prof. Somak Raychaudhury and IUCAA for the hospitality. I am also grateful and would like to acknowledge the support of all the staff at IUCAA. Thanks to the Canteen, Library, academic, non-academic, and computer center staff for their continued support. I would like to thank IUCAA for the infrastructure, computing facility, and resources. Several computing resources were bought using the contingency grant awarded to me through the Shyama Prasad Mukherjee Fellowship by CSIR. The author acknowledges their support.

Special thanks to my friends at IUCAA, who have made my stay memorable. I would like to mention the batch 2016: Bhaskar Biswas, Gitika Shukla, Pranoti Panchbhai, Somak Maitra, Sorabh Chhabra, Sunil Choudhary and Yash Bhargava. Ruchika Seth, Vidushi Sharma, Divya Rawat, Adarsh Ranjan, Rajorshi Chandra, Ajay Sharma, Aakash, Sayak Datta, just to name a few. The list includes many more friends of mine not mentioned here.

The author also thanks the Navajbai Ratan Tata Trust, and Prof. Ajit Kembhavi for funding partially the participation in Les Houches School on Gravitational Waves, 2018.

The numerical simulations in this thesis were performed on the Perseus and Pegasus cluster at The Inter-University Centre for Astronomy and Astrophysics, Pune, India (IUCAA). The author's Ph.D. was funded by the Shyama Prasad Mukherjee fellowship (CSIR). 

Finally, we thank Prof. K. G. Arun and Prof. Ivan Booth for their valuable comments and suggestions.

\endgroup

\cleardoublepage\pdfbookmark[1]{Publications}{Publications}

\bigskip

\begingroup
\let\clearpage\relax
\let\cleardoublepage\relax
\let\cleardoublepage\relax

\bigskip
\chapter*{Publications in Astronomy \& Astrophysics}
\section*{Included in this thesis}
\begin{enumerate}
 \item \textbf{Vaishak Prasad}. ``The Shear at the Common Dynamical Horizon in Binary Black Hole Mergers and its Imprint in their Gravitational Radiation." \\\emph{Published 28 April 2025.} \href{https://journals.aps.org/prd/abstract/10.1103/PhysRevD.111.084070}{Phys. Rev. D 111, 084070}, \href{https://arxiv.org/abs/2312.01136}{e-print: arXiv:2312.01136}.
\item \textbf{Vaishak Prasad}. ``Tidal deformation of dynamical horizons in binary black hole mergers and its imprint on gravitational radiation``. \emph{Published 16 February 2024}.\\ \href{https://journals.aps.org/prd/abstract/10.1103/PhysRevD.109.044033}{Phys.Rev.D 109, 044033}. Based on \href{https://arxiv.org/abs/2109.01193}{arXiv:2109.01193}.
  \item \textbf{Vaishak Prasad}, Anshu Gupta, Sukanta Bose, Badri Krishnan. ``Tidal deformation of dynamical horizons in binary black hole mergers." \\\emph{Published 9 February 2022.} \href{https://journals.aps.org/prd/abstract/10.1103/PhysRevD.105.044019}{Phys. Rev. D 105, 044019.}
 \href{https://arxiv.org/abs/2106.02595}{e-print: arXiv:2106.02595}.
 \item \textbf{Vaishak Prasad}, Anshu Gupta, Sukanta Bose, Badri Krishnan, and Erik Schnetter. ``News from Horizons in Binary Black Hole Mergers." \\\emph{Published 16 September 2020.} \href{https://journals.aps.org/prl/abstract/10.1103/PhysRevLett.125.121101}{Phys. Rev. Lett. 125, 121101.} \href{https://arxiv.org/abs/2003.06215}{e-print: arXiv:2003.06215}.
\end{enumerate}

\noindent Chapters 3, 4 and 5 are based on the articles above published in Physical Review Letters and Physical Review D, \textcopyright{} American Physical Society (2020, 2022, 2024 and 2025). They are reused here in accordance with the author rights of the American Physical Society.

\bigskip
\subsection*{Conference Proceedings}
\begin{enumerate}
    \item Dec 2023, \href{https://indico.cern.ch/event/1268737/}{The 10th International Conference on Gravitation and Cosmology (ICGC), IIT Guwahati}. \textbf{\emph{Talk on Tidal deformation of dynamical horizons in binary black hole mergers and its imprint in their gravitational radiation}}
    \item Feb 2022, \href{https://physics.iitm.ac.in/~csgc/events/csgc/csgc-2022.html}{Second Chennai Symposium on Gravitation and Cosmology} (Virtual). Invited talk on \textbf{\emph{Tidal deformability of dynamical horizons in binary black hole mergers.}}
    \item Jul 2021, \href{https://www.amaldi14.org/}{Amaldi 14}, July 2021 (Online), Poster on \textbf{\emph{Tidal deformation of dynamical horizons in binary black hole mergers.}}
    \item Dec 2019, \href{https://web.iisermohali.ac.in/web/icgc2019/index.html}{International Conference on Gravitational and Cosmology (ICGC)}, 2019, Mohali. Talk on \textbf{\emph{The relation between horizon shear and gravitational waves at null infinity.}}
    \item Jan 2019, 30th meeting of the \href{https://universe.bits-pilani.ac.in/hyderabad/iagrg/iagrg}{Indian Association for General Relativity and Gravitation (IAGRG)} 2019, Hyderabad.Talk on \textbf{\emph{Chronology protection problem in modified Kerr-Newman spacetimes.}}
    \item Jul 2018, \href{http://www.lkb.upmc.fr/gravitationalwaves2018/}{Les Houches Summer School on Gravitational Waves}, 4-27 July 2018, Les Houches, France. Talk on \textbf{\emph{The Chronology protection problem in modified Kerr-Newman spacetimes.}}
\end{enumerate}
\bigskip
\section*{Not included in this thesis}
\begin{enumerate}
\item Sashideep Gutti, Shailesh Kulkarni, \textbf{Vaishak Prasad}, Sheldon Suresh. ``Closed timelike curves and energy conditions in regular spacetimes." \emph{Published 14 January 2019.} \href{https://journals.aps.org/prd/abstract/10.1103/PhysRevD.99.024023}{Phys. Rev. D 99, 024023.}  \href{https://arxiv.org/abs/1803.06666}{e-print: arXiv:1803.06666}.
\item \textbf{Vaishak Prasad}, Rahul Srinivasan, Sashideep Gutti. ``The Chronology Protection Problem in modified Kerr-Newman space-times."  \emph{Published 14 January 2019.} \href{https://journals.aps.org/prd/abstract/10.1103/PhysRevD.99.024023}{Phys. Rev. D 99, 024023.}  \href{https://arxiv.org/abs/1803.06666}{e-print: arXiv:1803.06666}. 
\end{enumerate}

\endgroup

\cleardoublepage\pdfbookmark[1]{CodesDeveloped}{Codes developed / originated from this thesis}

\bigskip

\begingroup
\let\clearpage\relax
\let\cleardoublepage\relax
\let\cleardoublepage\relax

\bigskip
\chapter*{Codes independently developed \& originated from this thesis}
\section*{Included in this thesis}

\begin{enumerate}
\item \textbf{wfs\_match}: An MPI-enabled, parallel, scalable template-based best-match parameter estimation routine, primarily for exploring the correlation between horizon quantities and the observed gravitational wave signals. 
    \item \textbf{waveformtools}: A library of data handling functions for post-processing and handling Numerical Relativity simulation data and their waveforms. Includes tools for accurate computation of spin-weighted spherical harmonics (SWSH), decomposing 2D data into Spin-weighted spherical harmonic modes, carrying out center of mass corrections, BMS transformations, extrapolation to infinity on the gravitational waveforms, evaluation of polarizations, loading numerical relativity data from different NR catalogs (e.g. RIT, SXS, GT, etc) and their comparisons. Publicly available via PyPi/ git: \href{https://github.com/vaishakp/waveformtools}{https://github.com/vaishakp/waveformtools}
    \item \textbf{yumcee}: An MPI-enabled Markov-Chain and Nested Monte Carlo sampler. 
    \item \textbf{spectral}: A set of transforms and tools to aid in solving simple PDEs and ODEs numerically.
    \item \textbf{qlmtools}: A library of functions for the handling, analysis, and computation of quasi-local and 2-D dynamical horizon numerical data.
    \item \textbf{null\_solver}: An implementation of the dual-null characteristic numerical solver for studying scattering problems.
    \item \textbf{HPC utilities}: Large data handling, and syncing scripts across network storage devices. 
    \item \textbf{vistools}: Collections of tools to visualize NR simulations and dynamical horizons using Python, vtk, and visit.
    \item \textbf{quasinormal}: A library of functions for best-fit analysis of quasi-normal modes.
    \item \textbf{EinsteinToolkit}: Helped identify and reported certain I/O bugs.
\end{enumerate}
\bigskip
\section*{Codes developed that originated directly from this thesis}
\begin{enumerate}
\item \textbf{ahsolver}: A simple spectral apparent horizon solver
\item \textbf{pkginstaller}: Batch installation of software and their optimization to HPC hardware.
\end{enumerate}

\endgroup

\cleardoublepage\pdfbookmark[1]{Synopsis}{Synopsis}

\begingroup
\let\clearpage\relax
\let\cleardoublepage\relax
\let\cleardoublepage\relax

\chapter*{Abstract}

In a binary black hole merger, it is known that the
inspiral portion of the gravitational waveform from two distinct horizons
orbiting each other, and the merger and ringdown signals, correspond
to the final horizon being formed and settling down to equilibrium.  
However, we still lack a detailed understanding of the relation
between the horizon geometry in these three regimes and the observed
waveform. Here, in this thesis, after some preliminaries in the first two chapters (i.e. \ref{chap:intro}, \ref{chap:nr_sims}), we show in \ref{chap:waveform_rel} that the well-known inspiral
chirp waveform has a clear counterpart on black hole horizons,
namely, the shear of the outgoing null rays at the horizon.  
We demonstrate that the shear behaves very much like a compact binary coalescence waveform with increasing frequency and amplitude. Furthermore, the parameters of the system estimated from the horizon agree with those estimated from the waveform.  This implies
that even though black hole horizons are causally disconnected from us,
assuming general relativity to be true, we can potentially infer some of their detailed properties from gravitational wave observations.

An important physical phenomenon that manifests itself during the
inspiral of two orbiting compact objects is the tidal deformation of each under the gravitational influence of its companion.  In the case of binary neutron
star mergers, this tidal deformation and the associated Love numbers
have been used to probe properties of dense matter and the nuclear
equation of state.  Non-spinning black holes on the other hand have a vanishing
(field) tidal Love number in General Relativity.  
This pertains to the deformation of the asymptotic gravitational
field.  In certain cases, especially in the late stages of the
inspiral phase when the black holes get close to each other, the
\emph{source} multipole moments might be more relevant in probing
their properties and the No-Hair
conjecture; contrastingly, these Love
numbers do \emph{not} vanish. In \ref{chap:tddh}, we track the source
multipole moments in simulations of several binary black hole mergers and
calculate these Love numbers for axisymmetric modes.  We present evidence that, at least for
modest mass ratios, the behaviour of the source multipole moments is
Universal.

In the next part of the study, \ref{chap:gen_mul}, we uncover new features in the source multipole moments of the dynamical horizons in binary black hole mergers. We extend the previous study of the axisymmetric multipole moments and compute and analyze the non-axisymmetric moments. We show that the dynamical horizons encode detailed information about the dynamics of the binary black hole system. Additionally, the source multipole
moments of the dynamical horizon are shown to be strongly correlated with the gravitational field at
null infinity. Therefore the gravitational waves carried away from the system contain important information
about the geometrical structure of the dynamical horizons in the strong field regime. 

The final leg of the thesis (\ref{chap:flux_pm}) deals with a survey and analysis of the possibility of extending the new and important results found in the inspiral phase to the ringdown phase. We analyze the outer common dynamical horizon of the binary black hole system formed during the merger and study its dynamics and relationship with the outgoing gravitational waves from the system.  While we also find
that the gravitational field at the common dynamical horizon shares a similar multipolar structure as that at the future null infinity $\mathcal{I}^+$, we point out some limitations, inconsistencies, and anomalies that need to be addressed.

\manualmark
\markboth{\spacedlowsmallcaps{\bibname}}{\spacedlowsmallcaps{\bibname}} %
\refstepcounter{dummy}
\addtocontents{toc}{\protect\vspace{\beforebibskip}} %
\addcontentsline{toc}{chapter}{\tocEntry{\bibname}}

\vfill

\endgroup
\cleardoublepage

\vfill

\pagenumbering{arabic} %
\cleardoublepage %
\newpage

\cleardoublepage
\chapter{Introduction}
\label{chap:intro}
A black hole is one of the simplest and most symmetrical objects to be found in the universe. The mathematical equations that describe it are also simple. Though various ideas of a black hole existed since the 1700s \cite{Schaffer1783, LaplacenMichellBHs}, the first consistent mathematical description was the Schwarzschild solution given in 1916 \cite{SchwarzTrans}.  Predicted and described by the mathematical solution of the Einstein Field equations, these have been objects of central importance in gravitation and the cosmos. Their existence was corroborated by the discovery of gravitational waves that were consistent with that of black holes in general relativity \cite{Castelvecchi2016} at the existing sensitivity levels. %

To put it simply in words, it is believed that a black hole is a region of spacetime from which almost nothing, not even light, can escape. The mathematical description of stationary black holes in most gravitational theories is so simple that one does not need more than a few parameters to describe it \cite{Misner:1973prb}. Furthermore, those of Einstein's theory of gravity are vacuum solutions: except the singularity, no matter exists anywhere in the spacetime!

The most interesting aspect of the black hole is the boundary of the black hole region, called the event horizon \cite{Rindler1956,HawkingEllis:1973}. Significant work by mathematicians and physicists has paved the way to much of the unravelling of the features of this simple object. The theory of relativity also brought in new concepts, such as time dilation, length contraction, gravitational waves, gravitational lensing, etc which are currently being used to unravel the mysteries of the cosmos.  As the mathematical features of undisturbed black holes came to be understood, the study of quasi-static processes revealed the mysterious nature of black holes \cite{Bardeen:1973gs}. The laws governing these processes were found to have an uncanny resemblance with those of the fluids. Furthermore, the application of quantum theory to such objects resulted in the theoretical discovery of the Hawking radiation \cite{Hawking:1974sw}. Black holes thus became a testing ground for many of these concepts in relativity.

However, the formation and the origin of these objects were poorly understood. Rigorous mathematical description of the nature of black holes, in a differential and local form, began with the conception of the Raychaudhuri equation \cite{RaychEqn, Poisson:2009pwt}. Penrose used the study of congruences of geodesics to understand the trapping nature of black holes and the fundamental properties of the event horizon. These also led to insights into the formation of black holes and the singularity theorems \cite{PenroseSing}. %

The perturbation theory describes the dynamics of gravitational and other fields when the black hole is weakly disturbed %
and is a milestone in humanity's understanding of nature \cite{Chandrasekhar:1985kt}. 

Owing to the interest in the singularity theorem, no hair conjecture, and intrinsic properties of these objects, there came a need to understand the strong field dynamics of the black holes.  Although perturbation theory sheds light on the dynamics of black holes, the need to understand the dynamics of black holes actively absorbing matter, radiation, and energy was inevitable. Until recently, much of the research has been towards understanding black holes and their physical features in the static/stationary context, and at most infinitesimally perturbed from their equilibrium configurations. Owing to the complexities of the gravitational theories, one could only speculate on the possible nature of dynamical black holes. The strong field regime of black holes is not accessible as easily as their asymptotic weak filed regions, where gravity is weak and the gravitational wave detectors are placed. These are high curvature regions in the vicinity of the horizons that can host a range of interesting physical phenomena not easily known to the observers at infinity. Advancements in Numerical Relativity have opened up the possibility of studying the physical nature of black holes in this strong field, fully dynamical regimes and binary black hole mergers are the cosmic laboratories that present us with the very opportunity of witnessing this new physics. 

This thesis is an attempt along this direction, to understand the strong field dynamics of black holes. These have been possible with some of the important breakthroughs in numerical relativity \cite{Pretorius:2005gq,Campanelli:2005dd, Baker:2005vv}, due to the very dynamical, non-linear nature of these phenomena. 

The most interesting aspect of black holes is the presence of a horizon. As mentioned earlier, the notion of an event horizon that attempts to describe this hypersurface of no return is not suitable for describing black holes far away from equilibrium. These also include astrophysically important non-linear processes like binary black hole mergers. The limitations of the mathematical constructs used to describe black holes in or infinitesimally away from equilibrium are easy to understand. The event horizon, defined as the future boundary of the causal past of future null infinity, is teleological in its very definition due to the global constructs used therein. To locate the event horizon,  one would need to know the complete future and past of the spacetime. These were understood with time and the concepts of dynamical horizons took birth, building on the ideas of trapped surfaces and local constructs used by Penrose \cite{Penrose:1964wq}. Hayward used these to describe the trapped regions and the nature of their boundaries \cite{Hayward:2000ca}. These were furthered for use with numerical relativity by Ashtekar, Krishnan, and others \cite{Ashtekar:2004cn,Booth:2005qc}. %
Most of the physical laws known about black holes could be described as process laws even in fully dynamic scenarios. In the dynamical horizon formalism, it becomes very clear that it is the dynamical horizon, and not the event horizons, that are physically relevant to dynamic phenomena involving black holes. In this thesis, we aim to understand the fully dynamical nature and features of black holes and their dynamical horizons in the strong field regime, and their correlations with the dynamics far away in the radiation zone accessible to the detectors in fully dynamical and energetic scenarios. We do so using the dynamical horizon formalism and numerical relativity. In particular, we intend to probe these aspects when the black holes are close to merger and are in the mutual strong field regimes.

Starting with the first binary black hole detection in 2015 \cite{Abbott:2016blz}, numerous binary black
hole mergers have been observed to date
\cite{LIGOScientific:2018mvr,TheLIGOScientific:2016pea,Nitz:2018imz,Nitz:2019hdf,Venumadhav:2019lyq,Zackay:2019tzo}.
The parameters of these binary systems,
including the masses and spins of the individual black holes can be
inferred from the observed data \cite{TheLIGOScientific:2016wfe}. 
This inference relies crucially on gravitational waveform models
meant to represent, with sufficient accuracy, the gravitational 
wave emission from binary black hole mergers in general relativity
\cite{hannam:2014,schmidt:2015,bohe:2016gbl}.
The observations have so far been found to be consistent 
with standard general relativity \cite{Abbott:2018lct,LIGOScientific:2019fpa,Abbott:2018lct}. 
Deviations 
from general relativity might be revealed as detectors become more sensitive 
and as many more events are found, but we shall not consider alternate 
theories in this thesis.  Within general relativity, we also have detailed information
of the curved spacetime around a black hole merger from numerical simulations.
Since the first successful merger simulations
\cite{Pretorius:2005gq,Campanelli:2005dd,Baker:2005vv}, it is now
relatively straightforward to evolve black hole binaries through the
inspiral, merger, and ringdown regimes, at least for moderate 
mass ratios.  Indeed, the waveform models mentioned above are all based 
on, and ultimately verified by, comparisons with these numerical 
simulations.

The strong field regime of spacetime containing black holes is those regions of spacetime where the curvatures of space-time are much larger than the asymptotic flat regions. This region is thus in the vicinity of the black hole horizons.  As observers in the solar neighbourhood, we cannot directly observe the strong field region of space-time around black holes. We can only do so through the gravitational waves received in our detectors, which are placed in the asymptotically flat regions of the black holes (unless there is a flyby of a black hole and we are prepared, and survive it!). 
Since black hole horizons are very interesting objects of very high curvatures in spacetime, it is important to observe and understand the physics in their vicinity. The detection of gravitational waves by LIGO opens up the possibility to test our understanding of black holes and gravity, primarily in the weak field regime that we are in. But can it also reveal details of the dynamical and strong field regime? As the detectors improve in sensitivity, the need to understand the dynamical nature of black holes, their strong field properties and their effects on the gravitational radiation that reaches future null infinity is created. This would further enable us to test the no-hair conjecture and the gravitational theory.

One might be able to infer properties of spacetime regions
hidden behind the event horizon and causally disconnected from us.
The signal received at our observatories is generated by the
non-linearities and dynamics of the spacetime metric in the vicinity of the black holes.  These non-linear and dynamical fields are
responsible for both the signal seen by us, and also the
properties of spacetime inside the event horizon \cite{Jaramillo:2011rf,Jaramillo:2011re,Jaramillo:2012rr,Gupta:2018znn,Gunasekaran:2019jrq}.  

The important question that then arises is how can we probe the physics of strong field regions using the gravitational waves observed in the detectors Can we infer the strong field dynamics from these observations? In this thesis, we show clearly that this may indeed be possible. %
We run numerically accurate simulations of non-spinning binary black hole mergers using full numerical relativity and study the spacetime around black holes and the associated gravitational radiation. We uncover interesting relationships between the strong field and weak field dynamics of the systems for a range of mass-ratios.

These aspects were the main motivation for this thesis. The main theme of this thesis is to use numerical relativity to understand the strong field physics of black holes and their relationship with the gravitational radiation emitted from binaries involving them. Given the existence of binary black hole systems and the relevance to the data collected in the interferometers, we analyze the physics of binary black hole mergers. We study the gravitational field at the dynamical horizons of the binary black hole systems in the due course of inspiral, and that of the common dynamical horizon in the post-merger phase.

Some of the key results derived and observations made in this thesis are as follows. 
\begin{itemize}
	\item Inspiral phase: The gravitational radiation infalling at the dynamical horizons in a binary black hole merger is shown to have a familiar chirp-like waveform. They can be estimated given the gravitational radiation from the system recorded in detectors placed far away from the system. Furthermore, we also obtain an approximate relation relating them between the two black holes.
	
	The source multipole moments are found to have a universal behaviour in terms of a tidal expansion involving the parameters of the system. We define an alternate notion of a Love number of a black hole that may be more relevant close to the merger phase, and compute them using the source multipole moments.
	
	The generalized source multipole moments are defined and computed for the first time for dynamical horizons in binary black hole mergers. They are shown to carry rather precise imprints of the dynamics of, the outgoing gravitational radiation from the system.

	\item Post-merger phase: We show qualitatively that the shears of the common dynamical horizon formed in the due process of a merger have quasi-normal behaviour. We estimate the lowest QNM frequency and show that the remnant black hole parameters can be estimated from the post-merger shears. 
\end{itemize}
We also implement methods and develop codes to carry out waveform extrapolation and centre of mass corrections.

\section{Organization of the thesis}

The thesis is divided into six chapters. We begin by describing the numerical Infrastructure and simulations used for these studies. The second part details the study of the gravitational radiation at the dynamical horizons. The third part deals with deformations of the black holes in the strong field regime. We also attempt to address the important question of the tidal deformability of black holes. The fourth part contains a report on the preliminary results on the post-merger dynamics of the system and future directions.

The appendices contain information on methods for waveform preparations: numerical tools and codes developed and implemented to achieve the above results.  A brief description of them is as follows.

\begin{itemize}
	\item \ref{chap:nr_sims}: Numerical simulations. Describes the use and role of Numerical Relativity. Basic details about the numerical simulations and infrastructure used are described here.
	\item \ref{chap:waveform_rel}: Shear-News studies. Here we describe the relationship between the shear of the dynamical horizons and the outgoing gravitational radiation.
	\item \ref{chap:tddh}: Tidal deformability. Here, we use the source multipole moments to understand the tidal deformations of dynamical black holes in the strong field regime and show that the corresponding Love numbers do not vanish.
	\item \ref{chap:gen_mul}: Generalized multipole moment studies. In this chapter, we compute and study the source multipole moments of dynamical horizons, and discuss their correlation with the dynamics of the system and far the away wavezone.
	\item \ref{chap:flux_pm}: Post merger flux studies. Preliminary results and report on the post-merger energetics of the common dynamical horizon in the post-merger phase.
	\item \ref{chap:concl}: A brief summary and outlook. This chapter summarizes the results of the thesis, their implications, and the outlook for the future.
	\item The Appendices contain additional information on the tools and methods used. 
	\begin{itemize}
		\item Appendix \ref{appendix:dm}: Distance measures: Details of distance measures used in tidal deformability studies.
		\item Appendix \ref{appendix:waveform_prep}: Waveform preparations. The details of the implementation of preparations of NR waveforms.
		\item Appendix \ref{appendix:build_etk}: Installing the Einstein toolkit. A manual for building and compiling the toolkit on the Pegasus cluster.
	\end{itemize}
\end{itemize}

\cleardoublepage %
\flushbottom
\newpage

\chapter{Numerical Relativity Simulations}
\label{chap:nr_sims}

In this chapter, we give an overview of the use of Numerical Relativity simulations in the projects associated with this thesis.

\section{Numerical Simulations}

Our numerical simulations are performed using the publicly available Einstein Toolkit code \cite{Loffler:2011ay, EinsteinToolkit:web}, on the Perseus \cite{perseus:web} and Pegasus \cite{pegasus:web} HPC facility at IUCAA. The initial data is generated based on the puncture approach \cite{PhysRevLett.78.3606,Ansorg:2004ds}, which is then evolved through the Baumgarte - Shapiro - Shibata and Nakamura -  Oohara - Kojima (BSSNOK) formulation \cite{Alcubierre:2000xu, Alcubierre:2002kk, Brown:2008sb} using the $1+\log$ slicing and $\Gamma$-driver shift conditions.  The computational grid set-up is based on the multipatch approach using Llama \cite{PhysRevD.83.044045} and Carpet modules, which enable the mapping and coordinate transformation of multigrid set-up from curvilinear coordinates to Cartesian along with adaptive mesh refinement (AMR). It helps to optimally evolve 
the spacetime for a long time in a larger computational domain and to extract gravitational waves at faraway regions as compared to the Cartesian grid. Individual horizons and common horizons on the numerical grid are found via the method described in \cite{Thornburg:1995cp, Thornburg:2003sf}.

Gravitational waveforms are extracted \cite{Baker:2002qf} on coordinate spheres at various radii between $100M$ to $500M$. They are then extracted and prepared for use (i.e. transformed to the observer frame) by applying a series of transformations. These include the extrapolation to future null infinity $\mathcal{I}^+$, and the transformation to the center of mass frame. These operations are %
 carried out using a module developed by the author \cite{nrtools} in Python. This is further explained in Appendix \ref{appendix:waveform_prep}.

General quasi-local physical quantities are computed on the horizons following  \cite{Dreyer:2002mx, Schnetter:2006yt}. The implementation and the computation of the generalized multipole moments in  \eqref{mass_mult_lm} is new and does not exist in the QuasiLocalMeasures thorn of the Einstein Toolkit. They were computed in post-processing in Python. We developed and implemented methods to compute these generalized source multipole moments using numerical relativity data of dynamical horizons from the simulations. 

All the studies presented in this thesis consider non-spinning binary black hole systems with varying mass-ratio $q=M_2/M_1$, where $M_{1,2}$ are the component horizon masses ($M_1\geq M_2$). Furthermore, we use geometric units in which $G=c=1$. Thus, masses, lengths, and time can be expressed in terms of the total ADM mass of the system $M =M_1 + M_2$ as the total mass scales out of all these quantities.

We use the GW150914 parameter file available from \cite{wardell_barry_2016_155394} as
our template. For each of the simulations, as input parameters, we provide initial separation between the two punctures $d$, mass ratio $q$ and the radial and azimuthal linear momenta $p_r$, $p_{\phi}$ respectively, while keeping the total physical horizon masses $M=M_1+M_2=1$ fixed in our units. The parameters used in generating the initial data correspond to quasi-circular configurations. We run simulations whose initial parameters are also present in the RIT catalog \cite{Healy:2014yta,PhysRevD.95.024037, RITcatalog:web}
, so that a comparison can be made. These parameters are listed in \ref{tab:ic_qc0}. We compute the corresponding initial locations, the $x$, $y$, $z$ components of linear momentum for both black holes and grid refinement levels, etc., before generating the initial data and evolving it.

For computing the quasi-local quantities, we use a grid of size ($36, 74$) points on each of the dynamical horizons, unless and otherwise specified. This grid resolution allows us to safely study multipole moments of up to $\ell=2$. We also carry out a convergence test of the horizon data by running the simulation at two different grid resolutions of the horizon. We found that the errors between the neighbouring resolution runs decrease as the resolution increases.

Our simulations agree very well with the catalog simulations \cite{RITcatalog:web}, with merger time discrepancies of less than a few per cent. Our waveforms also agree well with the RIT catalog, and SpECTRE waveforms point-wise up to a few per cent in amplitudes and phases. See \ref{appendix:waveform_prep} for further details.  

In the \ref{tab:ic_qc0} below, we list all the simulation parameters used in the projects associated with this thesis.
\begin{table}
	\centering
	\begin{tabular}{|p{1.8cm}|p{1.8cm}|p{1.8cm}|p{1.8cm}|p{1.8cm}|p{1.8cm}|}
		\hline
		\multicolumn{6}{|c|}{Non-spinning BBH Simulations} \\
		\hline
		\hline
		Mass ratio &$d$ &$M_{1}$ &$M_{2}$ &$p_r$ &$p_t$\\
		\hline
		1.0     & 11.0        & 0.5       & 0.5       & -7.220e-04 & 0.09019  \\
		0.85    & 12.0        & 0.54051   & 0.4595    & -5.290e-04 & 0.08448  \\
		0.75    & 11.0        & 0.5714    & 0.4286    & -6.860e-04 & 0.08828  \\
		0.7     & 12.0        & 0.5882    & 0.4118    & -5.07e-04  & 0.08246  \\
		0.6     & 11.5        & 0.625     & 0.375     & -5.46e-04  & 0.08206  \\
		0.6667  & 11.75       & 0.6       & 0.4000    & -5.290e-04 & 0.08281  \\
		0.50    & 11.0        & 0.6667    & 0.3333    & -5.720e-04 & 0.0802   \\
		0.40    & 11.25       & 0.7143    & 0.2857    & -4.500e-04 & 0.07262  \\
		0.25    & 11.0        & 0.8       & 0.2       & -0.000308  & 0.05794  \\
		\hline
	\end{tabular}
	\caption[Initial parameters of all the simulations run and used in this thesis.]{Initial parameters for non-spinning binary black holes with quasi-circular orbits. Here, $q=M_2/M_1$ is the mass ratio, $d$ is the initial separation between the two holes, $p_r$ and $p_t$ are the radial and tangential (to the orbit) momenta in the initial data, respectively.}
	\label{tab:ic_qc0}
\end{table}

\begin{figure}
	\includegraphics[width=0.9\linewidth]{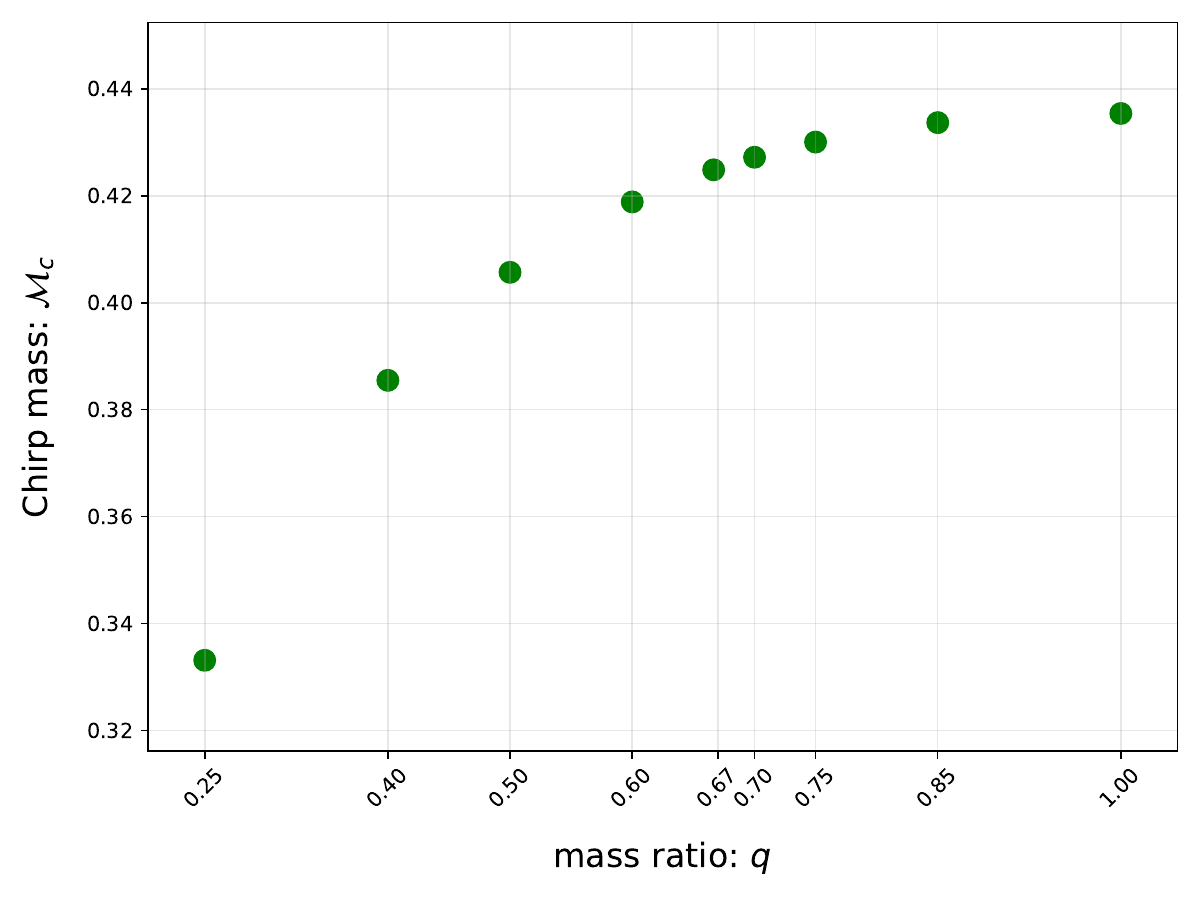}
	\caption[A graphical overview of the component masses of the numerical simulations run.]{The mass-parameters of the numerical simulations listed in \ref{tab:ic_qc0}.}
	\label{fig:all_sims_vis}
\end{figure}

\begin{figure}
	\includegraphics[width=0.9\linewidth]{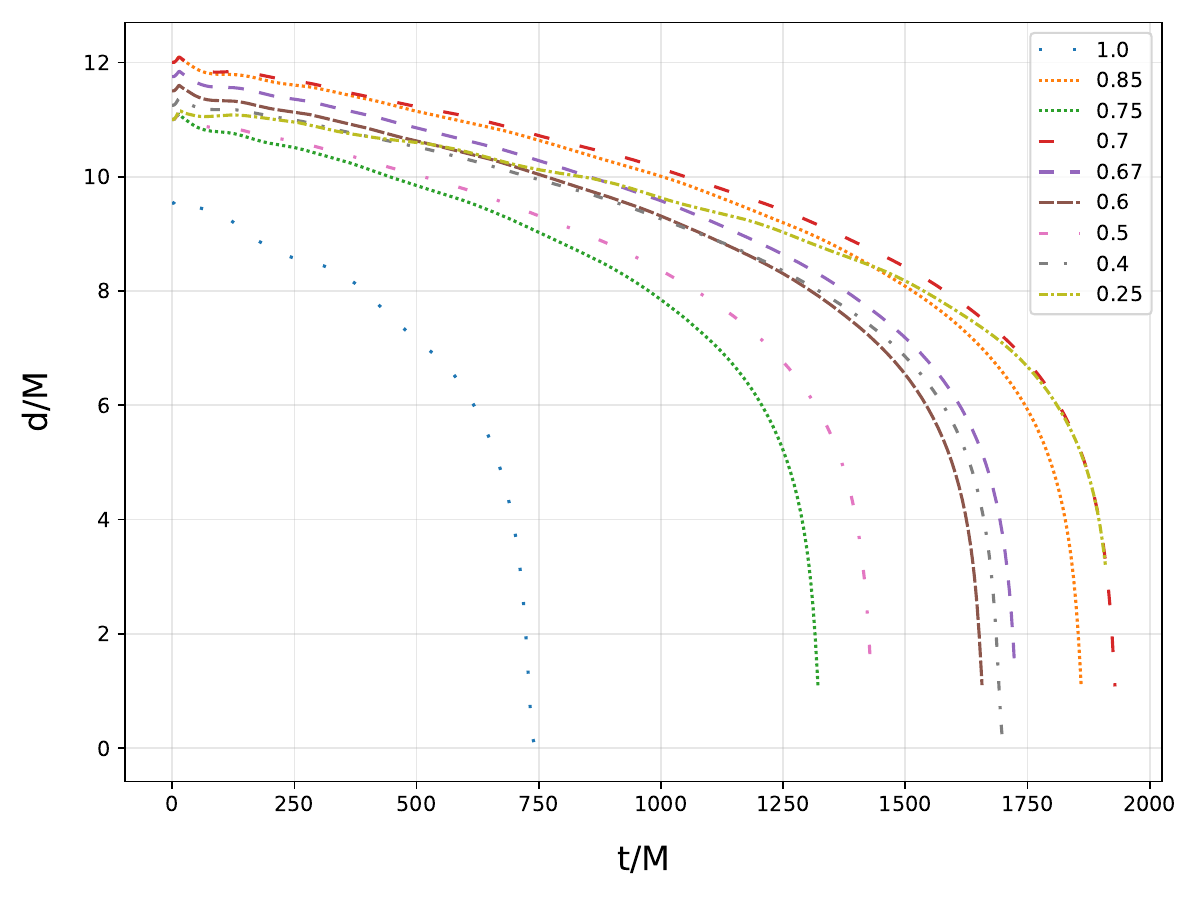}
	\caption[The coordinate distances of separation of BHs vs simulation time.]{The coordinate distances of separation between the two black holes for simulations of mass ratios $q$ listed in \ref{tab:ic_qc0}. The mass ratio of the respective curve is shown in the legend.}
	\label{fig:all_sims_distances}
\end{figure}

\begin{figure}
	\includegraphics[width=\linewidth]{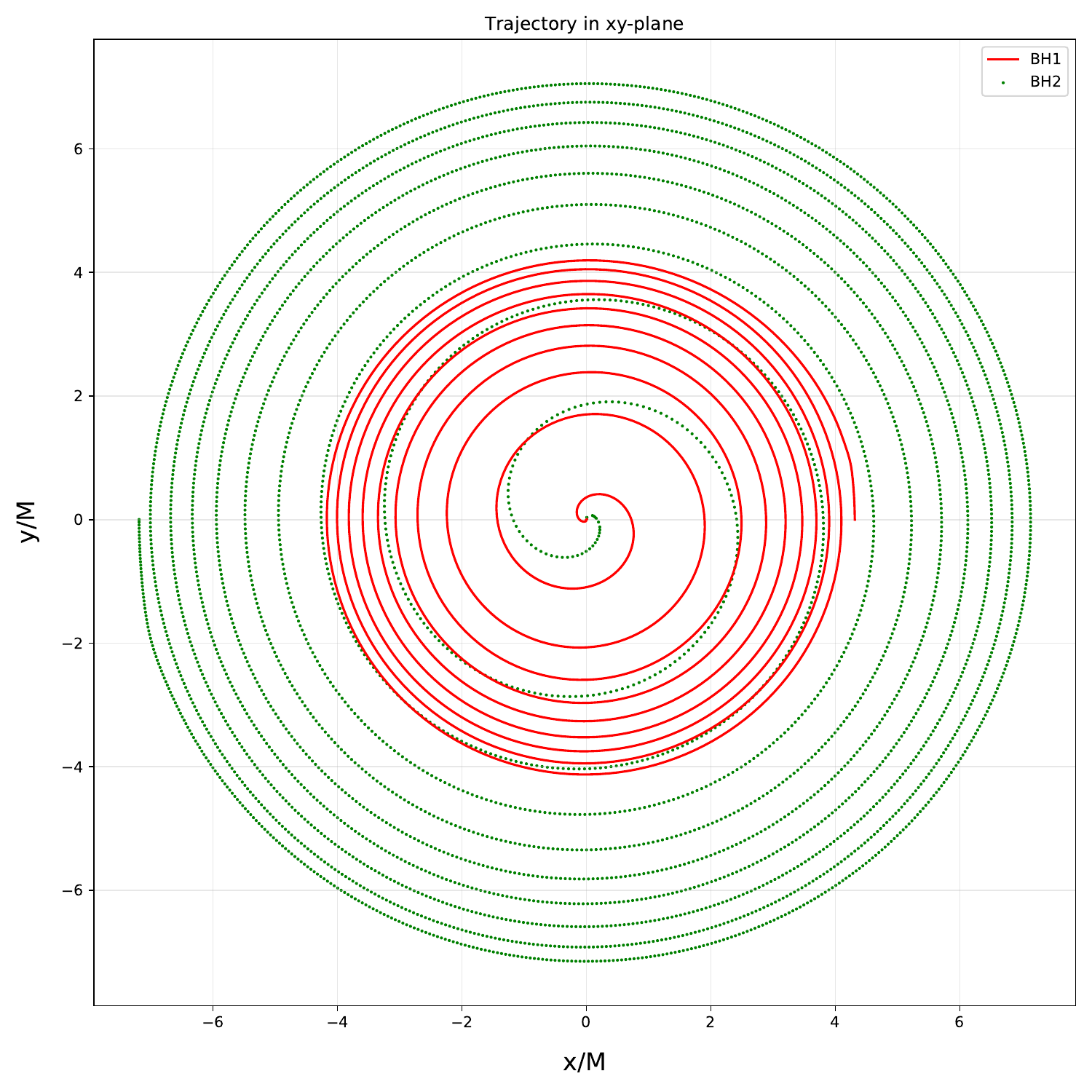}
	\caption{The coordinate trajectories of the two black holes in the inspiral of a BBH merger of a system with $q=0.6$.}
	\label{fig:trajs_q0p6}
\end{figure}
\begin{figure}
	\includegraphics[width=\linewidth]{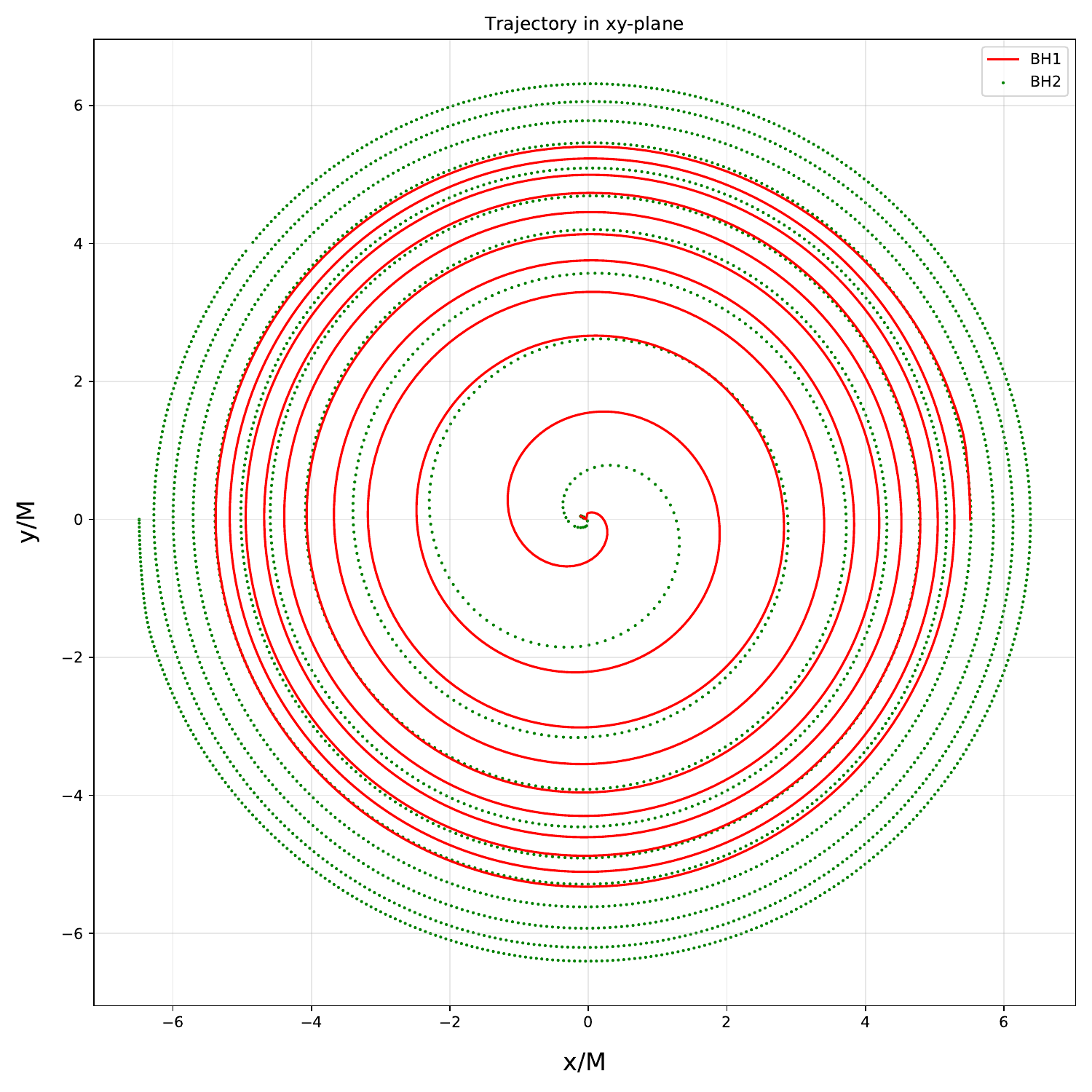}
	\caption{The coordinate trajectories of the two black holes in the inspiral of a BBH merger of a system with $q=0.85$.}
	\label{fig:trajs_q0p85}
\end{figure}

In the following section, we describe the basic details of the numerical simulations performed.

\begin{figure}
	\includegraphics[width=0.7\linewidth]{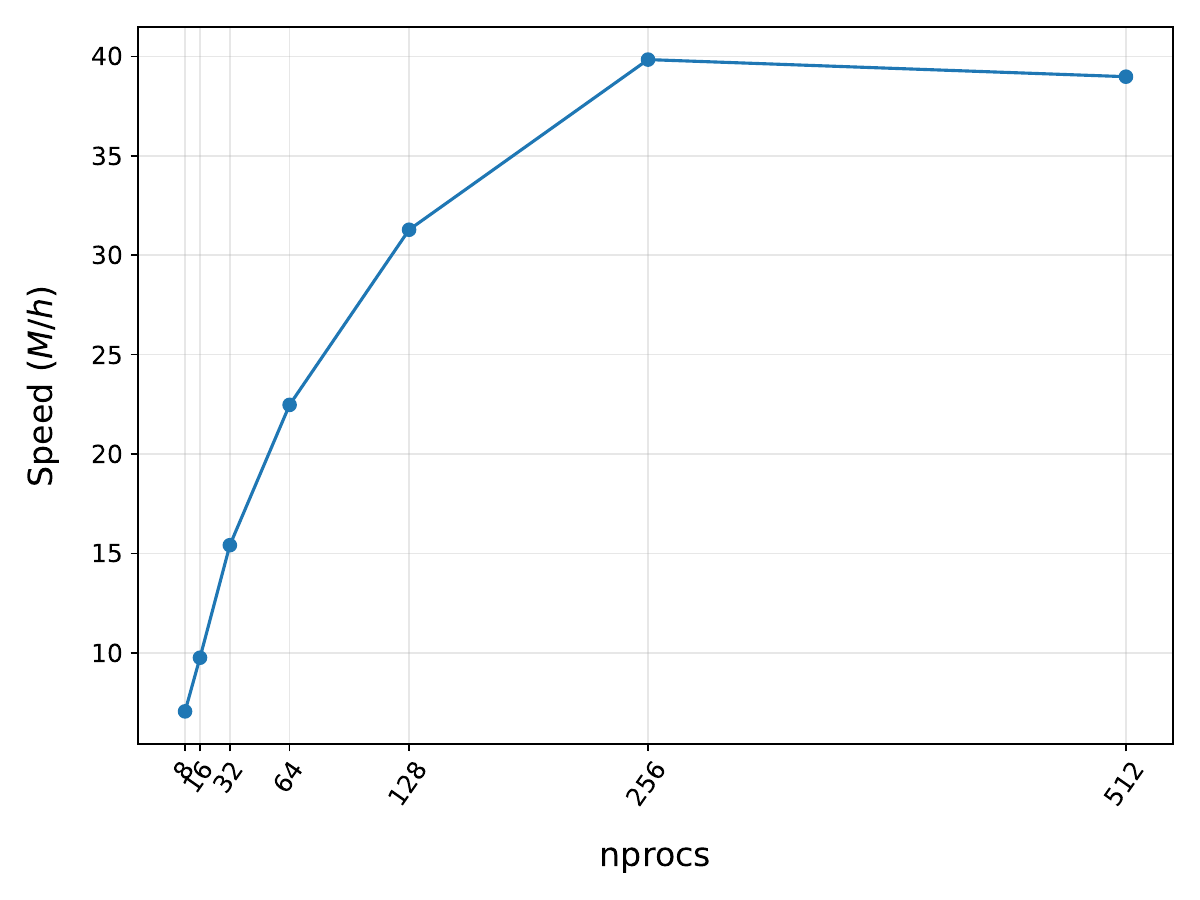}
	\caption[Strong scaling for ETK on Pegasus]{Shown here is the strong scaling for ETK on the Pegasus HPC for $q=1$ simulation. On the X axis, we show the number of MPI processors without hyperthreading and no OpenMP. On the Y axis, we show the number of physical time units per hour achieved 512 iterations since the start of the simulation. As can be seen, the performance improvement ceases as the number of MPI processors is increased to about 256 processors.}
	\label{fig:scaling}
\end{figure}

All the numerical simulations were performed at the Perseus \cite{perseus:web} and Pegasus \cite{pegasus:web} HPC at IUCAA, Pune India. In particular, the Einstein toolkit was built, compiled, and optimized to the specific hardware of the HPC at IUCAA by the author, and a simpler procedure is detailed in Appendix \ref{appendix:build_etk}. In \ref{fig:scaling} we show the strong scaling of the ETK code on Pegasus. As seen in the plot, the inter-node and MPI overhead starts to dominate the expenses and the computing gain as the number of MPI procs exceeds about 256.

Most of the numerical simulations were performed using 64 (CPU) processors in parallel using MPI, without OpenMP, and some with 128. The resolution of our numerical relativity simulations coincides with the gallery run of the GW150914 system\cite{wardell_barry_2016_155394}.

For analyzing, handling, and processing the numerical relativity data the author acquired and used various hardware resources, and developed various modules and codes. A brief description of this is given as under.

Hardware used:
\begin{enumerate}
	\item Workstation I: 6 core 12 threads CPU with 24 GB of RAM (IUCAA Resource)
	\item Workstation II: 8 core, 16 thread CPU with 32 GB of RAM and 1.25 TB SSD (Personal)
	\item Workstation III: 4 core, 8 thread CPU with 8GB of RAM and 500GB SSD (Personal).
	\item Storage I: 12 TB online NAS.
	\item Storage II: several 4TB external HDDs.
\end{enumerate}

All the codes used for analysis and processing were developed by the author in Python.
\begin{enumerate}
	\item Data transfer and syncing scripts: The NR data produced and saved at HPC storage was synced in chunks, with only required data being transferred to analysis network storage, accessible to the workstations used. 
	\item NRTools/ waveformtools: A numerical relativity package developed to handle NR data and waveforms. More details can be found at \cite{nrtools}.
	\item waveform\_match script: A scalable, MPI-based scheme implementing a new customized algorithm to carry out parameter estimation and data analysis on Numerical Relativity data. This includes the generation of template bank, and match routines to identify the best matching template.
	\item pkginstaller: A batch-like script for automatically downloading, compiling and building an optimized version of libraries for HPCs. These are used to install all prerequisites for NR codes \cite{pkginstaller}.
\end{enumerate}

Git was used for version control.

\section{Coordinate system on a Dynamical Horizon}
\begin{figure}
	\includegraphics[width=0.9\columnwidth]{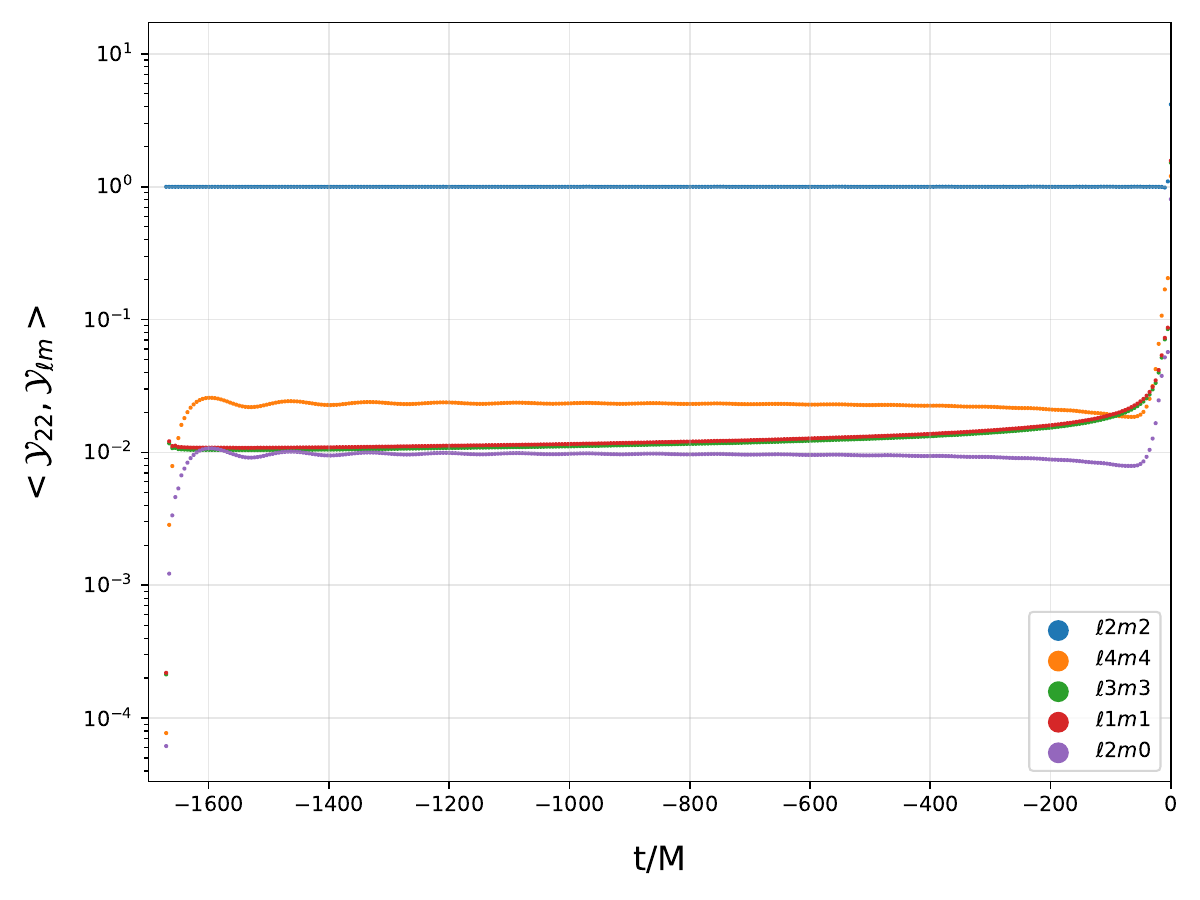}
	\caption[The (non) orthogonality of the spherical harmonic basis functions on the horizon.]{The (non) orthogonality of the spherical harmonic basis functions adapted to the coordinate system on the dynamical horizon, using the approximate Killing vector method of BH1 during the evolution of the $q=0.6$ system.}
	\label{fig:ylm_orthog}
\end{figure}

\begin{figure}[htpb]
	\includegraphics[width=0.8\columnwidth]{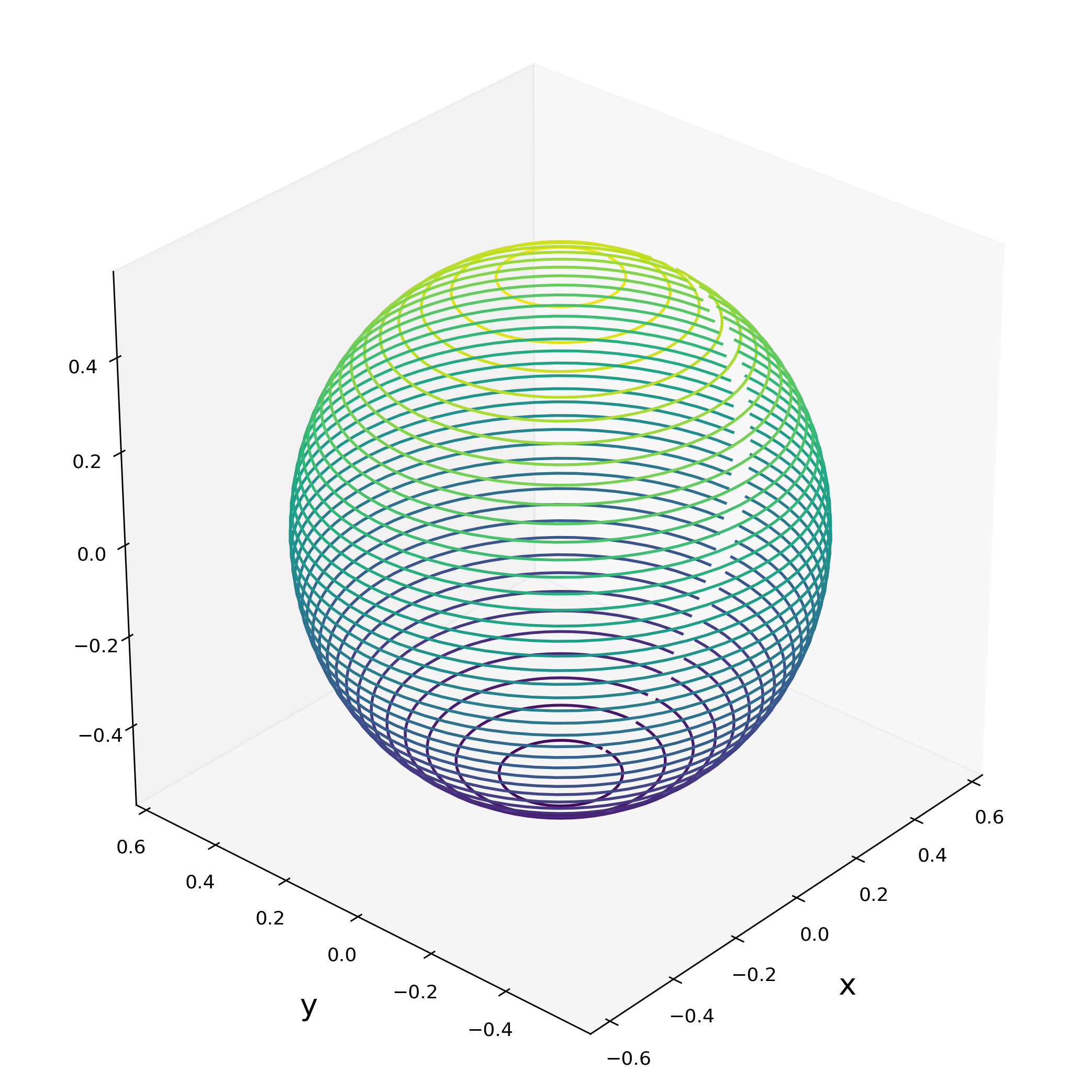}
	\caption[The level curves of the polar angle on a MOTS.]{The level curves of the polar angle $\zeta ~ \cos(\theta)$ on a MOTS of the dynamical horizon. The colour bar indicates the magnitude of $\zeta$.}
	\label{fig:zeta_level}
\end{figure}
\begin{figure}[htpb]
	\includegraphics[width=0.8\columnwidth]{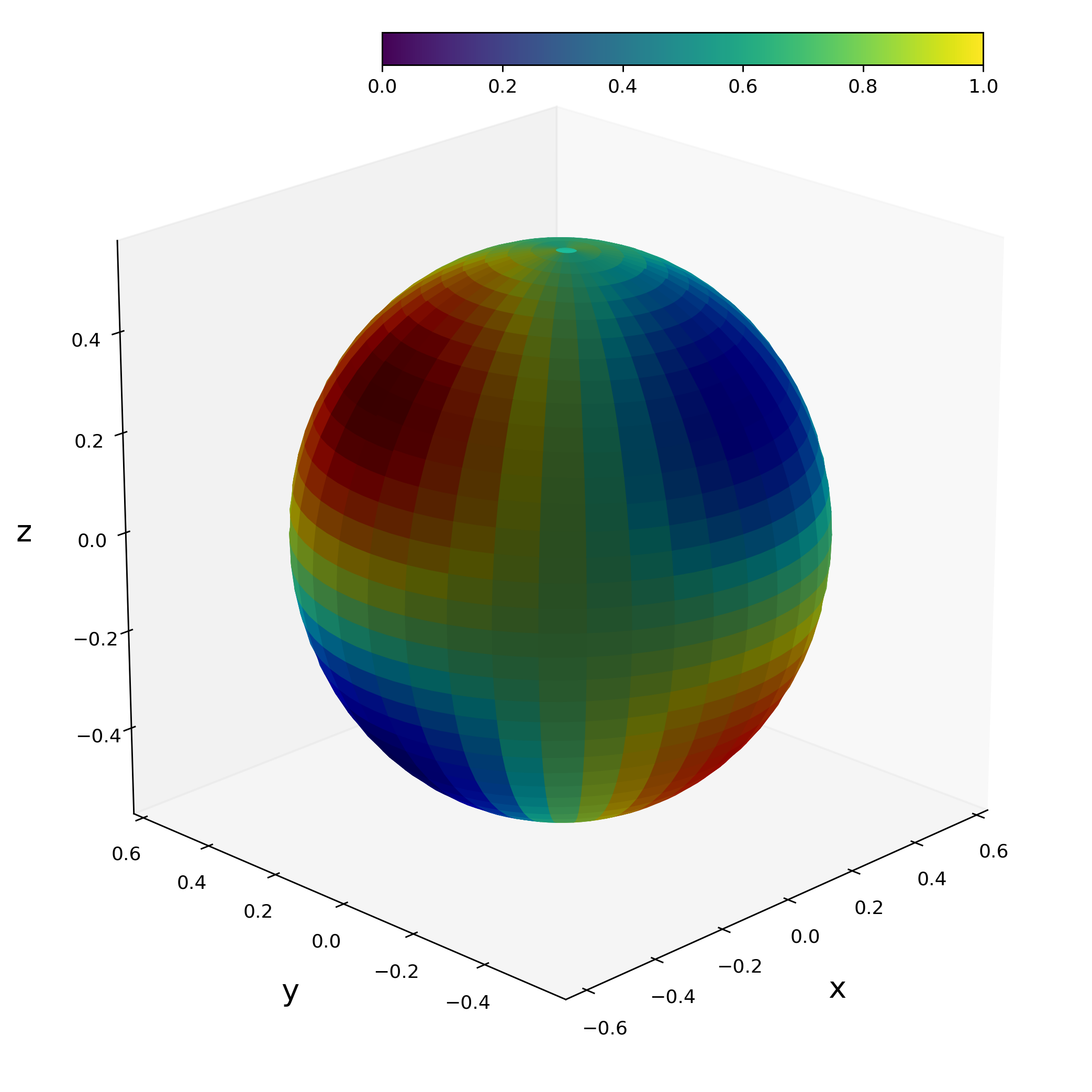}
	\caption[The deviation from sphericity of the polar angle.]{The deviation of the value of $\zeta$ from $\cos(\theta)$ at an instant of time, on a MOTS of the same dynamical horizon.}
	\label{fig:zeta_dev}
\end{figure}

\begin{figure}[htpb]
	\includegraphics[width=0.8\columnwidth]{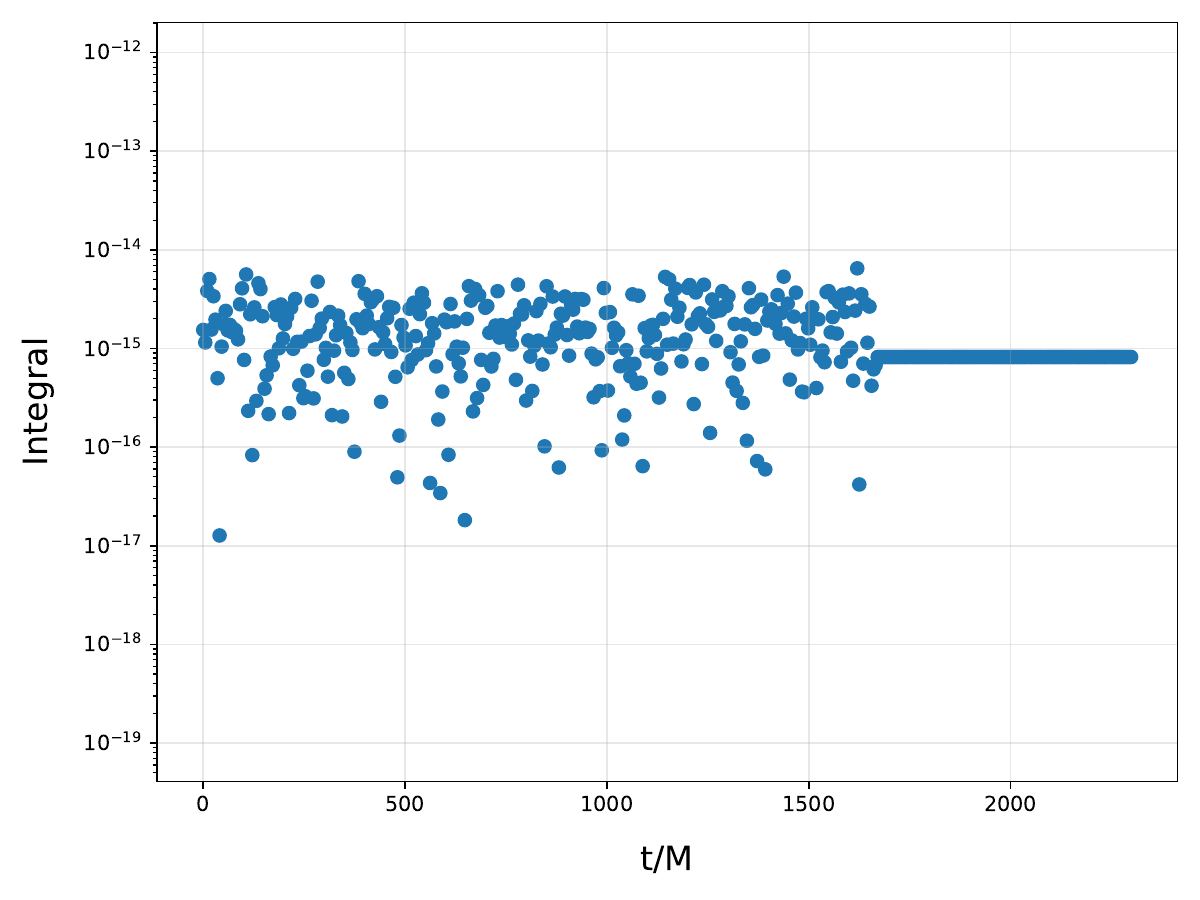}
	\caption[The surface integral of the polar angle on a MOTS.]{The integral of $\zeta$ on the MOTS of the dynamical horizon of the same dynamical horizon.}
	\label{fig:zeta_integ}
\end{figure}

The main object of study in this thesis is a dynamical horizon $\mathcal{H}$, \cite{Ashtekar:2004cn,Booth:2005qc} obtained by a time evolution of marginally trapped surfaces.

A dynamical horizon is located inside the event horizon, which marks the boundary of a trapped space-time region. 

We consider $\mathcal{H}$ to be foliated by 2-surfaces $S$ of spherical topology, with an intrinsic Riemannian metric $q_{ab}$. For the former, we obtain a marginally trapped surface and for the latter, we approximate it by large coordinate spheres in the numerical domain. For every cross-section $\mathcal{S}$, we assign outgoing and ingoing directions. Denote the outgoing future-directed null vector normal to $S$ by $n^+$, and the ingoing null normal as $n^-$ satisfying $n^+\cdot n^- = -1$.  Let $x$ be a complex null vector tangent to $S$
satisfying $x\cdot\bar{x}=1$ (the overbar denotes complex conjugation),
and $n^+\cdot x = n^-\cdot x = 0$.  

On the black hole, the basic object here is a marginally outer trapped surface (MOTS), again denoted by  $\mathcal{S}$. This is a closed space-like
2-surface with vanishing outgoing expansion $\Theta_{+}$: 
\begin{equation}
	\Theta_{+} = q^{ab}\nabla_a n^+_b = 0\,.
\end{equation}
Its scalar curvature is denoted by $\tilde{\mathcal{R}}$, the two-metric by $\tilde{q}_{ab}$, and its extrinsic curvature by $\tilde{K}_{ab}$.
The shear of the dynamical horizon is denoted by $\sigma = x^a \bar{x}^b \nabla_{a} n^+_b$. 

Most of the results in this thesis involve the computation of quasi-local quantities. These quantities are defined on the dynamical horizon $\mathcal{H}$. In order to carry out these computations, a preferred coordinate system needs to be erected on $\mathcal{H}$ and on every one of its slices (MOTS) $\mathcal{S}$, which we shall now describe. Note that such a coordinate system on $\mathcal{S}$ will be dynamical, as the system and the black holes in it evolve with time. This is partly achieved using the thorn QuasiLocalMeasures.

Consider a time slice of a numerical simulation in which a MOTS $\mathcal{S}$ embedded in the three-dimensional spatial hypersurface $\Sigma$ has been located. First, a preferred coordinate system on the leaves $S$ of the dynamical horizon is constructed using an appropriately defined axial vector field $\varphi^a$. The method of Killing transport is used to compute the axial field \cite{Ashtekar:2004gp, Schnetter:2006yt}. This method finds an approximate symmetry field on the MOTS from an initial guess. The field coincides with the direction field $\phi$ of the Boyer-Lindquist co-ordinates when the horizon is isolated e.g. when the separation between the black holes is large compared to their horizon sizes. Thus, even though this vector field $\varphi$ is not an exact Killing field on the MOTS, it furnishes a field that can be used to establish a dynamical coordinate system on the dynamical horizon and carry out the computations.  

Once the axial field is defined and computed on each slice $\mathcal{S}$, an invariant coordinate $\zeta$ (analogous to the polar coordinate variable $\cos{(\theta)}$ of the Boyer-Lindquist coordinates) can be defined and computed as:
\begin{equation}
	\tilde{D} \zeta = \dfrac{1}{R_S ^2} \tilde{\epsilon} \varphi^a
\end{equation}

Some ideal features required for a \emph{good} coordinate system are as follows:
\begin{enumerate}
	\item Integral curves length of $\varphi \approx 2 \pi$.
	\item Closed-ness of the integral curves of $\varphi$.
	\item Orthogonality of the ($\theta, \phi$) basis vectors on each $\mathcal{S}$.
	\item The coordinate $\zeta$ should integrate to zero on the sphere, and every $\phi=const$ meridian.
\end{enumerate}
The coordinate $\zeta$ is analogous to $\cos(\theta)$ on round unit spheres, which are the geometries of the cross sections of the isolated horizons in General Relativity. As a dynamical horizon geometry deviates from its isolated configurations, $\zeta$ deviates from $\cos(\theta)$ and this is depicted in \ref{fig:zeta_dev}.

For defining basis functions on every $\mathcal{S}$, the basis functions should be sufficiently orthogonal. In \ref{fig:ylm_orthog}, we show the extent of non-orthogonality of the spin-weight 0 (scalar) spherical harmonics used to compute the multipole moments in \ref{chap:gen_mul}. In \ref{fig:zeta_integ}, we show graphically that condition 4 above is satisfied by construction. In \ref{fig:zeta_level}, the level curves of $\zeta$ i.e. the curves on which $\phi$ varies are shown. 

Therefore, for the purposes of our study, we can safely use the coordinate system on the DH described above. More accurate methods exist for computing the axial vector fields \cite{Ashtekar:2013qta} which we reserve for future works.

\section{Units}

We work in geometric units where $G=c=1$. The total ADM mass of the numerical simulation $M$ is set to one and scales out of all quantities. Thus, all lengths, times, and masses are in units of $M$. We will define and explain mathematical notation when they are used for the first time.
\cleardoublepage %
\flushbottom
\newpage
\cleardoublepage

\newcommand{\hl}{\colorbox{gray}}
\newenvironment{changemargin}[2]{%
\begin{list}{}{%
\setlength{\topsep}{0pt}%
\setlength{\leftmargin}{#1}%
\setlength{\rightmargin}{#2}%
\setlength{\listparindent}{\parindent}%
\setlength{\itemindent}{\parindent}%
\setlength{\parsep}{\parskip}%
}%
\item[]}{\end{list}}

\chapter{Shear-News studies}
\label{chap:waveform_rel}

\section{Introduction}

In this chapter, the relations between the strong field and weak field dynamics of black holes are studied in the context of binary black hole mergers. \\ 

In a binary black hole merger, it is known that the
inspiral portion of the waveform corresponds to two distinct horizons
orbiting each other, and the merger and ringdown signals correspond 
to the final horizon being formed and settling down to equilibrium.  
However, we still lack a detailed understanding of the relation 
between the horizon geometry in these three regimes and the observed 
waveform. Here we show that the well known inspiral 
chirp waveform has a clear counterpart on black hole horizons, 
namely, the shear of the outgoing null rays at the horizon.  
We demonstrate that the shear behaves very much like a compact binary coalescence waveform 
with increasing frequency and amplitude. Furthermore, the parameters of the system estimated from 
the horizon agree with those estimated from the waveform.  This implies
that even though black hole horizons are causally disconnected from us, 
assuming general relativity to be true, we can potentially infer some of their 
detailed properties from gravitational wave observations.

The infalling flux of gravitational waves, representing tidal coupling, is a part of the energy balance governing the dynamics of the binary system.  Thus, any modifications of the infalling flux will also affect the overall dynamics of the system.  In situations when the black holes are spinning sufficiently rapidly, 
phenomena like super-radiance can play an important role (see e.g. \cite{Brito:2015oca}).   
In the regime when the effect of the companion can be treated 
perturbatively, these tidal effects can be calculated analytically \cite{Hawking:1972hy,Hartle:1974gy,Hartle:1973zz,OSullivan:2015lni,OSullivan:2014ywd,PhysRevD.70.084044,Chatziioannou:2016kem,Chatziioannou:2012gq}.  Here we shall go beyond these calculations, close to the merger where linear perturbation theory is not sufficient. All of these perturbative calculations refer to the event horizon for which, as a matter of principle, no generally valid non-perturbative quasi-local flux formula can exist.  As in almost all such numerical studies, we always work with dynamical horizons which are expected to differ significantly from the event horizon near the merger and do not suffer from the teleological properties of the event horizon.  Moreover, exact flux formulae, valid in full non-linear general relativity, are known for dynamical horizons (see e.g. \cite{Ashtekar:2004cn,Booth:2005qc}).  

Besides the effects on the orbital motion through tidal coupling, 
the infalling radiation must be very special for other reasons.  
The remnant black hole horizon is highly distorted on 
formation, and it loses its hair to reach its 
final equilibrium state represented by a Kerr black hole.  
However, the horizon, being a one-way membrane, cannot ``radiate away'' 
its hair.  Instead, it approaches its very special final 
state by absorbing just the right amount 
of infalling radiation 
that
precisely cancels any hair that it might have when
it 
is initially formed \cite{Gupta:2018znn}.  Thus, the infalling radiation 
also determines the highly non-trivial issue of the final state within 
general relativity.  Assuming that general 
relativity holds, are there strong correlations between the observed 
gravitational wave signal, and suitable fields on black hole horizons?  
If we have a sufficiently detailed understanding of these 
correlations, we might be able to understand the fundamental properties of 
dynamical strong field gravity and black hole horizons 
from gravitational wave observations.

The presence of such correlations might be difficult to
discern in analytic or numerical studies.  The relevant spacetime
regions could not be more different: The spacetime we inhabit is very
close to flat and extremely well described by linearized general
relativity.  On the other hand, the region where black
hole horizons live will have much higher curvature, and the
non-linear dynamics of the Einstein equations need to be taken into
account.  It is thus not immediately obvious
precisely which fields should be correlated in a gauge invariant manner, 
and how mathematical results might be proved.  

Despite these potential pitfalls, several authors have previously found
evidence for correlations between quantities on horizons and in the
wave zone
\cite{Jaramillo:2011rf,Jaramillo:2012rr,Rezzolla:2010df,Gupta:2018znn}.  These works have considered either the post-merger regime or subtleties regarding gravitational 
wave recoil.  Thus far none have considered what is one of the most well-known features of binary merger waveforms, namely the inspiral chirp with increasing frequency and amplitude. 
The evolution of the frequency and amplitude have been calculated to high orders in various
post-Newtonian approaches, while accounting for a variety of physical effects such as

precession and eccentricity  (see, e.g., \cite{Blanchet2014}).   Moreover, these post-Newtonian calculations have been combined with 
numerical relativity merger signals to construct complete waveform models including the inspiral, merger and ringdown regimes 
\cite{Ajith:2007kx,Ajith:2009bn,Santamaria:2010yb,Buonanno:1998gg,bohe:2016gbl,Khan:2018fmp,Khan:2019kot}.  We shall use these 
complete waveform models to quantitatively compare gravitational wave signals with horizon fields.

It has been speculated previously
\cite{Gupta:2018znn} if  the post-merger signal, the
gravitational wave news (in essence the time derivative of the
gravitational wave strain), is correlated with the shear
on the black hole horizons. Here we will extend this study
to the inspiral regime, and show quantitatively that the chirp
signal is also extremely well correlated with the shear.  
Remarkably, for reasons that we do not yet fully understand, we shall see
that very little effort is required to extract these correlations, 
and the gauge conditions employed in the simulations do not seem to play an important role.

\section{Basic notions} 

Our results deal with two
surfaces. The first is future null infinity $\mathcal{I}^+$, the endpoint of future null-geodesics which escape to infinity 
\cite{Bondi:1962px,Penrose:1964ge}.  The second
is a dynamical horizon $\mathcal{H}$ \cite{Ashtekar:2004cn,Booth:2005qc} 
obtained by a time evolution of marginally trapped surfaces. These 
two surfaces seem to be very different.  Future null
infinity $\mathcal{I}^+$ is an invariantly defined null surface 
where outgoing null geodesics end.
On the other hand, a dynamical horizon is located
inside the event horizon.  Nevertheless, both $\mathcal{I}^+$ and
$\mathcal{H}$ are one-way membranes and exact flux formulae hold for
both surfaces.

For both cases, we consider spacelike 2-surfaces $S$ of spherical
topology, with an intrinsic Riemannian metric $q_{ab}$.  
$S$ will be either a cross-section of $\mathcal{I}^+$ (approximated as a large coordinate sphere enclosing the
source), or a section of $\mathcal{H}$, i.e. a marginally trapped surface.  In
either case, we assign outgoing and ingoing
directions.   Denote the outgoing future directed null vector normal
to $S$ by $n^{+a}$, and the ingoing null normal as $n^{-a}$ satisfying $n^{+}\cdot n^{-} = -1$.  Let $m$ be a complex null vector tangent to $S$
satisfying $m\cdot\bar{m}=1$ (the overbar denotes complex conjugation),
and $n^{\pm}\cdot m= 0$.  

In the wave-zone, spacetime geometry is completely described by the 
Weyl tensor $C_{abcd}$.  In particular, outgoing transverse radiation 
is described by the Weyl tensor component \cite{Newman:1961qr}
\begin{equation}
 \Psi_4 = C_{abcd}n^{-a}\bar{m}^b n^{-c}\bar{m}^d\,.
\end{equation}
$\Psi_4$ can be expanded
in spin-weighted spherical harmonics ${}_{-2}Y_{\ell,m}$ of spin weight $-2$
\cite{Goldberg:1966uu}.  Let $\Psi_4^{(\ell,m)}$ be the mode component
with $\ell\geq 2$ and $-m\leq \ell \leq m$.  The $(\ell,m)$ component
of the news function $\mathcal{N}^{(\ell,m)}$ and its polarizations $\mathcal{N}_{+,\times}$ are defined as
\cite{Bondi:1962px}
\begin{equation}
  \mathcal{N}^{(\ell,m)}(u) = \mathcal{N}^{(\ell,m)}_+ + i\mathcal{N}^{(\ell,m)}_\times = \int_{-\infty}^u\Psi_4^{(\ell,m)}\,du\,.
\end{equation}
The outgoing energy flux is related to the integral of $|\mathcal{N}|^2$ over 
all angles.  In a numerical spacetime, it is in
principle possible to extract $\Psi_4$ going out all the way to
$\mathcal{I}^+$ \cite{Babiuc:2008qy}
to reduce systematic errors.  We shall however follow the common approach of
calculating $\Psi_4$ on a sphere at a finite radial coordinate $r$, extrapolating it to $\mathcal{I}^+$, and transforming it to the centre of mass frame. The integral in the previous equation is over time instead of the retarded time coordinate $u$. The lower limit in the integral is not
$-\infty$ but the earliest time available in the simulation.  The news
function is then a function of time at $\mathcal{I}^+$, starting
from the earliest time available in the simulation. A further time
integration of $\mathcal{N}$ yields the gravitational wave strain.  

Turning now to the black hole, the basic object here is a marginally
outer trapped surface (MOTS), again denoted $S$.  This is a closed spacelike
2-surface with vanishing outgoing expansion $\Theta_{(+)}$: 
\begin{equation}
  \Theta_{(+)} = q^{ab}\nabla_a n^{+}_b = 0\,.
\end{equation}
The shear of $n^{+a}$ is defined as
\begin{equation}
  \sigma = m^am^b\nabla_a n^{+}_b = \sigma_+ + i \sigma_\times\,. \label{shear_spin_coeff}
\end{equation}
Both $\mathcal{N}$ and $\bar{\sigma}$ have the same behaviour under
spin rotations $m\rightarrow me^{i\psi}$, i.e. they have the same spin
weight. Also, similar to the news function and the Bondi mass-loss
formula, $|\sigma^2|$ appears in the energy flux falling into the
black hole \cite{Ashtekar:2002ag,Ashtekar:2003hk}, though in this case
the flux also contains other contributions. It is shown
in \cite{Booth:2003ji} that for the case of a slowly evolving horizon,
which is what we are dealing with in the inspiral
phase, $ |\sigma|^2$ is the dominant part of the flux. Thus, as suggested in
\cite{Jaramillo:2011rf}, we will compare the shear at the horizon with
the news.    

\section{The numerical simulations}

The numerical simulations performed and used are described in the \ref{chap:nr_sims}.

For this work, we consider non-spinning binary black hole systems with varying mass ratios. This study uses a subset of the simulations listed in \ref{tab:ic_qc0}, whose parameters are listed in \ref{tab:ic_qc0_snr}. We chose 6 non-spinning cases ranging between $q=1.0$ to $0.25$. 

\begin{table}
\begin{center}
\begin{tabular}{|p{2cm}|p{2cm}|p{2cm}|p{2cm}|}
 \hline
 $q $ & $D/M$ & $p_r/M$& $p_{\phi}/M$\\
 \hline
1.0&  9.5332&0.0 &0.099322\\
0.85 &12.0 &-0.000529 &0.08448\\
0.75 &11.0 &-0.000686 &0.08828\\
0.667 &11.75 & -0.000529 &0.08281\\
0.5 &11.0& -0.000572&0.0802\\
0.25 &11.0&-0.000308 &0.05794\\
 \hline
\end{tabular}
\caption[Initial parameters of BBH simulations used in Shear-News correlation studies.]{Initial parameters for non-spinning binary black holes with quasi-circular orbits. $q=M_2/M_1$ is mass ratio, $D$ is the initial separation between the two holes, $p_r$ and $p_{\phi}$ are radial and azimuthal linear momenta respectively.}
\label{tab:ic_qc0_snr}
\end{center}
\end{table}

\section{Procedure}
\label{sec:shear_news_inspiral}
\begin{figure}
		\includegraphics[width=\columnwidth]{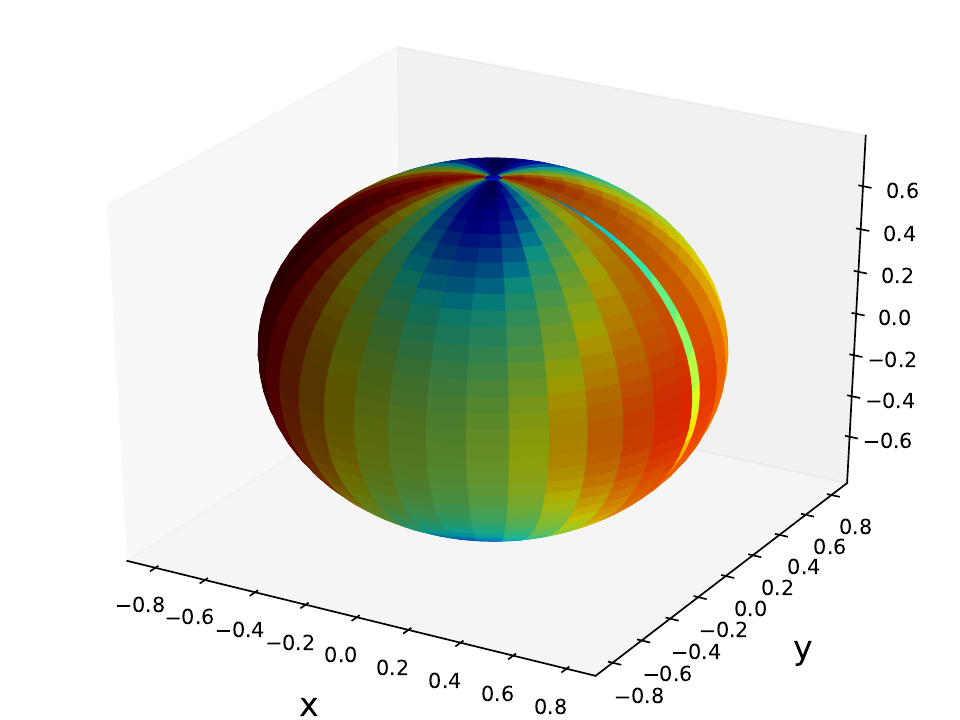}
		\includegraphics[width=\columnwidth]{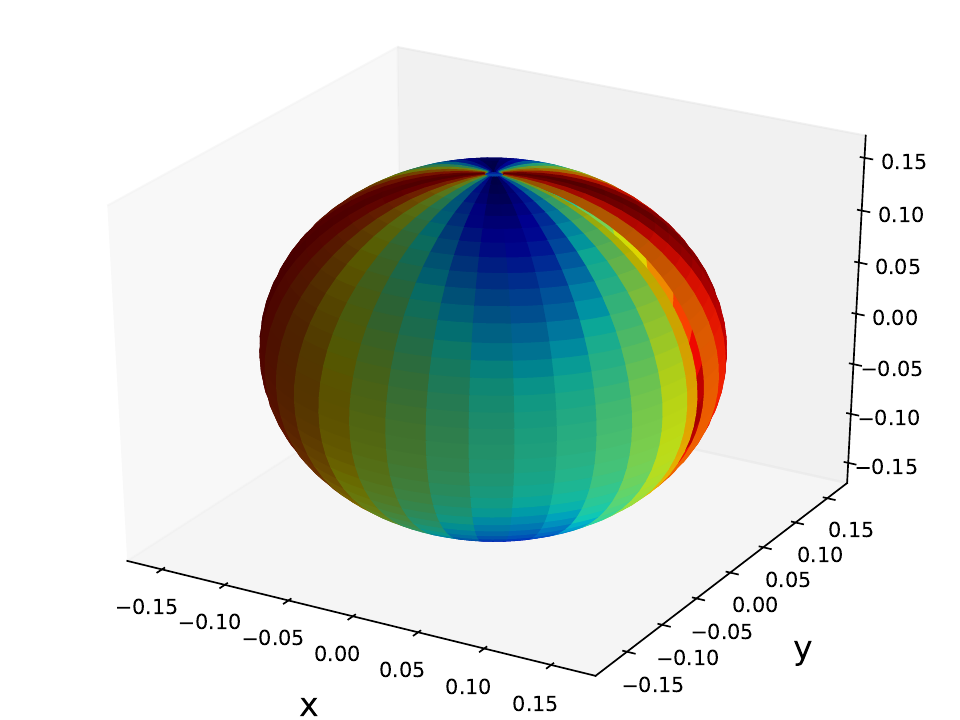}
	\caption[The angular pattern of the shear at the dynamical horizon during the inspiral.]{The 3D angular pattern of the shears. The real parts of the shears, $\sigma_+(t)$, for the two black holes $BH1$ (top) and $BH2$ (bottom) for the $q=0.25$ configuration have been plotted. This clearly shows the dominant quadrupolar patterns of the shears.}
	\label{fig:shear_news_3d}
\end{figure}
\begin{figure}
	\includegraphics[width=\columnwidth]{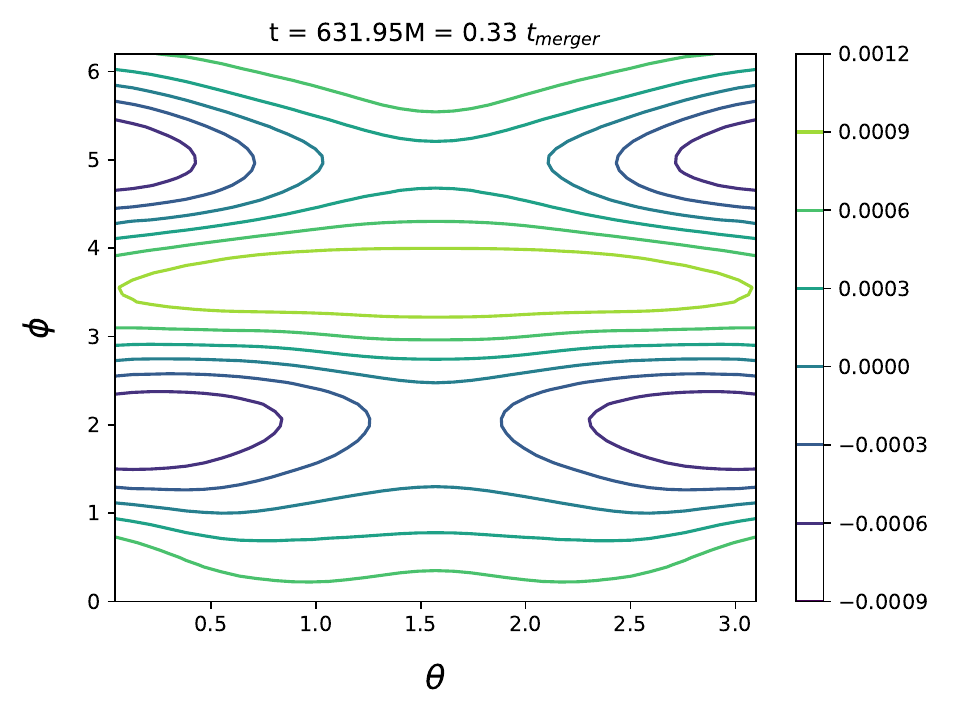}
	\includegraphics[width=\columnwidth]{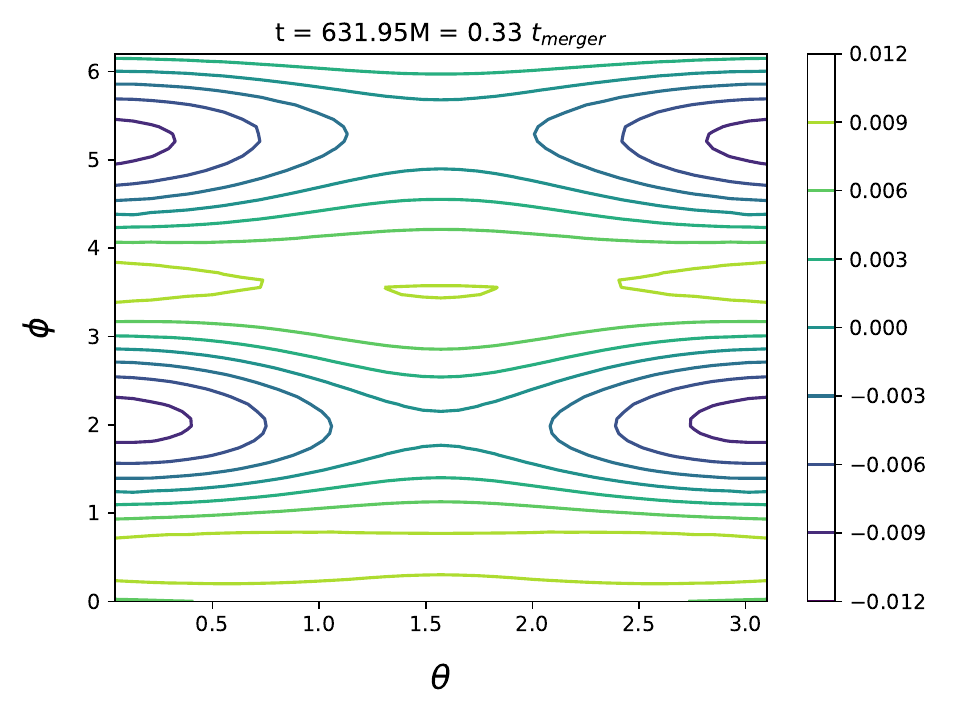}
	\caption[A 2D contour plot of the shears at the DH.]{The 2D contour plots of the shears. The contours of the shears which were plotted in 3D in the previous \ref{fig:shear_news_3d} shown here. The time instant at which the shears are plotted is depicted on top of the plots. The data corresponds to the MOTS at $t=631.95M$ units into the simulation and corresponds to one-third of the inspiral duration before the common horizon forms.}
	\label{fig:shear_news_cont}
\end{figure}

We begin by looking at the complex shears $\sigma_1$ and $\sigma_2$, of the outgoing null normal $\ell^a$ at the two individual horizons for the simulations, say $q=0.25$.   We write the shear as $\sigma = \sigma_+ + i\sigma_\times$.  As in \cite{Gupta:2018znn}, we introduce coordinates $(\theta,\phi)$ on the
horizons with the $z$-axes perpendicular to the orbital plane. Just like the waveform the angular distribution is mostly quadrupolar, i.e. $\sigma\propto {}_{-2}Y_{2,2}(\theta,\phi)$. This is evident from \ref{fig:shear_news_3d}, \ref{fig:shear_news_cont}. It will then be sufficient for our purposes to just look at the values of $\sigma_{1,2}$ near the north poles of the two horizons. This will not suffice for precessing spins or when higher modes become more important.  In these more complicated cases the approach suggested in \cite{Ashtekar:2013qta} can be followed.  

For each black hole, we first compare the 1-D shear data to the dominant (quadrupolar) news of the respective simulation. 

The news of the outgoing gravitational radiation is obtained by using the methods described in the Appendix \ref{appendix:waveform_prep}. This has been extrapolated to infinity, and corrected for the CoM drifts.

Given the shear, we now ask the question: How well do the shears behave in comparison to the outgoing gravitational wave news? To answer this question, we co-align each shear against the news for all the black holes and simulations by minimizing a least-squares order of merit for the `match' between the two waveforms. The minimum least-squares order of merit and the corresponding maximum match score is obtained by aligning the normalized waveforms over time and phase. A new algorithm was invented for this purpose. The functions for computing the match, and co-aligning the numerical waveforms are implemented in the \hl{waveformtools.waveformtools} module \cite{nrtools}.

\begin{figure}
  \includegraphics[width=\columnwidth]{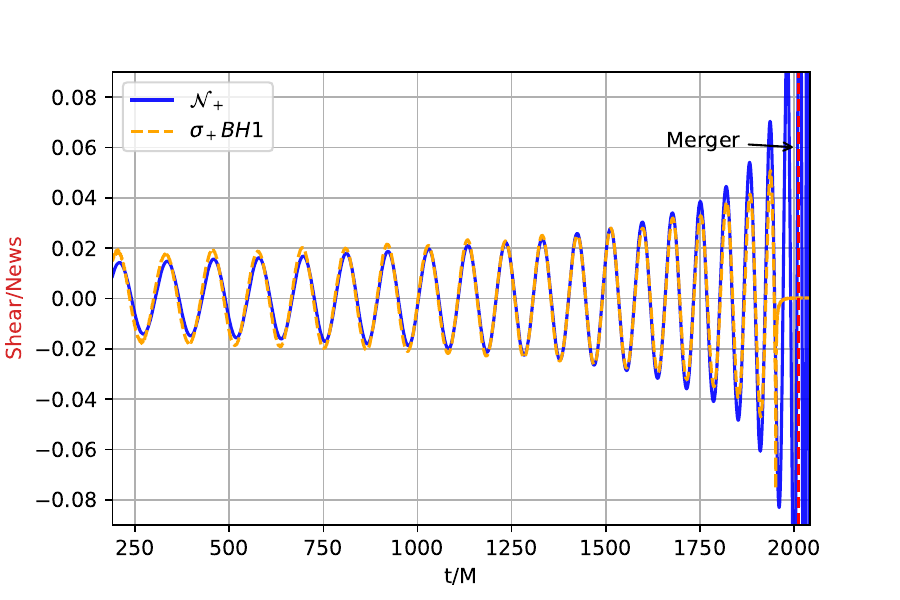}
  \includegraphics[width=\columnwidth]{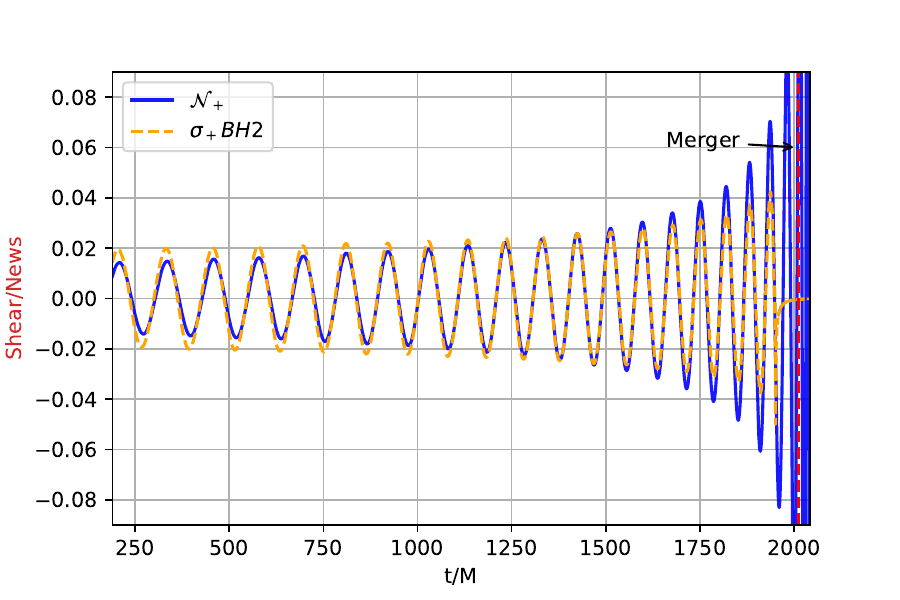}
  \caption[DH Shear - numerical news correlations for the $q=0.25$ BBH system.]{The real part of the shear, $\sigma_+(t)$, for the two black holes (BH1 and BH2) for the $q=0.25$ configuration, compared with the real part of the news function $\mathcal{N}_+^{(2,2)}$. The functions $\sigma_+$ and $\mathcal{N}_+^{(2,2)}$ have been suitably aligned and their amplitudes scaled so that they have unit norm over the time interval where the shear is defined. Similar results hold for the '$\times$' polarization and the other mass-ratios.}
  \label{fig:shear-news}
\end{figure}

\ref{fig:shear-news} shows the real parts $\sigma_+(t)$ for the two black holes and for the $q=0.25$ configuration.  The plot also shows the news function for the corresponding mode and polarizations, i.e. $\mathcal{N}_+^{(2,2)}$.
The time when the common horizon is formed is
indicated by a vertical line indicating "merger".
The shear is set to zero when the horizon can no longer be found reliably, which happens here somewhat before the merger.  We see clearly that the
shear is very similar to the well-known ``chirping''
gravitational waveform: it has increasing frequency and amplitude.  We
note also that since the shear is non-vanishing, the horizon is
not isolated and its area is increasing. However, the area
is not increasing rapidly and this area increase is not 
measured reliably in our simulations.  See 
\cite{PhysRevLett.123.171102,PhysRevD.100.084044} for a more 
accurate study of the area increase in a black hole merger.  

\begin{figure}[hbtp]
    \begin{tabular}{cc}
        \includegraphics[width=\myw, height =\myh]{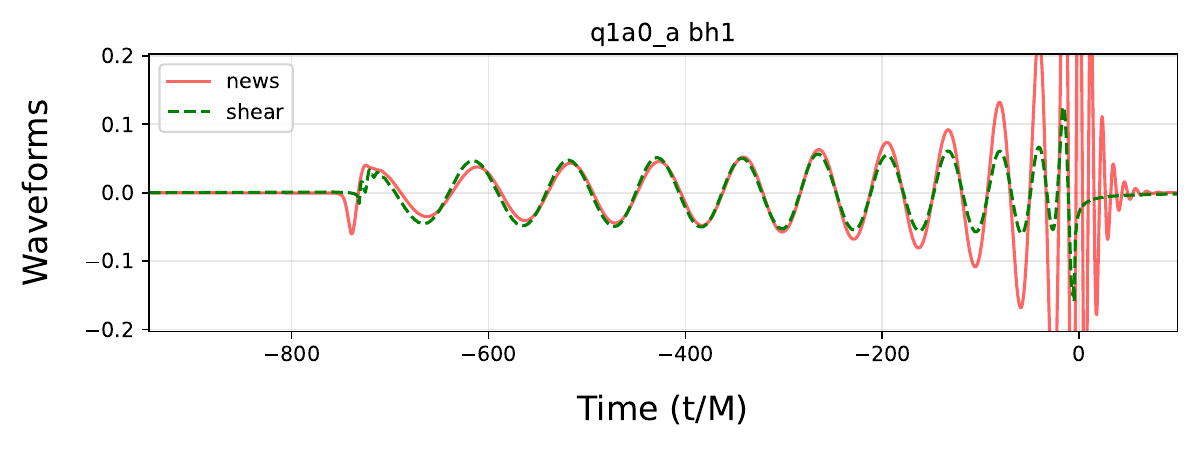} &
        \includegraphics[width=\myw, height =\myh]{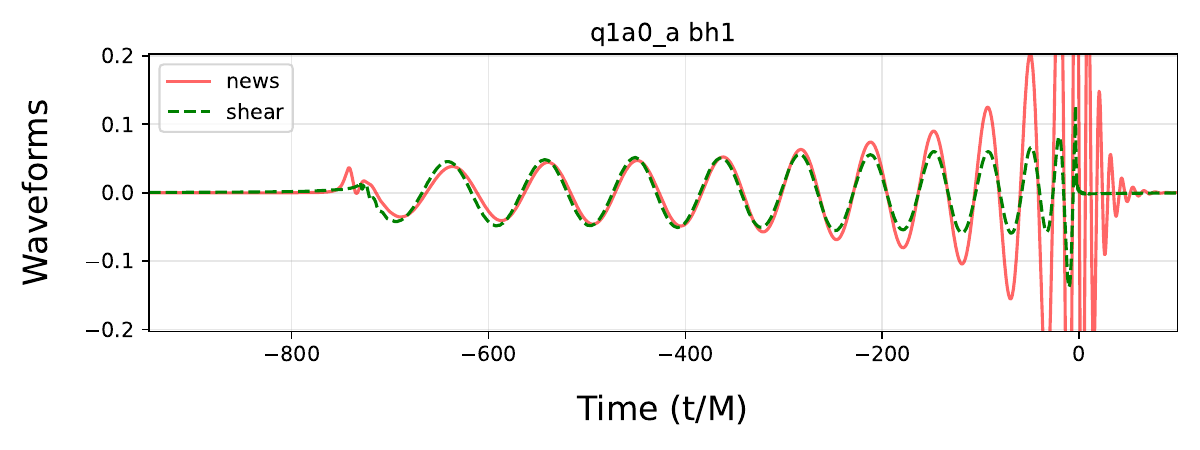} \\
    \end{tabular}
    \begin{tabular}{cc}
        \includegraphics[width=\myw, height =\myh]{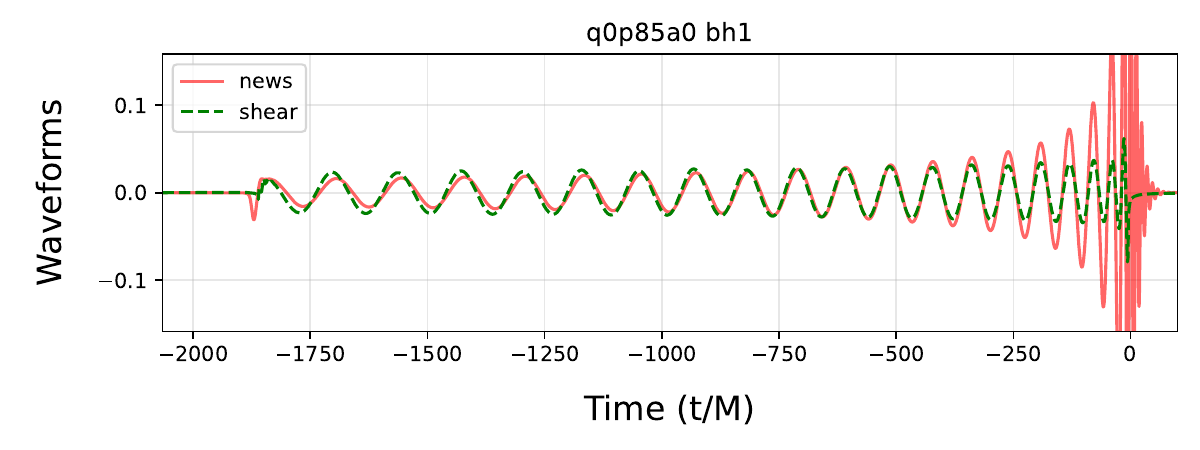} &
        \includegraphics[width=\myw, height =\myh]{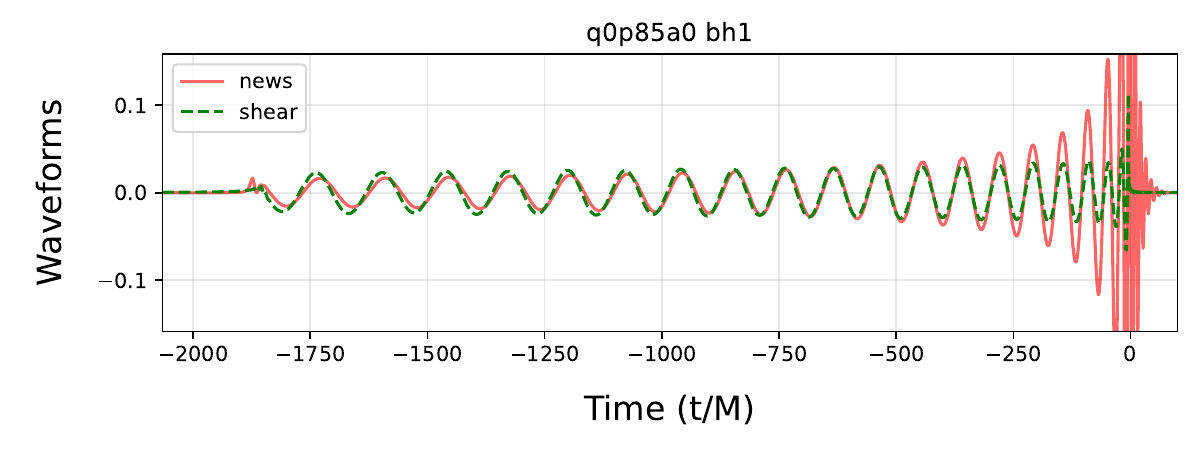} \\
         \\
    \end{tabular}
    \begin{tabular}{cc}
        \includegraphics[width=\myw, height =\myh]{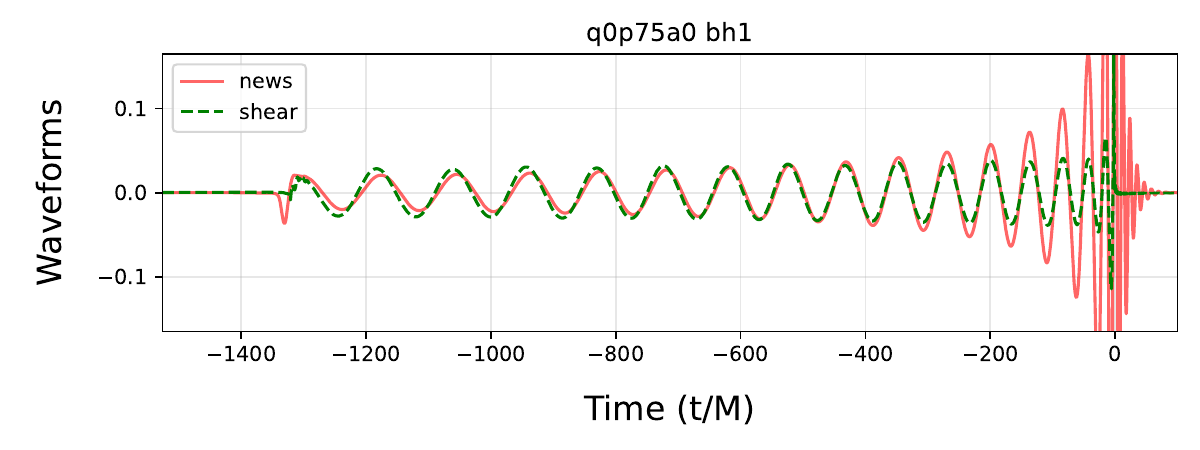} &
        \includegraphics[width=\myw, height =\myh]{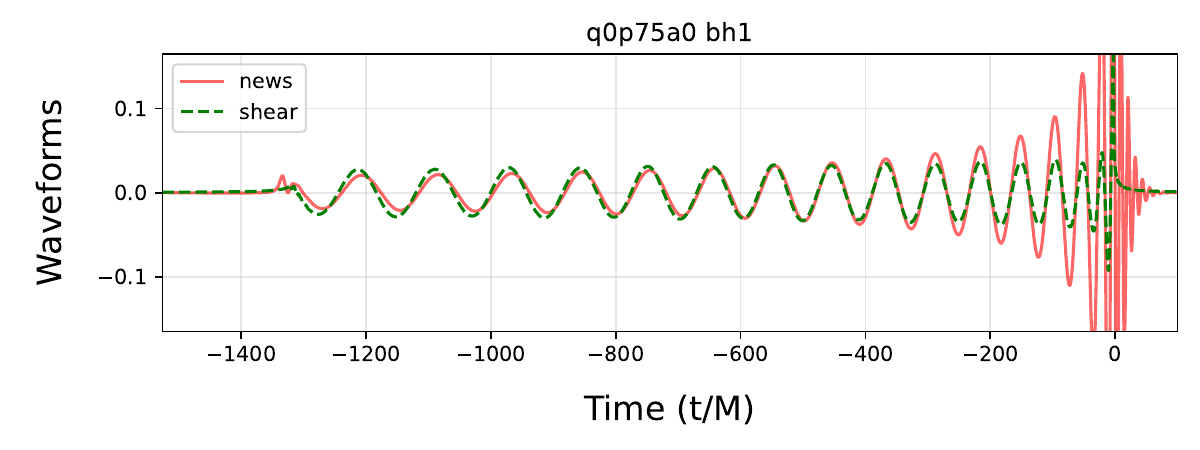} \\
         \\
    \end{tabular}
    \begin{tabular}{cc}
        \includegraphics[width=\myw, height =\myh]{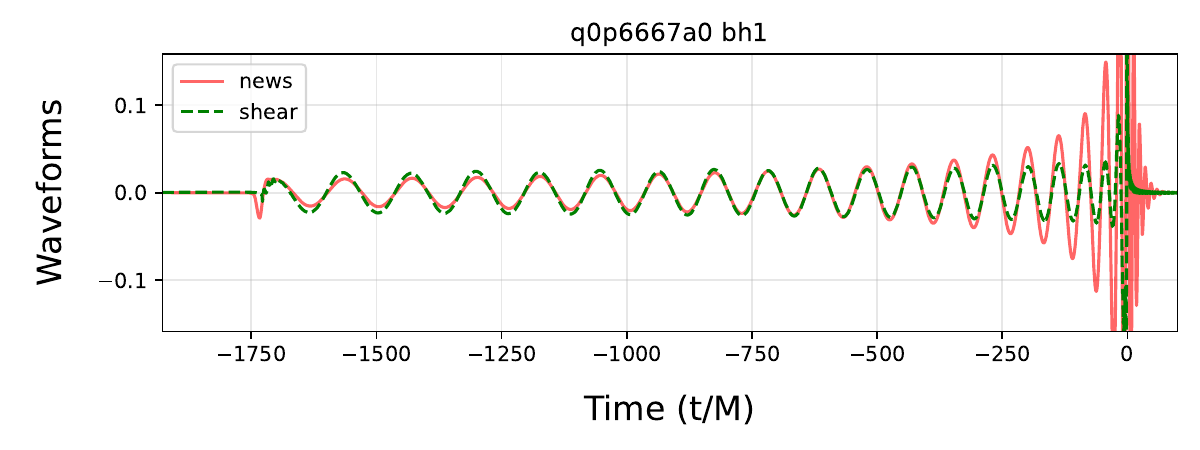} &
        \includegraphics[width=\myw, height =\myh]{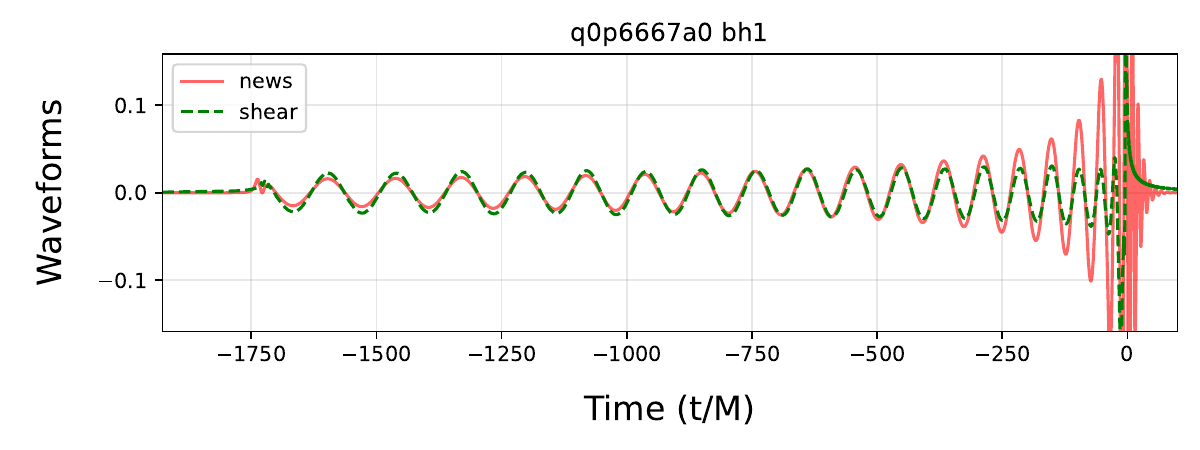} \\
         \\
    \end{tabular}
        \begin{tabular}{cc}
        \includegraphics[width=\myw, height =\myh]{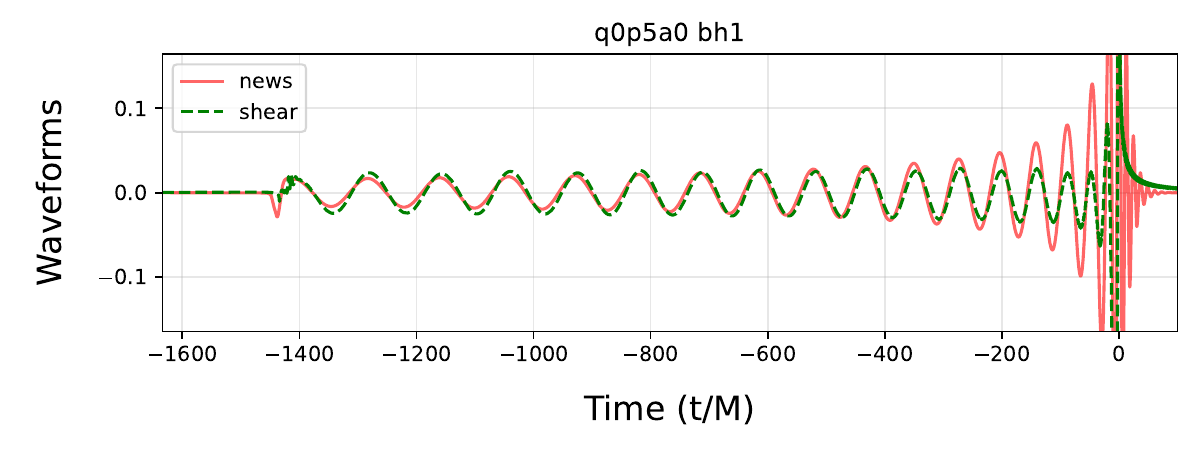} &
        \includegraphics[width=\myw, height =\myh]{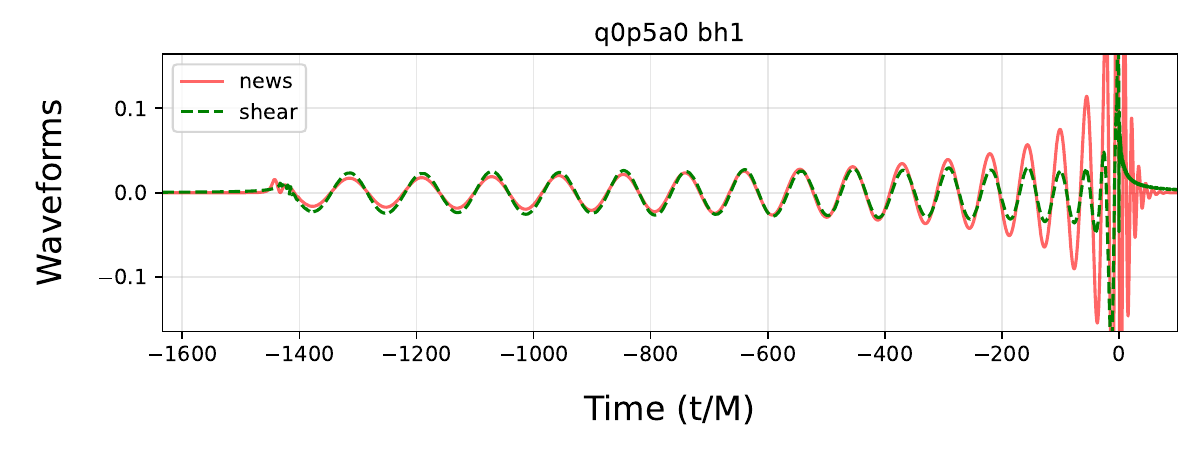} \\
         \\
    \end{tabular}
        \begin{tabular}{cc}
        \includegraphics[width=\myw, height =\myh]{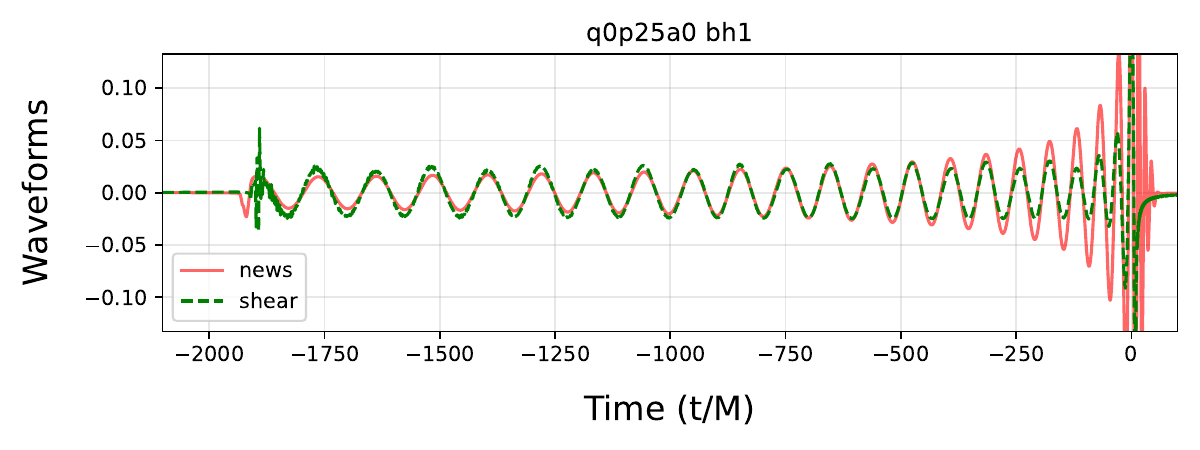} &
        \includegraphics[width=\myw, height =\myh]{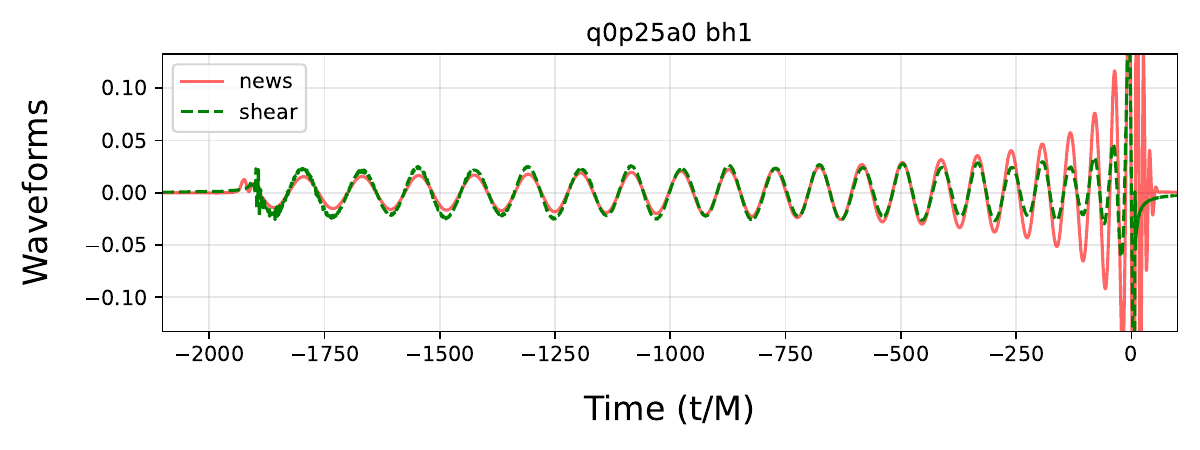} \\
         \\
    \end{tabular}
    \caption[Strong correlation between the DH shears and numerical news, BH1.]{The shear-news correlations for BH1. The co-aligned shears of the outgoing null vectors of the dynamical horizon of the larger black hole, against the News functions of outgoing gravitational radiation. The real part of the functions is shown on the left whereas the imaginary part is shown on the right. Each row represents the data from a numerical simulation of a particular mass ratio.} 
 \label{fig:shear_news_bh1}
\end{figure}

\begin{figure}[hbtp]
    \begin{tabular}{cc}
        \includegraphics[width=\myw, height =\myh]{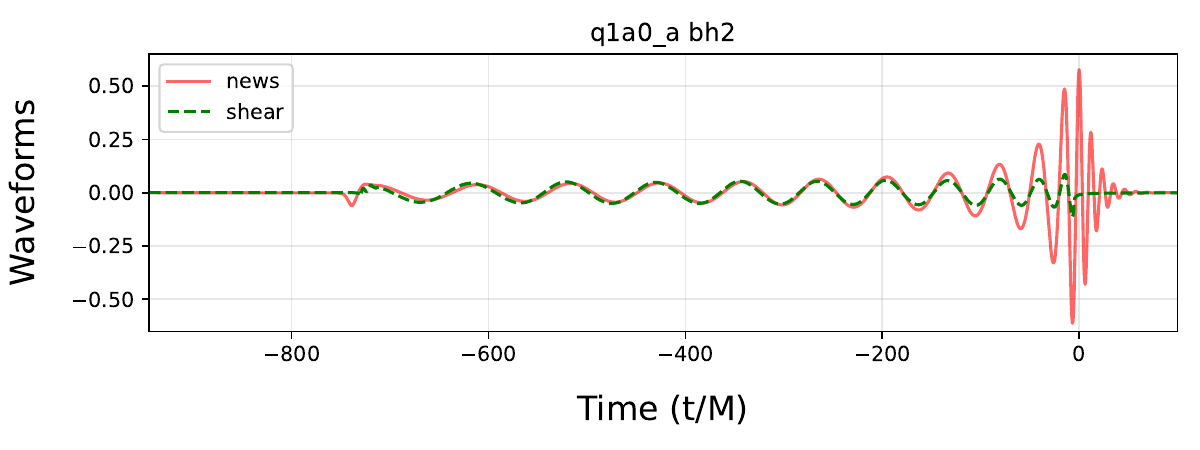} &
        \includegraphics[width=\myw, height =\myh]{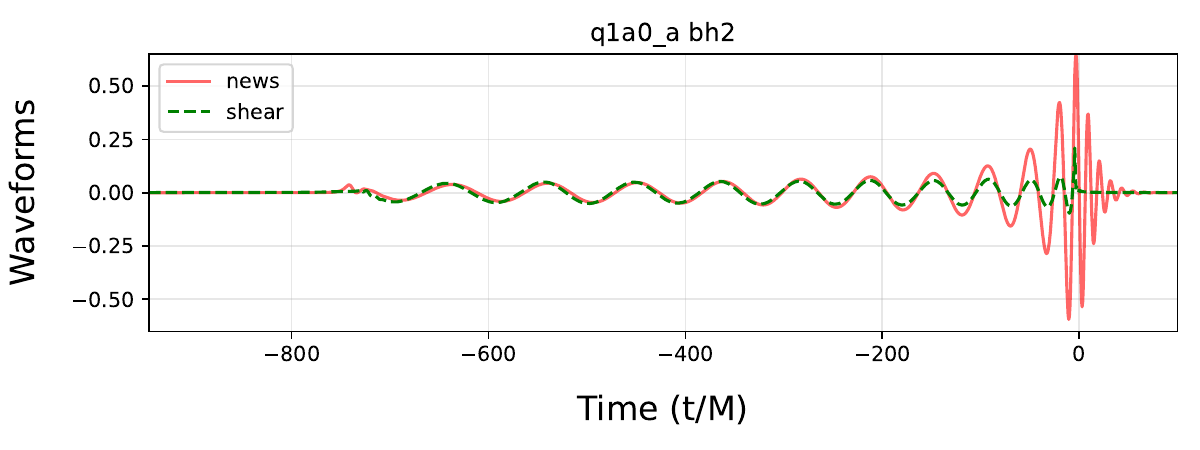} \\
    \end{tabular}
    \begin{tabular}{cc}
        \includegraphics[width=\myw, height =\myh]{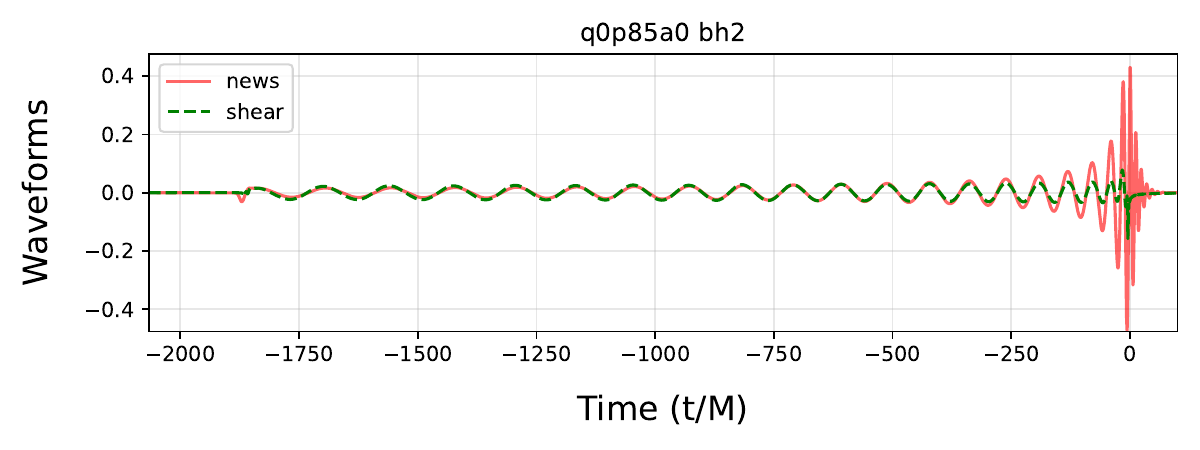} &
        \includegraphics[width=\myw, height =\myh]{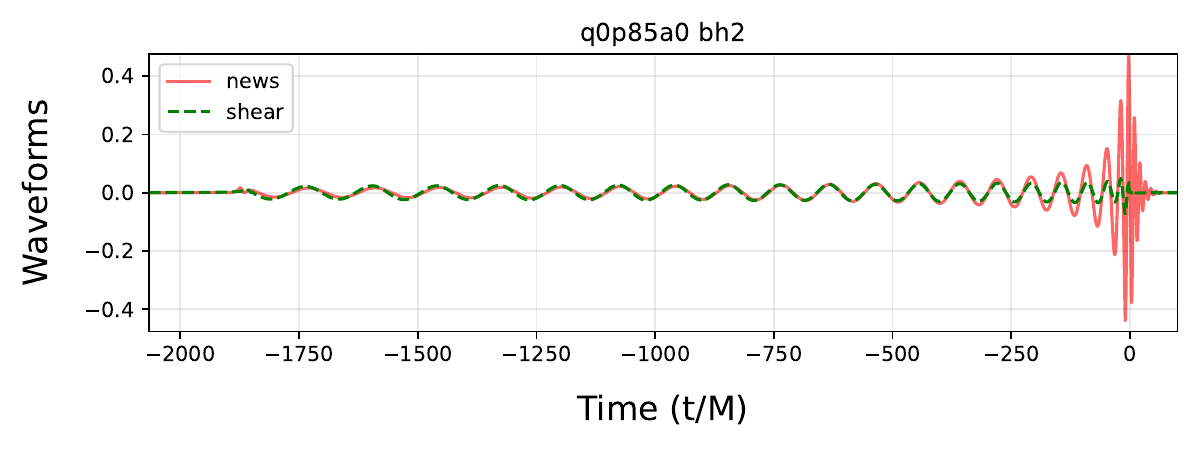} \\
         \\
    \end{tabular}
    \begin{tabular}{cc}
        \includegraphics[width=\myw, height =\myh]{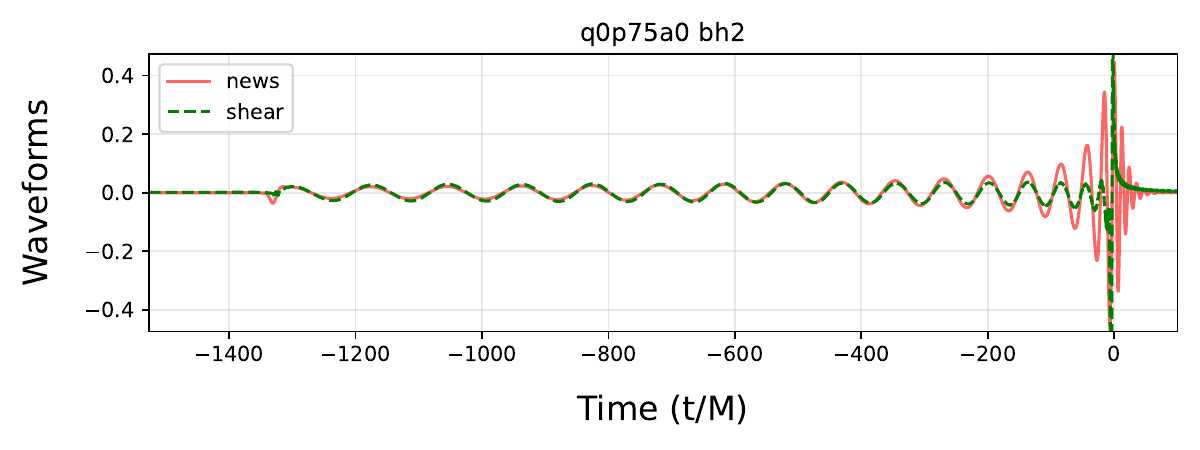} &
        \includegraphics[width=\myw, height =\myh]{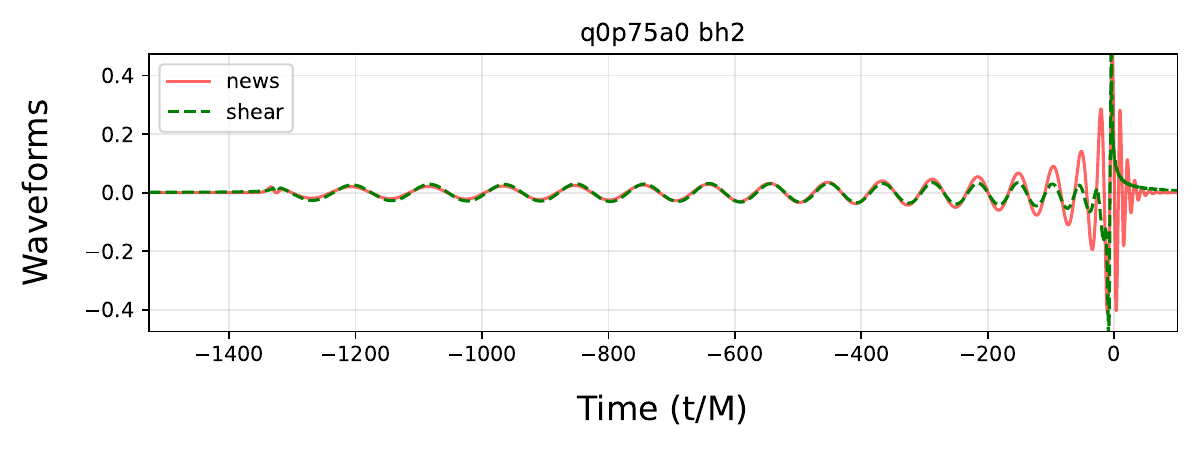} \\
         \\
    \end{tabular}
    \begin{tabular}{cc}
        \includegraphics[width=\myw, height =\myh]{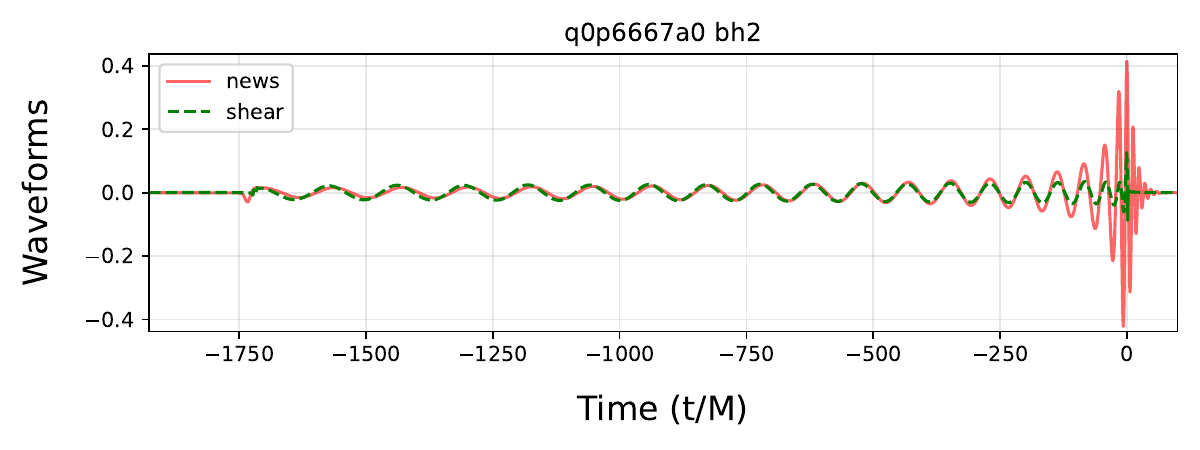} &
        \includegraphics[width=\myw, height =\myh]{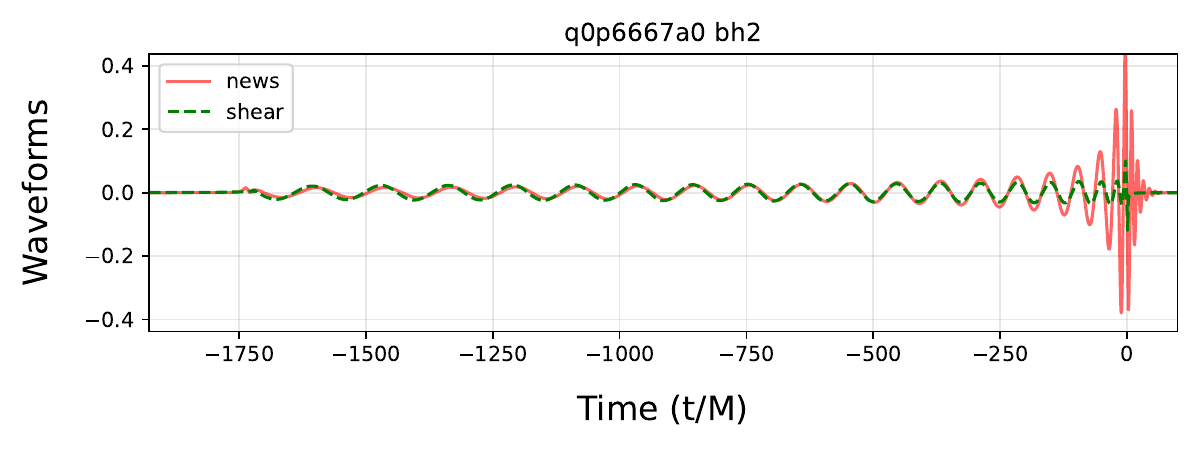} \\
         \\
    \end{tabular}
        \begin{tabular}{cc}
        \includegraphics[width=\myw, height =\myh]{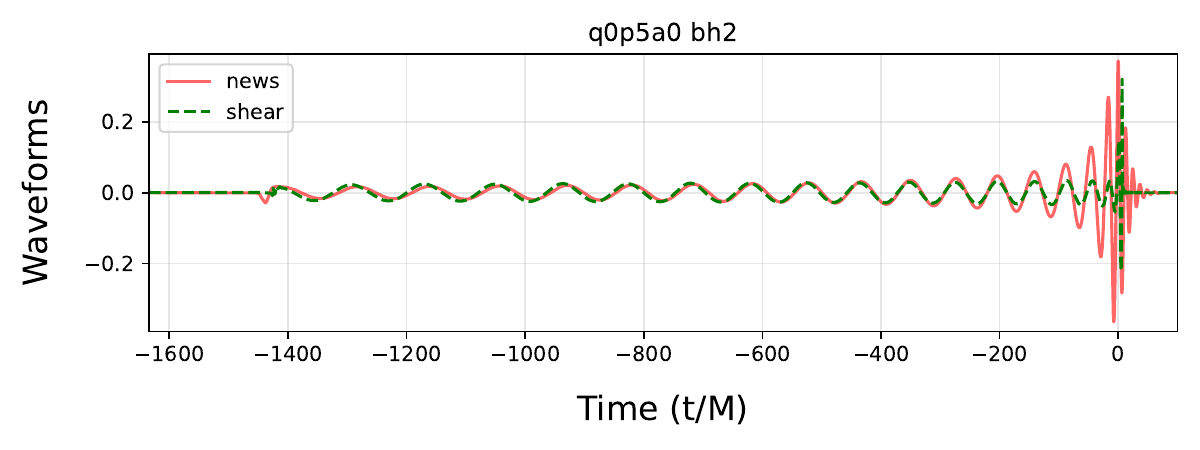} &
        \includegraphics[width=\myw, height =\myh]{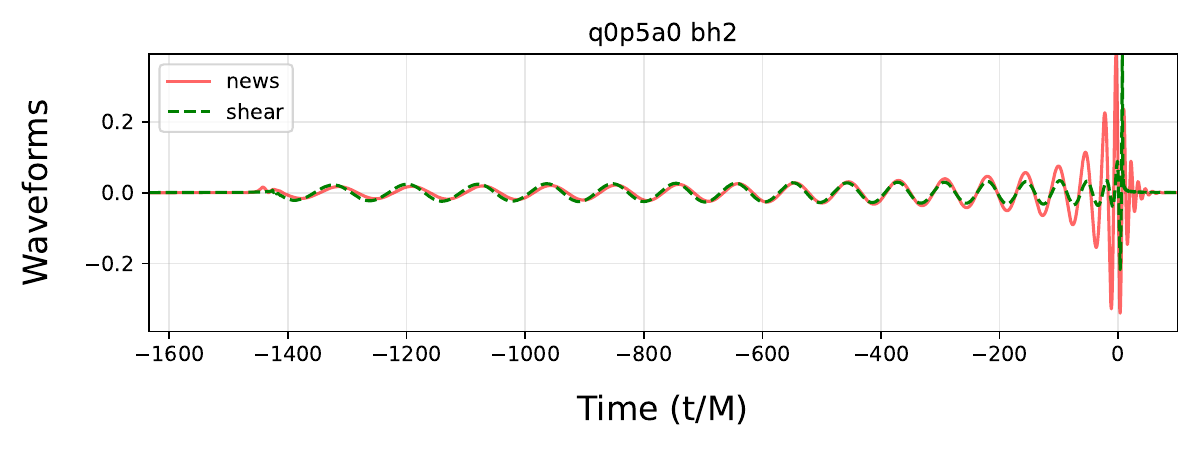} \\
         \\
    \end{tabular}
        \begin{tabular}{cc}
        \includegraphics[width=\myw, height =\myh]{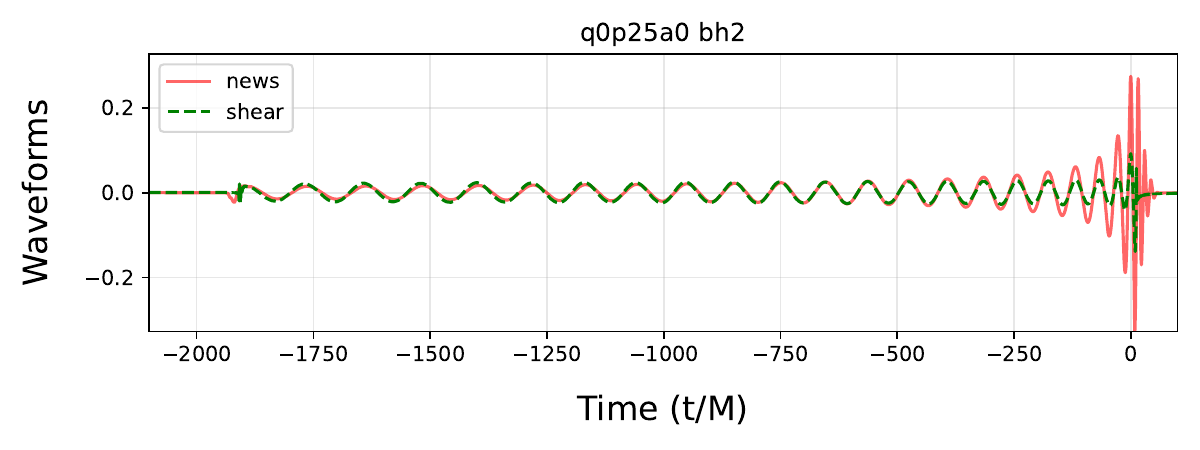} &
        \includegraphics[width=\myw, height =\myh]{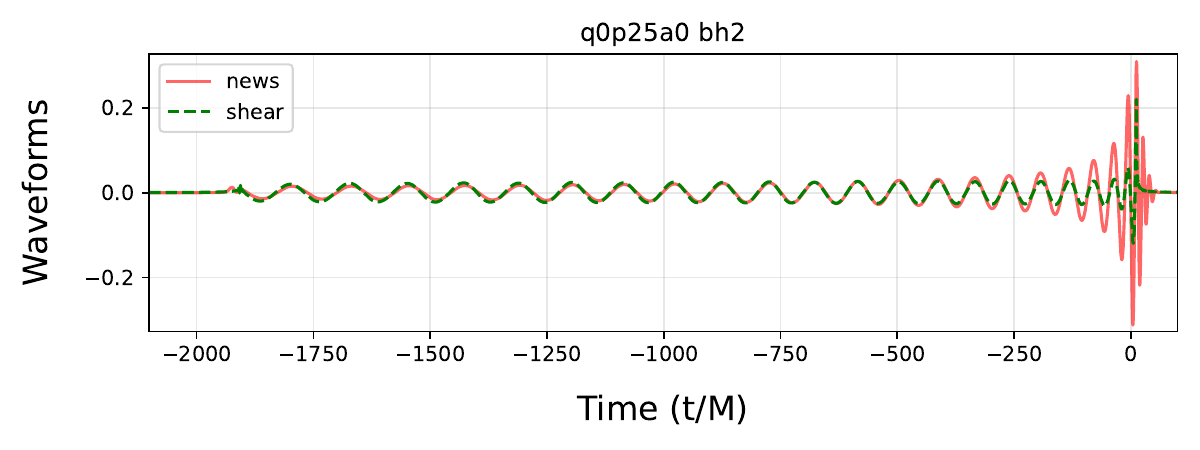} \\
         \\
    \end{tabular}
    \caption[Strong correlation between the shears and numerical news, BH2.]{The shear-news correlations for BH2. The co-aligned shears of the outgoing null vectors of the dynamical horizon of the smaller black hole, against the News functions of outgoing gravitational radiation. The real part of the functions is shown on the left whereas the imaginary part is shown on the right. Each row represents the data from a numerical simulation of a particular mass ratio.} 
 \label{fig:shear_news_bh2}
\end{figure}

\begin{figure}[hbtp]
    \begin{tabular}{cc}
        \includegraphics[width=\myw, height =\myh]{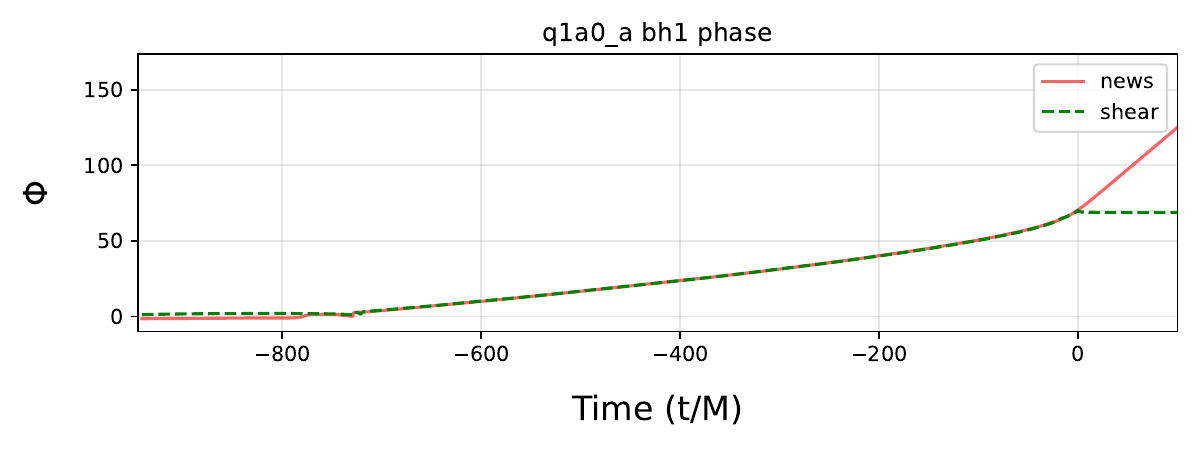} &
        \includegraphics[width=\myw, height =\myh]{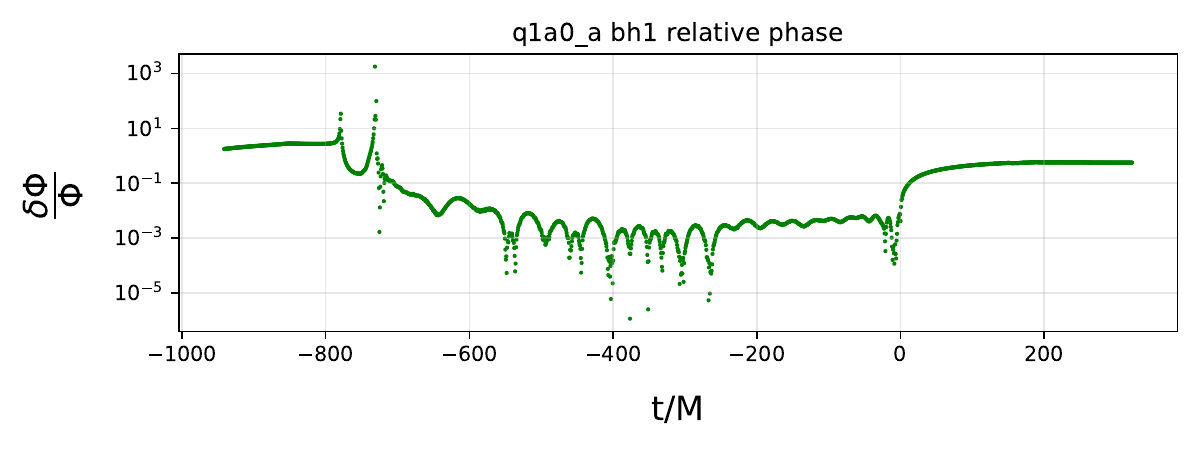} \\
    \end{tabular}
    \begin{tabular}{cc}
        \includegraphics[width=\myw, height =\myh]{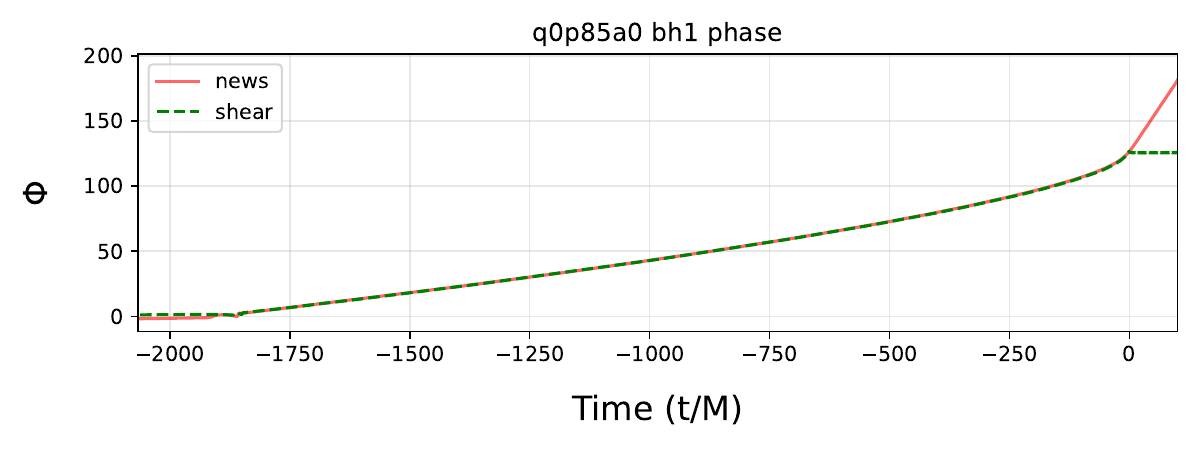} &
        \includegraphics[width=\myw, height =\myh]{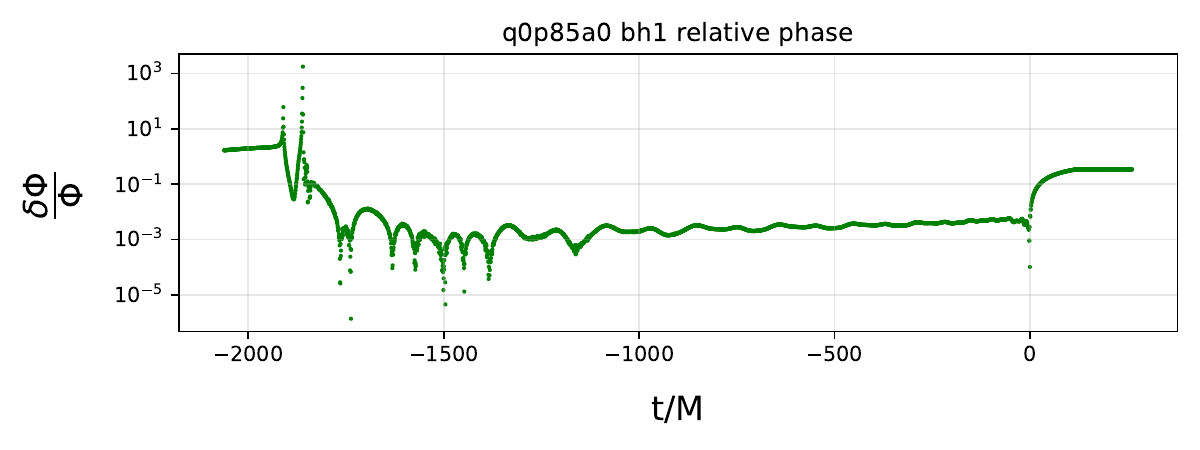} \\
         \\
    \end{tabular}
    \begin{tabular}{cc}
        \includegraphics[width=\myw, height =\myh]{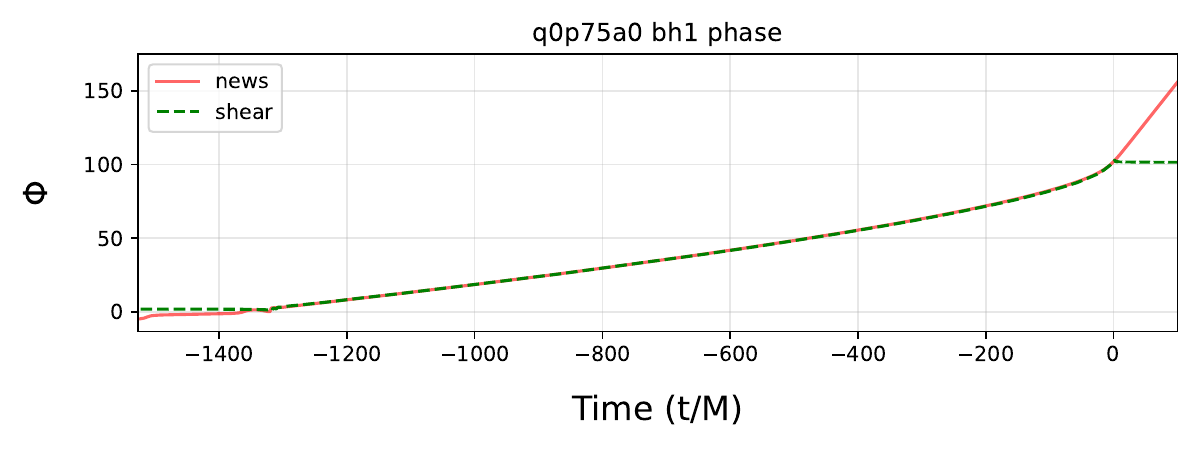} &
        \includegraphics[width=\myw, height =\myh]{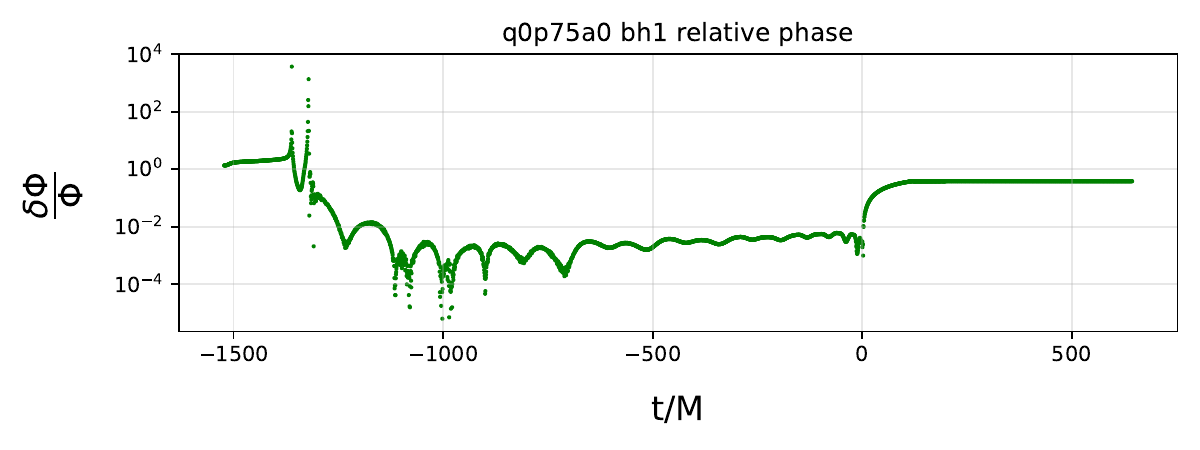} \\
         \\
    \end{tabular}
    \begin{tabular}{cc}
        \includegraphics[width=\myw, height =\myh]{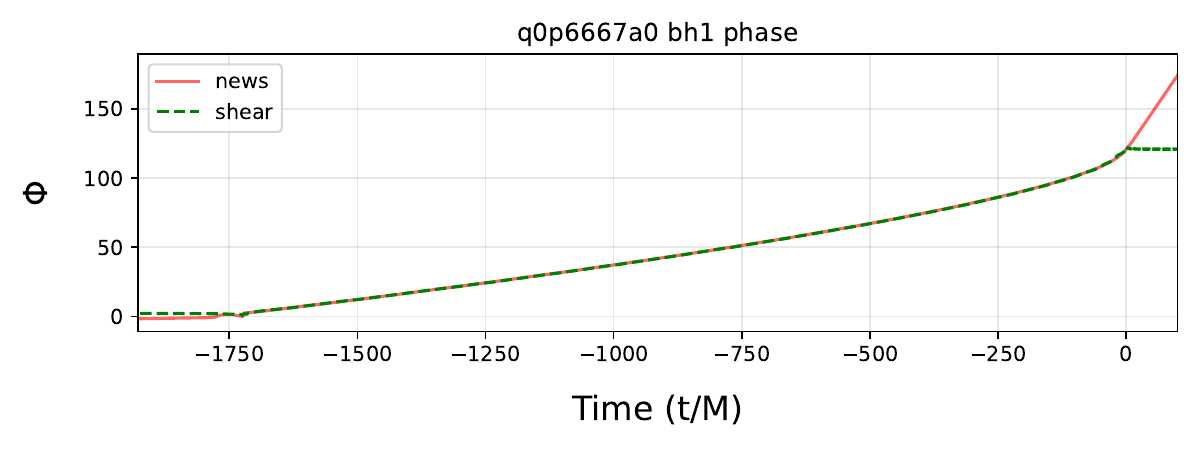} &
        \includegraphics[width=\myw, height =\myh]{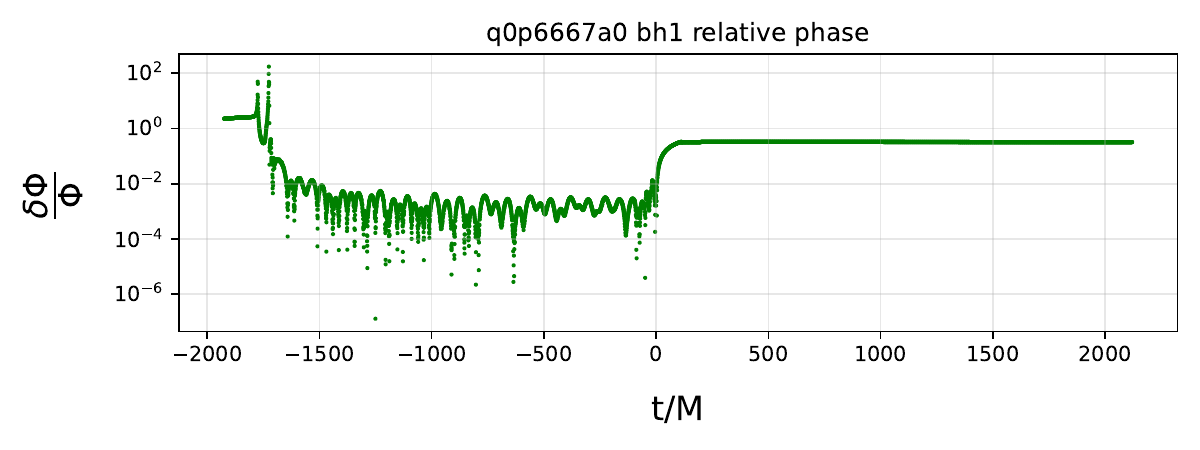} \\
         \\
    \end{tabular}
        \begin{tabular}{cc}
        \includegraphics[width=\myw, height =\myh]{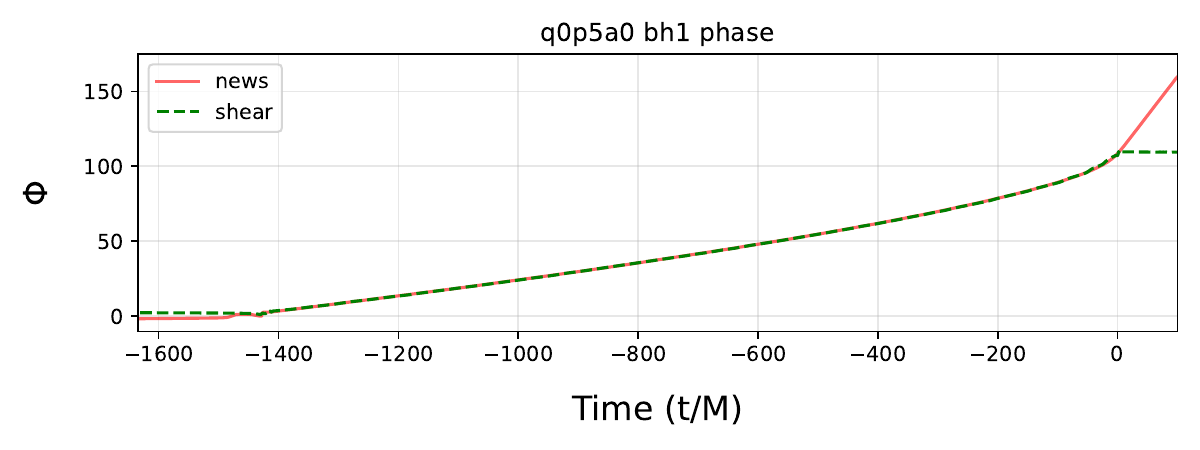} &
        \includegraphics[width=\myw, height =\myh]{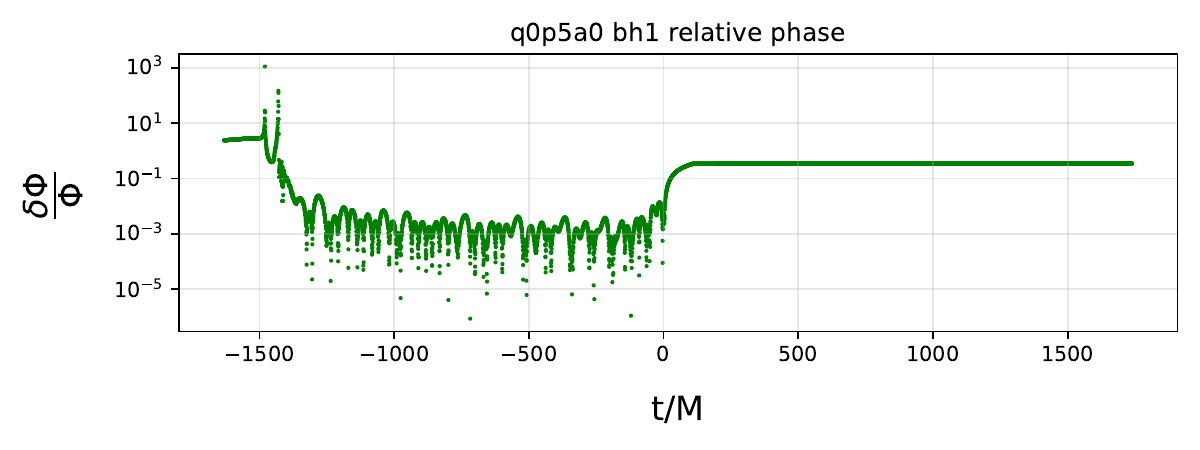} \\
         \\
    \end{tabular}
        \begin{tabular}{cc}
        \includegraphics[width=\myw, height =\myh]{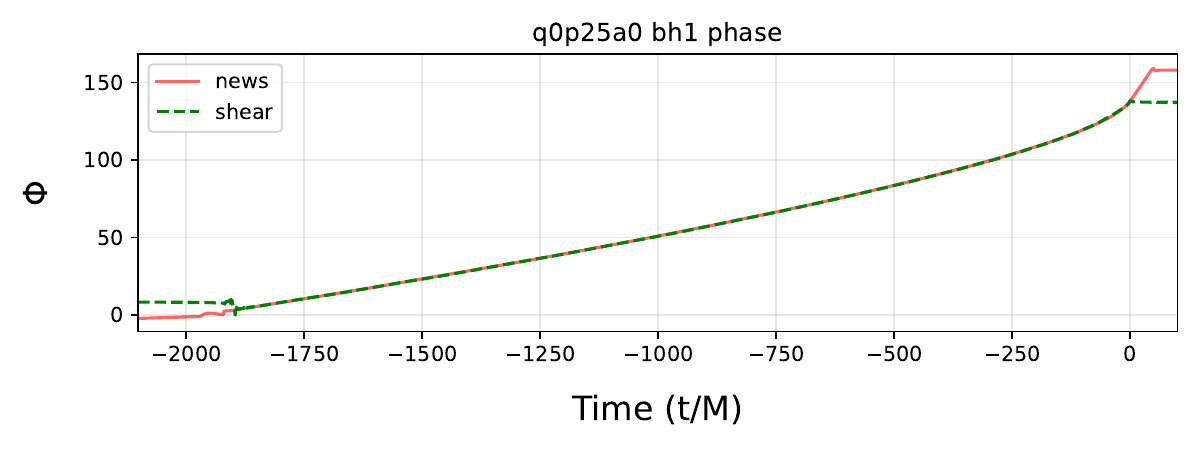} &
        \includegraphics[width=\myw, height =\myh]{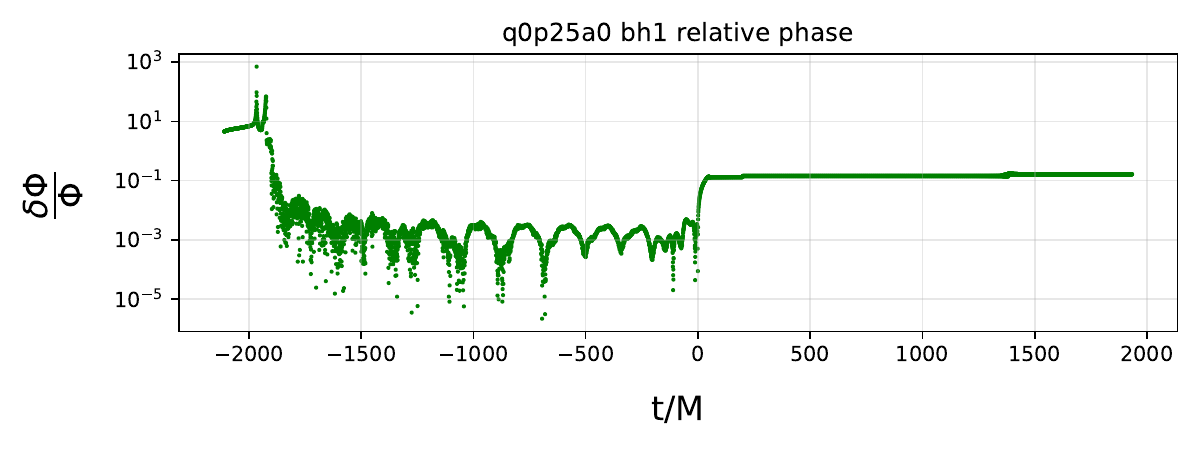} \\
         \\
    \end{tabular}
    \caption[Phase evolution differences between the DH shears and numerical news, BH1.]{The shear-news phase evolution for BH1. The phases of the co-aligned shears of the outgoing null vectors of the dynamical horizons of the larger black holes (BH1), against those of News functions of outgoing gravitational radiation are plotted here. The aligned phases are shown on the left whereas the relative deviation of the phases of the shear from those of the news functions are shown on the right. Each row represents the data from a numerical simulation of a particular mass ratio whose value is mentioned in the panel title.} 
 \label{fig:shear_news_phase_bh1}
\end{figure}

\begin{figure}[hbtp]
    \begin{tabular}{cc}
        \includegraphics[width=\myw, height =\myh]{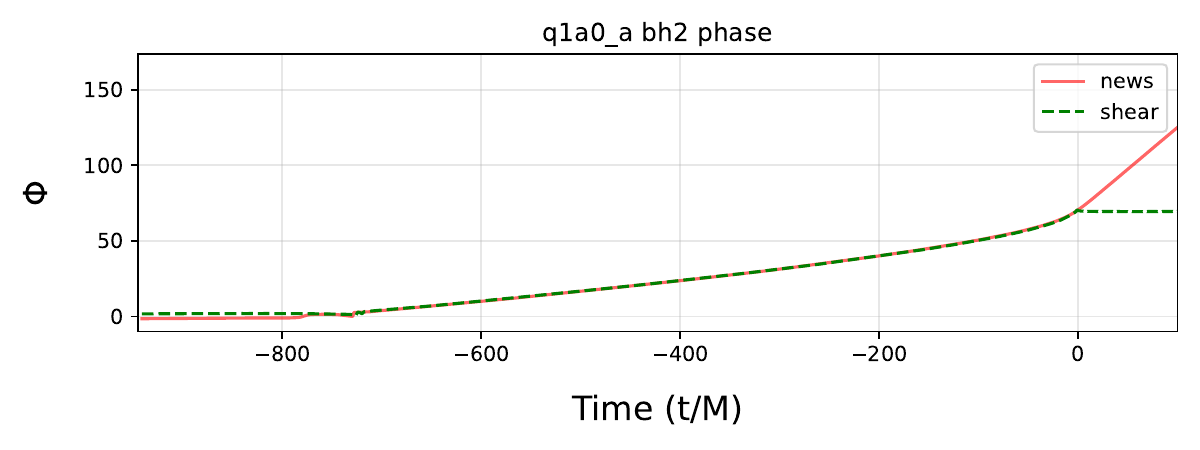} &
        \includegraphics[width=\myw, height =\myh]{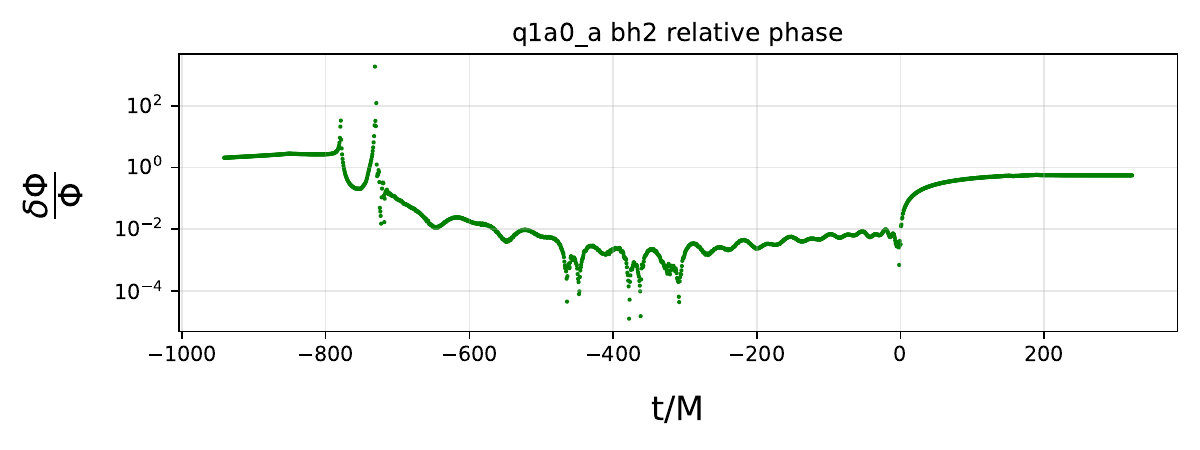} \\
    \end{tabular}
    \begin{tabular}{cc}
        \includegraphics[width=\myw, height =\myh]{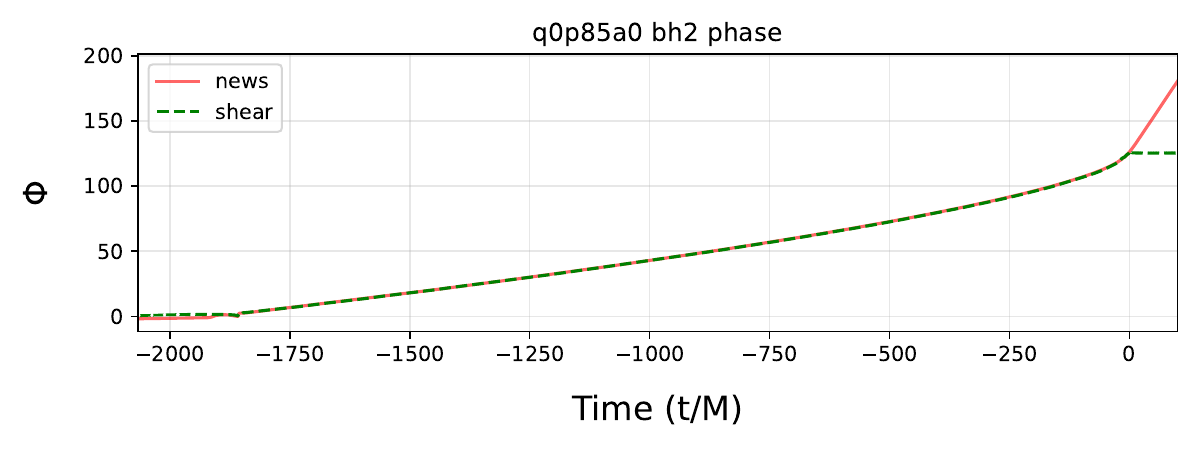} &
        \includegraphics[width=\myw, height =\myh]{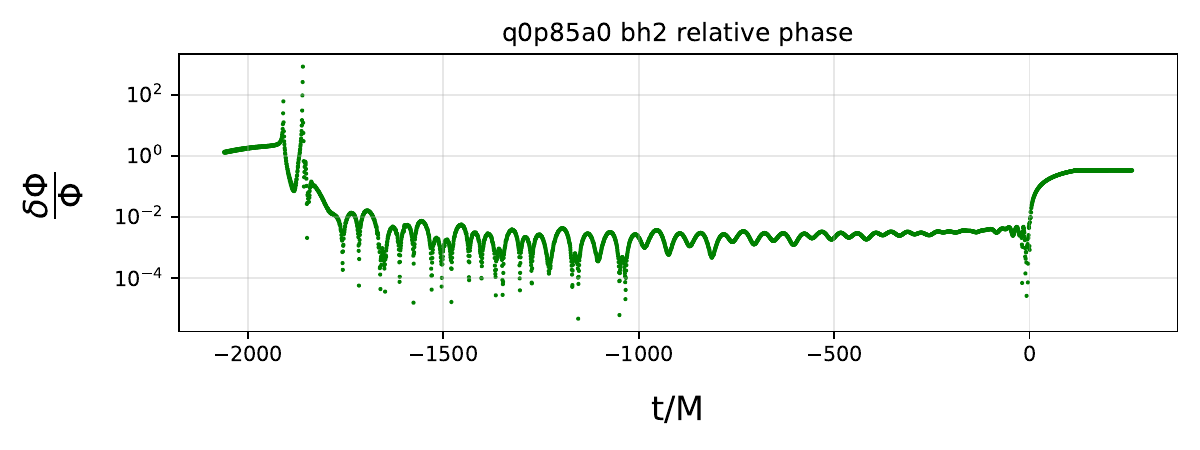} \\
         \\
    \end{tabular}
    \begin{tabular}{cc}
        \includegraphics[width=\myw, height =\myh]{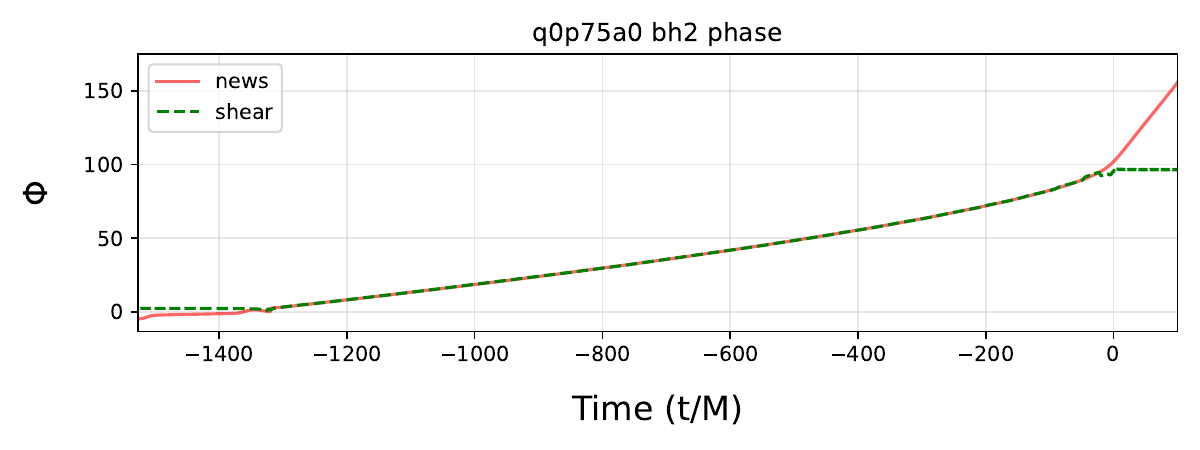} &
        \includegraphics[width=\myw, height =\myh]{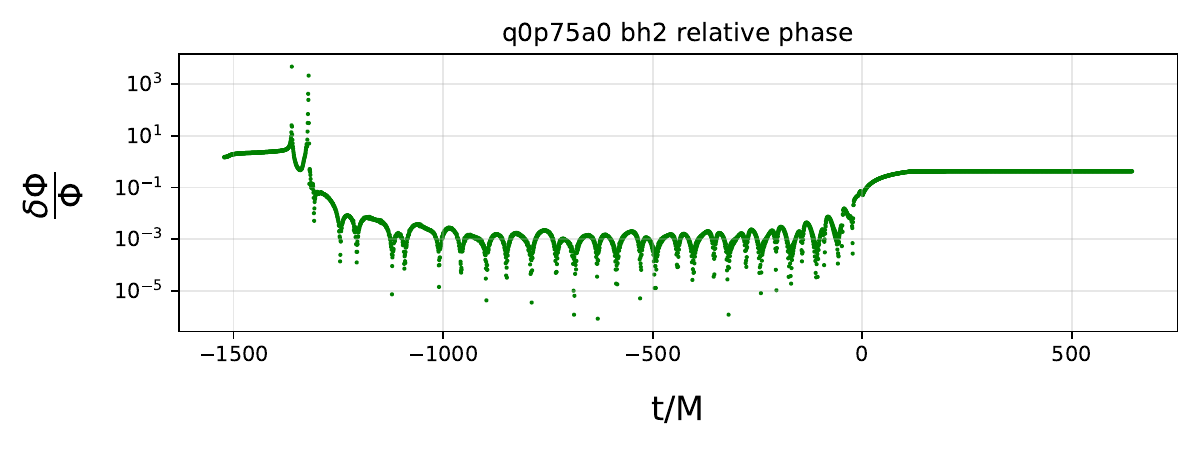} \\
         \\
    \end{tabular}
    \begin{tabular}{cc}
        \includegraphics[width=\myw, height =\myh]{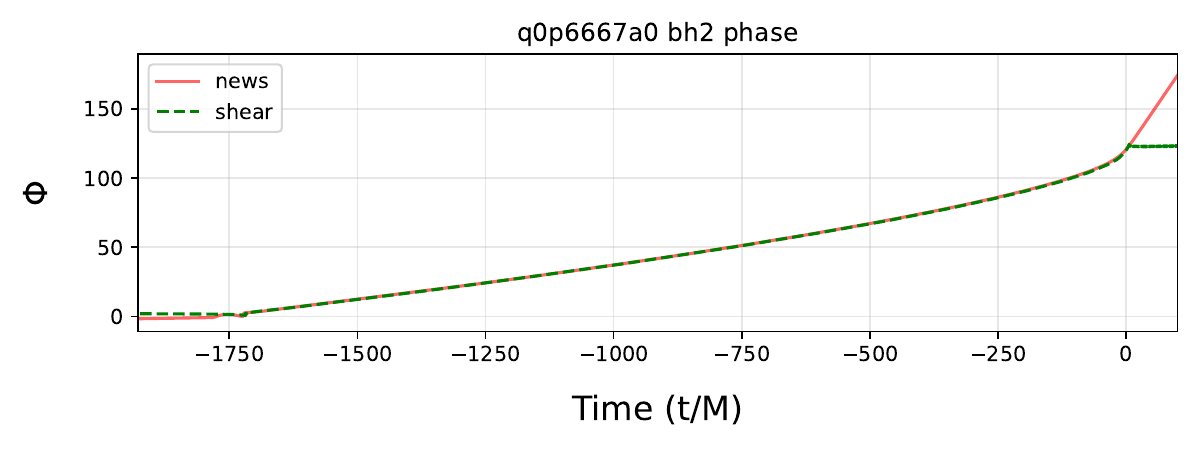} &
        \includegraphics[width=\myw, height =\myh]{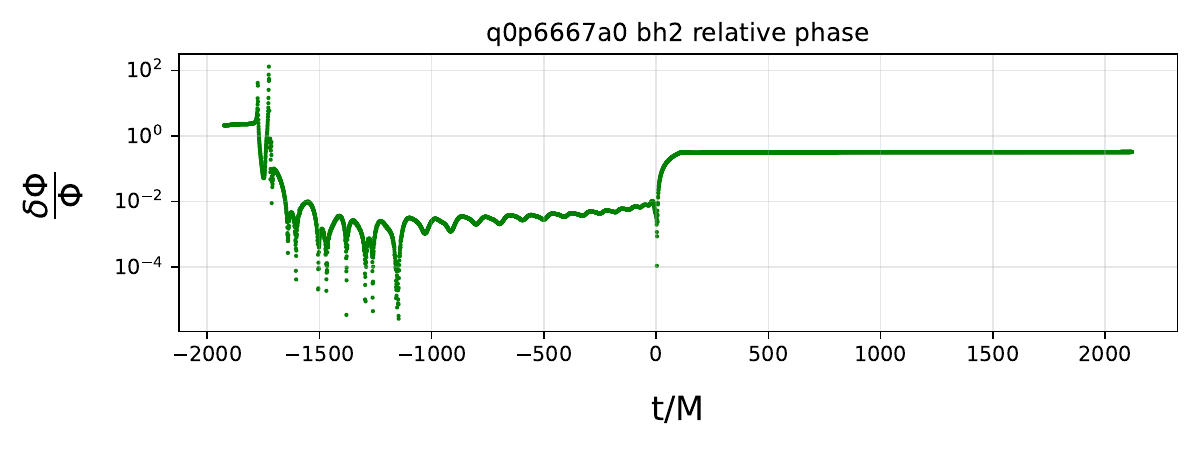} \\
         \\
    \end{tabular}
        \begin{tabular}{cc}
        \includegraphics[width=\myw, height =\myh]{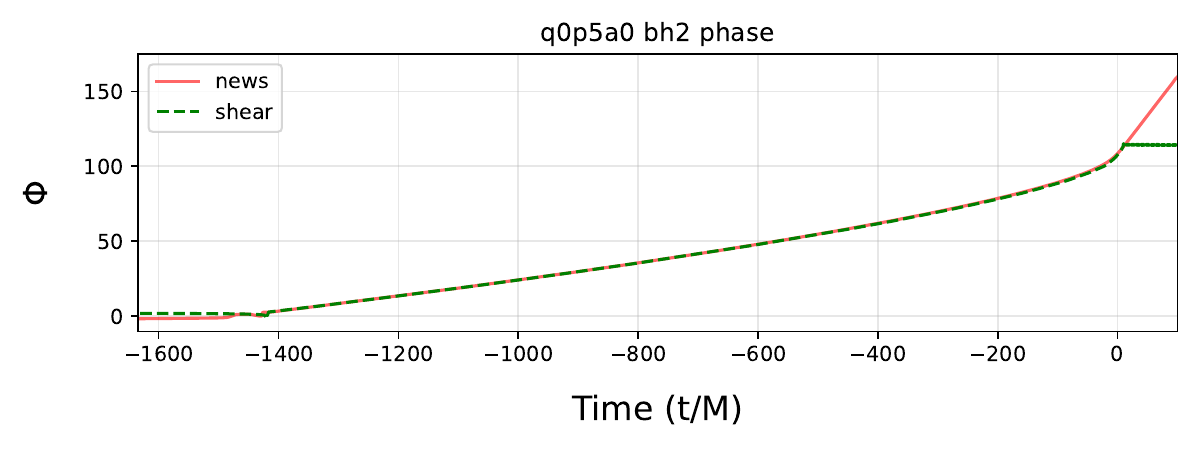} &
        \includegraphics[width=\myw, height =\myh]{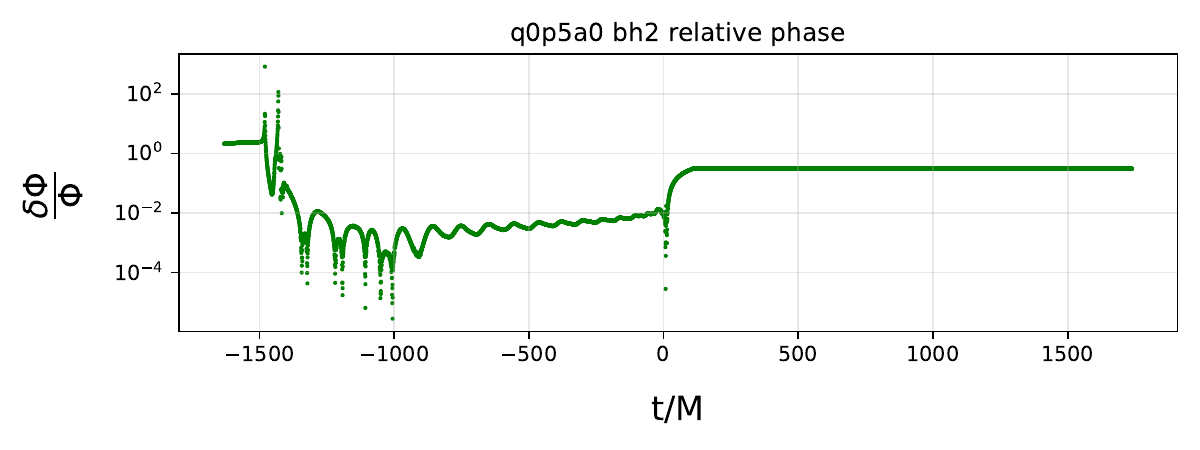} \\
         \\
    \end{tabular}
        \begin{tabular}{cc}
        \includegraphics[width=\myw, height =\myh]{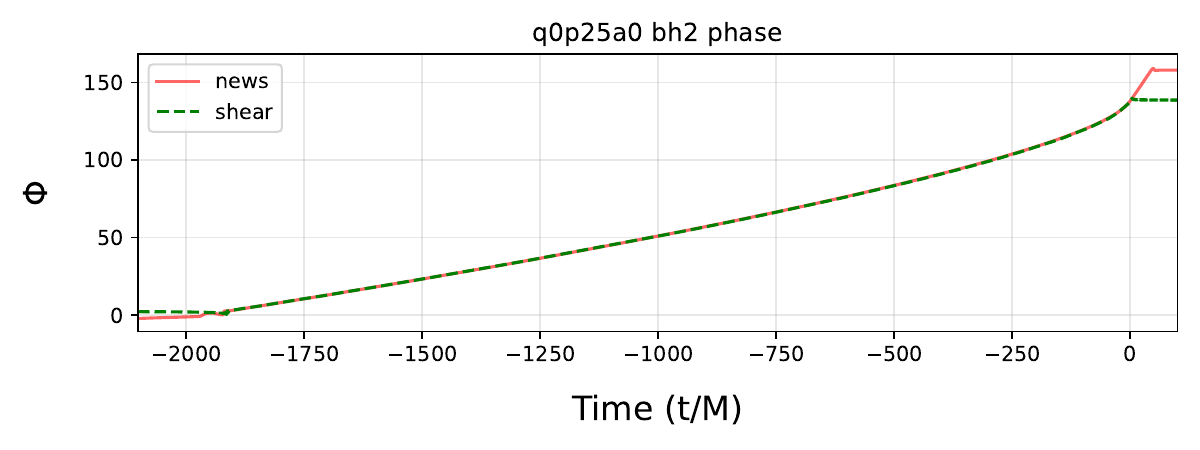} &
        \includegraphics[width=\myw, height =\myh]{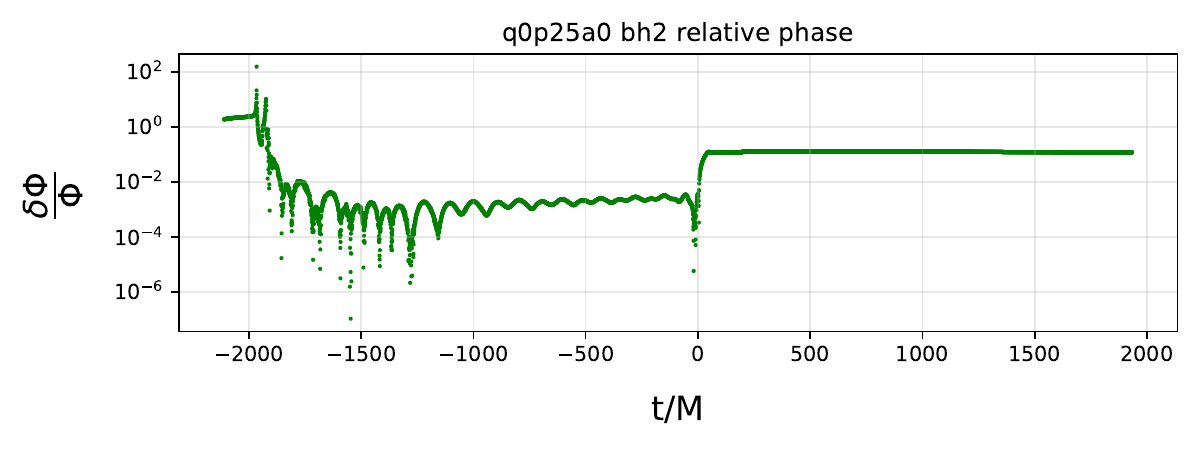} \\
         \\
    \end{tabular}
    \caption[Phase evolution differences between the DH shears and numerical news, BH2.]{The shear-news phase evolution for BH2. See \ref{fig:shear_news_phase_bh1} for a description.} 
 \label{fig:shear_news_phase_bh2}
\end{figure}

In \ref{fig:shear_news_bh1} and \ref{fig:shear_news_bh2}, we show a collage of the co-aligned shears of both polarizations of all the dynamical horizons and the news of the respective simulations. From these figures, it is evident that the shears of the DHs have similar behaviour to that of the news, and are strongly correlated with each other. In particular, the phase evolution of these waveforms is very close to each other. \ref{fig:shear_news_phase_bh1} and \ref{fig:shear_news_phase_bh2} show how well the phase evolutions of the shears and the respective news functions evolve. The difference is of the order of a per cent. However, it is noticeable that the amplitudes of the shears do not agree very well with those of the news functions. This discrepancy can be seen to be more towards the merger. This effect is present in the original data and is not the result of post-processing (like tapering). This needs to be investigated further.


Does this agreement between the shear and the gravitational wave
signals hold quantitatively?  To answer this question, we treat the
shear as a bona fide gravitational waveform and attempt to estimate its
parameters.  This was done using the algorithm used in the previous section and a template bank constructed from the gravitational waveforms of a million binary black-hole systems of varying mass ratios and chirp masses using a commonly used approximant. The routine was developed and implemented using  MPI to make the parameter estimation routine scalable to large computing systems. The parameters of the best-matching templates (using the match score defined above) were then returned as the best estimates of the parameters of the BBH system.

The chirp mass, $\mathcal{M} = Mq^{3/5}/(1+q)^{6/5}$, determines the frequency
evolution of the signal at the leading order. For any given simulation,
we have the three possible gravitational wave signals: the waveform
$h(t)$ extracted in the wave zone over a large sphere and extrapolated to $\mathcal{I}^+$ and transformed to CoM frame, and the shears $\sigma_{1,2}(t)$ calculated at the individual horizons.  

Here $t$ is the coordinate time used in the
numerical evolution.  For all three of these time series, we estimate 
$\mathcal{M}$ and $q$ using a well tested model for binary
mergers known as IMRPhenomPV2 \cite{PhysRevD.93.044006,PhysRevD.93.044007}.  This waveform model is
a development of the so-called phenomenological binary merger models
\cite{Ajith:2007kx,Ajith:2009bn,Santamaria:2010yb} and it includes also precession due to the misalignment of the individual spins
with the orbital angular momentum (though this is not relevant here). 
Other waveforms could also be used \cite{Buonanno:1998gg, Khan:2018fmp, Khan:2019kot} 
but we do not expect any significant differences for our purposes.  

We obtain three estimates of $(q,\mathcal{M})$ using, in turn, the
waveforms
$h_{+,\times}$ extracted in the wave-zone (on an extraction
sphere of radius $R_E= 500M$ for most cases, or at $R_E=300M$ for some) and extrapolated to infinity and transformed to CoM frame, and the shears $\sigma_{1,2}$ at the two
horizons. The waveform $h_{+,\times}$ is matched with the
model waveform itself, while $\sigma_{1,2}$ is matched with the news, i.e., the
time derivative of the model waveform. In each case, we use a
sufficiently fine grid of resolution $1/1001$ in $(q,\mathcal{M})$ and minimize a standard least-squares
figure-of-merit over the relative time-shift and initial phase (which will henceforth be referred to as alignment) as well as the mass parameters using the procedure described above.  \ref{tab:bestfit} shows the best fit values of
$\mathcal{M}$ and $q$ for the real part of the shears.  The chirp mass $\mathcal{M}$ is very well
measured, with typical errors of $\sim 0.5\%$ for the strain and 
$\sim  1-3 \%$ for the shears.  The mass ratio is more uncertain 
as expected, with errors of $\sim 9\%$ for the 
strain and $\sim 17-24\%$ for the shears.   We have chosen to 
present our results in terms of $(\mathcal{M},q)$ as independent 
parameters, though we could have used the total mass $M$ as well. 
The best-fit values of $M$ turn out to be 
very close to unity as they should. 

\begin{table}
\begin{center}
\begin{tabular}{|p{1.8cm}|p{1.8cm}|p{1.8cm}|p{1.8cm}|}
\hline
\multicolumn{4}{|c|}{Shear-news relations BH1} \\
\hline
\hline
$\hat{q}$ &plus &cross &mutual\\
\hline
1    &96.43 &98.09 &97.65 \\
0.85 &96.02 &96.02 &95.74 \\
0.75 &96.08 &96.63 &95.71 \\
0.67 &95.69 &96.42 &95.87 \\
0.5  &95.95 &96.80 &96.81 \\
0.25 &93.72 &95.08 &93.46 \\
\hline
\end{tabular}
\begin{tabular}{|p{1.8cm}|p{1.8cm}|p{1.8cm}|p{1.8cm}|}
\hline
\multicolumn{4}{|c|}{Shear-news relations BH2} \\
\hline
\hline
Simulation &plus &cross &mutual\\
\hline
1 &96.09 &97.93 &97.71 \\
0.85 &96.23 &96.23 &95.49 \\
0.75 &96.63 &97.18 &96.96 \\
0.6667 &95.51 &96.28 &95.54 \\
0.5 &95.70 &96.71 &96.83 \\
0.25 &95.51 &96.18 &95.45 \\
\hline
\end{tabular}
\end{center}
\caption[Overlap for Shear-News.]{The maximum match score between various waveform components for BH1 (top) and BH2 (bottom). The maximum match is computed by maximizing a least-squares figure of merit for a match of normalized waveforms over the timeshift and phase. The waveforms involved are the shears and news. The first column shows the mass ratio of the system. The second and third columns refer to the match between the respective polarizations of the shear and news. The third column shows the match between the two polarizations of the shear itself.}
\end{table}

Motivated by this excellent agreement, we postulate that given the news, it should be possible to predict the horizon shears.  Comparing the amplitudes and the phases of $\sigma_1$ and $\sigma_2$ (after aligning them with the news), we find that to a very good approximation the phase difference between them is very small, and 
\begin{equation}
    \frac{M_2\sigma_2}{M_1\sigma_1} \approx q^{-0.7}\,.
\end{equation}
We multiply the shears by the respective masses to make them dimensionless.  
Similarly, comparing the news and one of the shears, say $\sigma_1$, we find %
that the phase difference is again small and their amplitudes are related as follows:
\begin{equation}
    R_E\left|\mathcal{N}\right| \approx 0.5(1+q)M_1\left|\sigma_1\right|\,. \label{sn_rel}
\end{equation}
Here $R_E$ is the extraction radius.
With these relations, given the observed gravitational wave strain, one can estimate the amplitude and frequency content of the horizon shears for binaries consisting of non-spinning black holes. 

\begin{table}
\begin{center}
\begin{tabular}{|p{1cm}|c|p{1cm}|p{1cm}|p{1cm}|p{1cm}|p{1cm}|}
\hline
$\widehat{q}$ & $q$ & $\mathcal{M}$ & $q_1$ & $\mathcal{M}_1$ & $q_2$ & $\mathcal{M}_2$ \\
\hline
$1.0$ & $1.0$ & $0.432$ & $0.800$& $0.439$ & $1.200$& $0.439$ \\ 
$0.85$ & $1.0$ & $0.433$& $0.770$& $0.434$ & $0.770$& $0.434$ \\ 
$0.75$ & $0.78$ & $0.428$& $0.630$& $0.426$ & $0.870$& $0.426$ \\ 
$0.67$ & $0.779$ &$0.427$& $0.867$& $0.433$ & $0.747$& $0.429$ \\ 
$0.50$ & $0.498$ & $0.403$& $0.580$& $0.410$ & $0.580$& $0.410$ \\ 
$0.25$ & $0.222$ &$0.328$& $0.330$& $0.345$ & $0.250$& $0.333$ \\ 
\hline
\end{tabular}
\caption[Parameter estimation results using the DH shears.]{Best fit values of the mass ratio and chirp mass for
  i) the waveform extracted in the wave zone (denoted $q$ and
  $\mathcal{M}$), ii) the shear of the first black hole ($q_1$ and
  $\mathcal{M}_1$), and iii) the shear of the second black hole ($q_2$
  and $\mathcal{M}_2$).  The mass ratio is nominally $\widehat{q}$ for
  the puncture initial data.  }
\label{tab:bestfit}
\end{center}
\end{table}

\section{Post merger study}

In this portion, we explore whether the post-merger data of the numerical simulations used in the previous study (\ref{sec:shear_news_inspiral}) can be used to discover correlations between the shear of the common dynamical horizon and the gravitational news of the outgoing radiation in the post-merger phase. We find that, like in the inspiral phase, the shear of the dynamical horizon continues to be strongly correlated by outgoing gravitational radiation in the wave zone of the post-merger phase. 

\subsection{Introduction}

The study in \ref{sec:shear_news_inspiral} uncovered an interesting correlation between the shear of the dynamical horizon in the strong gravity regime of each black hole $\sigma$, and the gravitational wave news $\mathcal{N}$ recorded in a faraway detector in the inspiral phase. It was shown that the components of the shear of the individual dynamical horizons could be well described by the familiar chirp signal observed in gravitational waveforms.

In this section, we attempt to extend the investigation to the post-merger phase. In particular, we study the correlation between the shear of the common horizon of the system formed close to the merger and the news of the gravitational waves detected by a detector at some distance from the black hole center.

The system setup is exactly the same as before, consisting of a non-spinning binary black hole system in a quasi-circular configuration. In this study, instead of analyzing the two independent components of the shear tensor of each dynamical horizon in the inspiral phase, we analyze the two components of the shear of the outer common dynamical horizon formed as the two parent objects close in together. We use the same numerical simulations as in \ref{sec:shear_news_inspiral}.

In the study of dynamical horizons in a numerical evolution of binary black hole merger, it has been observed that a common envelope surrounding the two black holes forms close to the merger. %
The outer envelope (henceforth the common horizon) is a marginally outer trapped surface (MOTS) and when formed, is in a highly deformed state, far from spherical symmetry.  The outer common horizon of the system is a three-dimensional geometric object and is a dynamical horizon. Each of its slices is topologically $\mathcal{S}^2$ and can be regarded as the instantaneous apparent horizon of the remnant black hole. The common horizon settles down to the outer horizon of the remnant black hole which is expected to resemble the Kerr isolated horizon asymptotically (the Kerr hypothesis), by absorbing an appropriate amount of radiation and gaining axisymmetry in the process. However, we do not ascertain whether this indeed happens here.%

The shear of a dynamical horizon is a quasi-local tensorial quantity and can be described as a spin-2 field on each slice of the dynamical horizon. Alternatively, it can be described as a complex scalar in the spin-coefficient formalism. It has two independent components in the spin-coefficient formalism, like in the previous case \eqref{shear_spin_coeff}. %

Our previous study was focused on the inspiral phase of such systems. We found that the shear of the respective dynamical horizon of the black holes behaves similarly to the two independent components of the news of gravitational radiation received far away from the system. We expect such a relation to hold past the inspiral phase, for the common dynamical horizon. In this study, we find that it is indeed true and we quantitatively confirm this. In the next two sections, we present results that strongly suggest that the shear-news relationship that conveys the strong correlation between strong field and weak field dynamics is valid in all phases of the binary black hole mergers.

According to the perturbation theory, a remnant black hole's ringdown to an expected stationary, axisymmetric state can be described by perturbations. In the linear regime, these perturbations can be expressed as a superposition of damped-sinusoids, and are called the quasi-normal modes. The angular behaviour can be decomposed in terms of spin-weighted spheroidal harmonics and each quasi-normal mode can be labeled by three indices ($n, \ell, m$), for each spin weight $s$. Two different modes will in general have different amplitudes,  frequencies $\omega$, and damping rates $\gamma$. The temporal behaviour of these functions (at a particular spatial coordinate location) can be written down as:

\begin{equation}
    f(t) = \sum_{(n, \ell, m)} A_{n \ell m} e^{(-\gamma_{nlm} t)} e^{(i \omega_{n\ell m} t)}
\end{equation}
Or, alternatively:
\begin{equation}
    f(t) = \Sigma_{n, \ell, m} A_{n \ell m} e^{(-\gamma_{n \ell m} t)} \sin{(\omega t + \phi)} \label{qnm_f}
\end{equation}
Each mode ($n, \ell, m$) will have five free parameters.

In this work, we will analyze the correlation between the shears of the outer common dynamical horizon and the outgoing news function by:
\begin{enumerate}
	\item Attempting to fit the shears with damped sinusoids, and checking if they possess a quasi-normal behaviour. 
	\item Quantifying the deviation of the best-fit frequencies from expected quasi-normal modes of linearly perturbed Kerr black holes.
	\item Estimating the parameters of the remnant, and the parent black holes using matched filtering and a template bank.
\end{enumerate}

\subsection{Fitting procedures}

The fitting procedure followed was the following:

\begin{enumerate}
    \item $N$ number of damped sinusoids were fit to the data without any restriction on the parameters $(1 < N < 15)$. The obtained best-fit QNM frequencies are then compared against the theoretical results of QNM frequencies of a BH with mass and spin matching the remnant's masses and spins. 
    \item A least-squares minimization procedure to obtain the fit, with parameter guess values at the theoretical (expected QNM) values and small deviations of bounds about them was used.
\end{enumerate}

The latter procedure is computationally expensive and dis-favourable as we do not know beforehand which modes to expect.

\subsection{Challenges in QNM analysis of spinning remnants:}

\begin{enumerate}
\item For spinning black holes, there are $(2\ell+1)$ modes for each $\ell$ and $n$, corresponding to different azimuthal quantum mode number $m$. As opposed to the inspiral phase, we cannot always assume that ($2,2$) is the dominant mode here. %
\item Usually, at least the $n=0$ mode and one overtone are necessary to obtain a decent fit. This means that even if we restrict ourselves to $\ell=2$, we need to search for $5 \times 2 = 10$ modes. On the one hand, not all $10$ modes will be necessary for obtaining a good fit, as some will be subdominant. On the other hand, we do not know which/ how many of these $10$ modes will be necessary. This adds to the limitation of analysis and procedure, and to the computational cost.
\item Limitation of NR simulations performed: The overtone fitting depends upon the first few data points from the start of the simulation as the higher damping of the overtones will become subdominant quickly. Therefore, finding overtones requires that there be a sufficient number of data points before the oscillations visibly start in the data.
\item The damping is spaced closely for different $m$ modes given $n, \ell$. This causes difficulty in the identification and estimation of the modes.
\end{enumerate}

To investigate this correlation between the shear of the common dynamical horizon of the binary system (which is also the MOTS / dynamical horizon of the remnant black hole) and the news of the gravitational radiation as recorded in a detector, we carry out the following studies using the shear of the common horizon at its pole and the gravitational news of the system:

\begin{enumerate}
    \item Parameter estimation to infer the spin of the remnant black hole.
    \begin{enumerate}
        \item Using a template match-based approach and the lowest order QNM.
        \item Fit to multiple damped sinusoids.
    \end{enumerate}
    \item Parameter estimation to infer the mass-ratio and the chirp mass of the parent black holes of the binary system using the waveform approximant SEOBNRv4\_PHM.
\end{enumerate}

using the Newman-Penrose complex scalar, the shear of the common horizon

\begin{equation}
 \sigma = Re(\sigma) + i Im(\sigma) = \sigma_{+} + i \sigma_{\times} \label{eqn:sc-shear}
\end{equation}

We carry out this study on six simulations of non-spinning, quasi-circular configurations whose initial data parameters of the punctures are in \ref{tab:ic_qc0_snr}.

\subsection{Matching procedure}
\begin{figure}
\includegraphics[width=\columnwidth]{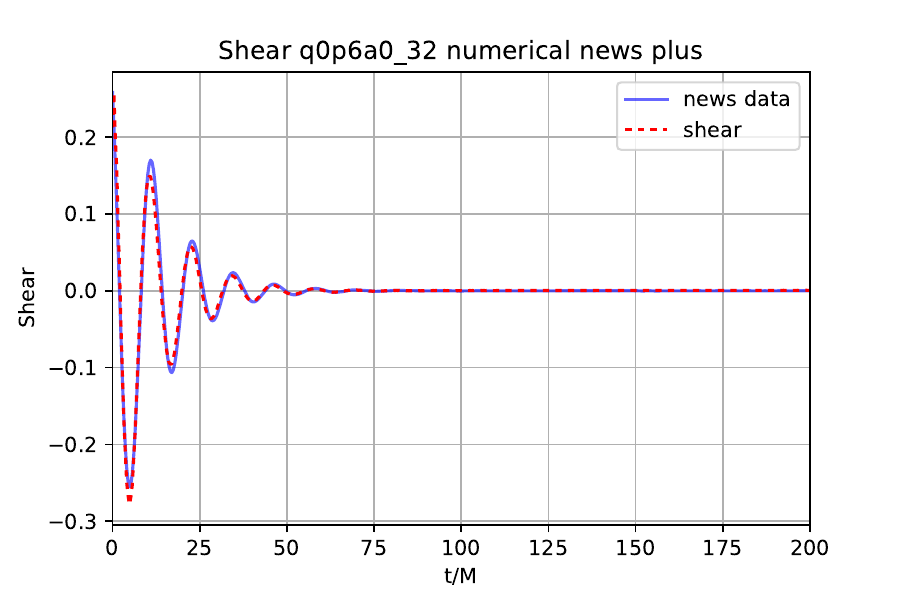}
\includegraphics[width=\columnwidth]{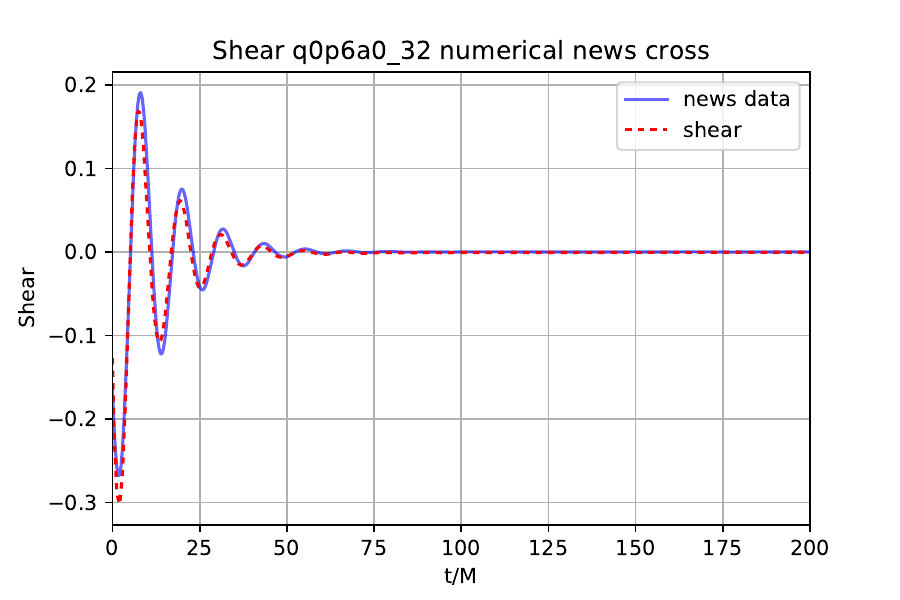}
\caption[Strong correlations between the common DH shear - News.]{The shears of the common horizon aligned in time and phase with the news, following the procedure described previously for $q=0.6$ simulation. These waveforms have been resampled for the purpose of presentation. Top: The real and imaginary parts of the shear co-aligned with the respective outgoing news functions. It can be seen that the shear and news data reach numerical noise floors after about $t=90M$ units from the merger $t=0M$.}
\label{fig:pmshear_news}
\end{figure}
\begin{figure}
	\includegraphics[width=\columnwidth]{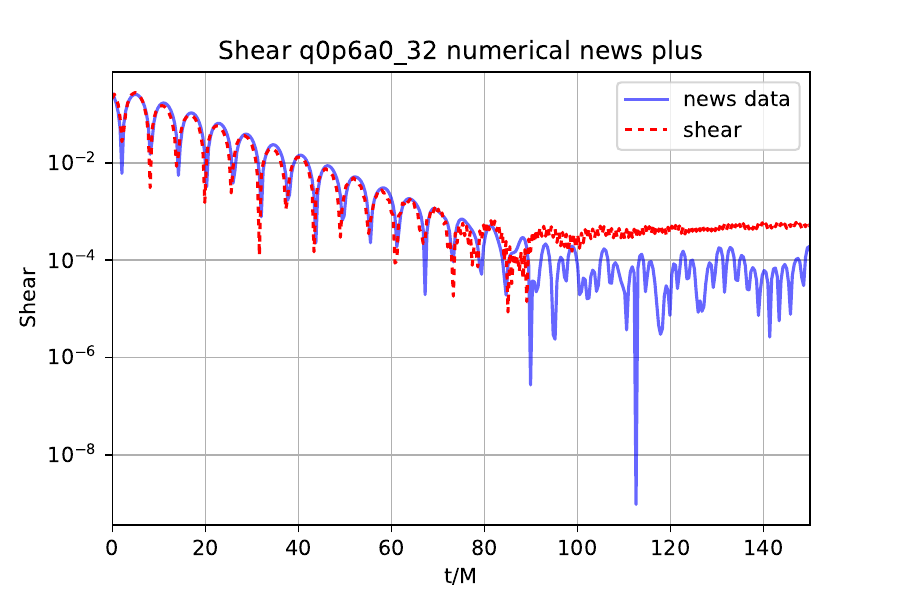}
	\includegraphics[width=\columnwidth]{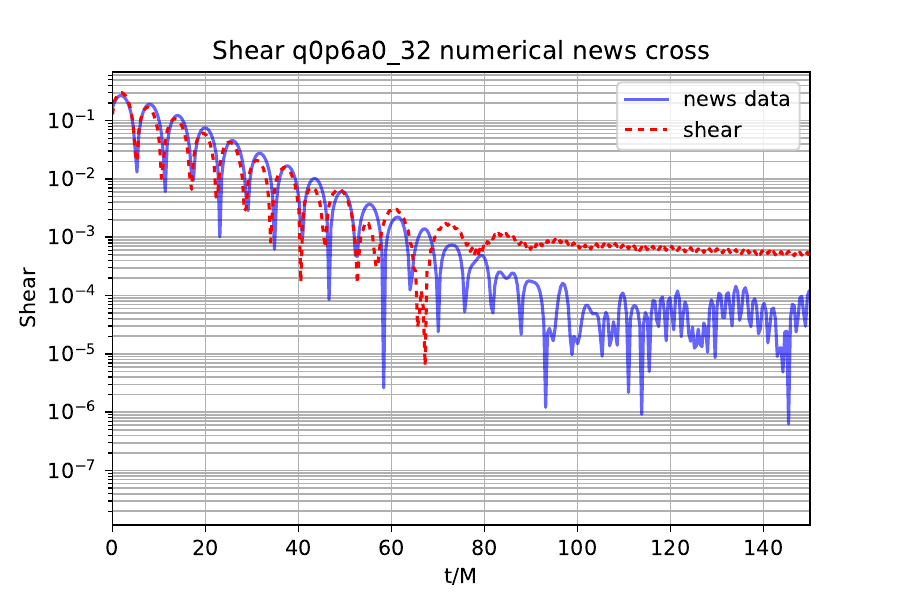}
	\caption[Strong correlations between the common DH shear - News, in logarithmic scale.]{The waveforms in the previous \ref{fig:pmshear_news} plotted on a logarithmic scale on the y-axis.}
	\label{fig:pmshear_news_log}
\end{figure}

We estimate the parameters of the parameters of the system to pin this relationship between the shear of the common dynamical horizon and the gravitational news. From black hole perturbation theory, we know that the waveform past the maximum intensity location can be well approximated by a superposition of quasi-normal modes of the final black hole. As per the no-hair conjecture, each QNM frequency corresponds to a unique set of physical parameters of the final black hole (mass, spin). The gravitational news can be obtained through a single derivative of the strain and is therefore also quasi-normal in behaviour. Therefore, if the shear-news correlation holds in the post-merger phase, one may expect the same quasi-normal behaviour for the shear of the common outer horizon as with the gravitation news, and therefore the strain of the system. It is difficult to decide what is meant by the post-merger/ringdown phase in binary black hole mergers. For this study, we use the maximum intensity location as the demarcating epoch. This is supported by the evidence that in numerical simulations, the temporal location of the peak intensity of the gravitational radiation can be mapped to the formation time of the common horizon of the system. %
Therefore, we analyze the shear and gravitational wave data post the maximum intensity location of their waveforms. To test this hypothesis, we explore the question: \emph{can we infer the parameters of the final black hole and/or the parent black holes from the post-merger shear of the common dynamical horizon? \label{q1}} %
We use two quantities for this analysis: the shear of the common outer horizon of the system $\sigma$ and the news of the system in the post-merger phase, recorded in a detector stationed at a faraway distance of $r = 100M$ units from the center of the system. We discuss these in the next two subsections.

\subsection{Remnant parameter estimation}

Firstly, we compare the shear of the common horizon and the gravitational news from the system, i.e. we align the shears with the news suitably in phase and time by maximizing the least squares figure of merit. When aligned in time and phase, we find that these two quantities are related through:
\begin{equation}
    \left\lvert \dfrac{\sigma}{\mathcal{N}} \right \vert = 24.81 q^2 -30.64 q +  10.62
\end{equation}
These coefficients are roughly the same for both the components of the shears and we postulate them to be universal. This is an interesting parallel with the results for the inspiral phase as presented in \eqref{sn_rel}%
The figure  \label{fig:shear_news} shows the same graphically. 

Secondly, we attempt to estimate the final spin of the remnant from the shear of the common horizon and the final mass.

\begin{table}
	\begin{center}
		\begin{tabular}{|p{1.8cm}|p{1.8cm}|p{1.8cm}|p{1.8cm}|p{1.8cm}|p{1.8cm}|}
			\hline
			mass-ratio q  &   Spin        &   $\omega_{(0)}$   &   $\gamma_{(0)}$  &   $\omega_{(1)}$   &   $\gamma_{(1)}$  \\
			\hline
			0.85     &   5.622e-01   &   4.819e-01       &   -8.457e-02      &   4.667e-01       &   -2.57e-01       \\
			0.75     &   5.588e-01   &   4.809e-01       &   -8.463e-02      &   4.656e-01       &   -2.57e-01       \\
			0.667   &   5.539e-01   &   4.794e-01       &   -8.473e-02      &   4.640e-01       &   -2.57e-01       \\
			0.5      &   5.329e-01   &   4.732e-01       &   -8.511e-02      &   4.573e-01       &   -2.58e-01       \\
			0.25     &   4.319e-01   &   4.471e-01       &   -8.654e-02      &   4.288e-01       &   -2.63e-01       \\
			0.167   &   3.515e-01   &   4.296e-01       &   -8.733e-02      &   4.095e-01       &   -2.66e-01       \\
			\hline
		\end{tabular}
		\caption[Expected values of QNM frequencies.]{Theoretical QNM frequencies of a black hole of masses and spins corresponding to those of the final black hole of the numerical simulations used in the analysis.}
		\label{tab:actual_qnms}
	\end{center}
\end{table}

\begin{table}
	\begin{center}
		\begin{tabular}{|p{1.8cm}|p{1.8cm}|p{1.8cm}|p{1.8cm}|p{1.8cm}|}
			\hline
			Simulation  &   $\omega_{(0)}$   &   $\sigma$    &   $\gamma_{(0)}$  &   $\sigma$   \\
			\hline
			0.85        &   5.236e-01       &   8.354e-04   &   7.805e-02       &   7.662e-04    \\
			0.75        &   5.226e-01       &   4.245e-04   &   8.076e-02       &   4.879e-04    \\
			0.667      &   5.258e-01       &   1.088e-03   &   8.263e-02       &   9.433e-04   \\
			0.5         &   5.102e-01       &   3.510e-03   &   8.055e-02       &   3.834e-03    \\
			0.25        &   1.198e+01      &   3.187e-03   &   8.059e-02       &   3.264e-03     \\
			0.167    &   4.371e-01       &   6.928e-04   &   8.383e-02       &   7.394e-04   \\
			\hline
		\end{tabular}

		\begin{tabular}{|p{1.8cm}|p{1.8cm}|p{1.8cm}|p{1.8cm}|p{1.8cm}|}
			\hline
			Simulation  &   $\omega_{(1)}$   &   $\sigma$    &   $\gamma_{(1)}$  &   $\sigma$    \\
			\hline
			0.85        &   3.979e-01       &   1.567e-02   &   3.839e-01       &   7.76e-03    \\
			0.75        &   2.081e-01       &   2.636e-02   &   4.963e-01       &   7.50e-03    \\
			0.6667      &   1.216e+01      &   2.939e+01   &   7.661e-01       &   3.28e-02    \\
			0.5         &   4.626e-03       &   3.387e+01   &   5.999e-01       &   2.56e-01    \\
			0.25        &   1.123e+01       &   7.798e-02   &   7.835e-01       &   1.32e-01    \\
			0.1667a0    &   6.624e-01       &   5.103e-03   &   9.307e-02       &   4.41e-03    \\
			\hline
		\end{tabular}
		\caption[Estimated QNM behavior of the real part of DH shears.]{The estimates of the first two quasi-normal frequencies that best fit the imaginary part of the shear $Re (\sigma)$ of the common horizon of the system to two damped sinusoid in \eqref{qnm_f}.}
		\label{tab:shearp_rem_fits}
	\end{center}
\end{table}

\begin{table}
	\begin{center}
		\begin{tabular}{|p{1.8cm}|p{1.8cm}|p{1.8cm}|p{1.8cm}|p{1.8cm}|p{1.8cm}|p{1.8cm}|p{1.8cm}|p{1.8cm}|}
			\hline
			Simulation  &   $\omega_{(0)}$   &   $\sigma$    &   $\gamma_{(0)}$  &   $\sigma$    \\
			\hline
			0.85        &   1.002e+01       &   7.647e-04   &   7.783e-02       &   6.812e-04  \\
			0.75        &   1.081e+01      &   5.972e-04   &   8.035e-02       &   6.459e-04   \\
			0.6667      &   5.230e-01       &   7.152e-04   &   8.206e-02       &   7.623e-04   \\
			0.5         &   5.090e-01       &   9.246e-04   &   8.198e-02       &   1.032e-03    \\
			0.25        &   4.621e-01       &   2.840e-03   &   8.490e-02       &   2.512e-03     \\
			0.167      &   3.468e+01       &   7.104e-04   &   8.238e-02       &   6.464e-04    \\
			\hline
		\end{tabular}
		\begin{tabular}{|p{1.8cm}|p{1.8cm}|p{1.8cm}|p{1.8cm}|p{1.8cm}|p{1.8cm}|p{1.8cm}|p{1.8cm}|p{1.8cm}|}
			\hline
			Simulation  &   $\omega_{(1)}$   &   $\sigma$    &   $\gamma_{(1)}$  &   $\sigma$    \\
			\hline
			0.85        &   9.973e+00       &   9.033e-03   &   3.690e-01       &   1.13e-02    \\
			0.75        &   2.863e-01      &   1.409e-02   &   3.140e-01       &   9.48e-03    \\
			0.667      &   1.392e-03       &   5.042e+00   &   3.718e-01       &   2.01e-02    \\
			0.5         &   2.067e+00       &   2.797e-01   &   2.773e+00       &   6.48e-01    \\
			0.25        &   1.187e+01       &   3.532e-01   &   1.131e+00       &   2.36e-01    \\
			0.167      &   3.446e+01      &   4.351e-03   &   8.456e-02       &   4.97e-03    \\
			\hline
		\end{tabular}
		\caption[Estimated QNM behaviour of the imaginary part of DH shears.]{The estimates of the first two quasi-normal frequencies that best fit the imaginary part of the shear $Im(\sigma)$ of the common horizon of the system.}
		\label{tab:shearx_rem_fits}
	\end{center}
\end{table}

\begin{table}
	\begin{center}
		\begin{tabular}{|p{1.8cm}|p{1.8cm}|p{1.8cm}|p{5cm}|}
			\hline
			Simulation  &      Overtone   &     \% error            &   Notes \\
			\hline
			0.85        &      (2, 2, 2)  &     -9.5, -11.9         &   other possibility is for m=1. Deviation is higher and $\omega_I > \omega_R$ \\                                      
			0.75        &      (2, 2, -2) &     -7.6,  -0.08        &   Quite close     \\ 
			0.6667      &       NA        &                         &   Spurious mode.  \\
			0.5         &       NA        &                         &   Spurious mode.  \\
			0.25        &       NA        &                         &   Spurious mode  of very high l \\
			0.167    &       (0, 2, 2) &    54.2, 6.57           &   Matches better with (0, 3, -2) \\
			\hline
		\end{tabular}
		\caption[Estimated QNM overtone behaviour of the real part of DH shears.]{The estimates of the ($n=1$) overtone quasi-normal frequencies that best fits the real part of the shear $Re(\sigma)$ of the common horizon of the remnant black hole of the respective simulation.}
		\label{tab:shearp_rem_pe_n1}
	\end{center}
\end{table}

\begin{table}
	\begin{center}
		\begin{tabular}{|p{1.8cm}|p{1.8cm}|p{1.8cm}|p{5cm}|}
			\hline
			Simulation  &      $n=0$ mode    (+, $\times$)    &  $n=1$ overtone       \\
			\hline
			0.85        &       0.224, 0.221    &  0.286, 0.208                \\                                      
			0.75        &       0.232, 0.231    &  0.534, 0.193                \\ 
			0.7         &       0.251, 0.262    &  0.297, 0.104                \\ 
			0.667      &       0.232, 0.231    &  34.14, 21.77                \\
			0.5         &       0.215, 0.197    &  0.032, 0.25                 \\
			0.25        &       0.065, 0.07     &  0.074, 0.133                \\
			0.167    &       0.043, 0.042    &  0.007, 0.006                \\
			\hline
		\end{tabular}
		\caption{The estimates of the amplitudes of the $n=0, 1$ modes obtained by fits.}
		\label{tab:shearpx_rem_pe_amps}
	\end{center}
\end{table}

\begin{table}
	\begin{center}
		\begin{tabular}{|p{1.4cm}|p{1.4cm}|p{1.4cm}|p{1.4cm}|p{1.4cm}|p{1.4cm}|p{1.4cm}|p{1.4cm}|}
			\hline
			$\widehat{q} $ & Best match mass & Actual mass& \% Error &Best match spin& Actual spin & \% Error& Match\\
			\hline
			0.85    &   0.973   &   0.952   &   2.132   & 0.716 &   0.620   &   15.600  &   0.999   \\ 
			0.75    &   0.962   &   0.954   &   0.825   & 0.693 &   0.614   &   12.798  &   0.999   \\ 
			0.6667  &   0.914   &   0.956   &   -4.386  & 0.657 &   0.607   &   8.292   &   0.988   \\ 
			0.5     &   1.030   &   0.962   &   7.164   & 0.758 &   0.576   &   31.470  &   0.953   \\ 
			0.25    &   1.013   &   0.978   &   3.543   & 0.475 &   0.451   &   5.274   &   0.989   \\ 
			\hline
		\end{tabular}
		\caption[Parameter estimation using the real part of the shear of the common DH.]{The results of a template-based parameter estimation carried out on the real part of the shear $\sigma_{+}$ of the remnant black hole using just the lowest/dominant quasi-normal mode.}
		\label{tab:shearp_rem_pe}
	\end{center}
\end{table}
\begin{table}
	\begin{center}
		\begin{tabular}{|p{1.4cm}|p{1.4cm}|p{1.4cm}|p{1.4cm}|p{1.4cm}|p{1.4cm}|p{1.4cm}|p{1.4cm}|}
			\hline
			$\widehat{q} $ & Best match mass & Actual mass& \% Error &Best match spin& Actual spin & \% Error& Match\\
			\hline
			0.85    &   0.837   &   0.952   &   -12.146     & 0.320 &   0.620   & -48.299   &   0.973   \\ 
			0.75    &   0.950   &   0.954   &    -0.349     & 0.639 &   0.614   &   4.009   &   0.996   \\ 
			0.6667  &   0.955   &   0.956   &    -0.032     & 0.634 &   0.607   &   4.435   &   0.998   \\ 
			0.5     &   0.962   &   0.962   &     0.009     & 0.621 &   0.576   &   7.736   &   0.999   \\ 
			0.25    &   0.997   &   0.978   &     1.907     & 0.599 &   0.451   &   32.788  &   0.990   \\ 
			\hline
		\end{tabular}
		\caption[Parameter estimation using the imaginary part of the shear of the common DH.]{The results of a template-based parameter estimation carried out on the imaginary part of the shear component $\sigma_{\times}$ of the remnant black hole using just the lowest/ dominant quasi-normal mode.}
		\label{tab:shearx_rem_pe}
	\end{center}
\end{table}

\begin{figure}
	\includegraphics[width=\columnwidth]{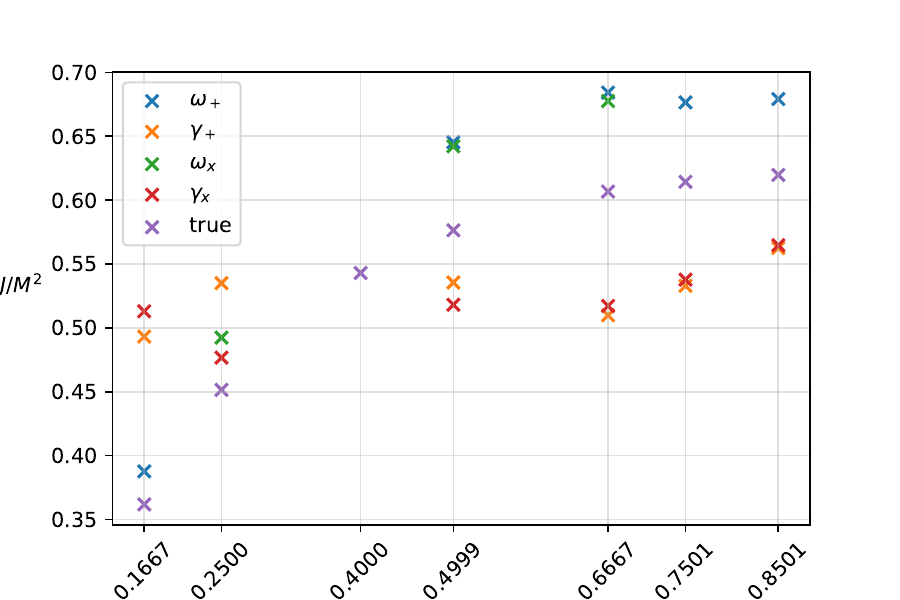}
	\includegraphics[width=\columnwidth]{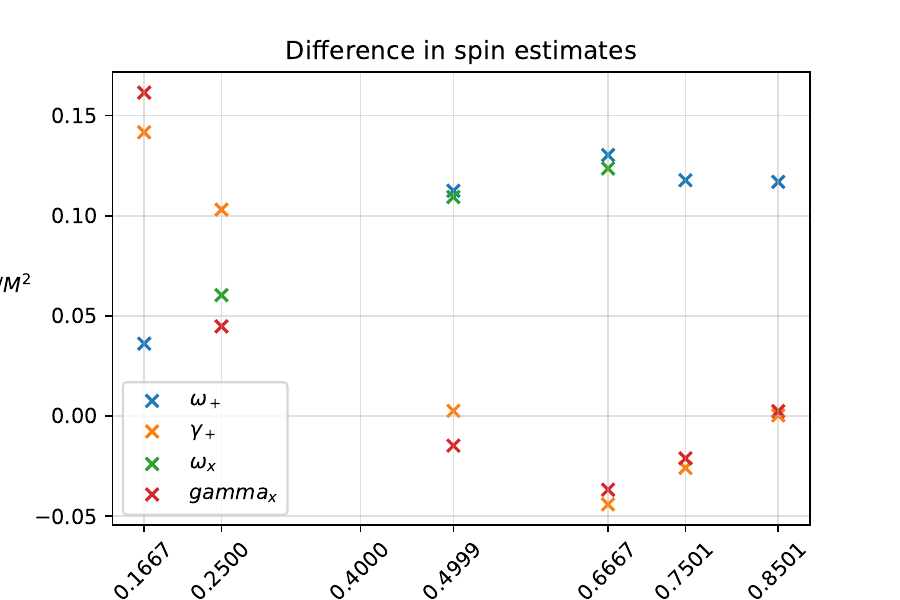}
	\caption[Parameter estimation of the remnant BH from the shear of the common DH.]{Top: Estimate of the spin of the remnant from the $n=0$ QNM frequencies. The data points in purple are the final quasi-local spin of the common horizon. Those in blue and orange are the spins obtained using the angular frequency and damping constant respectively of only the real part of the shear. The ones in green and red are the estimates obtained by using just the imaginary part of the shear. Bottom: Same as top, but the deviation from the quasi-local spins is shown instead. Please note that the last of the labels in the legend of the bottom figure in red that describes the damping of the imaginary part of the shear should read as $\gamma_\times$.}
	\label{fig:shear_spin}
\end{figure}

Here, we assume that the final spin of the black hole formed is aligned with the $z$ axis. This is expected and also confirmed by the numerical data. The two components of the shear of the common horizon in  \eqref{eqn:sc-shear} were fit to two damped sinusoids. The components of the shear of the outer common horizon of the system were found to fit well to two damped sinusoids. We could recover the $n=1, \ell= 2, m= 2$ and $n=2, \ell= 2, m=2$ quasi-normal modes.  The details of the first two QNMs are shown in \ref{tab:shearp_rem_fits} and \ref{tab:shearx_rem_fits}.  A table containing the QNM details from analytic computations is shown in \ref{tab:actual_qnms} for reference. Our estimates can be seen to agree well with these. Further details of the modes and amplitudes can be found in \ref{tab:shearp_rem_pe_n1} and \ref{tab:shearp_rem_pe_n1}.

Using the lowest QNM and the final mass of the remnant as reported by the numerical simulations, the spin of the remnant was computed. This was found to be close to the final spin as reported by the numerical simulation\footnote{The spin of the remnant can be computed from the numerical simulation using the $l=1$ spin multipole moment of the common horizon in the quasi-local formalism}. The details are represented in \ref{fig:shear_spin}. 

One may also expect a Price-like power law tail in the late time evolution \cite{Price:1972}. The late time portion in these figures may depict one or may correspond to numerical noise floors. The Price power law could not be confirmed owing to the resolution of data.

We also followed another approach to estimate both the masses and the spins of the remnant from the shear of the common outer horizon, using a procedure that minimizes a least squares figure of merit over the dominant and the lowest QNM. The recovered parameter values are shown in \ref{tab:shearp_rem_pe} and \ref{tab:shearx_rem_pe}.

Some observations on the fitting procedure:
\begin{enumerate}
\item Fits are good, provided we include at least one overtone. However, the estimated modes $(n, \ell, m)$ are not the same across all simulations. They correspond to different values of $m$. The \ref{tab:shearp_rem_pe} and \ref{tab:shearx_rem_pe} give details of the first two dominant modes estimated. These correspond to $n=1$ and $n=2$ but different $m$.

\item Fitting to more than two damped sinusoids results in the estimation of modes that do not correspond to any $(n, \ell, m)$ QNM frequencies.
\item In some cases, the real part of the shear gives significantly different results compared with the imaginary part. The imaginary part's data quality seems to be lower.
\item The frequency of shear data output is very small in the NR data. One cannot expect to find evidence of more than one overtone- preliminary conclusion.
\end{enumerate}

\section{Conclusions} 

\subsection{Inspiral phase} We have shown quantitatively that in a
black hole merger, the horizons
behaves just like gravitational wave signals seen
in the wave zone. The results here show that
there are strong correlations between gravitational wave signals seen
by gravitational wave detectors and suitable fields in the strong
field dynamical region near the black holes.  

Future work will extend this study in many directions, e.g. allowing spinning black holes, precession 
effects, possible super-radiance in the non-linear merger regime, and investigating the effect of different gauge conditions.
The horizon shears are related to the variation of the horizon multipole 
moments.  One might therefore be able to relate the radiative multipole
moments to the horizon moments.  On dynamical horizons, various balance
laws are known relating the change in the horizon multipole moments to 
fluxes across the horizon \cite{Ashtekar:2013qta}.  Combining this with 
the idea of slowly evolving horizons \cite{Booth:2003ji,Booth:2006bn} might provide 
an interesting route to further relate properties of waveforms with horizons ~\cite{Datta:2019euh,Datta:2019epe} 
and to perhaps build better waveform models.  

The correlation found here between the strong field and weak field regimes may give us a window into the strong field dynamics of black holes, a possible non-standard structure associated with the horizons, and possibly the physics of the interior region of black holes.

A deeper understanding 
of these observed correlations is still lacking. In particular, it is 
important to identify the precise spacetime region and the non-linearities 
that generate the gravitational waves seen at the horizons and in the 
wave zone; see e.g. \cite{Owen:2010fa,Nichols:2011pu} for work in this direction.  
Elucidating the precise relationship of our results with the 
perturbative calculations of tidal coupling \cite{Hawking:1972hy,Hartle:1974gy,Hartle:1973zz,OSullivan:2015lni,OSullivan:2014ywd} 
is of great interest as well.  In particular we mention the work by O'Sullivan \& Hughes 
\cite{OSullivan:2015lni,OSullivan:2014ywd} that studies high mass-ratio 
systems perturbatively, and especially the effect on the geometry of 
the event horizon. They find a strong correlation between the shear of 
the horizon with the particle orbit (and thus with the observed waveform) 
which is broadly consistent with our results.  We refer also to recent work \cite{Pook-Kolb:2020jlr,Pook-Kolb:2020zhm} which, among other results, confirms with high accuracy that the post-merger shear at the horizon corresponds to the quasi-normal ringing of the final black hole.  This is precisely what one sees in the waveform at infinity, and it thus lends additional support for the existence of correlations between the strong field region and the wave zone.

Note that in \ref{fig:shear-news} the difference between the shear and the news becomes larger near the merger. One needs to understand this better. This may be due to the additional terms in the horizon flux law that may start to matter.  It will be important to compare the full angular pattern of the shears at the horizon against the news.

These results suggest that the correlation between the strong field and the weak field dynamics of a BBH system carries past the inspiral phase for the systems studied. Also, there is evidence that relations hold for a wide range of mass ratios. Using these results, one can now understand the physics in the strong field, close to the horizons of the system using the gravitational wave information recorded in our detectors. These include, for instance, the amount of energy and angular momentum absorbed by the black holes. 

Some of the sources of errors in the parameter estimation results may be due to mode mixing in the waveform data, and the requirement of higher resolution numerical simulations. Furthermore, the errors were seen to be larger in the $n=2$ overtone results. Future directions include recovering and analysis of amplitudes of the various overtones, and higher overtones and extending the analysis to BBH systems with spinning black holes.

\subsection{Ringdown phase}

As the two individual black holes merge, a common envelope develops that surrounds the individual dynamical horizons of the two black holes. In this study, we computed the shears of this common dynamical and studied its relationship with the news of the outgoing gravitational radiation in the post-merger phase. The outgoing gravitational radiation has quasi-normal behaviour as expected. We found that the shear of the common dynamical horizon shares this behaviour as well. In particular, we found that the strong correlation between the shear and the outgoing news function continues through the merger into the post-merger phase. This is evidenced by the result that the shear matches very well with the news of the outgoing gravitational radiation. Also, the resolutions of the runs were a limitation in obtaining proper fits. %

\cleardoublepage %
\flushbottom
\newpage
\cleardoublepage

\chapter{Tidal deformability of dynamical horizons}
\label{chap:tddh}

\section{Introduction}
\label{sec:intro}

An important physical phenomenon that manifests itself during the
inspiral of two orbiting compact objects is the tidal deformation of each under the
gravitational influence of its companion.  In the case of binary neutron
star mergers, this tidal deformation and the associated Love numbers
have been used to probe properties of dense matter and the nuclear
equation of state.  Non-spinning black holes on the other hand have a vanishing
(field) tidal Love number in General Relativity.  
This pertains to the deformation of the asymptotic gravitational
field.  In certain cases, especially in the late stages of the
inspiral phase when the black holes get close to each other, the
\emph{source} multipole moments might be more relevant in probing
their properties and the No-Hair
conjecture; contrastingly, these Love
numbers do \emph{not} vanish. In this chapter, we track the source
multipole moments in simulations of several binary black hole mergers and
calculate these Love numbers using the $\ell=2, m=0$ multipole moments.  We present evidence that, at least for
modest mass ratios, the behaviour of the source multipole moments is
universal.

In this work, we study the relation between the geometry of the dynamical horizons and their tidal environment in a binary black hole scenario. We address the problem of tidal deformability of black holes in the strong field regime. 

The general treatment of the tidal deformation of compact objects has been perturbative. In such a treatment, an external tidal field is taken to induce a deformation of the compact object, and consequently, the potential or the asymptotic configuration of the gravitational field sourced by the compact object also changes \cite{binningtonpoisson2009}. In the non-relativistic theory, the deformation of the compact object can be quantified using its mass multipole moments, which are expressed as integrals over the source distribution. The external tidal field deforms the compact object (the source) and changes its distribution, i.e., the multipole moments.  At the linear order in the external tidal field, every $\ell^{\rm th}$ multipolar order of the external tidal field induces an $\ell^{\rm th}$ multipolar moment in the configuration of the compact object, and the constant of proportionality relating these is called a Love number. These Love numbers encode information about the constituent materials of the source. 

In General Relativity, the situation is more complex. The deformation of the compact object is now conveniently quantified using two sets of multipole moments: mass and current-type multipole moments, which quantify the configuration of the asymptotic gravitational field produced by the source. Although Love numbers can again be described as linearly relating the change in the field multipole moments and the external tidal field, the relation is not restricted to a simple proportionality. When the source object is spinning, the relation is tensorial and, hence, can involve the mixing of tidal fields at different multipolar orders. Thus, when the compact object is spinning, the $\ell^{\rm th}$ order multipole moment induced in the asymptotic field produced by the compact object can have contributions from any $\ell^{'{\rm th}}$ order component of the external tidal field with $\ell\neq \ell '$.

A sizeable literature exists on applications of tidal deformation and Love numbers~\cite{1955MNRAS.115..101B,Flanagan:2007ix,DamourNagar2009}, especially, in how they encode information about neutron star equation of state~\cite{Lattimer:2015nhk} in gravitational-wave signals~\cite{Abbott:2018exr,De:2018uhw,Capano:2019eae}.
On the other hand, tidal deformations of black holes and the associated Love numbers are discussed in, e.g., Refs.~\cite{Landry2014, Damour:2009va, binningtonpoisson2009} for the non-spinning case and Refs.~\cite{Pani:2015hfa, LeTiec:2020bos, LeTiec:2021} for the spinning case.

Since a black hole is a solution to vacuum field equations, the usual definition of the multipolar configuration of the system involving an integral of 
its mass density is irrelevant here. Furthermore, during a dynamical scenario such as a binary black hole merger, which is of great relevance to the recent gravitational-wave detections, it is difficult to define and study the change in the asymptotic gravitational field of each black hole as it inspirals towards the other and merges. Also, the situation is highly dynamical
as the two objects orbit under the mutual influence of their strong field gravity. 
Thus, the field Love numbers will pose a limitation to understanding this problem in the strong field regime.

Typically, past treatments of tidal deformability either start with the assumption that there exists a source distribution that is deformed due to an external tidal field \cite{DamourNagar2009} or involve the application of black hole perturbation theory~\cite{binningtonpoisson2009, Damour:2009va, Landry2014, Pani:2015hfa, LeTiec:2020bos, LeTiec:2021}. In the former treatment, a limit of the compactness is then taken to deduce the corresponding Love numbers for black holes. The problem with the ``limiting compactness'' approach is that black holes do not have a source distribution in the same sense as that of, e.g., a neutron star. Furthermore, in a true sense, the horizons of black holes evolve, cease to be null hypersurfaces and are no longer isolated in a dynamical tidal environment.

In the perturbative approach, the gravitational field far away from the black hole is studied. Often, simplifying assumptions, such as staticity/stationarity, slow variation in time, etc., are imposed. The perturbative approaches cannot be used to understand (a) the deformation of black holes when they are close to each other, and (b) the fully dynamical aspects of tides on black holes. Furthermore, in a numerical simulation, one does not have access to the asymptotic regions of the individual black holes to study the problem dynamically using the gravitational field far away from the system. Due to these reasons, it is desirable to employ an alternative treatment with a differently defined set of quantities that characterize the tidal deformability of black holes in the strong field regime.

Hence, more appropriately, one should address how the external tidal field induces a change in the geometry of the horizon. One has to study the tidal deformability of the quasi-local \emph{dynamical horizons} rather than the teleological event horizons, which only coincide with the former in the special case of stationarity. The information on the geometry of the dynamical horizon is encoded in its source multipole moments and can be used to study the problem of tidal deformability. One may expect a similar relation between the source multipoles and the $\ell^{\rm th}$ order multipolar external tidal field. In this sense, the dynamical horizon formalism is an ideally suited framework to treat the problem of the tidal deformation of black holes.

In this formalism, for black holes in a vacuum, a set of numbers -- called the source multipole moments -- fully characterize the horizon geometry.
In a tidal environment, these numbers can change, e.g., owing to an alteration in the horizon geometry in response to the external tidal field. Using this approach, one can directly study the deformation of the horizon geometry due to an external tidal field.
In this work, we address how the black holes are tidally deformed in a binary black hole merger scenario. We describe and use a convenient definition of tidal deformability suited to a binary black hole system and compute them numerically using full numerical relativity simulation data of binary black hole mergers of non-spinning black holes.
We use the framework of quasi-local horizons for our analysis; see e.g. \cite{Ashtekar:2004cn,Booth:2005qc} for reviews. We use the source multipole moments of the involved dynamical horizons to address the problem of their tidal deformability.  In particular, we use the mass multipole moments to characterize the deformation in the horizon geometry of non-spinning black holes and define a set of dimensionless tidal coefficients. We then calculate the leading order tidal coefficients that characterize the deformation of a black hole's multipole moments in the tidal environment of its companion in a binary black hole merger scenario. 

The plan for the rest of the chapter is as follows. \ref{sec:prelim} briefly reviews basic concepts and equations for dynamical horizons and describes the definition of the source multipole moments of horizons. The details of Numerical simulations are given in \ref{sec:nr_tddh}. In \ref{sec:com_tidal_coeffs} we compute the tidal coefficients and describe the fitting procedure. The results are summarized in \ref{sec:results} followed by \ref{sec:conclusions} with discussion and conclusions. Appendix \ref{appendix:dm} contains a comparison of distance measures, among the centroid distance and various orders of post-Newtonian approximations.

All the equations and quantities are expressed in geometric units, where $G = c = 1$. The masses of the horizons of the primary and secondary black holes in the initial data are $M_1$ and $M_2$, and their mass ratio is defined to be $q \equiv M_2/ M_1$, which is always $\leq 1$. Additionally, the total mass of the black hole horizons in the initial data, denoted by $M = M_1 + M_2$ is set to one. In a few places where tracking the dependence on $M$ is important, we show it explicitly. 

\section{Preliminaries}
\label{sec:prelim} 

Our calculation of the tidal deformations of black holes is based on the formalism of quasi-local horizons~\cite{Hayward:2000ca,Booth:2005qc,Ashtekar:2004cn,Gourgoulhon:2005ng,Visser:2009xp,Jaramillo:2011zw,Faraoni:2015pmn}. 
A detailed description of this formalism is beyond the scope of this article and we shall restrict ourselves to a brief overview of the most relevant concept, namely, that of black hole source multipole moments. 

The starting point is the notion of a marginally trapped surface, first introduced by Penrose in the context of the black hole singularity theorems \cite{Penrose:1964wq}. More specifically we need here the notion of marginally outer trapped surfaces (MOTS), which are closed space-like 2-dimensional surfaces of spherical topology, such that their outgoing null-normals $\ell^a$ have vanishing expansion $\Theta_{(n^{+})}$. Thus, if $\mathcal{S}$ is a MOTS, $\Tilde{q}_{ab}$ the Riemannian metric on $\mathcal{S}$, and $n^{+a}$ an outward-pointing null-normal to $\mathcal{S}$, then a MOTS has 
\begin{equation}
  \Theta_{(n^+)} := \Tilde{q}^{ab}\nabla_a n^+_b = 0\,.
\end{equation}
Under time evolution, a MOTS $\mathcal{S}$ traces out a 3-dimensional world tube sometimes referred to as a dynamical horizon \cite{Pook-Kolb:2020zhm,Pook-Kolb:2020jlr}, or a marginally trapped tube \cite{Andersson:2008up,Andersson:2007fh}. We shall not delve here into properties of this time evolution. We shall instead just consider various properties of $\mathcal{S}$ as functions of time. It turns out that the time evolution of MOTS is generally found to be smooth, which means that we end up with smooth functions of time. When two black holes collide, the process by how two distinct dynamical horizons merge to yield a single final dynamical horizon turns out to have interesting topological and dynamical properties \cite{PhysRevD.100.084044,PhysRevLett.123.171102,Pook-Kolb:2021gsh,Booth:2021sow}. Again, this is beyond the scope of this chapter. Here, we shall only consider the two dynamical horizons corresponding to the two individual black holes before they merge. 

For each of the two dynamical horizons, we shall calculate the source multipole moments. These were first introduced in \cite{Ashtekar:2004gp} for isolated horizons, and extended and used in \cite{Schnetter:2006yt} for dynamical horizons. These multipole moments have found applications, for example, in predictions of the anti-kick in binary black hole mergers \cite{Rezzolla:2010df} and for studying tidal deformations of black holes \cite{Cabero:2014nza,Gurlebeck:2015xpa}.

For defining these multipole moments, let $\mathcal{S}$ be a MOTS on a Cauchy surface $\Sigma$. Let $K_{ab}$ be the extrinsic curvature of $\Sigma$ embedded in spacetime, $r^a$ the unit space-like normal to $\mathcal{S}$ and tangent to $\Sigma$. Let $\mathcal{S}$ have the 2-Ricci scalar $\Tilde{\mathcal{R}}$. Also, let its areal radius be denoted by $R_{\mathcal{S}}$, its mass by $M_\mathcal{S}$ and its angular momentum by $J_S$. Furthermore, let $\mathcal{S}$ be axisymmetric and let $\varphi^a$ be the axial symmetry vector field on $\mathcal{S}$.  The symmetry vector $\varphi^a$ can be used to construct a preferred coordinate system $(\theta,\varphi)$ on $\mathcal{S}$ analogous to the usual spherical coordinates on a sphere; let $\zeta = \cos\theta$. We can then use spherical harmonics in this preferred coordinate system to construct multipole moments.  As expected we have two sets of moments $M_n$, $J_n$ such that $M_0$ is the mass $M_{\mathcal{S}}$ and $J_1$ is the angular momentum $J_{\mathcal{S}}$.  The expressions for the multipole moments  are the following
\begin{equation}
  \mathcal{M}_n = \frac{M_{\mathcal{S}}R_{\mathcal{S}}^n}{8\pi}\oint_S \Tilde{\mathcal{R}} P_n(\zeta) d^2S\,, \label{mass_mult}
\end{equation}
and 
\begin{equation}
  \mathcal{J}_n = \frac{R_{\mathcal{S}}^{n-1}}{8\pi} \oint_S P_n^\prime(\zeta) K_{ab}\varphi^a r^b d^2S \,, \label{spin_mult}
\end{equation}
where $P_n(\zeta)$ is the $n^{\mathrm{th}}$ Legendre polynomial and $P_n^\prime(\zeta)$ its derivative.

In a binary black hole merger scenario, the individual horizons of the black holes are dynamical and many of the above assumptions do not hold. For example, the Weyl scalar $\Psi_2$ of the two black holes are not time-independent, and neither are their areas, curvature, and other geometric quantities on $\mathcal{S}$. However, following Ref.~\cite{Schnetter:2006yt}, we shall continue to interpret the surface density and current in the same way so that the multipole moments share the same definitions as above.

The dynamical horizons are geometric objects that exist in spacetime independent of the spacetime foliation used to locate them. In this sense, the dynamical horizons are gauge invariant. On the other hand, the computation of the multipole moments in a numerical relativity simulation requires choosing a spacetime foliation. A different choice of spacetime slicing will give different values of the multipole moments $(\mathcal{M}_n, \mathcal{J}_n)$. For every choice of slicing the physical laws, like the flux/balance laws, will hold for the corresponding dynamical horizon. The foliations and gauges we choose (1+log slicing, $\Gamma$-driver shift \cite{alcubierre2003h}) are such that the horizon geometries are close to Kerr when the black holes are far apart (at the start of the simulation)  and they settle down close to Kerr at late times after the merger. In all numerical simulations done so far using these gauge choices, we find the same reasonable behaviour as expected \cite{Nielsen:2010wq}, just as in the case of waveform extraction. Although bad gauge choices are expected to affect the multipole moment values, we do not expect them to change significantly when the gauge choice is reasonable. We expect that our results will be the same qualitatively for other reasonable choices of gauge. A quantitative study of the dependence of these results on the choice of gauge is beyond the scope of this work.

In this work, we attempt to understand the tidal deformation of the horizons using the dynamical horizon formalism. The horizon geometry is characterized by the mass and spin multipole moments of the dynamical horizon. The presence of the companion induces a change in the horizon geometry of the black holes. For instance, the 2-Ricci scalars ($\Tilde{\mathcal{R}}$) of the two-dimensional slices of the dynamical horizons change from their equilibrium configuration,  i.e., from their isolated horizon values. During the inspiral phase, the two black holes influence each other and mutually change their horizon geometries. As they inspiral toward each other in the tidal environment of their companion the source multipole moments (as defined in \eqref{mass_mult} and \eqref{spin_mult}), which encode information about the deformed geometries of the MOTS of the individual dynamical horizons, evolve in time.

In this work, as we deal with non-spinning black holes, we attempt to understand the tidal deformability of black holes using mass multipole moments. Using the evolution of the mass multipole moment of the MOTS of the dynamical horizons of both the black holes, we define and compute dimensionless tidal coefficients (Love numbers)~\footnote{We use \emph{tidal coefficients} and \emph{Love numbers} interchangeably} that universally characterize the tidal deformability of black holes in General Relativity. Although our methods are generically applicable to all multipolar deformations, in this work we will study tidal deformation of black holes using $\mathcal{M}_2$ as defined in  \eqref{mass_mult}.

Turning now to our object of interest, namely a binary black hole system in the inspiral/pre-merger phase. Imagine starting with two Kerr black holes far apart, with the source mass and spin multipoles $(\mathcal{M}_n,\mathcal{J}_n)$ defined above exactly as for a Kerr black hole. As the black holes spiral in and approach each other, these multipole moments will vary under the influence of the gravitational field of the other. Let the variation in $\mathcal{M}_n$ and $\mathcal{J}_n$ be respectively $\delta\mathcal{M}_n$ and $\delta\mathcal{J}_n$ (these will be functions of time). In general, this deformation of the multipole moments will depend on the mass and spin of the companion, the separation $d$ between the black holes, and also possibly their relative velocity. However, for simplicity, we will not study the dependence on all these parameters in this work. The relative velocity would be important in describing the evolution of the multipole moments in the later stages of the inspiral and the merger and is not included here. Furthermore, we do not study the effect due to spinning black holes.

Note that as the horizon masses of the two black holes evolve during the inspiral, they deviate from the physical masses $M_{1,2}$ in the initial data. However, the fractional deviation turns out to be sub-percent in the analysis domain (i.e., $\lessapprox 0.1$\%). Similarly, the least upper bound for the spins acquired by the black holes was found to be $\lessapprox 10^{-2}$.

We assume the masses and spins to remain constant throughout the evolution. Therefore, in what follows, we restrict ourselves to the simpler situation of non-spinning black holes and also ignore their relative velocity.

We require that in the limit $d\rightarrow \infty$, $\delta\mathcal{M}_n$ and $\delta\mathcal{J}_n$ should vanish. From dimensional arguments, we are then naturally led to an expansion of the multipole moment of the black hole whose deformation is being studied. Denoting its mass by $M_{td}$ and that of its companion responsible for the tidal field by $M_{tf}$, the expansion can be written in the form:
\begin{align}
  \dfrac{\delta \mathcal{M}_n}{M_{td}^{n+1}} &= \sum_{i,j = 1}^{\infty} \alpha_{ij}^{(n)}\dfrac{M_{td}^i M_{tf}^j}{d^{i+j}}\,.
  \label{exp}
\end{align}
The deformation of the spin multipole moments $\mathcal{J}_n$ can also be expanded in a similar fashion. However, since we are dealing with non-spinning black holes in this work, we will restrict ourselves to the behaviour of mass multipole moments. Such an expansion has been previously used to understand the tidal deformations of black holes in the Bowen-York initial data set \cite{Cabero:2014nza}. Along similar lines, the following observations can be made on the above expansion. The expansion must start at $i = j =1$ since when we take the limit of either of the masses tending to zero, the left-hand side must vanish. Also, there should be no positive exponents on the distance measure $d$; i.e., $\delta\mathcal{M}$ not contain any terms that are proportional to $d^{\nu > 0}$ since the perturbation should vanish in the limit $d\rightarrow 0$. The coefficients $\alpha_{ij}^{(n)}$ are dimensionless coefficients independent of the parameters of the system as they have been scaled out by appropriate factors. These coefficients are therefore the same for all black holes and are thus universal. They should characterize the tidal deformability of the black holes. 

\section{Numerical simulations}
\label{sec:nr_tddh}

We ran numerical simulations of orbital mergers for a set of non-spinning binary black holes with varying mass ratios. The systems are evolved numerically using puncture data \cite{PhysRevLett.78.3606} describing two black holes in quasi-circular orbits, with typical initial separations of $10$-$11 M$. These simulations cover the dynamical behaviour starting at 5-6 orbits before the merger and going up to the merger phase. The gravitational waveform is extracted \cite{Baker:2002qf} at various distances between $100 M$ to $500 M$ from the merger location.

Since this work is focused on the study of individual horizon geometry of the black holes in the inspiral phase, we track the individual horizons up to merger and compute quasi-local quantities \cite{Dreyer:2002mx, Schnetter:2006yt} using the isolated and dynamical horizon formalism \cite{Ashtekar:2004cn}. 

We compute the quasi-local quantities on the horizon on an angular grid of size ($37, 76$) along the longitudinal and latitudinal directions, respectively
The multipole moments in  \eqref{mass_mult} are obtained from the numerical simulations run using the Einstein Toolkit, using the QuasiLocalMeasures thorn.

We consider non-spinning binary black hole systems with varying mass ratios 
$q=M_2/M_1$, where $M_{1,2}$ are the masses of the primary and secondary objects. The larger black hole with mass $M_1$ will be denoted by BH1 and the smaller one with mass $M_2$ by BH2. We will be studying the tidal deformation of 
each of the black holes BH1 and BH2 due to the influence of their companion. 

We use the GW150914 parameter file available from Ref.~\cite{wardell_barry_2016_155394}, as a template for our simulations. For each case, as input parameters, we provide initial separation between the two punctures $D$, mass ratio $q$ and radial and azimuthal linear momentum $p_r$, $p_{\phi}$ respectively, while keeping the total horizon mass $M=M_1+M_2$ of the system to be $1.0$ in units of $c=G=M=1$. Parameters are listed in \ref{tab:ic_qc0_tddh}. We compute the corresponding initial locations, the $x$, $y$, $z$ components of linear momentum for both black holes and grid refinement levels, etc., before generating the initial data and evolving it. We chose non-spinning cases ranging between $q=1.0$ to $0.4$, based on the initial parameters listed in \cite{Healy:2014yta,PhysRevD.95.024037}. Our simulations match very well with the catalogue simulations \cite{RITcatalog:web}, having merger time discrepancies of less than a few per cent.
\begin{table}
	\centering
\begin{tabular}{|p{1.8cm}|p{1.8cm}|p{1.8cm}|p{1.8cm}|p{1.8cm}|p{1.8cm}|}
\hline
\multicolumn{6}{|c|}{Non-spinning BBH Simulations} \\
\hline
\hline
Mass ratio &$d$ &$M_{1}$ &$M_{2}$ &$p_r$ &$p_t$\\
\hline

1.0     & 11.0        & 0.5       & 0.5       & -7.220e-04 & 0.09019  \\
0.85    & 12.0        & 0.54051   & 0.4595    & -5.290e-04 & 0.08448  \\
0.75    & 11.0        & 0.5714    & 0.4286    & -6.860e-04 & 0.08828  \\
0.6667  & 11.75       & 0.6       & 0.4000    & -5.290e-04 & 0.08281  \\
0.50    & 11.0        & 0.6667    & 0.3333    & -5.720e-04 & 0.0802   \\
0.40    & 11.25       & 0.7143    & 0.2857    & -4.500e-04 & 0.07262  \\
\hline
\end{tabular}
\caption[Initial parameters of the NE simulations used in the strong field tidal deformability studies.]{Initial parameters for non-spinning binary black holes with quasi-circular orbits. Here, $q=M_2/M_1$ is the mass ratio, $d$ is the initial separation between the two holes, $p_r$ and $p_t$ are the radial and tangential (to the orbit) momenta in the initial data, respectively.}
\label{tab:ic_qc0_tddh}
\end{table}

\section{Computing the tidal coefficients}
\label{sec:com_tidal_coeffs}

We now describe a method to compute the leading and sub-leading order tidal coefficients appearing in the expressions for the expansion of the perturbed multipole moments numerically using the aforementioned simulations of BBH mergers. 

To distinguish between the black hole that is tidally deformed and the black hole that sources the tidal field we use $M_{td}$ for the former and $M_{tf}$ for the latter. Thus, $M_{td}$ can be $M_1$ and $M_{tf}$ can be $M_2$ or vice-versa depending on whether the tidal deformation of $M_1$ is being studied or that of $M_2$. We denote an alternate-mass ratio defined using $M_{td}$ and $M_{tf}$ by $\gamma$:
\begin{equation}
  \gamma = \dfrac{M_{tf}}{M_{td}}
\end{equation}
As opposed to the mass ratio defined earlier in \ref{tab:ic_qc0_tddh}, $\gamma$ takes values in the range $[0.4, 2.25]$.

\subsection{Identification of the leading order term}

We begin by noting that  \eqref{exp} is a power series in the inverse of the distance of separation of the two orbiting black holes. We postulate that there would be no term at order less than $1/d^3$ (i.e., no term involving a lower exponent on $1/d$, such as $1/d^2$ or $1/d$) since the Newtonian tidal force enters at the order of $1/d^3$. Therefore, the leading term is ${\cal O}(1/d^3)$. The sub-leading order term is ${\cal O}(1/d^4)$ and is responsible for post-Newtonian tidal influences. Henceforth, we will use \emph{third-order} and \emph{fourth-order} to refer to the exponent on the distance, with the third-order term being the leading and the fourth-order being the sub-leading term.  

The expression for the perturbative expansion for $n=2$ mass multipole moment at the leading (Model~A), and leading and sub-leading (Model~B) order perturbations due to $M_{tf}$ are, respectively, 
\begin{equation}
  \mathbf{Model \quad A:}\quad \dfrac{\mathcal{M}_2}{M_{td}^3} = \dfrac{a_3 ^{(2)}}{d^3} + {\rm const.}\,
  \label{mass_mult_a3}
\end{equation}
and 
\begin{equation}
  \mathbf{Model \quad B:}\quad \dfrac{\mathcal{M}_2}{M_{td}^3} = \dfrac{a_3 ^{(2)}}{d^3} + \dfrac{a_4^{(2)}}{d^4} + {\rm const.} \label{mass_mult_a34}
\end{equation}
Here, the tidal coefficients are related to $a_3^{(2)}$ and $a_4^{(2)}$ through:
\begin{align}
a_3^{(2)} =& M_{td} M_{tf}^2 \alpha^{(2)} _{12} + M_{td}^2 M_{tf} \alpha^{(2)} _{21}\,, \label{a3_exp}  \\
a_4^{(2)} =& \alpha^{(2)} _{13} M_{td} M_{tf} ^3 + \alpha^{(2)}_{22} M_{td}^2 M_{tf}^2 + \alpha^{(2)}_{31} M_{td}^3 M_{tf} \,.
\label{a34_exp}  
\end{align}
Note, that in the case of an isolated non-spinning black hole, the horizon geometry is spherically symmetric, therefore, the $n=2$ mass multipole moment is zero, resulting in $\delta{\mathcal{M}}_2 = \mathcal{M}_2$. However, in the above model, we include an overall constant in the fits to allow for any systematic errors in the numerical computation of the multipole moment and to possibly take into account the fact that the initial data is described by punctures and not real black holes. We use the distance measure which is computed by the simple Euclidean separation between the geometric centroids of the two black holes at every time step in the simulation. This agrees well with the physical distance measures between the black holes. More details can be found in Appendix \ref{appendix:dm}.

We will now detail the procedure for computing the tidal coefficients $\alpha^{(2)} _{12}$, $\alpha^{(2)} _{21}$, $\alpha^{(2)} _{13}$, $\alpha^{(2)}_{22}$ and $\alpha^{(2)}_{31}$ using these models. Thereafter, we compare the fits obtained. The superscript $(2)$ is used to denote the fact that we are analysing $n=2$ mass multipole moment; since this chapter deals exclusively with $n=2$ mass multipolar deformations, it will be dropped henceforth. Resolving higher $n$ multipole moments would require a sufficiently fine grid on the horizon. Due to computational limitations, we restrict our analysis to $n=2$ multipolar deformations in this work.

\subsection{The Fitting procedure}

\subsubsection{The parameter estimation routine}
To estimate the tidal coefficients we follow a two-step fitting procedure:
\begin{enumerate}
  \item Fit of the multipole moment data to distance data, separately to each of the simulations and black holes, and estimating the coefficients $a_3$ and $a_4$ using the model  \eqref{mass_mult_a34}.
  \item The re-fitting of all values of the best-fit parameters from the above step to the masses $M_{td}$ and $M_{tf}$, in a combined manner, to the models in \eqref{a3_exp} and \eqref{a34_exp} to obtain the tidal coefficients $\alpha_{ij}$.
\end{enumerate}
We carry out linear regression of the numerical data in the two steps to fit the numerical data to the models. We minimize a least-squares objective function in the process. We directly compute a least-squares objective function on a grid in the parameter space and locate the minimum. Since this method is computationally expensive, we follow the two-step procedure mentioned above. 

For carrying out the regression, we relate the models and the data in a matrix form as
\begin{equation}
  Y = X A \,,
\end{equation}
where $Y$ is a matrix of dimensions ($K$, $1$) of the left-hand side of the respective model, $X$ is the matrix of data points of dimensions ($K$, $L$) and $A$ the matrix of the parameters of dimensions ($L$, 1). For instance, in the fitting procedure of step one, $K$ is equal to the number of data points in the time-series data of the multipole moment/distance and $L$ is the number of parameters in the model (which is 3 for Model~B). 

We use the following least-squares objective function in the minimization procedure (the summation convention is assumed on repeating indices):
\begin{equation}
  \mathcal{L}(A) = \sum_i (Y_i - X_{ij} A_j)^2 \,,
  \label{lsof}
\end{equation}
Here $X_{ij}A_{j}$ is the prediction from the respective linear models. The linearization of the above models in \eqref{mass_mult_a3}, \eqref{mass_mult_a34}, \eqref{a3_exp}, and \eqref{a34_exp} is done by assuming the fitting parameters as coefficients of linear variables. To exemplify this, let us consider the two steps of the fitting procedure for Model~B. In step one, the linearized model corresponding to  \eqref{mass_mult_a34} would be written as:
\begin{equation}
  \dfrac{\mathcal{M}_2}{M_{td}^3} = a_3 x_3 + a_4 x_4 + {\rm const.}\,,
\end{equation}
where $x_3 = 1/d^3$ and $x_4=1/d^4$ are the linear fitting variables. For re-fitting, \eqref{a3_exp} and \eqref{a34_exp} would be used. The re-fitting model corresponding to  \eqref{a34_exp} would be written as:
\begin{equation}
  a_4 = \alpha_{13} \mu_{13} + \alpha_{22} \mu_{22} + \alpha_{31} \mu_{31}\,,
\end{equation}
where $\mu_{13} = M_{td} M_{tf}^3$, and so on.

\subsubsection{Error estimation}

The best-fit parameter values (denoted by $\hat{A}$)
are those that minimize the least-squares objective function  \eqref{lsof} for the respective models. We compute the errors on the best-fit parameter values from the diagonal components of the variance-covariance matrix $\mathcal{C}$ of the data, and the fit residue $\sigma_{fit}$:
\begin{equation}
  \mathcal{C} = X^T X
\end{equation}
and
\begin{equation}
  \sigma_{fit} = \sqrt{\mathcal{L}(\hat{A})}
\end{equation}
as
\begin{equation}
  \sigma(\hat{A}) = \sigma_{fit} \times Diag(\mathcal{C})\,.
\end{equation}
Here, the matrix of best-fit parameters is denoted by $\hat{A}$ and their standard error of estimates by $\sigma(\hat{A})$. We will refer to the individual elements of these matrices by $\hat{A}_j$ and $\sigma_j$ respectively. Since these parameters can correspond to the fitting procedure of either step one or step two, we will explicitly mention which step we are referring to during their usage.

\section{Results}
\label{sec:results}
\begin{table}
\begin{tabular}{|p{4.5cm}|p{1.8cm}|p{1.8cm}|p{1.8cm}|p{1.8cm}|p{1.8cm}|}
\hline
\multicolumn{6}{|c|}{Tidal coefficients} \\
\hline
Model                           &   $\alpha_{12}$   &   $\alpha_{21}$    &   $\alpha_{13}$   &   $\alpha_{22}$   &  $\alpha_{31}$   \\
\hline
Third-order model               &   $-0.15 \pm 0.21$  &   $-3.43 \pm 0.21$   &   NA              &   NA              &   NA             \\
Third and fourth-order model    &   $-0.89 \pm 0.21$  &   $-5.45 \pm 0.21$   &   $1.79 \pm 1.72$   &   $5.3 \pm 3.65$    &   $4.68 \pm 1.72$  \\
\hline
\end{tabular}
\caption[Estimates of the tidal deformability coefficients.]{Tidal coefficient values estimated from a re-fit of the fit coefficients $a_3$ and $a_3, a_4$ in  \eqref{mass_mult_a3} and  \eqref{mass_mult_a34}, respectively.}
\label{tab:fit_coeffs_com}
\end{table}
\begin{figure}[htp]
    \centering
	\includegraphics[width=\columnwidth]{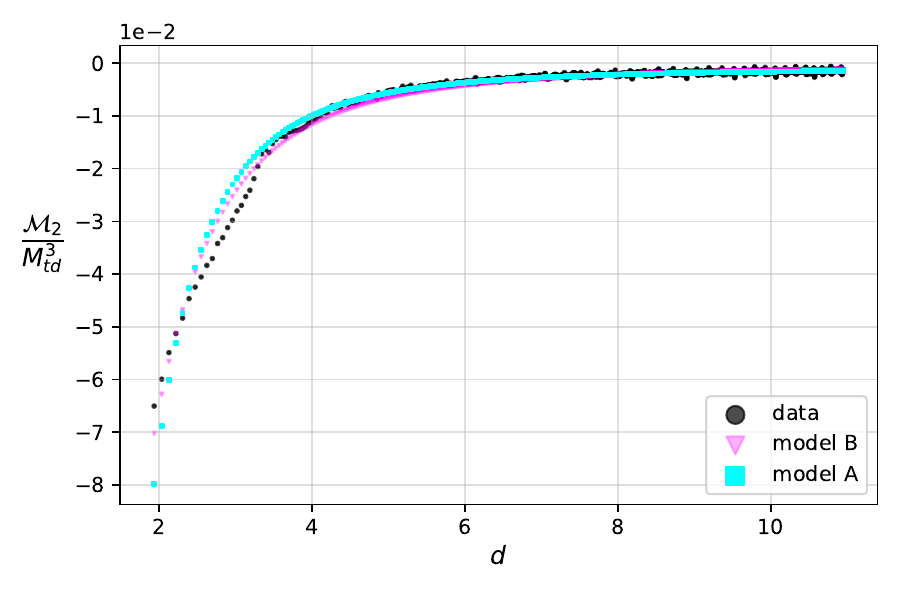}
    \includegraphics[width=\columnwidth]{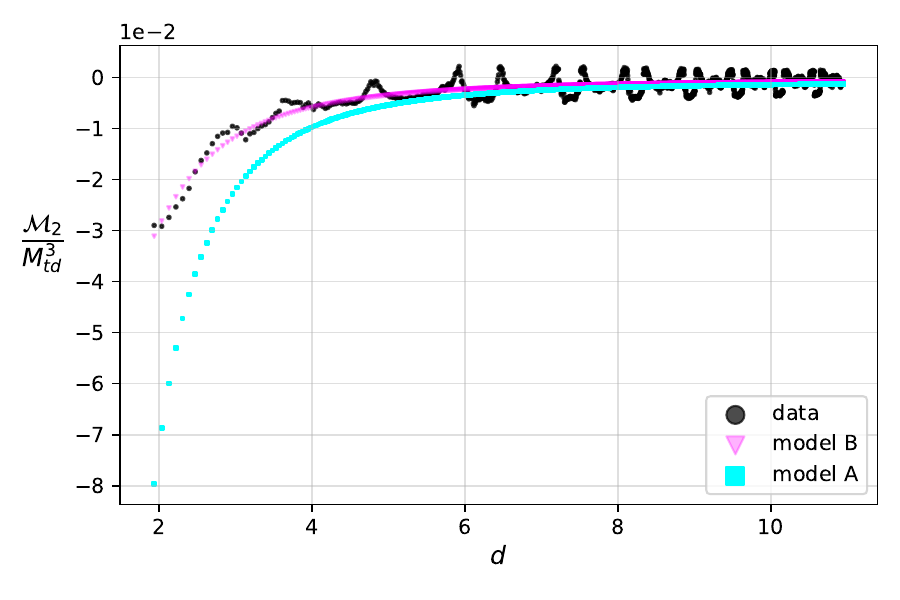}
\caption[The fit of the source multipole moment of the DHs to the tidal expansion.]{The fit of the $n = 2$ mass multipole moment of BH1 (top) and BH2 (bottom) for the simulation $q=0.4$. The multipole moment data is plotted against the distance $d$.  
In these two figures, the model where only the leading (third) order term is included, namely, Model~A of  \eqref{mass_mult_a3}, is shown in cyan (squares) whereas the model where the leading and sub-leading order terms have been included, namely, Model~B of  \eqref{mass_mult_a34} is shown in magenta (triangles). The data is in black (dots).}
\label{fig:mlfits_34_q0p4}
\end{figure}
\begin{figure}[htp]
	\centering
	\includegraphics[width=\columnwidth]{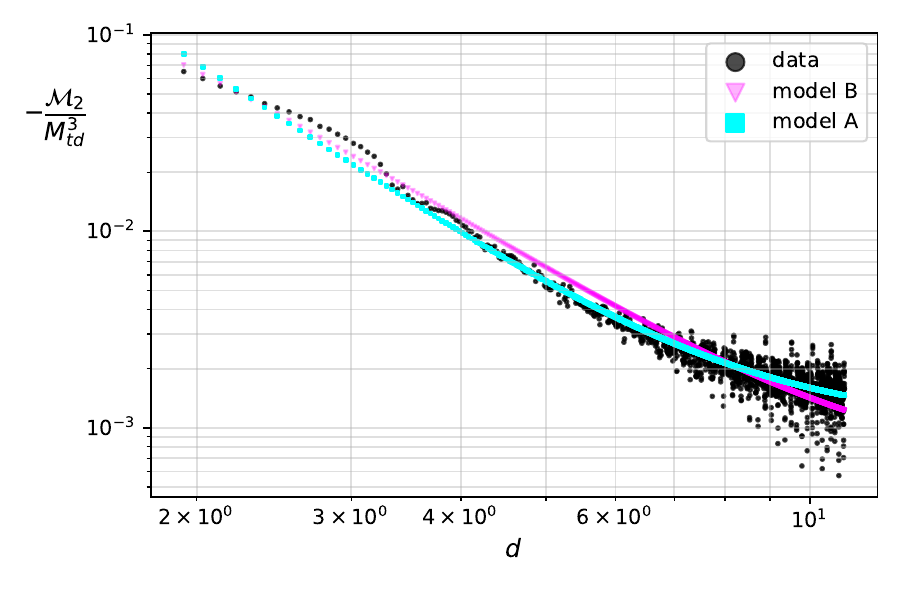}
	\includegraphics[width=\columnwidth]{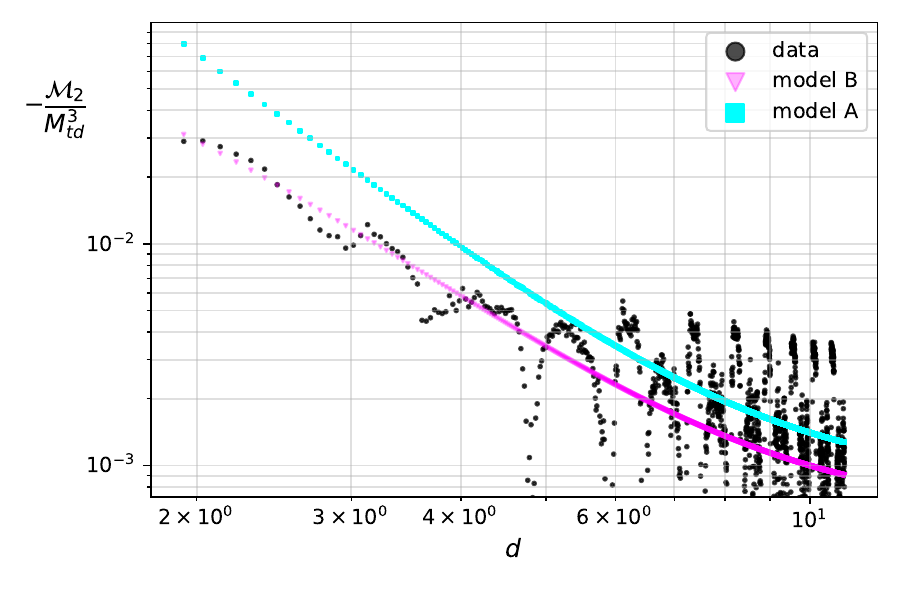}
	\caption[The fit of the source multipole moment of the DHs to the tidal expansion, shown in logarithmic scale]{The fit of the $n = 2$ mass multipole moment of BH1 (top) and BH2 (bottom) for the simulation $q=0.4$. The multipole moment data is plotted against the distance $d$ in the logarithmic scale.  
		In these two figures, the model where only the leading (third) order term is included, namely, Model~A of  \eqref{mass_mult_a3} is shown in cyan (squares) whereas the model where the leading and sub-leading order terms have been included, namely, Model~B of  \eqref{mass_mult_a34} is shown in magenta (triangles). The data is in black (dots).}
	\label{fig:mlfits_34_q0p4_log}
\end{figure}
\begin{figure}[htp]
   \centering
       \includegraphics[width=\columnwidth]{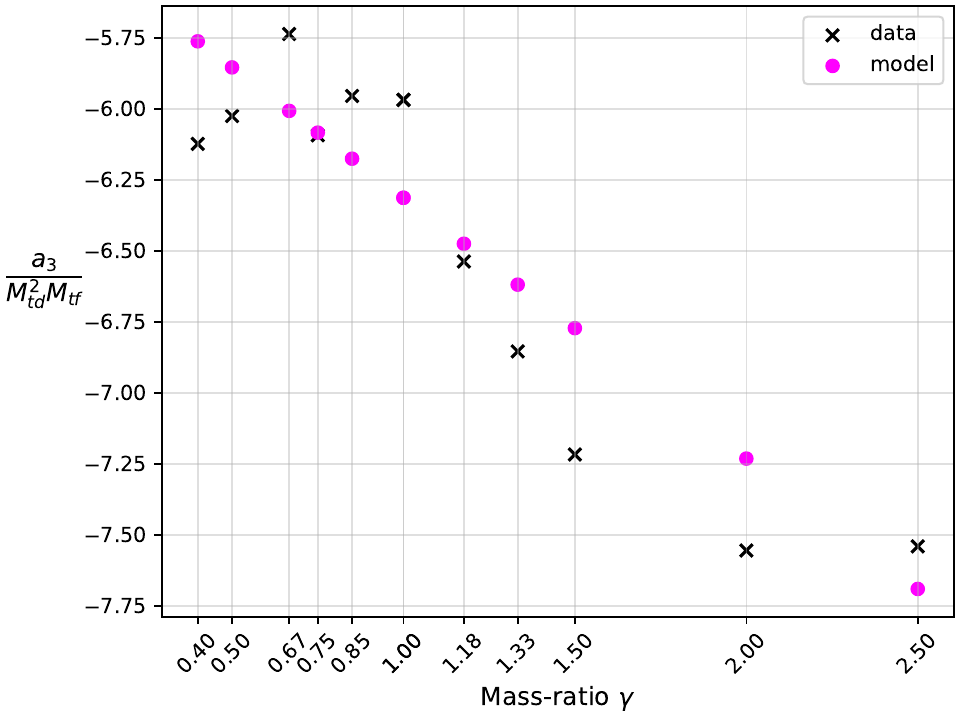}
\caption[The universality in the expansion coefficients and the estimation of the leading order tidal Love numbers.]{The re-fit of the coefficient $a_3$ to obtain the tidal coefficients in  \eqref{a3_exp} (Model~B). Here 
 $a_3/(M_{td}^2 M_{tf})$ is plotted against $\gamma$. The data points are individual best-fit parameters of Model~B in  \eqref{a34_exp} for both black holes in each simulation. The data points for the fit are in black (crosses) and the best-fit model values are in magenta (dots).}
\label{fig:refits_4_3order_12_com}
\end{figure}
\begin{figure}[htp]
	\centering
       \includegraphics[width=\columnwidth]{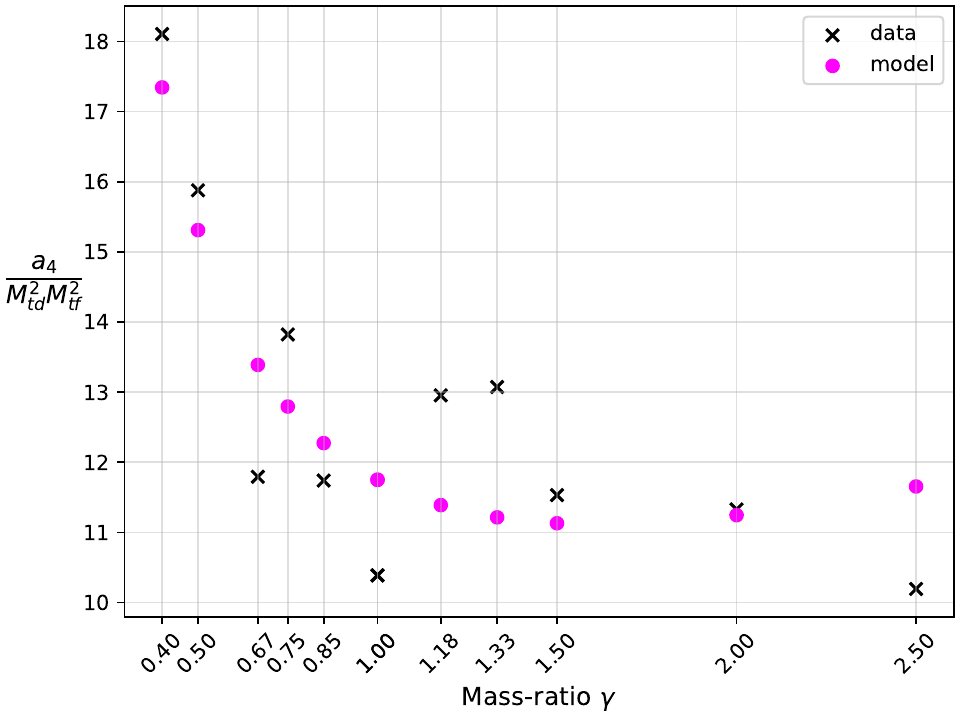}
\caption[The universality in the expansion coefficients and the estimation of the sub-leading order tidal Love numbers.]{The re-fit of $a_4$ to the model of  \eqref{a34_exp}, with three terms, 
for the simulations $q\geq 0.4$ (Model~B). Here,
$\dfrac{a_4}{M_{td}^2 M_{tf}^2}$ is plotted against $\gamma$. The data points are individual best-fit parameters of Model~B in  \eqref{a34_exp} for both black holes in each simulation. The data points for the fit are in black (crosses) and the best-fit model values are in magenta (dots).}
\label{fig:refits_4_4order_123_com}
\end{figure}
We obtain the best-fit parameters for the two models  \eqref{mass_mult_a3} and  \eqref{mass_mult_a34} using the two-step procedure mentioned in the preceding section and tabulate the results in \ref{tab:fit_coeffs_com}. In this table, the final results of step two of the fitting procedure (i.e. re-fits to the models in \eqref{a3_exp} and \eqref{a34_exp}) are denoted in the format $\hat{A}_j \pm \sigma_{j}$: the best-fit parameter value and the respective standard error. 

In \ref{fig:mlfits_34_q0p4} and \ref{fig:mlfits_34_q0p4_log}, we show the fit of the $n=2$ mass multipole moment to the two models  \eqref{mass_mult_a3} (Model~A) and  \eqref{mass_mult_a34} (Model~B) for $q=0.4$ simulation data. In these plots, we show the data points and the best fitting model i.e. the values of the $n=2$ multipole moment $\mathcal{M}_2$ predicted by the models in \eqref{mass_mult_a3} and \eqref{mass_mult_a34} for the best fit parameter values $\hat{A}$. In the top panel, we plot the multipole moment data {\it vs} the distance of separation for the more massive black hole (left) and the less massive black hole (right). In the bottom panel, we plot the same in logarithmic scale on both the axes.

In \ref{fig:refits_4_3order_12_com} and \ref{fig:refits_4_4order_123_com}, we show the results of step two of the fitting procedure (re-fitting results) of the parameters $a_3$ and $a_4$ of Model~B appearing in  \eqref{a3_exp} and  \eqref{a34_exp}, respectively, which were obtained from step one of the fit for the set of binary black hole configurations as listed in \ref{tab:ic_qc0_tddh}. In these plots, we show the data points and the best fitting model i.e. the values of $a_3$ and $a_4$ predicted by the models in \eqref{a3_exp} and \eqref{a34_exp} for the best fit parameter values $\hat{A}$.

Some comments are in order regarding these fits. 

First, as can be clearly seen, Model~B, consisting of both $1/d^3$ and $1/d^4$ terms, fits the data better in comparison to  Model~A (which has only the $1/d^3$ term).

Second, although the overall fit to the data of Model~A is worse than that of Model~B, nevertheless the former model fits the data relatively well in the early inspiral phase, i.e., at large distances of separation between the black holes, especially, for the heavier black hole (the top panel of \ref{fig:mlfits_34_q0p4}).

Since the system spends more time in the early inspiral phase (at large distances of separation), there are more points at large $d$ to which the model fits well. However, this model fails to explain the data as the black holes close in. This is expected because the sub-leading term in  \eqref{mass_mult_a34} is expected to become increasingly important at low separation distances $d$. This is 
why we expect Model~B to be more accurate.

Third, the quality of fit of the leading order model to the data: the fit of Model~A is visibly worse, especially, for the smaller black hole (right panels of \ref{fig:mlfits_34_q0p4}). From a perturbative point of view, we can intuitively explain this as follows. The change in the geometry of the smaller black hole due to the presence of the larger black hole is more than the change in geometry, as quantified by the multipole moments, of the larger black hole due to the smaller one, as the larger black hole is more massive leading to more variations in its multipole moment. 
The sub-leading (fourth) order term contains higher powers of the perturbing black hole's mass $M_{tf}$ than at the leading (third) order ($M_{tf} ^3$ as opposed to $M_{tf} ^ 2$ in  \eqref{a3_exp}~and~ \eqref{a34_exp}).
Therefore, the effect of the fourth-order term will be easily visible in the multipole moment of the smaller black hole $M_2$ (see right panels in \ref{fig:mlfits_34_q0p4}) and the fit between data and Model~A is visibly worse in the plots for the smaller black hole.

Fourth, the fourth-order term has the opposite sign to that of the third. This is visible in the same plots \ref{fig:mlfits_34_q0p4} at small values of $d$. The effective slope of the data points deviates from $1/d^3$ behaviour and reduces in magnitude as the distance between the black holes reduces. Therefore, the fourth-order term is responsible for reducing the overall tidal force involving the leading and sub-leading term, consistent with the velocity-independent post-Newtonian correction to the Newtonian tidal force (see Ref.~\cite{xupaik2016} and references therein):
\begin{equation}
    \vert\vert \mathbf{F}_{1PN} \vert\vert \sim \dfrac{M_{td}^2 M_{tf}}{d^3}\,.
\end{equation}
One feature of the model in  \eqref{a34_exp} is that there are three independent terms at the fourth-order in distance whereas there seems to be just one term in the expression for the 1PN correction to the Newtonian gravitational force. It was verified that all three terms were necessary to explain the data points, which have a quadratic dependence on the mass-ratio $q$ -- as can be seen in the bottom panel of \ref{fig:refits_4_4order_123_com}, and obtain a good fit to the data. This needs to be probed further in the future.

Fifth, there is room for some error in the fourth order in $d$. At this 1PN order, terms involving the relative velocities of the black holes would enter into the expansion of  \eqref{mass_mult_a34}. These terms might be contributing more to the tidal effects as the mass ratio decreases.

Therefore, in order to analyse the multipole moment data to infer the tidal coefficients at further smaller mass ratios than $q=0.4$, one may need to take into account new terms involving velocities in the perturbative multipole expansion. 

To further check the precision of these fits, we considered multiple sets of data having various combinations of our simulations, with additional configurations for intermediate $q=\{0.6, 0.7\}$;  excluding a few configurations randomly; or even adding/removing a few lower $q \leq 0.6$. Across these sets, we find $\alpha_{12}$ to be $\approx -1.0$ and $\alpha_{21} \approx -5.3$. Based on the results, we conclude that the tidal Love numbers at the leading order ($\alpha_{12}$ and $\alpha_{21}$) are estimated at a higher precision than for those at the sub-leading order ($\alpha_{13}$, $\alpha_{22}$ and $\alpha_{31}$). This is reasonable since we are attempting to estimate the coefficients of the sub-leading order term, to which there are three contributions ( \eqref{a34_exp}). The contribution of the fourth-order term is not visible until the onset of the late inspiral phase of the evolution -- i.e., at smaller values of $d$ (as can be seen in \ref{fig:mlfits_34_q0p4}). It has the largest impact during this phase, where the number of data points is significantly less than that in the early inspiral phase, thus making it difficult to estimate it at the same precision.

The estimates for the tidal Love numbers at the fourth order may be expected to be computed with more accuracy by increasing the number of simulations, and the resolution of the runs.

\section{Discussion and Conclusions}
\label{sec:conclusions}

The primary goal of this chapter was to study the tidal effects on horizons during the inspiral phase of binary black hole mergers. Past studies have examined this problem using the perturbations of the asymptotic gravitational fields, but this is the first time that tidal effects have been explored for a binary system comprising similar masses using full numerical relativity simulations, to deduce a black hole's tidal coefficients.

We used source multipoles computed on the marginally outer trapped surfaces of the space-like dynamical horizons of both black holes for this analysis. We simulated non-spinning BBH with varying mass ratios $q \in [0.4, 1]$. We defined a convenient set of tidal coefficients to characterize the tidal deformability of black holes in a binary system using the source mass multipole moments of a dynamical horizon. This approach does not involve the field multipole moments that are defined far away from the system, nor do they involve assuming a matter distribution whose compactness limit is taken to describe the black hole case. 
Assumptions regarding the stationarity/ slow evolution of the tides were not made either.  By definition, these numbers are dimensionless and should be independent of the system considered, i.e., the same for all black holes (of any mass and spins) in General Relativity. We computed five tidal Love numbers which characterize the $n=2$ mass multipolar deformations of the dynamical horizon geometry: two at the leading order and three at the sub-leading order.

These results show in explicit detail, as noted in existing literature, that, although the Love numbers that characterize the tidal deformations of the gravitational fields of non-spinning black holes far away from the system are zero, the corresponding Love numbers that characterize the strong field tidal deformation of the horizon geometry in the strong field regime do not vanish.

The relations in \eqref{mass_mult_a34},~ \eqref{a3_exp}~and\eqref{a34_exp}, together with the best-fit parameters listed in \ref{tab:fit_coeffs_com}, show how the mass multipole moment of the horizon evolves in relation to the dynamics of the system in a binary black hole merger scenario, apart from describing the evolution of tidal deformations. We find that the evolution of the multipole moments can be described quite accurately by the model in  \eqref{mass_mult_a34} up to the merger. 

Here, it must be noted that in simulations of lower mass ratios, we notice an increase in the error of our fits to numerical data. This can be attributed to the increasing importance of the terms involving relative velocities of the black holes, and terms involving higher powers of $1/d$. In principle, more tidal Love numbers at successively higher orders can be computed in the manner described here, given a sufficient number of numerical simulations, and are carried out at sufficiently high resolution. It will be important to examine those cases with higher resolution runs in the future.

Analogous to the treatment of neutron star tidal deformability, a source-independent definition of the tidal deformation of a black hole horizon using the source multipole moments of the horizon must also be possible. This would involve studying the perturbation of the geometry of an otherwise isolated horizon due to the external perturbing fields, and finding a relation between them -- a quest that may be pursued in a future work.

\cleardoublepage %
\flushbottom
\newpage
\cleardoublepage

\newcommand{\eS}{\mathcal{S}}

\chapter{Generalized source multipole moments of dynamical horizons}
\label{chap:gen_mul}

\section{Introduction} 

 When black holes exist in a tidal environment such as another black hole, they deform under mutual gravitational forces. In a BBH merger scenario, this tidal environment is highly dynamical and the black holes exist in the strong field regime of each other. The existing perturbative approach is therefore not feasible. We look at a different approach to studying the tidal deformation of the black holes, using their dynamical horizons. In this work, which is a continuation of the previous chapter,  we find interesting features in the evolution of the deformations of the dynamical horizon geometry in a binary black hole merger scenario in General Relativity, using numerical relativity. First, in the inspiral phase, owing to the deformations, the dynamical horizons of the two black holes are found to steadily acquire 'gravitational hairs' i.e. multipole moments as they evolve, that vanish when the horizons are isolated. The dominant deformation is shown to be of a quadrupolar nature. Second, we show that they encode detailed information about the dynamics of the binary black hole system. Third, the deformations are shown to be strongly correlated with the gravitational waves received from the system at infinity. The dominant quadrupole multipole moment is particularly shown to be strongly correlated with the gravitational field of the system at future null infinity. Therefore, the gravitational waves carried away from the system contain information about the geometrical structure of the black holes in the strong-field regime. Fourth, we also find that, in the post-merger phase, the multipolar structure of the outer common dynamical horizon of the system is strongly correlated with that of the individual horizons just before the merger. The outer common horizon then settles down to equilibrium as suggested by the decay of its multipole moments gained by the system through the inspiral phase.

Numerous binary black hole mergers have been observed to date starting with the first detection in 2015 \cite{Abbott:2016blz,LIGOScientific:2018mvr,TheLIGOScientific:2016pea,Nitz:2018imz,Nitz:2019hdf,Venumadhav:2019lyq,Zackay:2019tzo,theligoscientificcollaboration2021gwtc21,theligoscientificcollaboration2021search}. The parameters of these binary systems, including the masses and spins of the individual black holes, can be inferred from the observed data \cite{TheLIGOScientific:2016wfe}. These observations have been used to extract information about the overall dynamics of the binary black hole system. The observations have so far been found to be consistent with standard general relativity \cite{Abbott:2016apu, LIGOScientific:2019fpa, Abbott:2018lct}. 

The dynamics of the gravitational radiation in the far i.e. weak field regime, far away from the horizons of the merging black holes where the GW detectors are placed is fairly well understood. In particular, the multipole moments of gravitational radiation are relevant here \cite{thorne1980, Bondi:1962px, Sachs:1962wk, Newman:1962, Janais1965}. Various waveform models exist that can be used to approximately construct the gravitational wave signals given the parameters of the system \cite{Abbott:2016apu} for most quasi-circular systems, and some eccentric systems. On the other hand, in the strong field regime, the physical boundaries of the black hole regions are described by their dynamical horizons, which are located inside the event horizons and are thus outside of the domain of outer communications, and are only beginning to be explored. In recent times, advances in numerical relativity have paved the way for understanding the dynamics of the binary black hole system in the strong field dynamical regime, which have previously been inaccessible by analytical methods. In particular, the evolution and properties of dynamical horizons of black holes are of interest here. There have been efforts in the recent past directed towards understanding of the dynamics of the horizons, but mostly of axisymmetric scenarios like head-on collisions \cite{Owen:2009sb, Pook-Kolb:2020zhm, Pook-Kolb:2020jlr}. However, understanding of the dynamics of the horizon geometry in the inspiral, merger and ringdown phase of generic and realistic non-axisymmetric dynamical scenarios, like that of a binary black hole merger, is still in its infancy \cite{Gupta:2018znn}.

In binary black hole systems, the dynamical horizons of the black holes exist in a tidal environment under the influence of mutual gravitational forces. They are tidally coupled, possess an influx of energy and momentum into them and are hence highly dynamical entities. In such a scenario, in this chapter, we take an alternative approach to studying the tidal deformations of black holes that may be more relevant as the black holes close in towards the merger. To address this problem, we use the \emph{source} multipole moments of the dynamical horizons to study their deformations, as opposed to the field multipole moments used in the literature (see \ref{chap:tddh} for a discussion). We find that the evolution of these deformations shows a universal behaviour, and can be described by a generic expansion in terms of the distance of separation between the black holes. Such an expansion was previously used to study the deformations of initial data in \cite{Cabero:2014nza}, and to study the axisymmetric deformations during the evolution in fully dynamical scenarios in \cite{prasad2021tidal}. Here, we extend the study to more general non-axisymmetric deformations of the dynamical horizons.

We then show that the general features of the dynamics of the binary black hole system such as their orbital phasing, relative location of the black holes, etc. are imprinted on the deformed geometry of their dynamical horizons. In our previous work (\ref{chap:waveform_rel}), the correlations between the gravitational fields at two surfaces: the dynamical horizon and future null infinity were studied. In particular, the infalling radiation at the dynamical horizons was found to be strongly correlated with the news function of the outgoing gravitational radiation from the system \cite{shear-news2020}. In another recent study, the axisymmetric deformations of the dynamical horizons in binary black hole scenarios were studied and were used to quantify the strong field tidal deformability of black holes \cite{prasad2021tidal} (\ref{chap:tddh}). In this work, we further add to the understanding of the deformations and the strong field dynamics of black holes in general relativity (i.e. at the dynamical horizons) and their correlations with the weak field dynamics (i.e. at future null infinity) by studying the evolution of the horizon geometry of dynamical horizons, described by its source multipole moments, for the first time in a binary black hole scenario through the inspiral, merger and ringdown phases. We then show that not just the infalling radiation at the horizon, but the deformed geometry of the gravitational field in the strong field regime i.e. that of the dynamical horizons themselves is strongly correlated with the spacetime geometry at null infinity. Thus, although future-directed causal curves from the dynamical horizons cannot reach $\mathcal{I}^+$, one can potentially use these results to reconstruct the multipolar structure of the deformed horizon geometries using the information at $\mathcal{I}^{+}$. 

\section{Basic notions} 

The main quantities of interest are (a). the generalized source multipole moments of the dynamical horizons, which are defined at a dynamical horizon $\mathcal{H}$, \cite{Ashtekar:2004cn,Booth:2005qc} obtained by a time evolution of marginally trapped surfaces, and (b). the multipolar structure of the gravitational news function at the future null infinity $\mathcal{I}^+$, the endpoint of future null-geodesics which escape to infinity \cite{Bondi:1962px,Penrose:1964ge}. 

A dynamical horizon is located inside the event horizon, which marks the boundary of a trapped space-time region. Future null infinity $\mathcal{I}^+$ is an invariantly defined null surface where outgoing null geodesics end. 

We consider $\mathcal{H}$ and $\mathcal{I}^+$ to be foliated by 2-surfaces $\mathcal{S}$ of spherical topology, with an intrinsic Riemannian metric $\Tilde{q}_{ab}$. For the former, we obtain a marginally trapped surface and for the latter, we approximate it by large coordinate spheres in the numerical domain. For every cross-section $\mathcal{S}$, we assign outgoing and ingoing directions. Denote the outgoing future-directed null vector normal to $S$ by $n^+$, and the ingoing null normal as $n^-$ satisfying $n^+\cdot n^- = -1$.  Let $x$ be a complex null vector tangent to $\eS$
satisfying $x\cdot\bar{x}=1$ (the overbar denotes complex conjugation),
and $n^+\cdot x = n^-\cdot x = 0$.

\subsection{Dynamics at $\mathcal{I}^+$}

In the weak-field regime, the vacuum %
spacetime geometry is completely described by the 
Weyl tensor $C_{abcd}$.  In particular, outgoing transverse radiation 
is described by the Weyl tensor component \cite{Newman:1961qr}
\begin{equation}
	\Psi_4 = C_{abcd}n^{-a}\bar{x}^bn^{-c}\bar{x}^d\,.
\end{equation}
$\Psi_4$ can be expanded
in spin-weighted spherical harmonics ${}_{-2}Y_{\ell,m}$ of spin weight $-2$
\cite{Goldberg:1966uu}.  Let $\Psi_4^{(\ell,m)}$ be the mode component
with $\ell\geq 2$ and $-m\leq \ell \leq m$.  The $(\ell,m)$ component
of the news function $\mathcal{N}^{(\ell,m)}$ and its polarizations $\mathcal{N}_{+,\times}$ are defined as
\cite{Bondi:1962px}
\begin{equation}
	\mathcal{N}^{(\ell,m)}(u) = \mathcal{N}^{(\ell,m)}_+ + i\mathcal{N}^{(\ell,m)}_\times = \int_{-\infty}^u\Psi_4^{(\ell,m)}\,du\,.\label{news}
\end{equation}
The outgoing energy flux is related to the integral of $|\mathcal{N}|^2$ over 
all angles.  In a numerical spacetime, it is in
principle possible to extract \cite{Babiuc:2008qy} or evolve $\Psi_4$ going out all the way to
$\mathcal{I}^+$ \cite{SpECTRE_CCE_2021} to reduce systematic errors and gauge changes. We shall however follow the common approach of
calculating $\Psi_4$ on a sphere at a finite radial coordinate $r$ and then extrapolate it to $\mathcal{I}^{+}$ in post-processing using Python. We will now describe this procedure.

First, we extrapolate the waveform to $\mathcal{I}^{+}$ using an improved second-order perturbative procedure that also corrects for the angular momentum of the system
up to first order detailed in   29 of \cite{Nakano2015}. All time integrations are handled using fixed frequency integration \cite{Reisswig_2011}, and differentiations using an 11-point finite difference stencil. For all the simulations, we use the gravitational waveform extracted at coordinate spheres of radius $r = 500M$ from the centre of the spatial domain for extrapolation. 

Second, we transform the waveform to the centre of mass frame. We compute the coordinate centre of mass motion of the BBH system (see \cite{CBW_waveform_correction_2019} for details). We then use the methods described in \cite{transf_asym_wavef_2016} it to translate and boost the gravitational waveform $\Psi_4$ that had been extrapolated to $\mathcal{I}^{+}$. In particular, for translation, and transformation we retain terms unto the leading order. These transformations are implemented using a staggered angular grid on a sphere with a size of ($120 \times 240$). 

After these transformations, our waveforms are found to agree quite well with those of the RIT catalogue \cite{RITcatalog:web}, with phase evolutions deviating at most by $1 \%$ through the inspiral phase. The numerical simulations themselves agree quite well with those in the RIT catalogue e.g. the merger time discrepancies are less than a few per cent.

Given the waveforms in the center of the mass frame at $\mathcal{I}^+$, the integral in the previous equation is carried out over the simulation time $t/M$ instead of the
retarded time coordinate $u$. The lower limit in the integral is not
$-\infty$ but the earliest time available in the simulation, after discarding the initial junk radiation phase.  

The news function is then a function of time, starting
from the earliest time available in the simulation. A further time
integration of $\mathcal{N}$ yields the gravitational wave strain.

\subsection{Dynamics at $\mathcal{H}$}

On the black hole, the basic object here is a marginally outer trapped surface (MOTS), again denoted by  $\mathcal{S}$. This is a closed space-like
2-surface with vanishing outgoing expansion $\Theta_{+}$: 
\begin{equation}
	\Theta_{+} = q^{ab}\nabla_an^+_b = 0\,.
\end{equation}
Its scalar curvature is denoted by $\tilde{\mathcal{R}}$, the two-metric by $\tilde{q}_{ab}$, and its extrinsic curvature by $\tilde{K}_{ab}$.
The shear of the dynamical horizon is denoted by $\sigma = x^a \bar{x}^b \nabla_{a} n^+_b$. 

At $\mathcal{H}$, there are two sets of moments, the mass and current or angular momentum multipoles:  $\mathcal{M}_{lm}$, $\mathcal{J}_{lm}$, although defined quasi-locally on the foliations of the three-dimensional dynamical horizon, can be used to reconstruct the horizon geometry in a gauge invariant manner. These moments were first defined for isolated horizons \cite{Ashtekar:2004gp} and extended to axisymmetric \cite{Schnetter:2006yt} and non-axisymmetric dynamical horizons \cite{Ashtekar:2013qta}. These multipole moments can be used to study the intrinsic geometry of dynamical horizons. They have been used in predictions of the anti-kick in binary black hole mergers \cite{Rezzolla:2010df}, in the study of the no-hair conjecture in general astrophysical environments \cite{Gurlebeck:2015xpa}, and for studying tidal deformations of black holes \cite{Cabero:2014nza, Gurlebeck:2015xpa, prasad2021tidal}. In a more recent work \cite{prasad2021tidal}, the axisymmetric tidal deformations of dynamical horizons were studied by using the source multipole moments of the dynamical horizon in binary black hole scenarios.

We now describe how we compute these moments. Consider a time slice of a numerical simulation in which a MOTS $\eS$ embedded in the three-dimensional spatial hypersurface $\Sigma$ has been located. First, a preferred coordinate system on the leaves $\eS$ of the dynamical horizon is constructed using an appropriately defined axial vector field $\varphi^a$. The method of Killing transport is used to compute the axial field \cite{Ashtekar:2004gp, Schnetter:2006yt}, as it is readily available. This method finds an approximate symmetry field on the MOTS from an initial guess. The field coincides with the direction field $\phi$ of the Boyer-Lindquist co-ordinates when the horizon is isolated e.g. when the separation between the black holes is large compared to their horizon sizes. Thus, even though this vector field $\varphi$ is not an exact Killing field on the MOTS, it furnishes a field that can be used to establish a coordinate system on the dynamical horizon.  

Once the axial field is defined and computed, an invariant coordinate $\zeta$ (analogous to the polar coordinate variable $\cos{(\theta)}$ of the Boyer-Lindquist coordinates) can be defined as:
\begin{equation}
	\tilde{D} \zeta = \dfrac{1}{R_S ^2} \tilde{\epsilon} \varphi^a
\end{equation}
Here $\tilde{\mathcal{D}}$ is the derivative operator and $\tilde{\epsilon}_{ab}$ the volume form on $\mathcal{S}$ compatible with the two-metric $\tilde{q}$, and $R_{\mathcal{S}}$ is the areal radius of $\mathcal{S}$. The freedom to add a constant to $\zeta$ is removed by requiring that $\oint_{\mathcal{S}} \zeta d^2 S =0$. In order to construct the source multipole moments, one then needs to construct a harmonic basis on each $\mathcal{S}$. This is defined by effectively mapping $\mathcal{S}$ to a unit sphere, and using the spherical harmonic functions $\mathcal{Y}_{\ell m}$ of that unit sphere whose axial field is $\varphi$ and polar angle $\cos(\theta) = \zeta$. Using these, one can define the mass multipole moments as:
\begin{equation}
	\mathcal{M}_{\ell m} = \frac{M_{\mathcal{S}}R_{\mathcal{S}}^l}{8\pi} \sqrt{(2l+1) \dfrac{(\ell-m)!}{(\ell+m)!}} \oint_S \Tilde{\mathcal{R}} \mathcal{Y}_{\ell m} d^2S\,, \label{mass_mult_lm}
\end{equation}
Similarly, the spin-multipole moments can be defined using the same procedure:
\begin{equation}
	\mathcal{J}_{\ell m} = \frac{R_{\mathcal{S}}^{\ell + 1}}{8\pi} \sqrt{(2\ell+1) \dfrac{(\ell-m)!}{(\ell+m)!}} \oint_S   K_{bc} \varphi^b r^c \, _{-1}\mathcal{Y}_{\ell m} d^2S\,, \label{spin_mult_lm}
\end{equation}
Expressed in functional form, $\mathcal{Y}_{\ell m}(\zeta, \phi) = P^m _\ell(\zeta) e^{-im\phi}$. Here $K_{ab}$ is the extrinsic curvature of $\Sigma$, $P^m _\ell$ are the associated Legendre polynomials corresponding to the eigenfunctions of the Laplacian on the unit sphere described above, and $\phi$ is an affine coordinate on the integral curves of the vector field $\varphi^a$. $_{-1}\mathcal{Y}_{\ell m}$ denotes harmonics of spin-weight $-1$. In this notation, $\mathcal{M}_{00}$ is the mass $M_{\mathcal{S}}$ of the slices $\mathcal{S}$ of the dynamical horizon, and $\mathcal{J}_{10}$ is its angular momentum $J_{\mathcal{S}}$. $n!$ denotes the factorial of an integer $n$. These moments are defined for all positive $\ell$, and for each $\ell$, the magnetic mode number $m$ takes integer values ranging from $(-\ell, \ell)$. It is to be noted that modes for which $m\neq0$ are complex in general and we denote their strength as absolute magnitude or a quadratic sum $\vert \mathcal{M}_{\ell m} \vert = \sqrt{Re(\mathcal{M}_{\ell m})^2 + Im(\mathcal{M}_{\ell m})^2}$. We will only study the mass-multipole in this chapter, and address the spin-multipoles in a later work.

The choice of normalization of the spherical harmonics are as follows. For a given mode $(\ell, m)$),
\begin{equation}
	\oint \mathcal{Y}_{\ell m} \mathcal{Y}_{\ell m} ^* d^2 S= 4\pi  
\end{equation}
We have not included the Condon-Shortley phase in the definition of our harmonics. 

Given that these are not exact eigenfunctions of the Laplacian operator on $\mathcal{S}$, the harmonics may not exactly be orthogonal to each other. i.e. for modes $(\ell, m) \neq (\ell^{\prime}, m^\prime)$,
\begin{equation}
	\oint \mathcal{Y}_{\ell m} \mathcal{Y}_{\ell^\prime m^\prime} ^* \neq 0 
\end{equation}
However, we found that the departure from orthogonality was minimal of the order of $\approxeq 10^{-3}$, and can be safely neglected for the purposes of this study. As a simplifying assumption, we use the grid coordinate angle $\phi$ in place of the affine coordinate on the integral curves of $\varphi$. We refer the reader to \ref{chap:nr_sims} for further details.

In this work, we compute and study the evolution of the mass dipole ($\ell=1$) and the quadrupole ($\ell=2$) moments, $\vert m \vert \leq \ell$ moments of the individual dynamical horizons of the black holes during the inspiral phase.
We also study their relationship with the dynamics of the binary black hole system and the gravitational news function at $\mathcal{I}^+$.

\section{The numerical simulations}
\label{sec:nr_gen_mult} 

Numerical simulations were performed and used, whose details can be found in \ref{chap:nr_sims}.

General quasi-local physical quantities are computed on the horizons following 
\cite{Dreyer:2002mx, Schnetter:2006yt}. The framework to compute the generalized multipole moments defined in  \eqref{mass_mult_lm} does not exist in the QuasiLocalMeasures thorn of the Einstein Toolkit. The new definitions in  \eqref{mass_mult_lm} were implemented and computed using a post-processing script developed in Python. Using numerical relativity data as inputs, we compute the generalized source multipole moments. The axisymmetric source multipole moments were cross-checked against those computed in the ETK. 

We consider non-spinning binary black hole systems with varying mass ratio 
$q=M_2/M_1$. We chose 3 non-spinning cases with mass-ratios $q=0.6, 0.7, 1$ for the purposes of this study. For computing the quasi-local quantities, we use a grid of size ($36, 74$) on each of the dynamical horizons. This grid resolution allows us to safely study multipole moments of up to $\ell=2$. We also carry out a convergence test of the horizon data by running the simulation at two different grid resolutions of the horizon. 

Our simulations agree very well with the catalogue simulations \cite{RITcatalog:web}, with merger time discrepancies of less than a few per cent.  The results presented here are general features that were seen across the simulations $q=1, 0.6$, and $0.7$ described in the \ref{tab:ic_qc0_gen_mul}. We present the results using $q=0.6$. The outer common horizon of the configuration $q=0.6$ appears at $t=1656.045M$, which we designate as merger time. When the common horizon is found, it has an areal radius of $R_{c} = 1.708$. 3D visualizations were performed using python and \emph{VisIt} \cite{VisIt}.
\begin{table}
	\begin{center}
		\begin{tabular}{|p{2cm}|p{2cm}|p{2cm}|p{2cm}|}
			\hline
			$q $ & $D/M$ & $p_r/M$& $p_{\phi}/M$\\
			\hline
			1   &   9.5332  &  0            &   0.09932\\
			0.6 &   11.5    &  -5.46e-04    &   0.08206\\
			0.7 &   12.0    &  -5.07e-04    &   0.08246\\
			\hline
		\end{tabular}
		\caption[Initial parameters of the NR simulations used in the generalized multipole moment studies.]{Initial parameters for non-spinning binary black holes with quasi-circular orbits. $q=M_2/M_1$ is a mass ratio, $D$ is the initial separation between the two holes, and $p_r$ and $p_{\phi}$ are the linear momenta in the radial and azimuthal directions respectively.}
		\label{tab:ic_qc0_gen_mul}
	\end{center}
\end{table}

\section{Results}

\subsection{General behaviour}
\begin{figure}[htp]
	\includegraphics[width=\columnwidth]{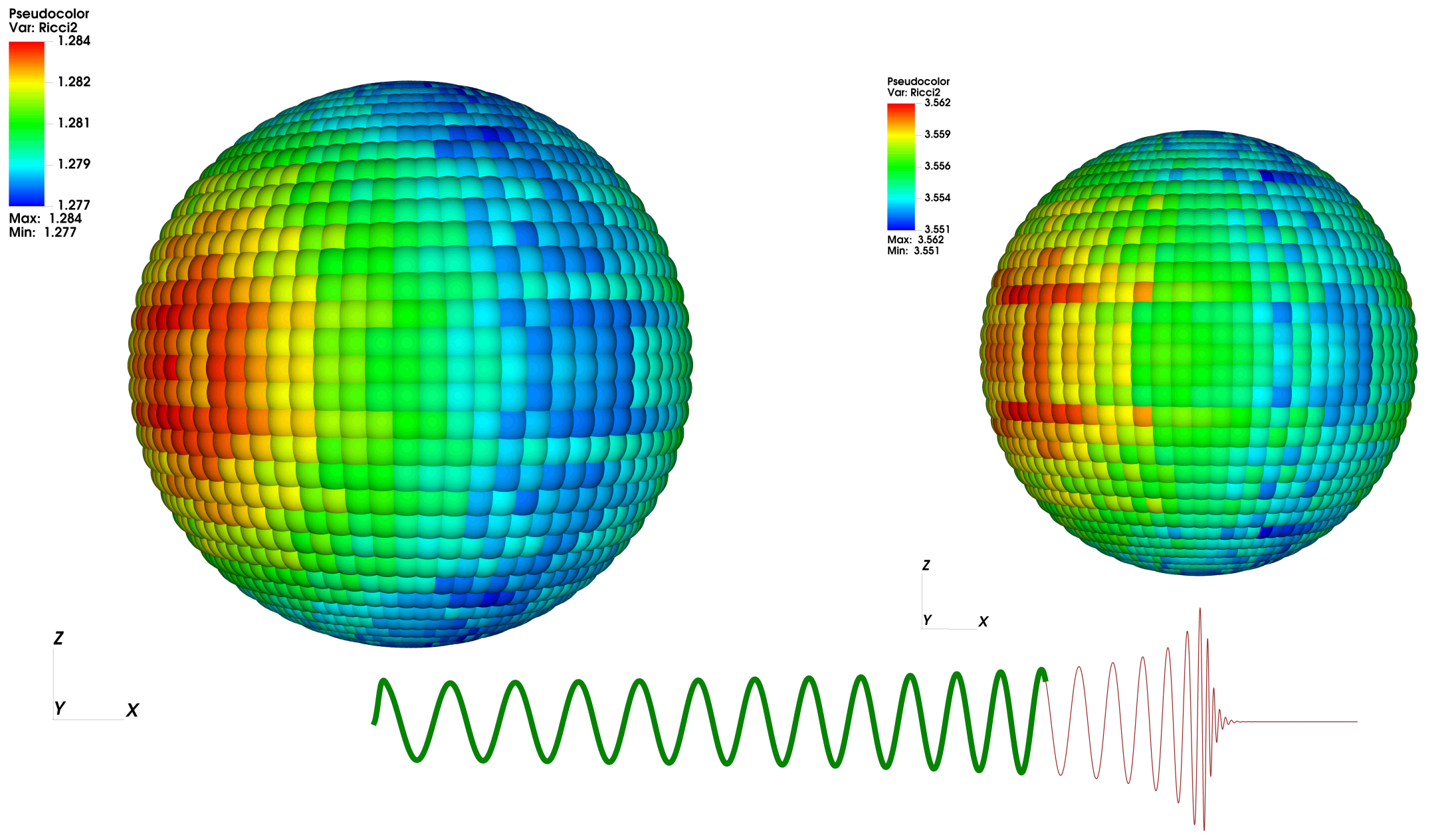}
	\caption[The tidal deformation of the geometries of the DHs in a BBH system.]{The deformation of the dynamical horizons of the black holes can be directly visualized in terms of the 2-Ricci scalar $\Tilde{\mathcal{R}}$ of their two-dimensional slices $\mathcal{S}$. Here the 2-Ricci scalars of $\mathcal{S}$ are visualized for the $q=0.6$ system when the black holes are at a separation of $d\approx 7.75M$ (about 67.4\% of the initial separation), approximately 5 orbits after the start of the simulation, as shown by the thick green line in the waveform plot below the figure. The total number of orbits before the merger is around $9$ and the corresponding complete waveform cycles in the simulation are shown by the thin red line. The more massive black hole BH1 is on the left. The values of the Ricci scalar are shown on the colour bars to the left of each black hole. A movie for $q=0.6$ can be viewed \href{https://drive.google.com/file/d/10Tc3itNioIBwhJnc1fRp2KV_L1UbmRGm/view?usp=sharing}{\color{blue} here} ~\cite{td_movie:2021}. This movie shows the mutual deformations of the leaves $\eS$ of the individual dynamical horizons as the system evolves in time. Importantly, it shows that the deformation of the horizon geometry is mutual for both the horizons and the Ricci scalar distribution patterns approximately face each other at all points on the orbit. The dominant quadrupolar structure can be seen, which is reflected in the numerical values of the strengths of the multipole moments in \ref{fig:Mlm_time_q0p6}, \ref{fig:Mlm_time_q0p7} and \ref{fig:Mlm_time_q1}.}
	\label{fig:R2_bh12_q0p6}
\end{figure}

\begin{figure}[htp]
	\includegraphics[width=\columnwidth]{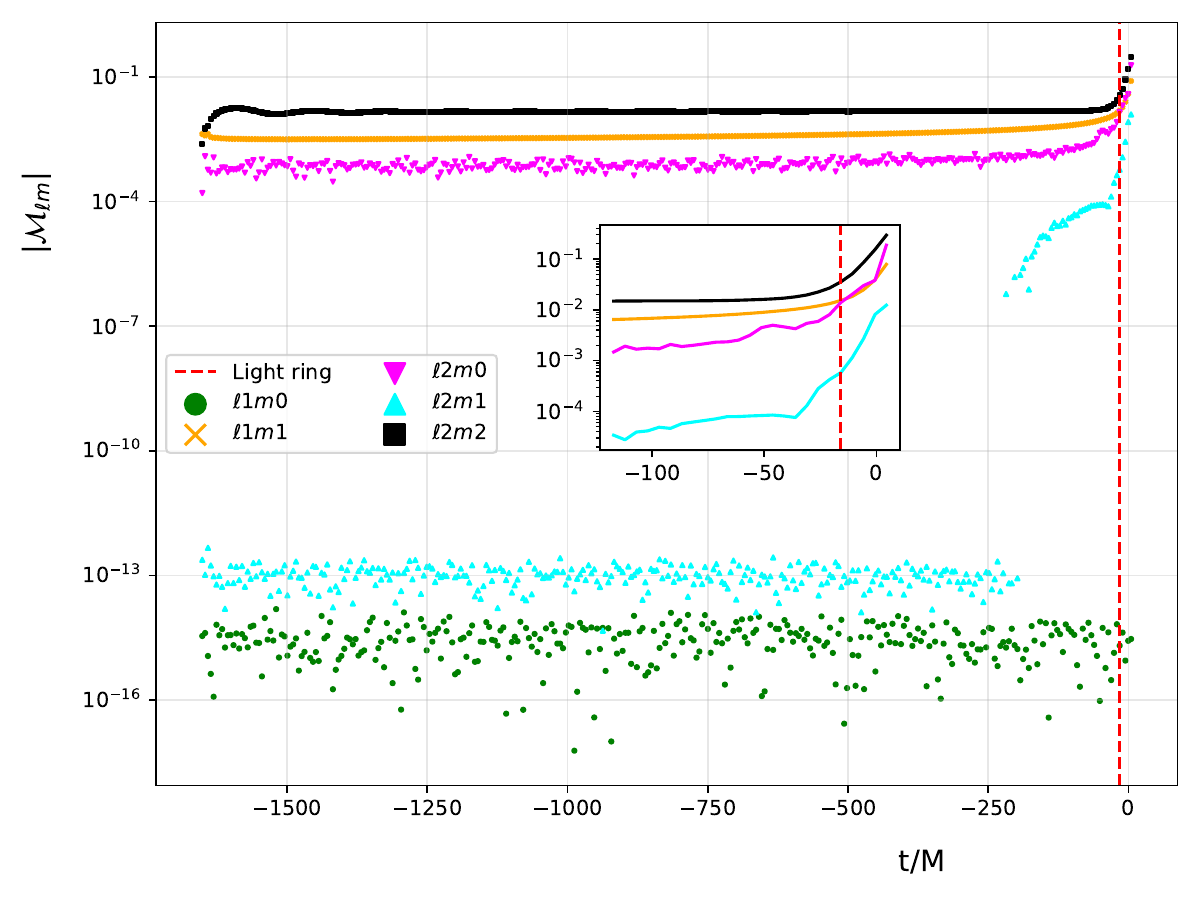}
	\includegraphics[width=\columnwidth]{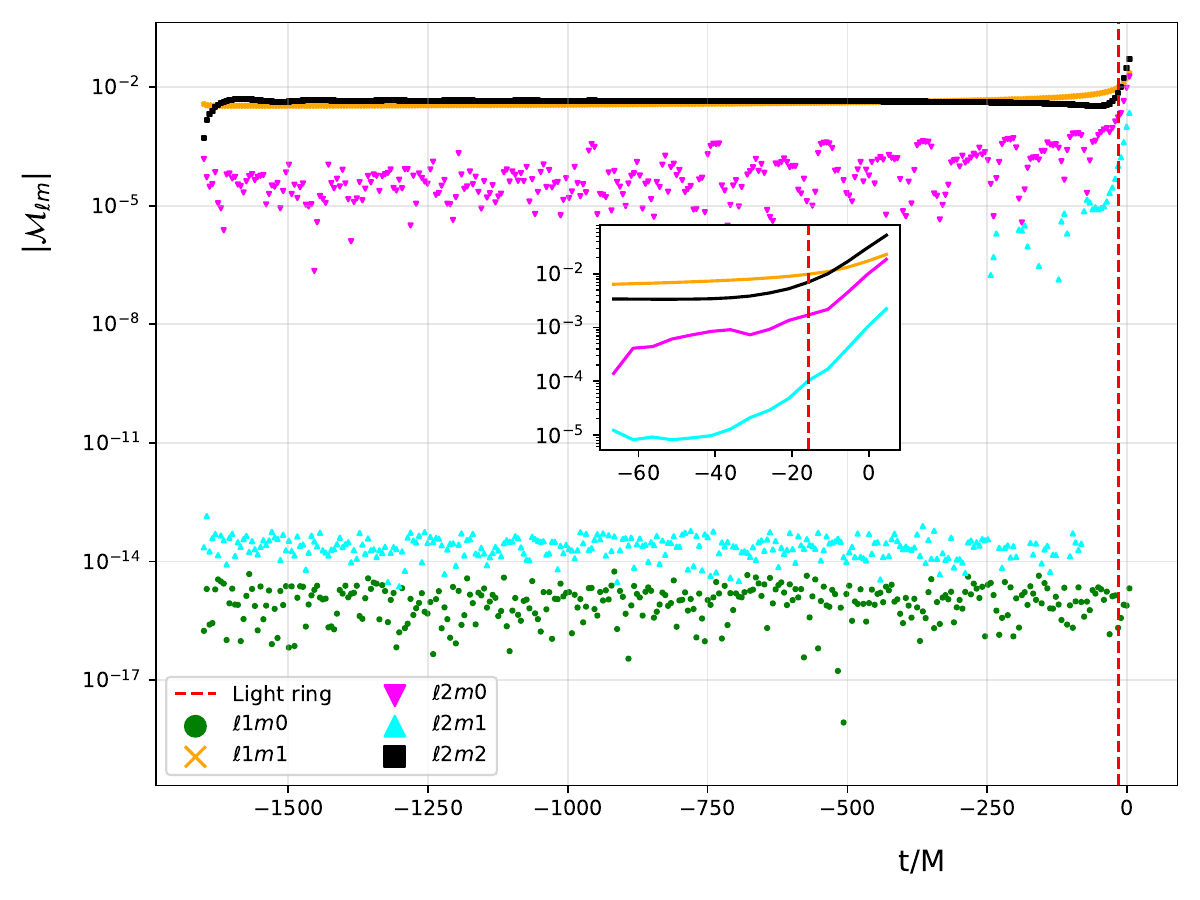}\\
	\caption[The time evolution of the strengths of the various multipolar components of the deformation of the horizon geometry of BHs for $q=0.6$ BBH system.]{The time evolution of the strengths (absolute magnitudes) of the multipole moments $\ell=1, \ell=2$ $\vert m \vert < \ell$ for the larger black hole (BH1, left) and smaller black hole (BH2, right), for mass ratio $q=0.6$ system. Here, the values below $10^{-8}$ depict the numerical noise floors of the respective moments. The time of crossing of the light ring of the system is denoted as a dotted line in red.}
	\label{fig:Mlm_time_q0p6}
\end{figure}

\begin{figure}[htp]
	\includegraphics[width=\columnwidth]{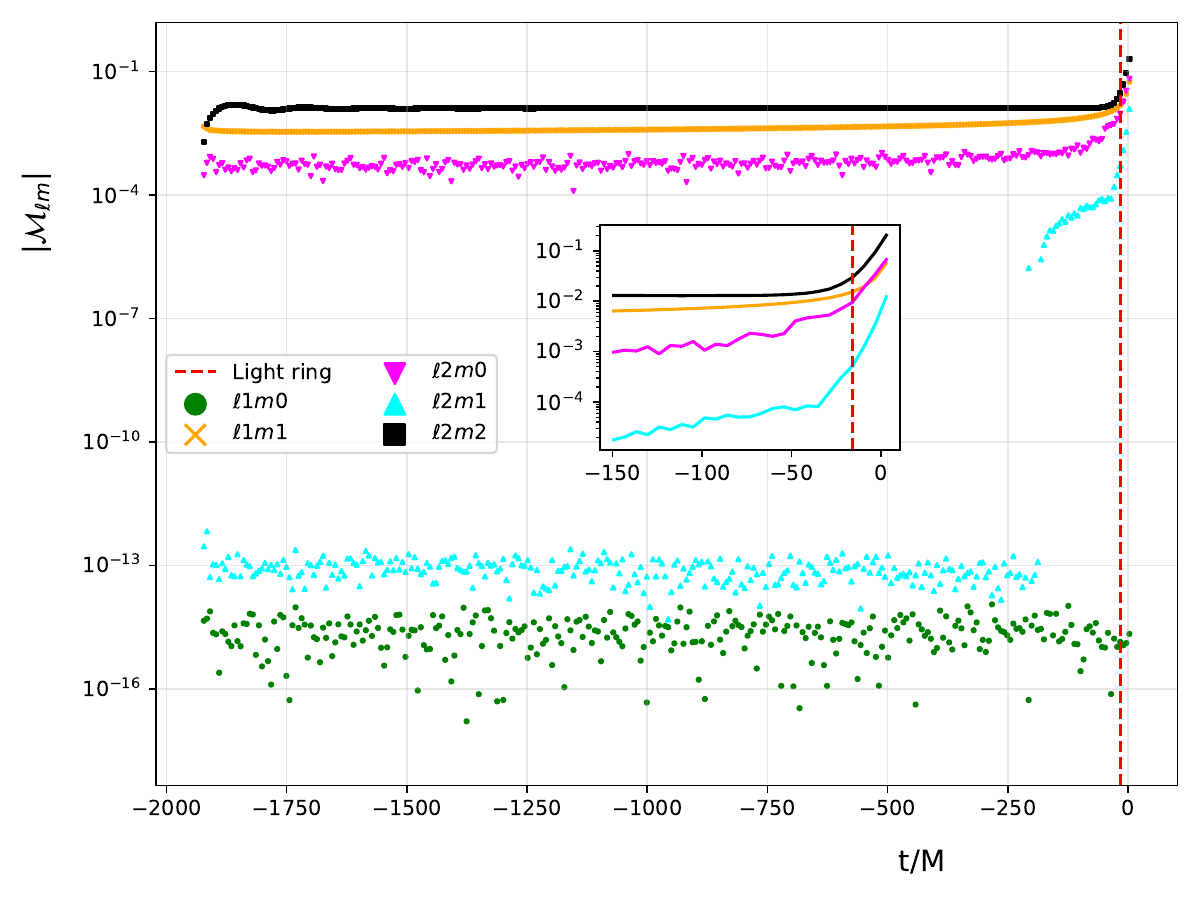}
	\includegraphics[width=\columnwidth]{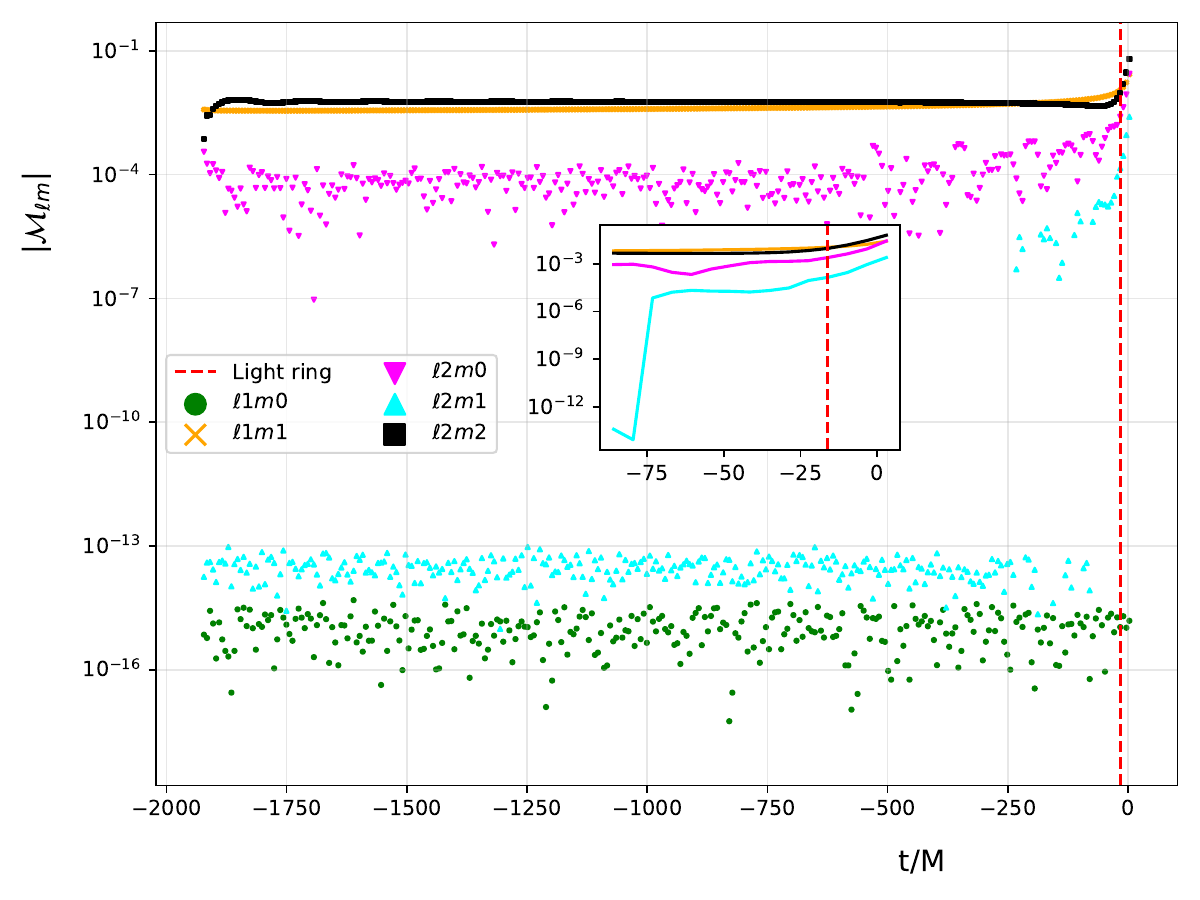}\\
	\caption[The time evolution of the strengths of the various multipolar components of the deformation of the horizon geometry of BHs for $q=0.7$ BBH system.]{The time evolution of the strengths (absolute magnitudes) of the multipole moments $\ell=1, \ell=2$ $\vert m \vert < \ell$ for the larger black hole (BH1, left) and smaller black hole (BH2, right), for mass ratio  $q=0.7$ system. Here, the values below $10^{-8}$ depict the numerical noise floors of the respective moments. The time of crossing of the light ring of the system is denoted as a dotted line in red.}
	\label{fig:Mlm_time_q0p7}
\end{figure}

\begin{figure}[htp]
	\includegraphics[width=\columnwidth]{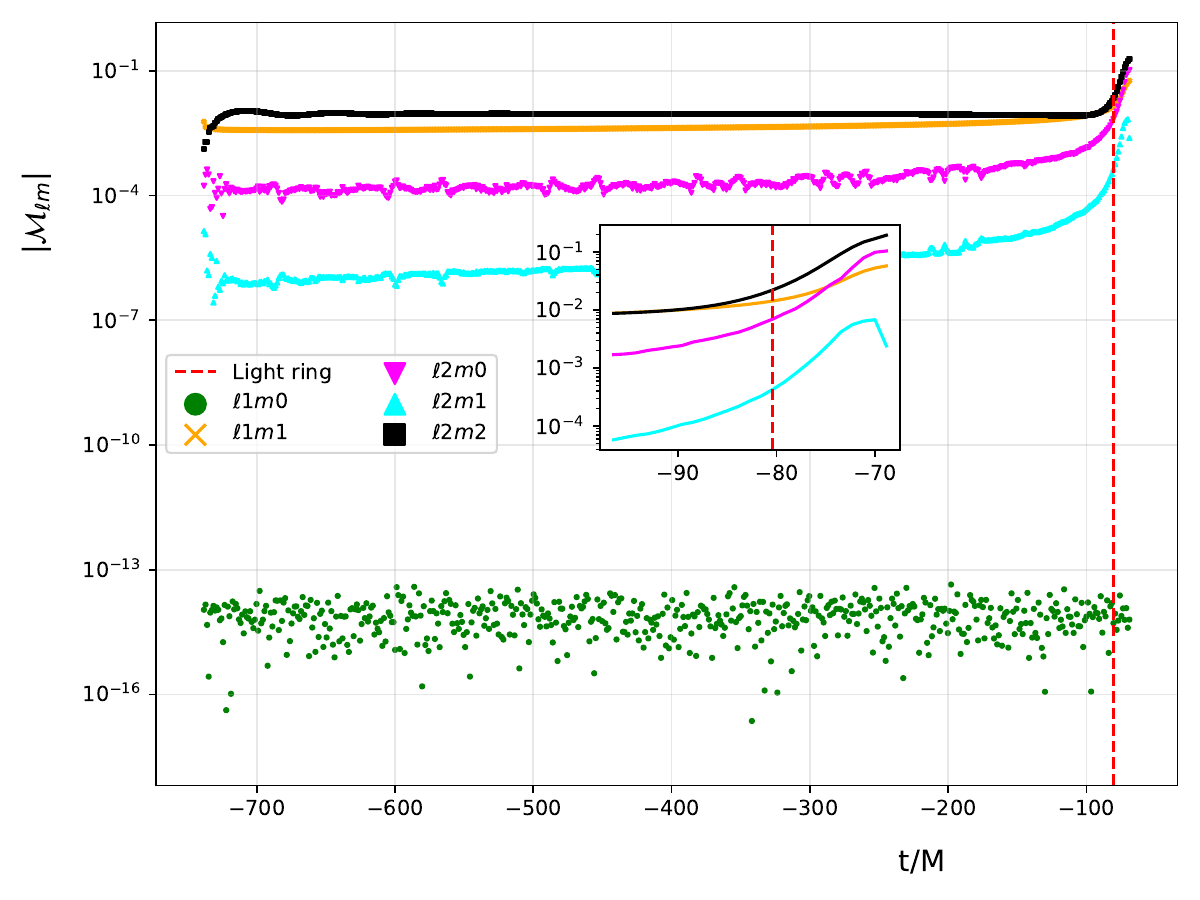}
	\includegraphics[width=\columnwidth]{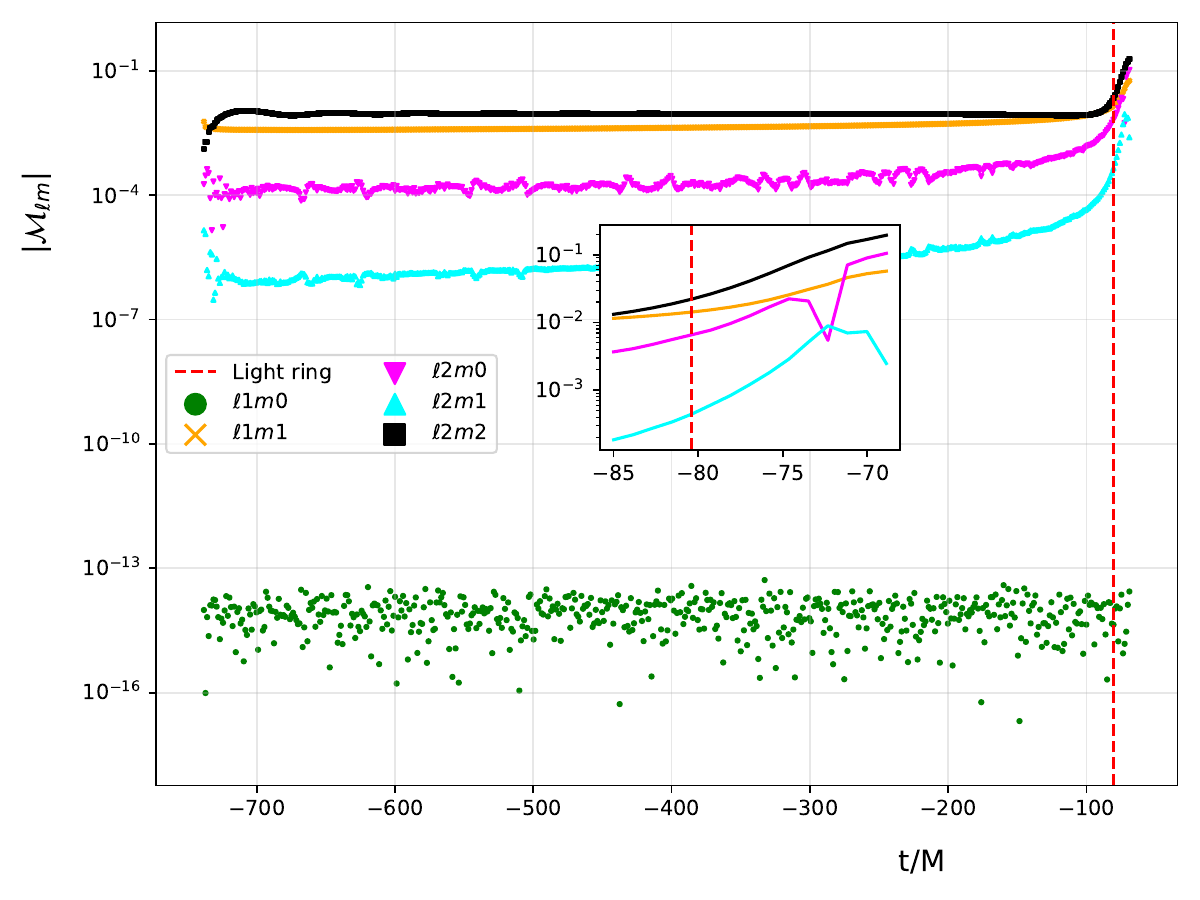}
	\caption[The time evolution of the strengths of the various multipolar components of the deformation of the horizon geometry of BHs for $q=1$ BBH system.]{The time evolution of the strengths (absolute magnitudes) of the multipole moments $\ell=1, \ell=2$ $\vert m \vert < \ell$ for the larger black hole (BH1, left) and smaller black hole (BH2, right), for mass ratio $q=1$ system. Here, the values below $10^{-8}$ depict the numerical noise floors of the respective moments. The time of crossing of the light ring of the system is denoted as a dotted line in red.}
	\label{fig:Mlm_time_q1}
\end{figure}

We first discuss the relative strengths of the various multipole moments $\vert \mathcal{M}_{\ell m}\vert$. For both the black holes of all the simulations, amongst all the moments at $\ell=1, 2$ multipolar order, the multipole moment $\mathcal{M}_{22}$ was found to have the largest and monotonically increasing amplitude in the inspiral phase, followed by $\mathcal{M}_{20}$ and $\mathcal{M}_{1\pm 1}$ (note that for an isolated Kerr horizon, apart from its mass, only $\ell=2$, $m=0$ mass moment is non-zero at the $\ell=2$ quadrupolar order). This can be seen in the 3D \ref{fig:R2_bh12_q0p6}, the time evolution of the strengths of the multipole moments in \ref{fig:Mlm_time_q0p6}, \ref{fig:Mlm_time_q0p6}, and \ref{fig:Mlm_time_q0p6}, and the \href{https://drive.google.com/file/d/10Tc3itNioIBwhJnc1fRp2KV_L1UbmRGm/view?usp=sharing}{\color{blue} movie} \cite{td_movie:2021} in which the evolution of the 2D-Ricci scalars of the dynamical horizons in the inspiral phase have been visualized. In \ref{fig:R2_bh12_q0p6}, a snapshot of the movie is presented. The multipolar deformations of the horizon geometries are mutual and are dependent on the location of the black holes in the binary system. The quadrupolar i.e. $\ell=2, m=2$ pattern can be clearly seen in the movie.

\begin{figure}[htp]
	\includegraphics[width=\columnwidth]{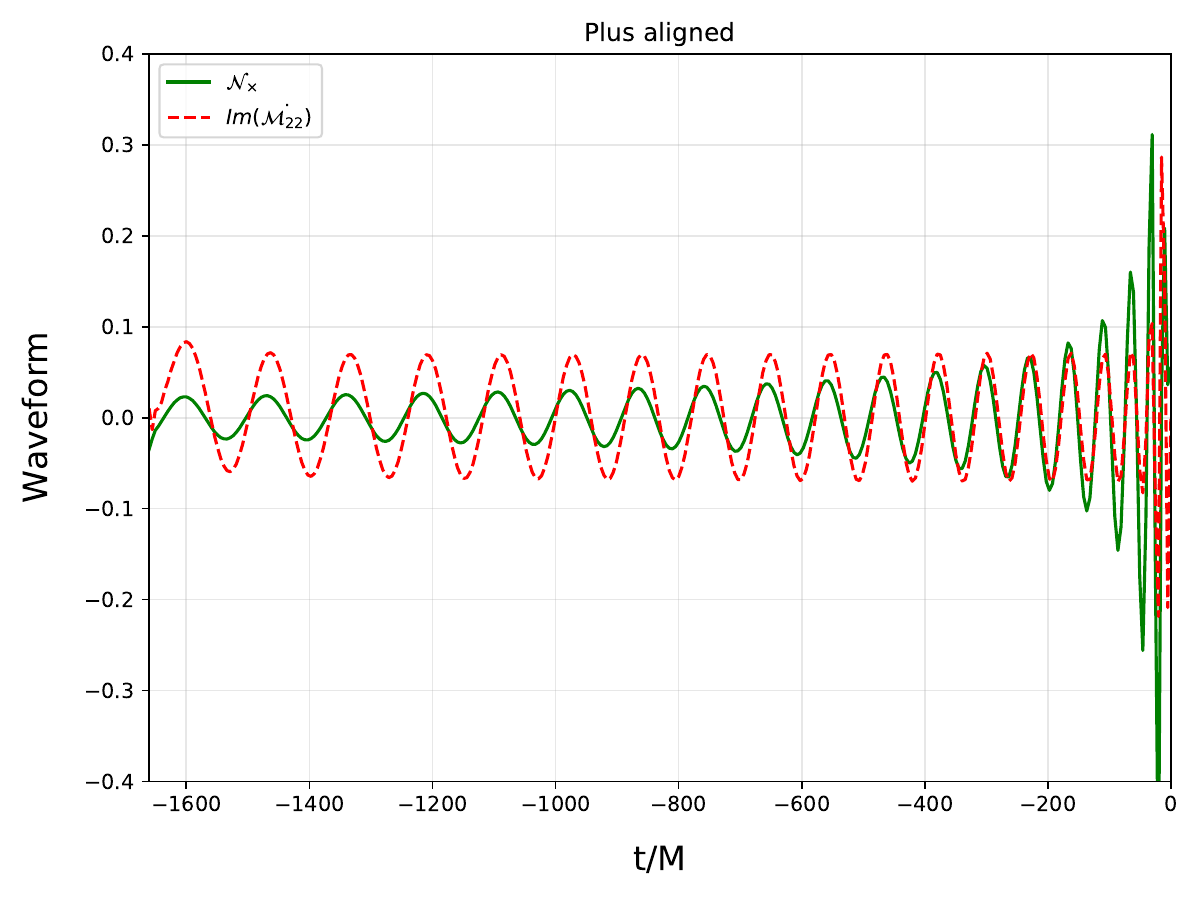}
	\includegraphics[width=\columnwidth]{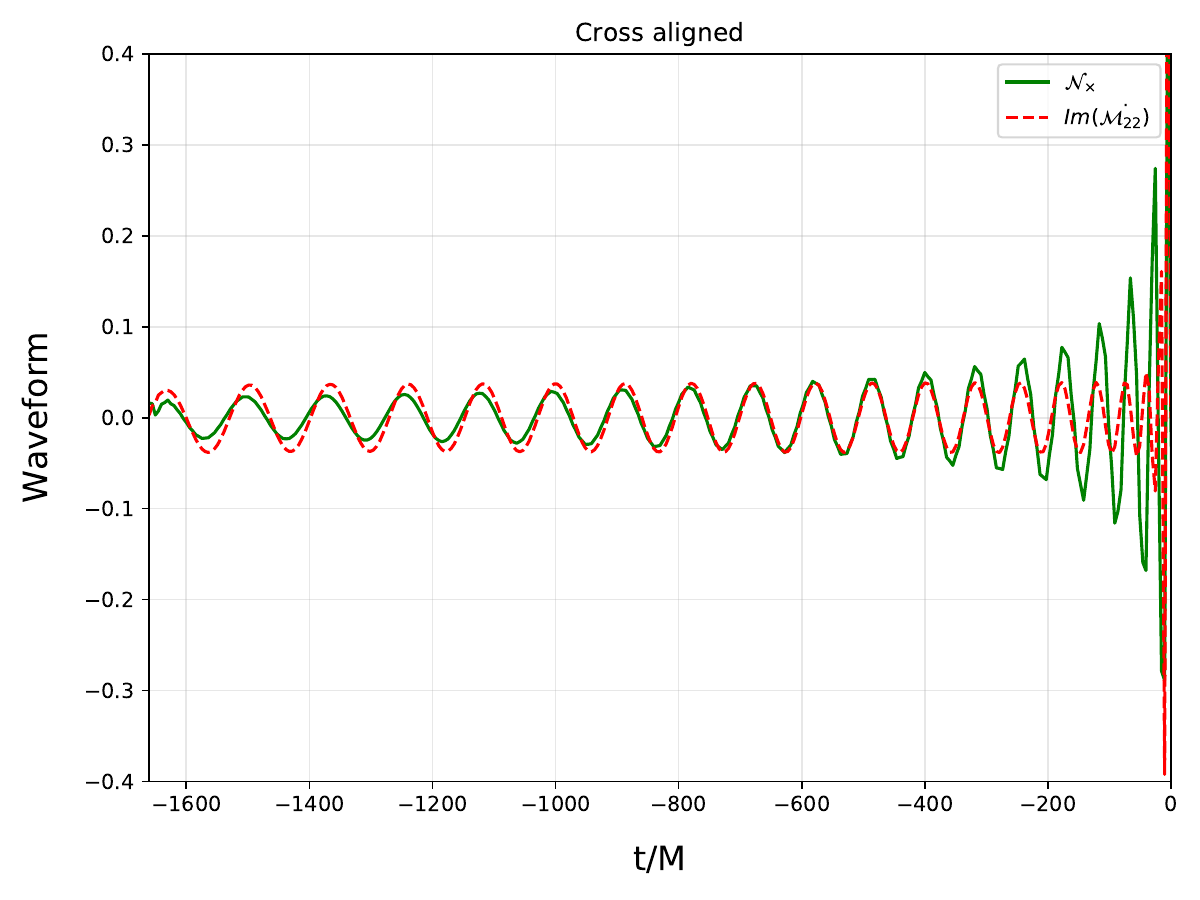}
	\caption[Strong correlations between the geometry of the deformed horizons and numerical news from the system.]{The co-aligned $\ell=2, m=2$ quadrupole moment and news waveforms. The top and the bottom panels are for the real and imaginary parts of the waveforms.  Note that the phasing of the multipole moment closely follows that of the gravitational news waveform.}
	\label{fig:M22_px_news}
\end{figure}

\begin{figure}[htp]
	\includegraphics[width=\columnwidth]{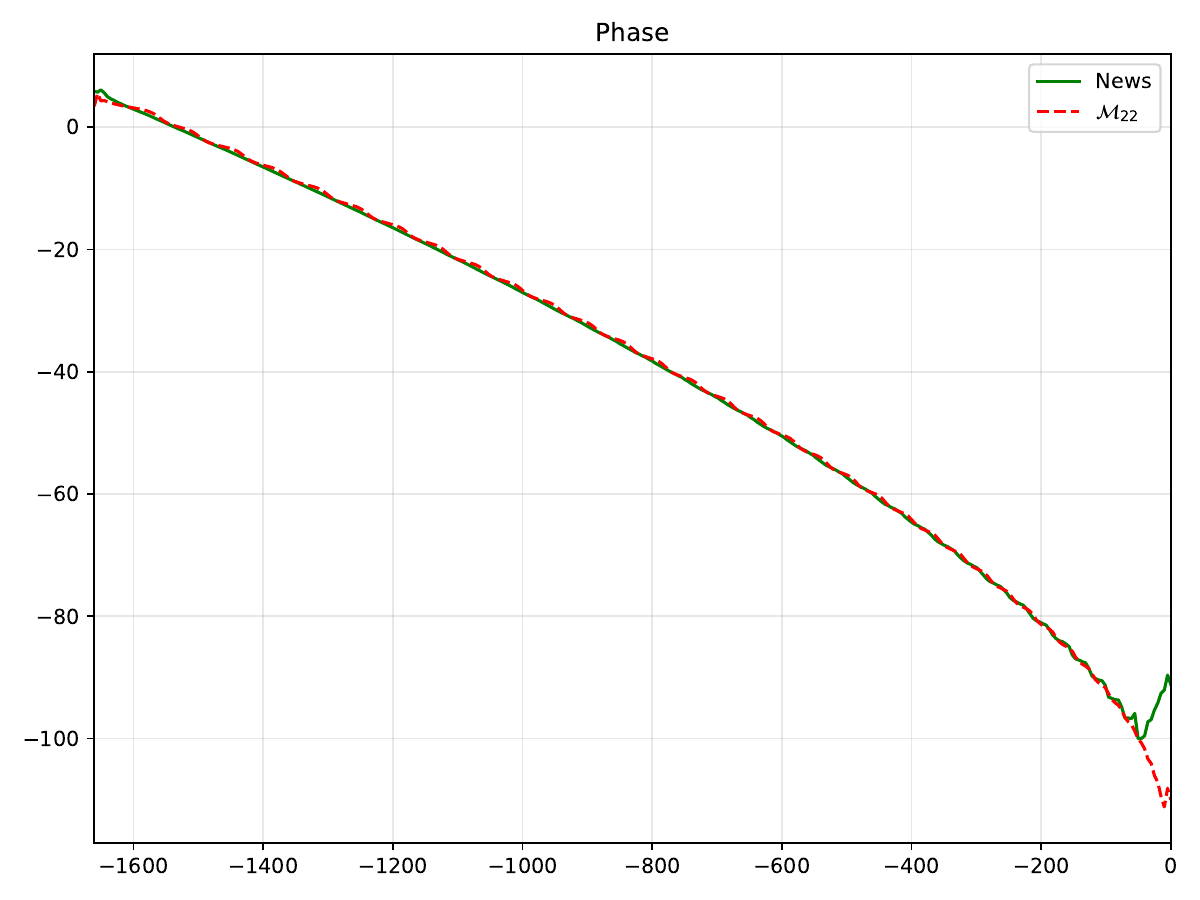}
	\includegraphics[width=\columnwidth]{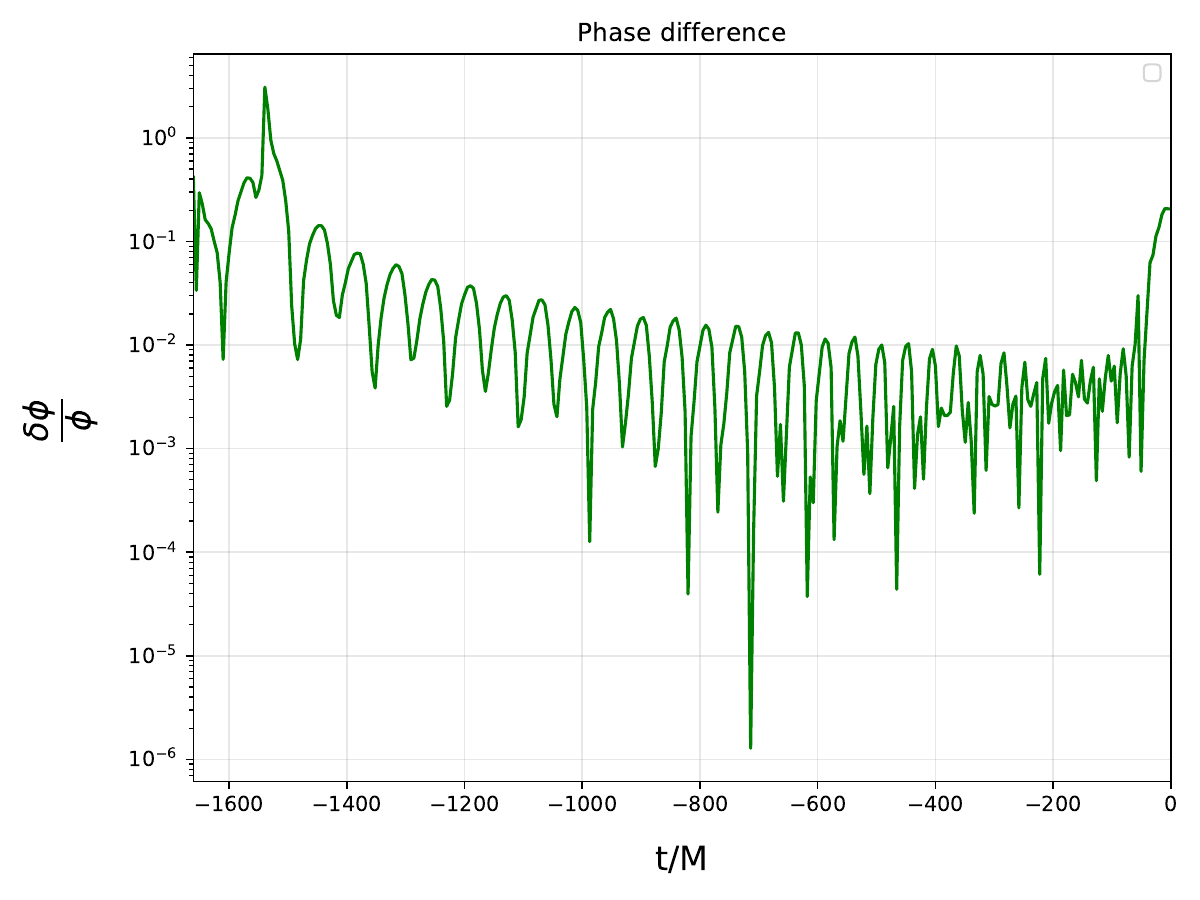}
	\caption[The phase evolutions of the horizon deformation and the numerical news from the system.]{The phases of the $\ell=2, m=2$ multipole moment and the news shown above in \ref{fig:M22_px_news}. On the top panel, we plot the phases and on the bottom we plot the fractional deviation of the phase of the multipole moment from that of the news. This shows that the multipole moment and the news are in phase, up to less than a few per cent.}
	\label{fig:M22_phase_news}
\end{figure}

\begin{figure}[htp]
	\includegraphics[width=\columnwidth]{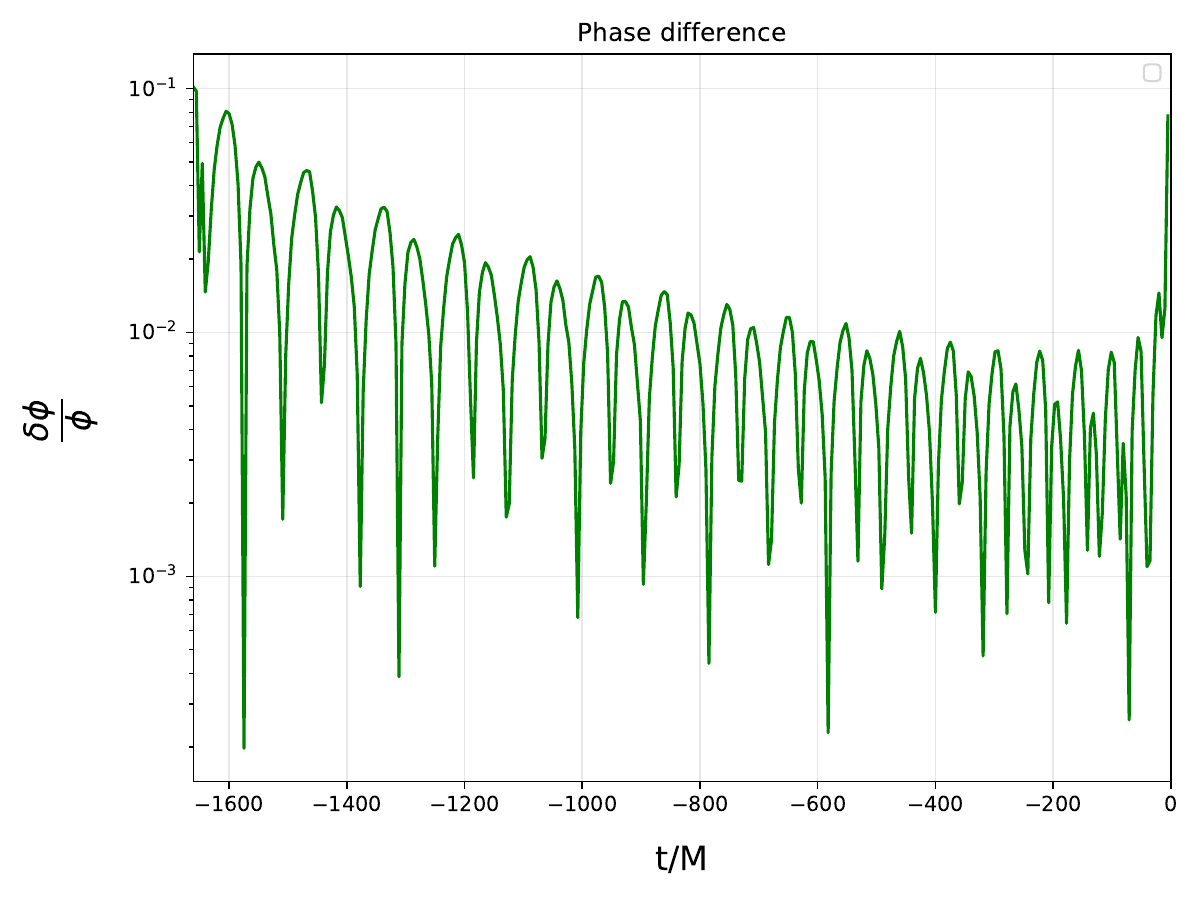}
	\caption[Strong correlations between the orbital dynamics and the dipolar horizon deformations.]{The fractional deviation of the phasing of the dipolar mode $\mathcal{M}_{11}$ from the orbital phasing of the binary system.}
	\label{fig:M11_phase_orb}
\end{figure}

The non-axisymmetric multipole moments (i.e., $m\neq 0$) of the individual dynamical horizons of the black holes are oscillatory in nature. It was found that the multipole moments $\mathcal{M}_{22}$ of the two black holes were in-phase with each other while the moments $\mathcal{M}_{1\pm1}$ differed by a phase of $\pi$ radians.

\subsection{Relationship with the orbital dynamics and $\mathcal{I}^+$}
We found the multipole moments of dynamical horizons of the two black holes to be strongly correlated with each other. Furthermore, the dominant multipole moment $\mathcal{M}_{22}$ of the dynamical horizons is found to be strongly correlated with the dominant $(\ell=2, m=2)$ mode of the gravitational wave strain extracted at a very large distance $r=100M$ from the system. The movie \cite{td_movie:2021} aids in the visualization of some of these results. 

In \ref{fig:M22_px_news}, we plot the strongest quadrupolar multipole moment $\mathcal{M}_{22}$ vs. the news function of the gravitational waves emitted from the system (\ref{news}), suitably normalised and aligned in time and phase. The timeshift was found to be $\approx 501M$, approximately consistent with the light travel time corresponding to the extraction radius for the news. In \ref{fig:M22_phase_news}, we compare the phasing of the multipole moment against the news. In \ref{fig:M11_phase_orb}, we compare the phasing of the dipole moment with the orbital phasing of the system. The orbital phasing was obtained by using the centroids of the apparent horizons of the two black holes. It was found that the quadrupole mass moment $\mathcal{M}_{22}$ encodes accurate information about the phasing of the gravitational waveform from the system, whereas the dipole moment $\mathcal{M}_{1\pm 1}$ quite accurately reflects the orbital phasing of the system. Thus like with the waveform received at $\mathcal{I}^+$, these multipole moments contain accurate information about the evolution of the binary system.

To demonstrate this further, we attempt to estimate the parameters of the binary system using the source multipole moments $\mathcal{M}_{22}$ of the individual dynamical horizons, and a standard least squares figure of merit, using a procedure detailed in \cite{shear-news2020}. To carry out this, we used a template bank of gravitational wave strain in the mass-ratio, chirp-mass parameter space constructed using the well-known phenomenological waveform model IMRPHenomPv2. We find that the parameters of the binary system could be estimated quite accurately (with an error of $0.12\%$ in the mass ratio and $0.01\%$ in the chirp-mass of the binary system). Therefore the dynamical horizon horizon carries accurate information on the dynamics of the system and can be used to extract information about the binary black hole system like their masses, velocities, orbital angular momentum, etc. These results also show that the source multipole moments are strongly correlated with the multipole moments of the gravitational field at null infinity \cite{Bondi:1962px, Sachs:1962wk, Newman:1962, Janais1965}, bringing out the interesting tight correlations that exist between the strong and weak field dynamics in a BBH merger.
\begin{figure}[htp]
	\includegraphics[width=0.7\columnwidth]{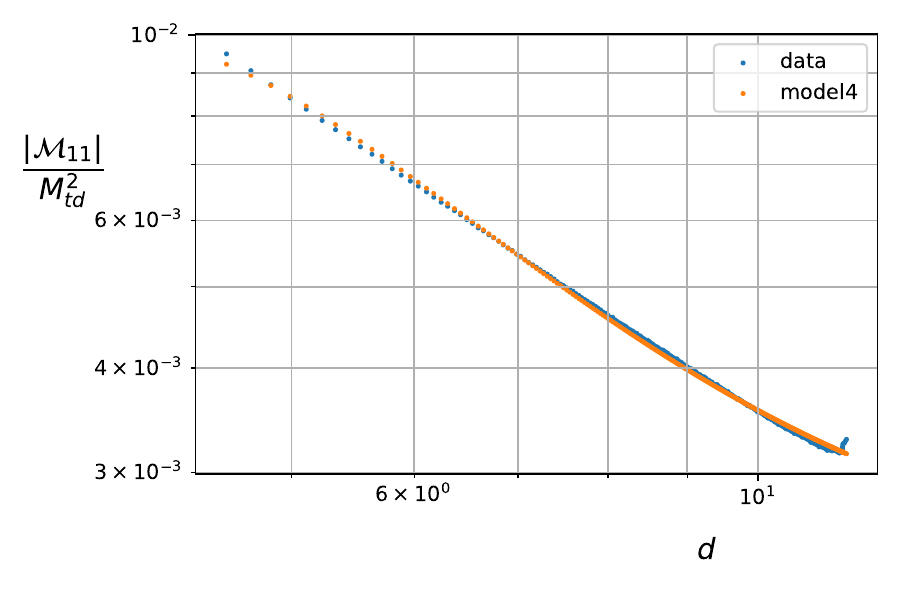}
	\includegraphics[width=0.7\columnwidth]{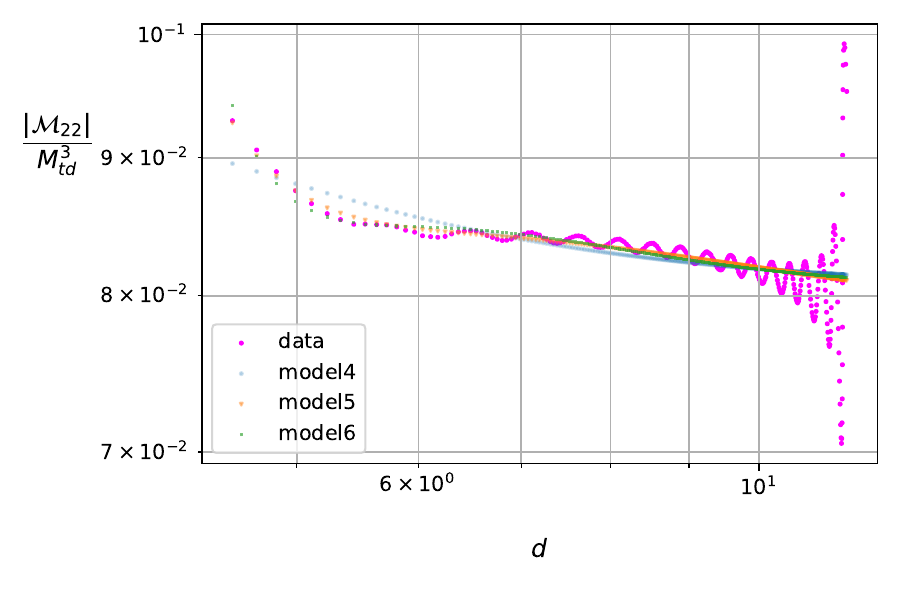}
	\includegraphics[width=0.7\columnwidth]{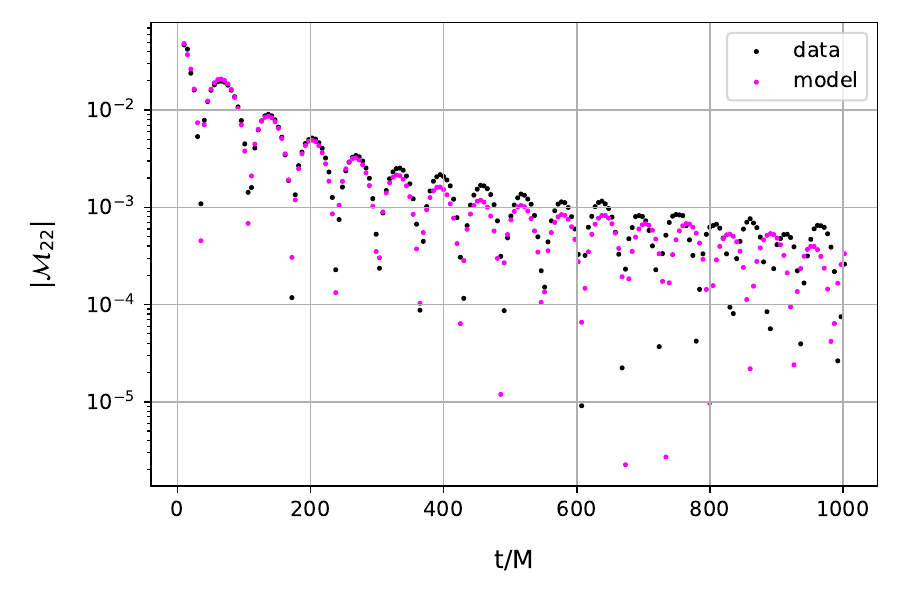}
	\caption[Fit of the generalized source multipole moments to tidal expansions.]{Left: The fit of the $\ell=1, m=\pm 1$ multipole moments to the relation in  \ref{exp_gen_mult}, with two terms at the third and fourth order in the separation $d$, i.e. $1/d^3$ and $1/d^4$. Center:  the fit of the $\ell=2, m=\pm 2$ multipole moments to the relation in  \ref{exp_gen_mult}, with terms upto fourth, fifth and sixth order in $d$. Right: the isolated oscillations of the overall amplitude of the multipole moment $\vert \mathcal{M}_{22} \vert$ seen in the figure at the center. }
	\label{fig:Mlm_dist_osc}
\end{figure}
\subsection{Tidal deformations}
The evolution of the strengths of the multipole moments was also found to display a generic behaviour. By means of maximizing a least-squares figure of merit, we found that the evolution of the multipole moments of both the black holes and across the two simulations, can be described by a generic tidal expansion of the form: 
\begin{equation}
	\dfrac{\delta \lvert \mathcal{M}_{\ell m} \rvert}{M_{\mathcal{H}}^{\ell+1}} = \sum_{i=3} ^{\infty} \dfrac{a_i}{d^i} \label{exp_gen_mult}
\end{equation}
where $M_{\mathcal{H}}$ is the mass of the black hole that is being discussed, and $d$ is a measure of the distance of separation between the holes. We use the coordinate distance measure, which is also close to the physical distance measures. Please see \ref{appendix:dm} for a discussion. In particular, the dipole moment was found to be well described by the above expansion that includes terms upto the order of $1/d^4$ and the quadrupole moment required terms upto the order $1/d^6$ (see \ref{fig:Mlm_dist_osc}). These are consistent with the results of \cite{prasad2021tidal}. These moments can therefore be used to study the tidal deformability of black holes and compute their corresponding Love numbers in a manner described there. 
\begin{figure}[htp]
	\includegraphics[width=\columnwidth]{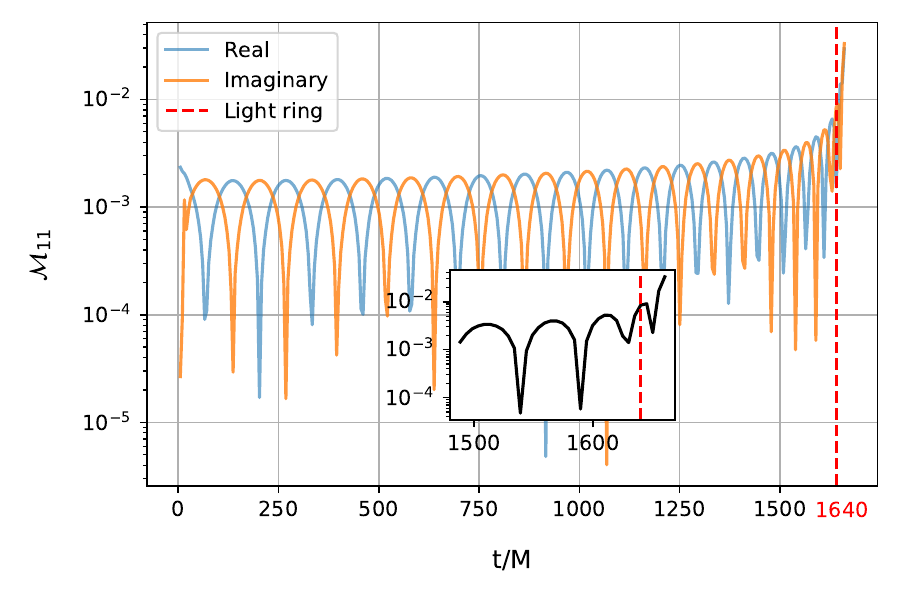}
	\includegraphics[width=\columnwidth]{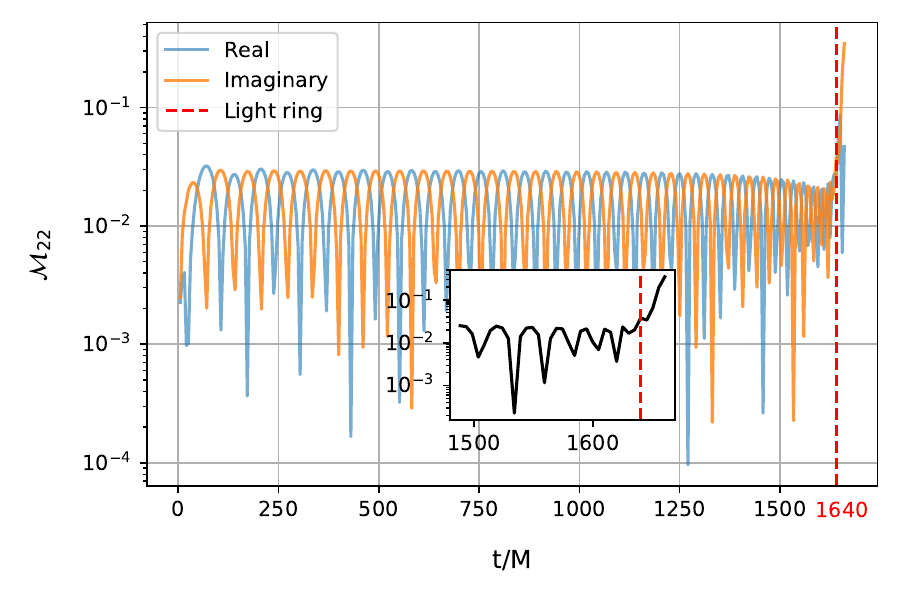}
	\caption[The time evolution of the dipolar deformations of the DH.]{The time evolution of some of the $\ell=1 m =1$ (top), $\ell=2, m=2$ (bottom) multipole moments. The $\ell=1, m= \pm1$ moments are identical to each other. The red line shows the temporal location of the light ring $r\approx2.856$. The close-up of these plots shows that the location where the change in the growth pattern of the multipole moments occurs is consistent with the epoch of light ring crossing by the system.}
	\label{fig:M11_Ml22_LCO}
\end{figure}

Apart from the real and imaginary parts of the moment, its magnitude also displays oscillatory behaviour as shown in \ref{fig:Mlm_dist_osc}. These oscillations are decaying with time, and exist in the multipole moments of both the dynamical horizons. To understand this further, we decomposed the evolution of the magnitude of the moment into a non-oscillatory portion as described by  \eqref{exp_gen_mult}, and an oscillatory part. This was done by fitting the multipole moment strength to  \eqref{exp_gen_mult} (we call this the non-oscillatory part), and then computing the normal distance of the data point from the best-fit curve (the oscillatory part). The fit was carried out on the data up to $t=1000M$. We found that the oscillatory part can be described by a superposition of power-law damped sinusoids of the form:
\begin{align}
	\dfrac{\vert\mathcal{M}_{22} \vert}{M_{\mathcal{H}}^{3}} = \sum_i A_i t^{(-\gamma_i)} \sin(\omega_i t + \phi_i)
\end{align}
with power law indices $\gamma = 1.47$ and $2.29$ respectively. The time periods of oscillations of these modes were found to be at $T_1 = 161.56M$ and $T_2 = 124.71M$ respectively. It is worth noting that the latter is close to half the average orbital time period (i.e. twice the orbital frequency)in the domain considered. These values are expected to be dependent on the mass ratio of the system. As this feature was observed in both the simulations $q=1, 0.6, 0.7$, we are led to speculate if these correspond to dynamic tides akin to the $w-$ mode oscillations of a star, quasi-normal modes of the tidally coupled dynamical horizons excited in the inspiral phase (e.g. due to residual eccentricity), or are mere numerical artefacts.

\subsection{Plunge} The multipole moments were found to display two distinct behaviours in the pre-merger phase, as can be seen in \ref{fig:Mlm_time_q0p6}, \ref{fig:Mlm_time_q0p7}, \ref{fig:Mlm_time_q1} and \ref{fig:M11_Ml22_LCO}. While the majority of the portion of the evolution of these moments reflected the adiabatic, quasi-circular dynamics of the system, their behaviour changed to a steeper increase in strength with the seizure of oscillations closer to the merger. We found that this epoch is very close to the time of crossing of the light ring of the binary black hole system. The light ring of the system is defined as the last circular (unstable) photon orbit. We estimate the light ring radius $r_{L}$ of the system using the adiabatic re-summed 1PN Hamiltonian \cite{buonanno2000}:
\begin{equation}
	r^3 -3 r^2 + 5 \nu = 0 
\end{equation}
where $\nu = m_1 m_2/ (m_1 + m_2)^2$ is the symmetric mass ratio. It was found that a sharp change in behaviour occurs when the black holes cross the light ring of the system. 

The multipole moments ($\ell=2, m=\pm1$) which were zero (at the numerical noise floor) for the majority of the inspiral phase, were found to gain in magnitude with oscillatory behaviour as the black holes approached each other. At these times close to the merger, the dynamical horizons will be boosted to high velocities. Thus, during these last moments of the merger of the individual horizons of the black holes and before the common horizon appears, the various multipole moments of the horizons grow to comparable strengths as the dynamical horizons strongly deform under each other's influences. 

A distinctive feature was observed for the $q=1$ case. The $\ell=2, m=2$ moment, although the strongest, does not monotonically increase in strength closer to the merger. This may be due to the enhanced symmetry of the equal mass binary system.

Owing to these and the previous results in \cite{shear-news2020}, we can conclude that the multipole moments of a dynamical horizon are strongly correlated with the dynamics of the system. It thus is also strongly correlated with the shear of its outgoing null normal, and the outgoing gravitational radiation emitted from the system.

\subsection{Post-merger}
\begin{figure}[htp]
	\includegraphics[width=0.9\columnwidth]{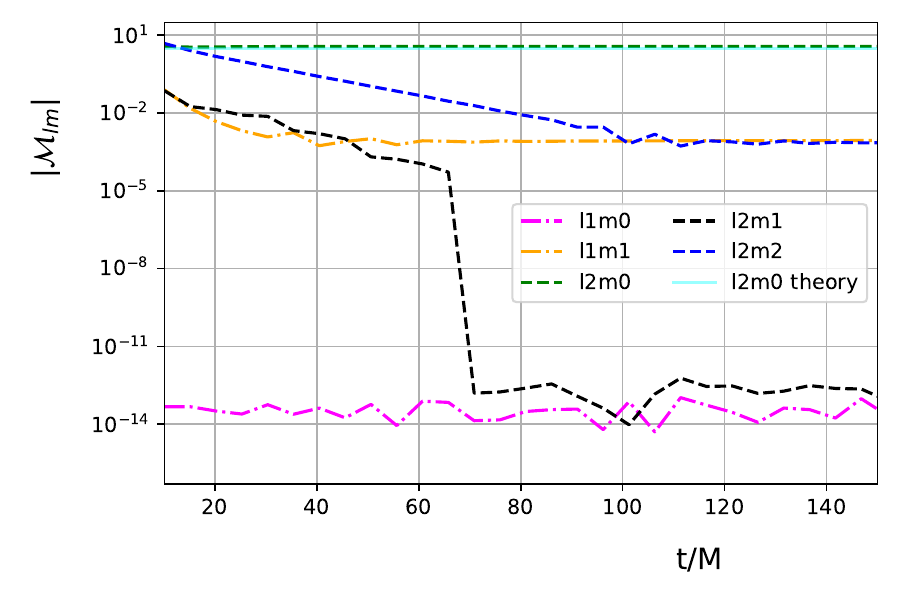}
	\includegraphics[width=0.9\columnwidth]{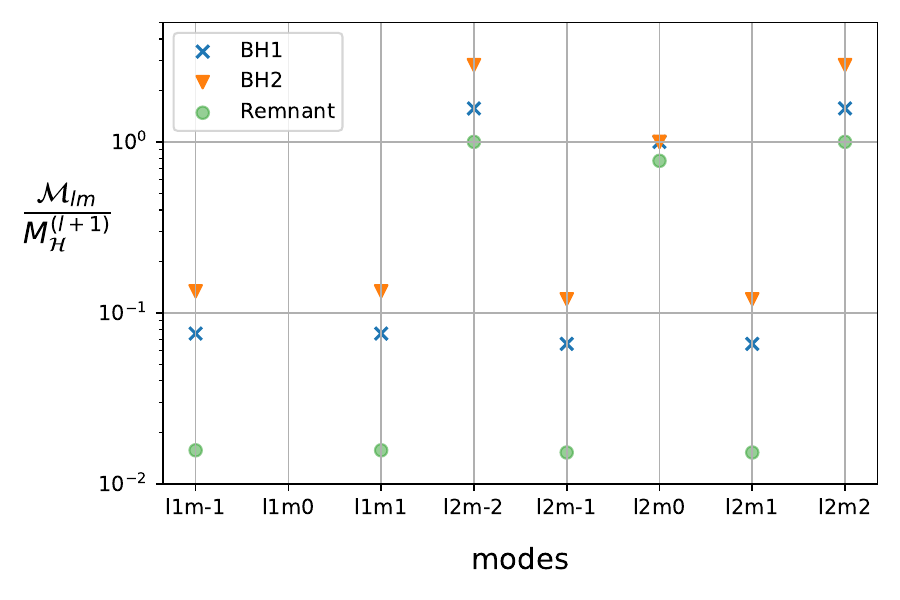}
	\caption[The deformation of the dynamical horizon in the post-merger phase.]{Top: The time evolution of the $\ell=1,2$ (top) multipole moments of the outer common horizon. Here, their strengths (absolute magnitudes) are plotted against time. Each mode has its own numerical noise floor. The theoretical value of the moment $\mathcal{M}_{20}$ of an unperturbed black hole of the same mass is plotted in cyan. Bottom: the strengths of the multipole moments of the individual horizons of the black holes moments before the formation of the common outer horizon of the system, and that of the common horizon just when it is formed. The moment $\ell=1, m=0$ is below machine precision as expected for all three horizons.}
	\label{fig:ml_pm}
\end{figure}
Extending the correlations of the multipole moments of the dynamical horizons with the outgoing gravitational waves further, we analyze the multipole moments of the outer common horizon in the post-merger phase, which is formed late into the inspiral phase, immediately after it is found. 

As the two black holes cross the light ring of the system, common envelopes surrounding the individual black hole horizons form. The outer common horizon is a dynamical horizon of the remnant that settles down to the Kerr isolated horizon, which is spinning and highly distorted when formed. Thus, one expects the multipole moments of the common horizon formed to be different from that of an isolated Kerr horizon. The dynamical horizon then proceeds to equilibrium by absorbing radiation and losing `hairs', i.e. the multipole moments are expected to decay to the corresponding isolated Kerr values. Fig.~\ref{fig:ml_pm} shows graphically that this is indeed the case. Here the time evolution of the mass multipole moments of the common horizon are plotted. The multipole moments display quasi-normal behaviour. The damping rate of the strengths of the moments $\vert \mathcal{M}_{lm} \vert$ were found to be close to the theoretical estimate of a Kerr black hole with the same mass and spin as that of the remnant. E.g, damping rate of $\mathcal{M}_{22}$ was found to be consistent with the $n=0, \ell=2, m=2$ mode with a deviation of $2.74\%$. Since the remnant black hole is spinning, the coordinate system established on the dynamical horizon can also rotate along with it and thus real part of the quasi-normal frequencies of one mode can only be directly estimated relative to another. By transforming the coordinate system using the spin of the horizon, one can obtain the correct quasi-normal frequency. Therefore the estimation of the real part of the quasi-normal frequencies requires more care and better resolution, which we will pursue in future work. For an isolated Kerr horizon, the only non-zero mass multipole moment at $\ell=1,2$ order is  $\mathcal{M}_{20}$. The moment $\mathcal{M}_{20}$ approaches the value of the corresponding Kerr black hole with a final deviation of $2.73\%$ from the expected theoretical estimate. The strengths of the moments $\mathcal{M}_{1\pm1}$ and $\mathcal{M}_{22}$, which are expected to go to zero, decay to the order of $10^{-4}$. We suspect that this is due to systematic errors arising from the rotation of the coordinate system on the common horizon, and its limited grid resolution and is to be investigated further. As expected from reflection symmetry, the mass multipole moment $\mathcal{M}_{10}$ is practically zero, whereas the moment $\mathcal{M}_{2-2}$ falls below the machine precision.

What decides the deformed state of the common horizon once it is formed? The initial configuration of the parent black holes is expected to decide the deformed state of the common horizon once it is formed. Interestingly, we found that the relative strengths of the multipole moments of each of the individual horizons of the black holes just before the formation of the common horizon are similar to that of the common horizon when it is formed (more so for the horizon of the more massive black hole). This can be seen in Fig.~\ref{fig:ml_pm}. Thus the deformed state of the individual horizons just before the common horizon appears plays a role in the setting up of initial conditions of the common horizon for the post-merger dynamics. The common horizon thus formed roughly inherits the multipolar structure of the horizon geometry of the black holes at the end of the inspiral phase and then loses them as it settles down to equilibrium in the post-merger phase.

\section{Conclusions}
In this work, the gravitational fields at the dynamical horizons in the strong-field regime and their correlation with the gravitational waves emitted from the system at future null infinity were studied in a binary black hole scenario.
The horizons of black holes in a binary environment are not isolated; their horizon geometries are dynamical and distorted due to their mutual tidal interactions, and details of the strong field dynamics of the system. In this work, these distortions were studied using the source multipole moments of the dynamical horizons. This was computed for the first time, separately using the numerical relativity data. In a binary black hole merger scenario, they are found to encode accurate information about the dynamics of the system. In particular, they display a chirp-like behaviour in the inspiral phase. These moments show a distinctive behaviour towards the late inspiral phase as the dynamical horizons plunge towards each other, and are roughly consistent with the epoch of crossing of the light ring of the system. The evolution of the strengths of the source multipole moments was also found to follow a universal behaviour: they can be described by a series expansion in the inverse of the distance of separation between the holes. 

These correlations allow one to study the strong field regime using gravitational wave observations. The correlations of the source mass multipole moments with the dynamics of the system allow the analogous, indirect interpretation that the changing source multipole moments are related to the emission of gravitational waves from the system.  We also may expect these correlations to extend to all multipolar orders, which will be studied in future work. 

Potential applications of these results are numerous. The estimation of non-axisymmetric tidal deformability coefficients following \cite{prasad2021tidal}, can be studied. In the modelling of gravitational waveforms from binary black hole systems (e.g. in the post-Newtonian and effective one-body approach), the deformation of the dynamical horizons is not usually taken into account. As shown here, the dynamical horizons are strongly deformed in the late inspiral phase, and this can potentially have effects on the late-inspiral, plunge waveforms. On the numerical side, the universal behaviour discovered in the deformations of the horizons could be used to inform the numerical evolution system of numerical relativity codes, thus accelerating the evolution system. This might prove to be useful and enable us to resolve the spacetime region, especially around the smaller black hole, and allow us to extend the parameter space for the numerical simulations of extreme mass ratio inspirals. This is currently being explored.

Owing to these and the previous results in \cite{shear-news2020}, we can conclude that the multipole moments of a dynamical horizon is strongly correlated with the dynamics of the system. It thus is also strongly correlated with the shear of its outgoing null normal, and the outgoing gravitational radiation emitted from the system.

These results urge us to re-state the following conjecture:  \emph{In a dynamical scenario involving binary black holes, the source multipole moments associated with charges of the dynamical horizon will be correlated with the multipolar structure of the Bondi flux of the outgoing gravitational radiation from the system received at future null infinity $\mathcal{I^+}$}

Understanding these correlations further can be useful in discerning the mass multipolar deformations of the horizons of black holes in fully dynamical scenarios using gravitational wave observations. They may also be used to test and understand further important aspects of the no-hair conjecture and probe possible non-standard structures of gravity (e.g. quantum gravitational corrections) in the strong field regime. 

The results presented here also allow for the following simple interpretation: In a binary black hole scenario, the individual dynamical horizon geometries of black holes gain a structure i.e., ``gravitational hairs'', away from their isolated Kerr geometries in the inspiral phase through the mutual tidal interactions, and lose them in post-merger dynamics of the common horizon. 

On the numerical side,  the computation of multipole moments requires a choice of an axial vector field on the horizon. In this work, this has been achieved by using the method of Killing transport, which may be inaccurate very close to the merger. Accurate descriptions of the merger phase would require a better choice of the axial vector field and is reserved for a future study. 
 
\cleardoublepage %
\flushbottom
\newpage
\chapter{A Post merger study: Flux absorbed by the common dynamical horizon}
\label{chap:flux_pm}

\section{Introduction}

After the merger of the black holes in a binary, a common dynamical horizon develops that envelopes the individual dynamical horizons of the black holes. This horizon grows by absorbing gravitational radiation in the post-merger dynamics and is eventually expected to settle down to that of a Kerr black hole. The shear of the outer common dynamical horizon was studied in the preceding chapter, using the data at the north pole of the common dynamical horizon. In this chapter, we study the shears and the associated energetics of the common dynamical horizon using the full 2D shear data.  %

Let the 2 slices of the dynamical horizon parameterized by $x$ have the null normals $(n^\pm)$ with  $n^+ \cdot n^- = -1$.

\begin{figure}[b]
	\centering
	\includegraphics[width=0.8\columnwidth]{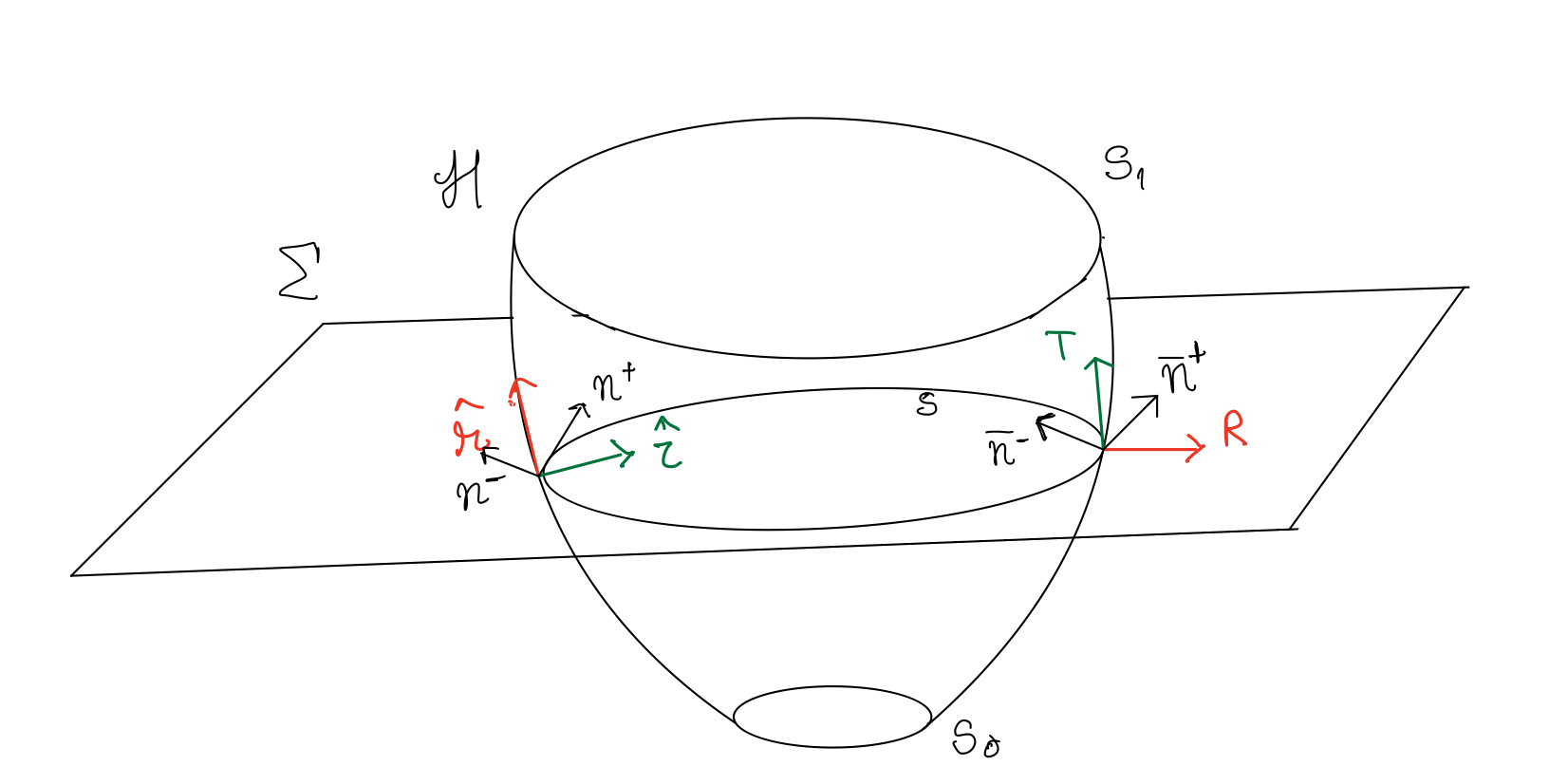}
	\caption{The null normals to a growing dynamical horizon $\mathcal{H}$.}
	\label{fig:dhfig}
\end{figure}

The null normals to a set of MOTS $\mathcal{S}$ that represents the slices of a DH can be constructed from the fiducial set of normals to $S$ embedded in a space-like three-hypersurface $\Sigma$. These can be written as:
\begin{align}
	\bar{n}^{\pm a} = \dfrac{1}{\sqrt{2}} (T^a \pm R^a)
\end{align}

Another set of null normals can be constructed using the normals to $\mathcal{S}$ that are naturally adapted to the dynamical horizon $\mathcal{H}$:
\begin{align}
	n^{\pm a} = \dfrac{1}{\sqrt{2}} (\hat{\tau}^a \pm \hat{r}^a)
\end{align}

Here $\hat{\tau}$ is the time-like normal to and $\hat{r}$ is the space-like normal to the leaves $\mathcal{S}$ but tangential to the dynamical horizon $\mathcal{H}$.

The null normals to a DH have a rescaling freedom. Thus these two sets of null normals must be related by a boost transformation:
\begin{align}
	n^{\pm a} = f \bar{n}^{\pm a}   
\end{align}
In a numerical relativity simulation, the set $(\bar{n}^{\pm})$ are easily computable and available. The shears corresponding to these null normals will be labelled as $\sigma_{\bar{n}^+}$ and $\sigma_{n^+}$. We will now attempt to study the energetics of the dynamical horizon using the shear and describe some of the obstacles involved.

\subsection{General formulae}

In this exercise, we attempt to analyse the amount of gravitational radiation absorbed by the black hole. The amount of radiation absorbed at a cross-section $\mathcal{S}$ of a space-like future outer trapping horizon $\mathcal{H}^+$ of a black hole in dual null foliation is given by the expression [25] of Ref.~\cite{Ashtekar:2004cn}:
\begin{equation}
	\dfrac{1}{2G} (R_2 - R_1) = \int_{\Delta_H} T_{ab} \hat{\tau}^a \xi^b _{(R)} d^3 V + \dfrac{1}{16 \pi G} \int_{\Delta H} N_R (\vert \sigma \vert^2 + 2 \vert \varsigma \vert^2) d^3V 
	\label{flux}
\end{equation}
Here, $\sigma$ and $\zeta$ are tensors that describe the longitudinal and transverse components of the gravitational radiation infalling at the MOTS $\mathcal{S}$. These are defined by:

\begin{equation}
	\sigma_{ab} = -\tilde{q}^{ac} \tilde{q}^{bd} \nabla_a \ell_b, \quad and \quad \varsigma^a = -\tilde{q}^{ab} \hat{r}^c \nabla_c \ell_b
\end{equation}

Here $\tilde{q}_{ab}$ is the two-metric on the MOTS, $(\ell, n, m, \bar{m})$ is a null tetrad on the DH, and $R$ is the areal radius of the MOTS.

In a BBH merger, $T_{ab} = 0$. The choice of the lapse function $N_R = \vert \partial R \vert$ on $\mathcal{H}$ simplifies the expression for the flux to:
\begin{equation}
	\dfrac{1}{2G} (R_2 - R_1) = \dfrac{1}{16 \pi G} \int_{\Delta H} \dot{R} (\vert \varsigma \vert^2 + 2 \vert \zeta \vert^2) d^3V 
	\label{flux2}
\end{equation}
Here, the time parameter $t$ is used to label the surfaces. However, in a simulation, one has convenient access to the null normals $n^\pm$. The integrand with respect to $n+$ and $\bar{n}^+$ can be related through the boost factor $f$:
\begin{align}
	(\vert \sigma_{n^+} \vert^2 + \vert \varsigma \vert^2) \dot{R} = f^4 \vert \bar{\kappa \vert^2 \dot{R}^2} + f^2 \vert \sigma_{\bar{n}^+} \vert^2 - \bar{\omega} \cdot \bar{\kappa} \dot{R} \vert \bar{\omega} \vert^2 
\end{align}

Presently, the unavailable quantities are the boost factor $f$ and the vector $\varsigma$. Therefore, we now study the energy flux absorbed by the common dynamical horizon using the shear $\sigma_{n^+}$ in place of $\sigma_{\bar{n}^+}$.

The Hawking energy of the two dimension slice $\mathcal{S}$ of the DH $\mathcal{H}^+$ is given by:
\begin{equation}
	E = \dfrac{R}{16\pi} \oint * (\mathcal{R}) = \dfrac{R}{2}
\end{equation}
By virtue of the Gauss-Bonnet theorem for surfaces of spherical topology. Here $*$ represents the Hodge dual with respect to the metric $\tilde{q}_{ab}$ of $\mathcal{S}$. This is called the quasi-local mass and also represents the irreducible mass of the horizon.

The equation  \eqref{flux2} above gives the variation of the Hawking energy along the null vector $\xi = N_R n^+$, where $N_R$ is the norm of the vector $\partial R$, formed by taking the three-gradient of the areal radius $R$ on the dynamical horizon $\mathcal{H}^+$.  
The absorption of radiation at the horizon must result in an increase in the horizon mass of the MOTS. We report on the attempts to study this using available numerical simulation data from previous studies.

\subsection{Implementation}

Procedure for computing the flux in \eqref{flux}:
The shear spin coefficient $\sigma$ was decomposed into modes using spin weight $(-2)$ spherical harmonics. The SWSHs are defined using the intrinsic coordinate system on the MOTS:
\begin{equation}
	C_{\ell m} = \oint_S \sigma(\theta, \phi)  \; _{-2}Y_{lm}(\theta, \phi) d^2 S
\end{equation}
For more information on this decomposition on the MOTS, please see \ref{chap:gen_mul}. Strictly, the modes need to be decomposed in Spin Weighted Spheroidal Harmonics. Development of the associated techniques is underway. However, as a first estimate, we proceed with the Spin Weighted Spherical Harmonics.

On decomposition, it was found that the $m=0$ mode has the highest (DC-like) contribution to the shear $\sigma$. This is unphysical as we expect the shear to vanish as the black hole approaches its end state, and the horizon tends to an isolated one. The origin of this is unknown. It could, e.g., be the result of a numerical artefact/inaccuracies in the established coordinate system. The full integrity of the SWSH functions on the sphere (e.g. orthogonality, etc) needs to be carefully investigated.  To handle this, the $m=0$ contribution to the shear was discarded and the flux was reconstructed.

The flux due to just the $\ell=2$ moments at each instant of time can be reconstructed as:
\begin{equation} 
	\mathcal{F}_{\sigma} = \sum_{m=-2, \\ m\neq 0} ^{m=2} (\vert Re(C_{\ell=2, m}) \vert^2 + \vert Im(C_{\ell=2, m}) \vert^2)
\end{equation}
In this exercise, the contribution from $\zeta$ has been neglected. There is some evidence that this term is subdominant. This term is important (and non-zero) when the MOTS has angular momentum. %

The vector $\partial R$ lives in the dynamical horizon, and is orthogonal to every vector in $S$. It therefore lies in the direction perpendicular to $S$. The norm can be parameterized in terms of $t$ and the derivative $\frac{dR}{d t}$ where $t$ is the CCTK time coordinate available from the data.

\section{Results}

\subsection{The shear modes}

\begin{figure}
	\centering
	\includegraphics[width=0.8\columnwidth]{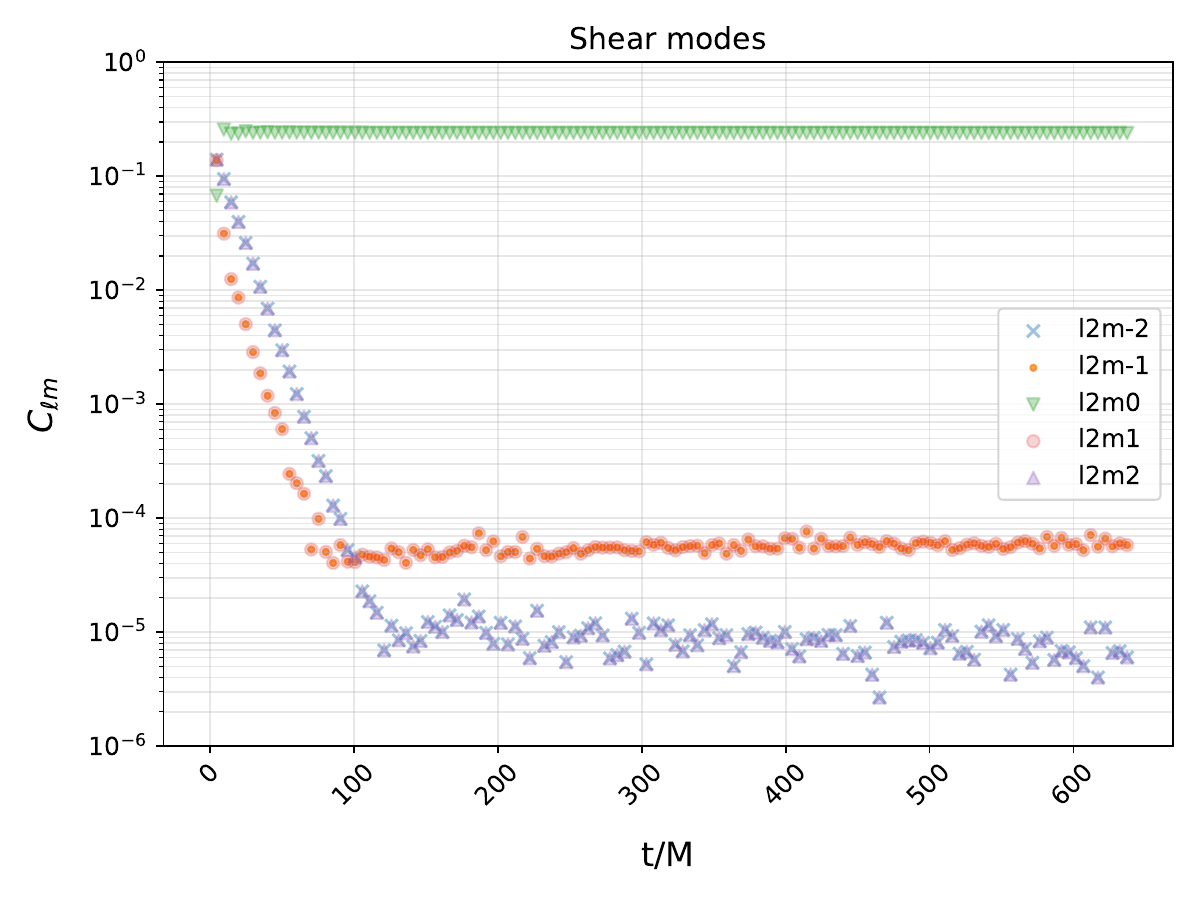}
	\caption[The quadupolar modes of the shear at the common DH.]{The $\ell=2$ mode strengths $\vert C_{\ell m} \vert$ of the shear $\sigma$ of the common dynamical horizon. The final mode strength values have been subtracted from these values to remove DC shifts.}
	\label{fig:modes1}
\end{figure}

\begin{figure}
	\centering
	\includegraphics[width=0.8\columnwidth]{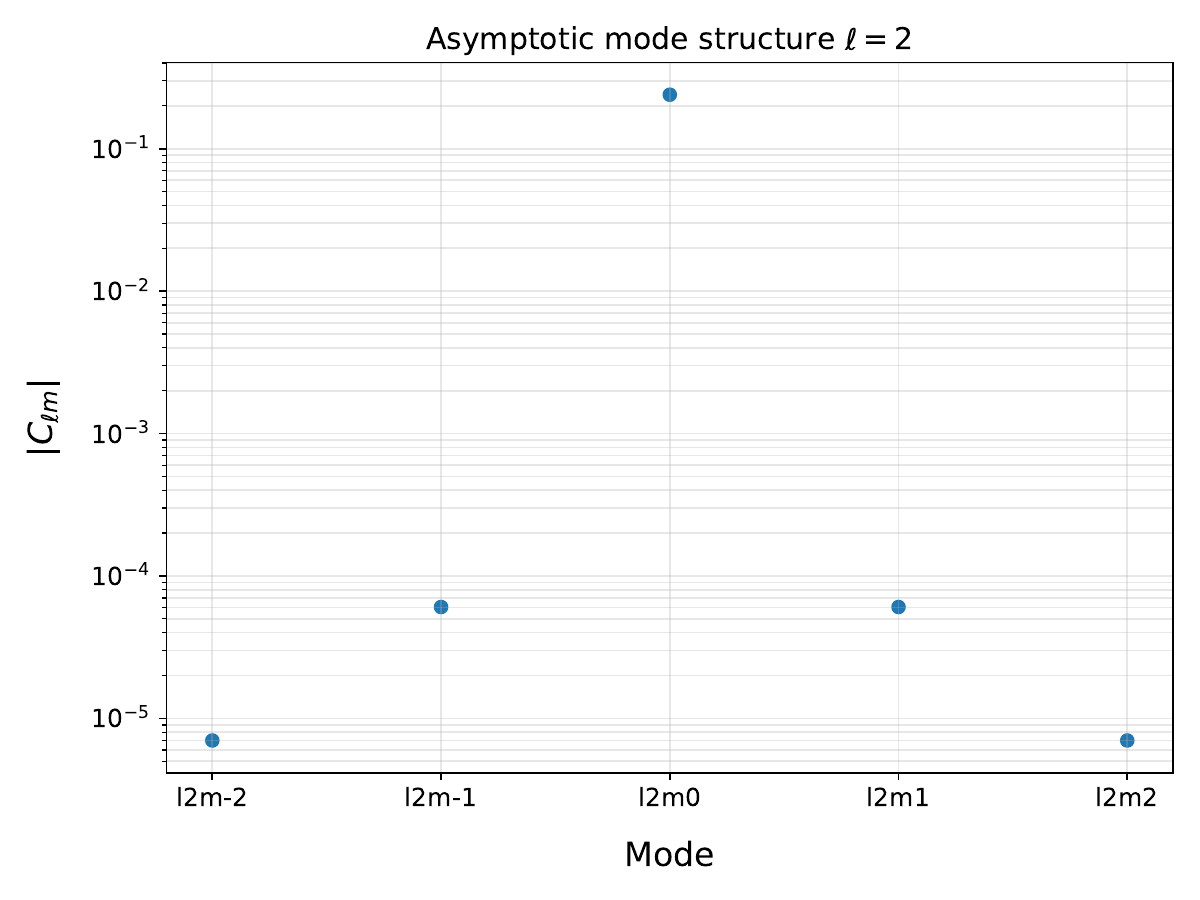}
	\caption[The asymptotic mode strengths of the shears at the common DH.]{The asymptotic (final) $\ell=2$ mode strengths $\vert C_{\ell m} \vert$ of the shear $\sigma$ of the common dynamical horizon. These values have been subtracted from the mode strengths to remove DC shifts (numerical noise floor) to produce \ref{fig:modes1} above.}
	\label{fig:asm_modes}
\end{figure}

The modes of the shear display quasi-normal with some anomalous behaviour. These observations are associated to the \ref{fig:modes1}, \ref{fig:asm_modes}. Some observations are

\begin{enumerate}
	\item There is a symmetry between the positive and negative $m$ modes as expected.
	\item The mode $\ell=0$ does not decay to the numerical noise floor. It reaches a magnitude of around $0.24$.
	\item The modes $\ell=2, m \neq 0$ reach their respective numerical noise floors of order $10^{-5}$
\end{enumerate}

\subsection{The flux due to shear}

The shear of the dynamical horizons is expected to account for the major portion of the energy absorbed. We wish to quantify this.

We compute the flux due to shear as:
\begin{equation}
	F_{t} = \dfrac{1}{16 \pi G} \int_{S} \dot{R} \, \vert \sigma \vert^2  \, d^2 S \label{flux_sub}
\end{equation}

Where $S$ is a MOTS and $d^2 S$ is the proper area element on it.

\begin{figure}
	\centering
	\includegraphics[width=0.8\columnwidth]{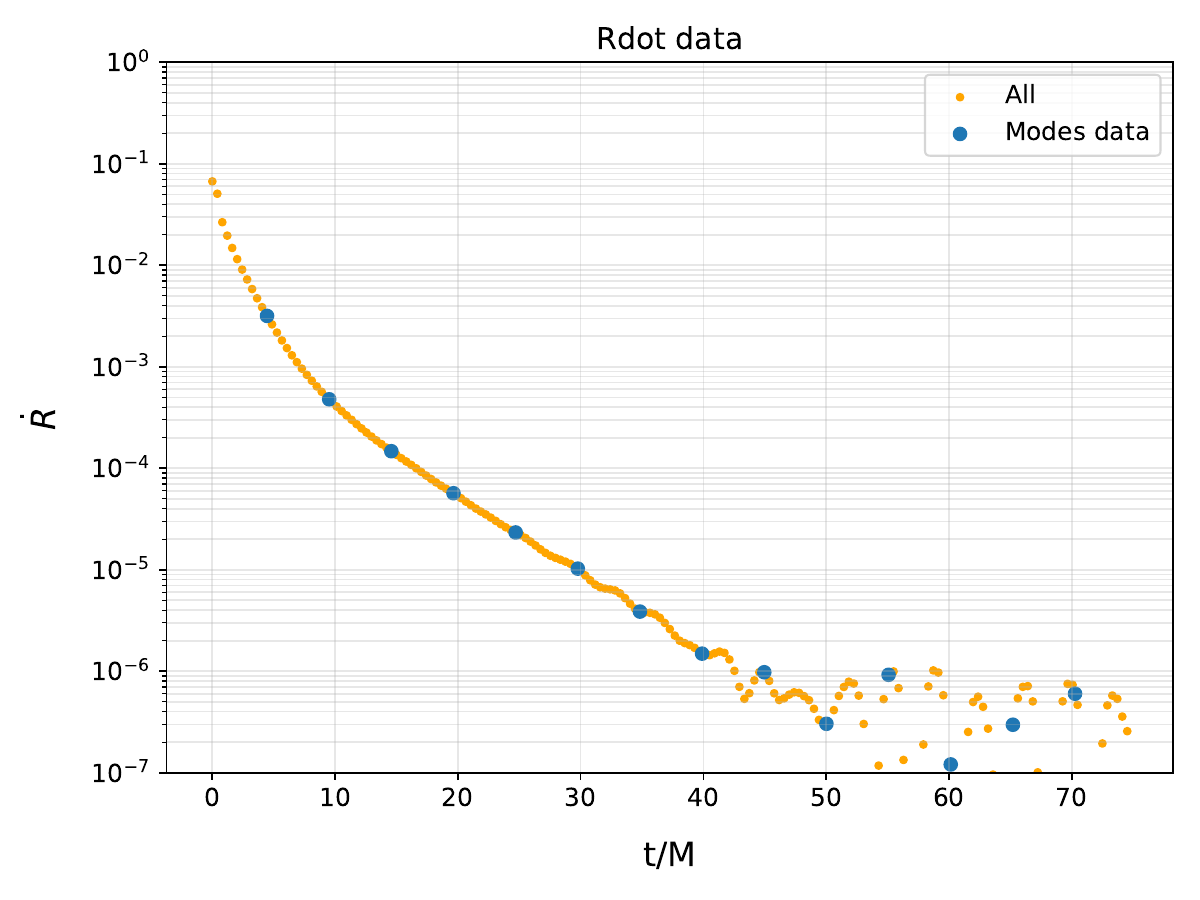}
	\caption[The time rate of change of the areal radii $\dot{R}$ of the MOTS]{The time rate of change of the areal radii $\dot{R}$ of the MOTS. Superposed, are points for which the shear mode data is available.}
	\label{fig:rdot}
\end{figure}
\begin{figure}
	\centering
	\includegraphics[width=0.8\columnwidth]{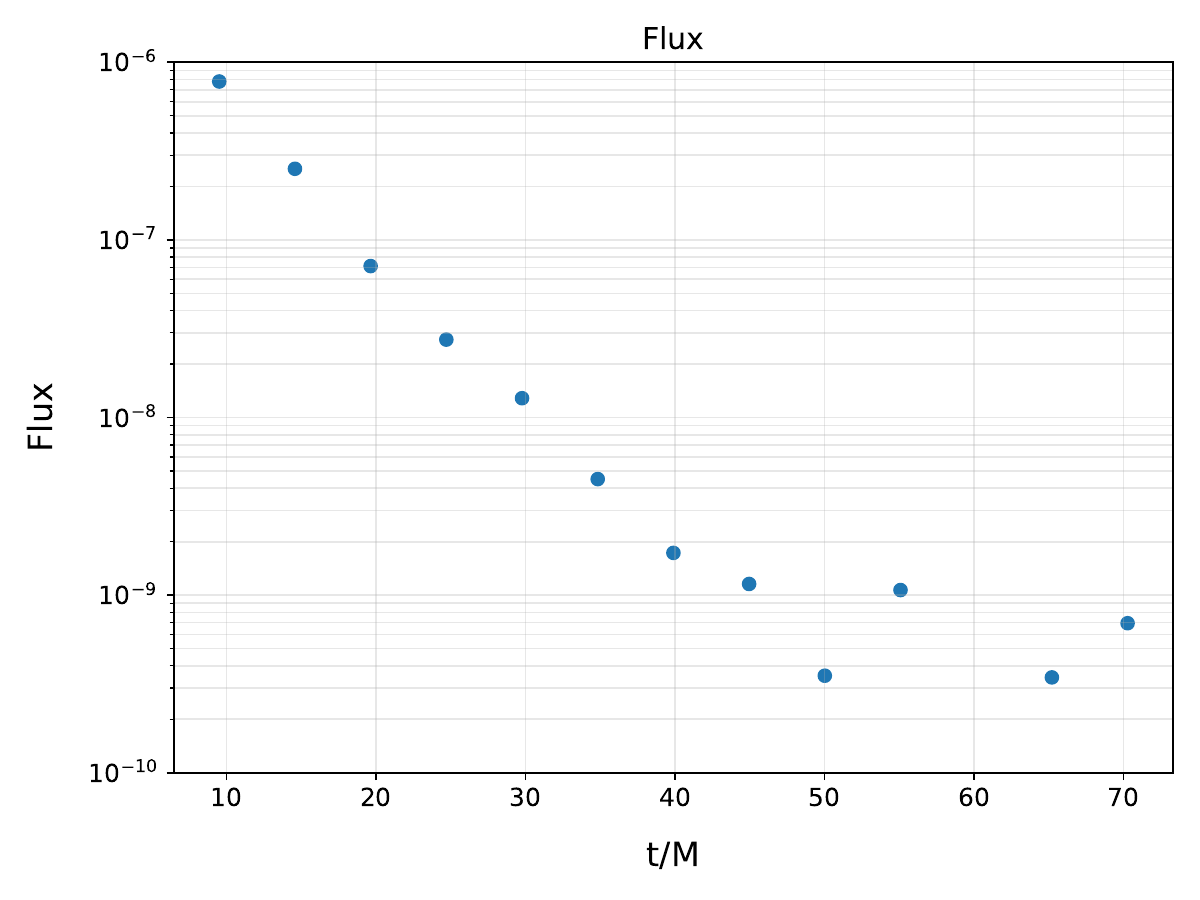}
	\caption[The flux of the infalling gravitational radiation]{The flux of the infalling gravitational radiation measured by the $\ell=2$ components of the shear in \eqref{flux_sub}.}
	\label{fig:flux}
\end{figure}
\begin{figure}
	\centering
	\includegraphics[width=0.8\columnwidth]{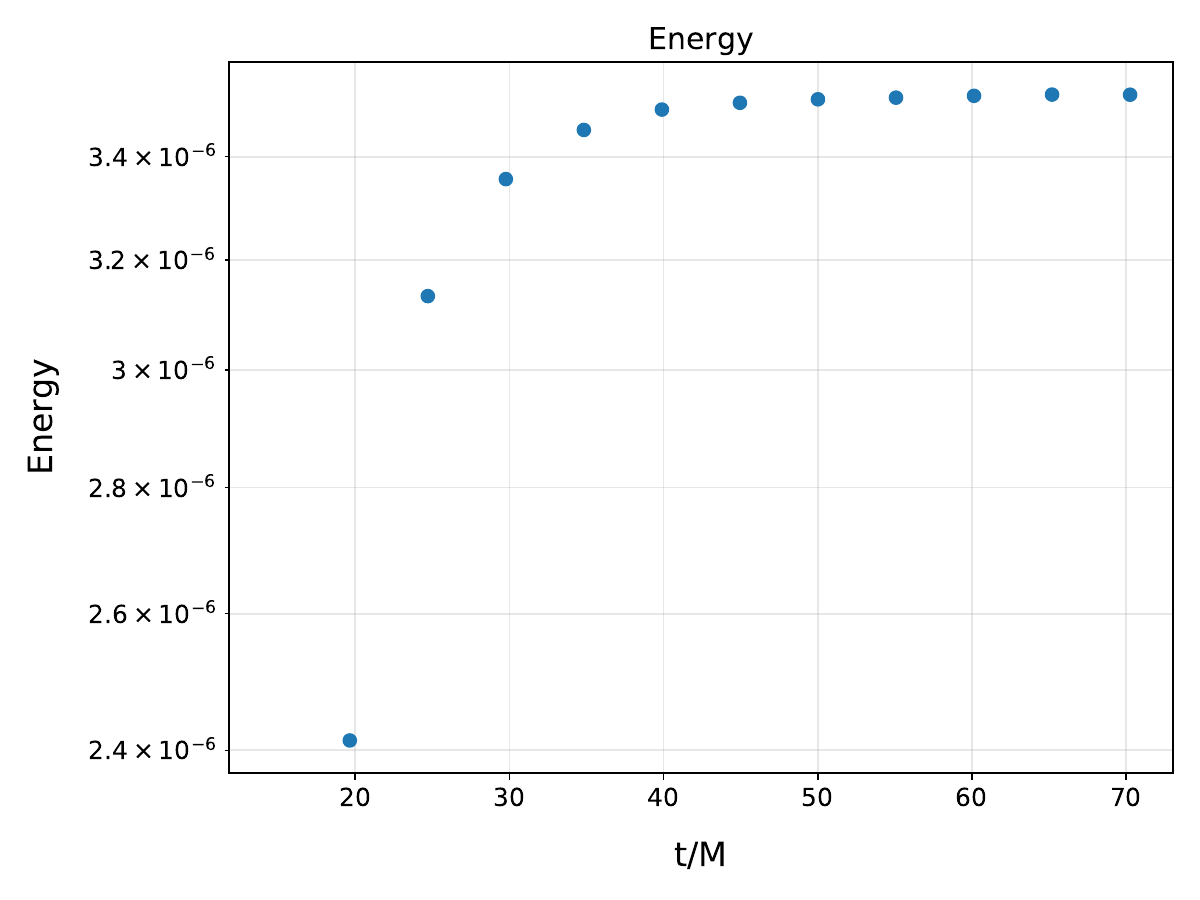}
	\caption[The energy of infalling gravitational radiation computed using the shears.]{The energy of infalling gravitational radiation measured by the $\ell=2$ components of the shear due to the flux in \eqref{flux_sub}.}
	\label{fig:mass_inc}
\end{figure}
\begin{figure}
	\centering
	\includegraphics[width=0.8\columnwidth]{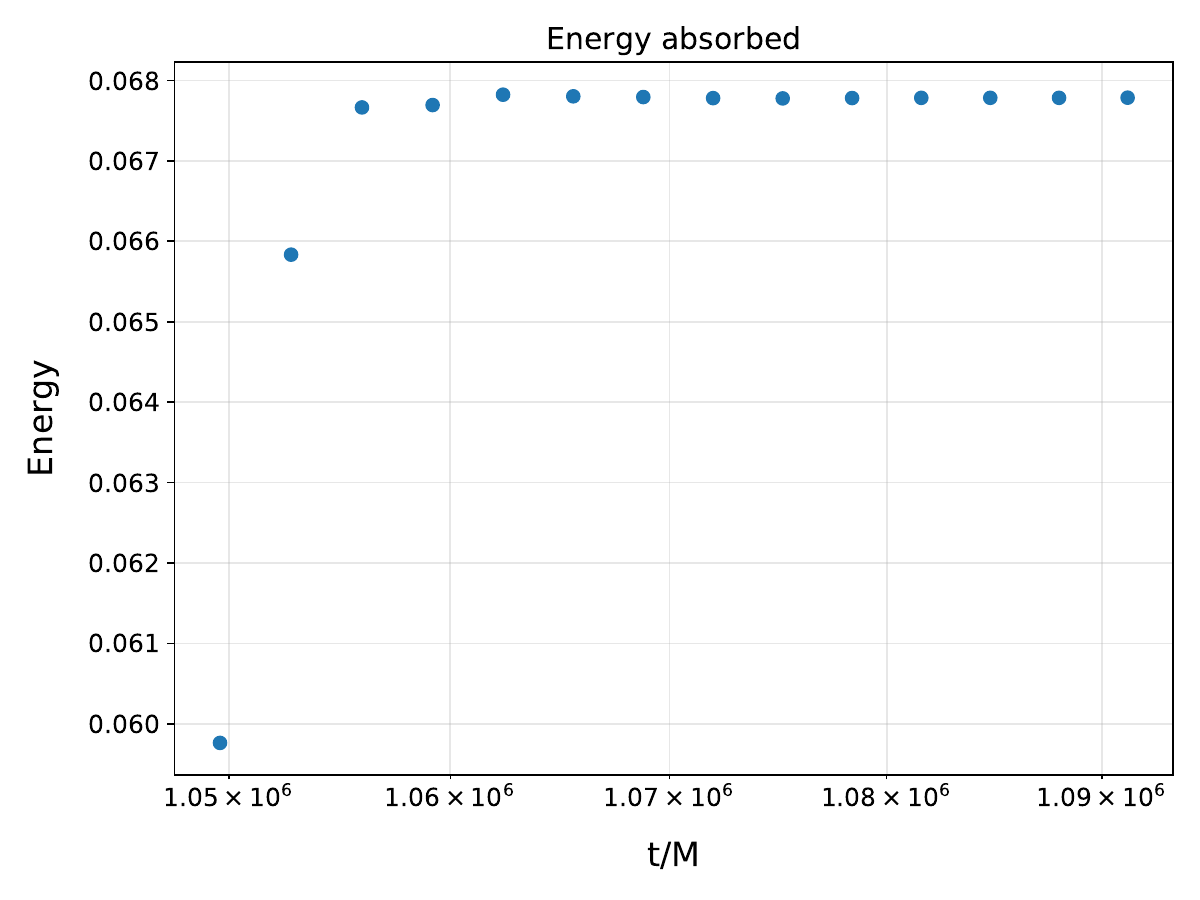}
	\caption[The energy of infalling gravitational radiation computed from the mass increase.]{The energy of infalling gravitational radiation computed using the increase in the common horizon mass as reported by QLM.}
	\label{fig:mass_inc_qlm}
\end{figure}

Some observations can be made from \ref{fig:flux}, \ref{fig:mass_inc} and \ref{fig:mass_inc_qlm}:

\begin{enumerate}
	\item As per the flux due to shear (with incorrect factor $N_R$) shown in these plots, the total energy absorbed by the common horizon is around $3 \times 10^{-6}$.  
	\item As per the horizon mass increase from QLM, the energy absorbed is around $6.8 \times 10^{-2}$
\end{enumerate}
However, more work is required to confirm whether the factors $f$ can account for this discrepancy. 

\section{Total energy absorbed}

The total energy absorbed can be estimated by taking the difference in the initial ($M_i$) and final mass $M_f$ of the common horizon of the system:
\begin{equation}
	\hat{E} = M_{f} - M_i
\end{equation}
Here the masses refer to the quasi-local Hawking masses of the common horizon.

This was computed for a small subset of the simulations with $q\geq 0.5$ with trustworthy post-merger data. It was found that the total energy absorbed can be described in terms of a simple fit across the mass ratios $q$:
\begin{align}
	E(q) = (-7.34 \times 10^{-2} \pm 9.083 \times 10^{-3}) \times q +  (1.14 \times 10^{-1} \pm 6.27 \times 10^{-3})
	\label{Minc_fit}
\end{align}

\begin{figure}[htpb]
	\centering
	\includegraphics[width=0.8\columnwidth]{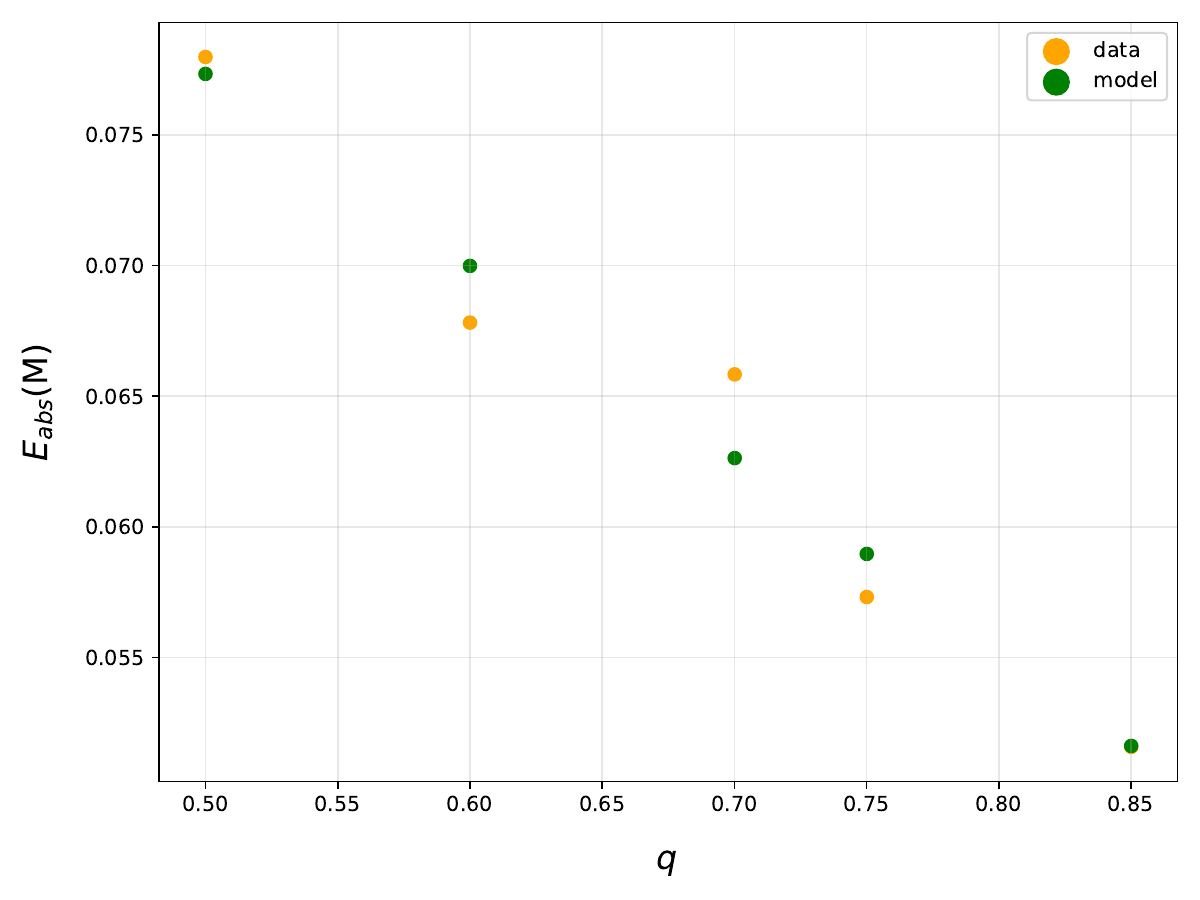}
	\caption[The total energy absorbed by the common dynamical horizon]{The total energy absorbed by the common dynamical horizon. The points in orange are computed from NR data. Those in green are the values predicted by the fit in \eqref{Minc_fit}}
	\label{fig:pm_mass_inc}
\end{figure}
This fit corresponds to a p-value of $1.26\times 10^{-6}$. The \ref{fig:pm_mass_inc} describes the fit graphically.

\section{Conclusions}

The mass increase of the common dynamical horizon was found using the horizon masses of its foliation. However, to relate this to the flux of radiation at the horizon, the computation of the boost scalar $f$ is required to verify the energy balance laws numerically.

\chapter{Conclusions}
\label{chap:concl}

In this thesis, the strong field dynamics of binary black hole mergers, and their associated gravitational radiation have been studied. The dynamical horizon formalism, numerical relativity and novel application of data analysis techniques have been used. There are two sets of boundaries in a BBH spacetime, one containing the future null infinity in the vicinity of which gravitational wave detectors are placed, and the other in the strong field regime of the spacetime where the black holes live. The latter can be appropriately described as dynamical horizons, one for each of the black holes. These dynamical horizons live in a different regime, where spacetime curvatures are high relative to that at future null infinity. The dynamical horizons pose an important boundary of the spacetime: like at future null infinity, gravitational wave energy can leave the spacetime, into these surfaces in the strong field regime. In this thesis, we studied the aspects of the dynamics of these surfaces, the elements of the dynamical horizon, the gravitational fields at the dynamical horizons and their relationship with the dynamics of the BBH system. 

\subsection{Numerical aside}   
We run and use full numerical relativity simulations of BBH mergers using the publicly available Numerical Relativity code, the Einstein Toolkit (ETK) at IUCAA, Pune. I have set up and compiled the Einstein Toolkit on the Pegasus supercomputing cluster. This involved compiling and tuning the external libraries to the architecture of the HPC hardware. The mutual compatibilities and dependencies of the ETK and its external libraries were identified before their compilation. Then, the Einstein Toolkit was configured and optimized to run on Pegasus HPC using compilers and libraries tuned and optimized to the HPC hardware. A simpler version of this procedure has been documented in the Appendix\ref{appendix:build_etk}.

We then run several numerical simulations of non-spinning, quasi-circular BBH mergers and track their dynamical horizons. The individual horizons were tracked through the inspiral and the common horizon in the post-merger phase. Several quantities associated with the dynamical horizon were then computed and used in the analysis. Python was used to compute the generalized source multipole moments and several other quantities associated with the dynamical horizon, that are outside of the capabilities of the ETK.  

In the analysis of the quantities, python was used along with the Message Parsing Interface (MPI), vectorization, and numba parallelization schemes. The latter also allows for extension to GPU-based computations. These are useful owing to tensorial data of large dimensional quantities involved. Several libraries have been developed to achieve this and we briefly summarize their capabilities below:
\begin{enumerate}
	\item waveformtools: Organization, parsing and information retrieval system for Numerical Relativity simulation data, waveform extrapolation to infinity, waveform corrections and tools (centre of mass, BMS transformations, frequency domain integration, comparisons). Their usage has been described in \hyperref[appendix:waveform_prep]{Appendix~B}.  
	\item qlmtools: Computation of various quasi-local quantities associated with the dynamical horizon.
	\item quasinormal: QNM analysis. A collection of fitting functions and procedures for unconstrained and constrained estimation of quasi-normal modes.
	\item nullsolver: characteristic initial value solver for studying scattering problems.  
	\item pkginstaller: A batch-like script for automatically downloading, compiling and building an optimized version of libraries for HPCs. These are used to install all prerequisites for NR codes \cite{pkginstaller}.
\end{enumerate}
These are being maintained in their respective git repositories.

In \ref{chap:waveform_rel}, I show that the shear of the dynamical horizons, which represents a part of the gravitational radiation infalling at the horizon, is strongly correlated with the news of the outgoing gravitational radiation from the system. The shear gives a measure of the flux of the infalling transverse gravitational radiation at the horizon and forms an important part of the tidal coupling between the black holes. Using these strong correlations, we show that one can predict the shears of the dynamical horizons, which are causally disconnected from observers, using the gravitational radiation escaping from the system at future null infinity. We also give fitting formulas relating the shear of the dynamical horizons of the two black holes to each other in the inspiral phase. Thus, in classical general relativity, the strong field dynamics of BBH mergers are strongly correlated with the weak field dynamics. During the inspiral phase, the component black holes accrete mass and angular momentum by absorbing gravitational radiation. Since the shear measures this infalling gravitation radiation, we plan to study numerically this accretion process and contribute to the improvement of waveform modelling. How is the gravitational radiation at the dynamical horizon strongly correlated with the outgoing radiation, although they are causally disconnected? How does the infalling gravitational radiation affect the gravitational radiation received at future null infinity, even though they are causally disconnected?  

The analysis and results in this thesis suggest the following explanation. In the inspiral phase, the nonlinear gravitational dynamics of the black holes in spacetime are responsible for the generation of gravitational radiation. A part of this radiation is absorbed at the dynamical horizons, which acts like a feedback mechanism, affecting its dynamics i.e. of the black holes. The observed gravitational radiation that escapes to infinity (especially its phasing) crucially depends on the dynamics of the black holes and the balance laws at their horizons, thus changing, while respecting causality.

In \ref{chap:tddh}, we study the tidal deformability of black holes in the strong field regime. In the approach, we use the source multipole moments of the gravitational field at the dynamical horizons of the individual black holes in the inspiral phase of the BBH merger. The source multipole moments deviate from their isolated values due to the mutual tidal interaction. Using a simple tidal expansion of this deviation, we show that one can compute the associated tidal deformability coefficients, which are universal. This approach may be more relevant and important to understand aspects of the no-hair conjecture in dynamical and strong field regimes, especially when the black holes are close to each other.

In \ref{chap:gen_mul}, we implement and compute for the first time, the full spectrum of deformations of the dynamical horizon geometry in the inspiral merger and ringdown phases. This was achieved using generalized source multipole moments computed via the approximate Killing vector method. Although dynamical horizons need not admit an axial symmetry vector field in full BBH merger scenarios, this method provides a simple way to compute the tidal deformations. We show that these deformations are themselves strongly correlated with the dynamics of the BBH system, and to the gravitational radiation received at infinity. We show that several important features of the BBH orbital dynamics are imprinted on the dynamical horizons. In particular, we show that the behaviour of these moments is universal. The dynamical horizons grow during the inspiral phase accreting not just energy and angular momentum from its surroundings, but also gravitational hair deposited by the infalling radiation. The individual dynamical horizons of the BBH system are shown to acquire these deformations during the inspiral phase, and the common horizon loses them during the post-merger phase through quasi-normal decay. These results pave the way for further studies of the no-hair conjecture in the strong field regime and the effect of the deformations of the dynamical horizons on the gravitational waves received in the detectors. These studies are in progress. Furthermore, the computations of spin-multipole moments are also underway. 

In the final \ref{chap:flux_pm}, we attempt to compute the energy accreted by the common dynamical horizon in the post-merger dynamics. We lay out the procedure and gauge transformations necessary to implement these computations. This is currently in the planning stages.

We are also studying the effect of the accretion of energy and angular momentum on the gravitational waveform emitted from the binary system and how it can enable us to improve the existing waveform models. 

The main takeaway from this thesis is that strong correlations hold between the dynamical horizons in the strong field regime and the future null infinity. These correlations, previously conjectured in the post-merger phase, have been shown to hold in all three phases of BBH mergers. Apart from the possibility of pursuing analytical evidence to the correlations, one may wonder, if these correlations hold in more general scenarios like modified horizon dynamics and modified gravity.

\setcounter{figure}{0}
\renewcommand\thefigure{\thesection.\arabic{figure}}    
\setcounter{table}{0}
\renewcommand{\thetable}{A\arabic{table}}%

\begin{appendices}

\cleardoublepage %
\flushbottom
\newpage
\cleardoublepage\chapter{Distance Measures}
\label{appendix:dm}

\section{Introduction}

An estimate of the distance between the two black holes in a binary is required to calculate the tidal coefficients. We now discuss various choices for this distance measure and describe how we compute it from the numerical simulations. In the simulations we carried out, the numerical evolution uses a coordinate system described in \cite{Thornburg_2003}. We define and compute the following measures of distances of separation between the black holes:

\begin{itemize}
	\item The Euclidean distance between the geometric centroids of the individual apparent horizons of the two black holes. 
	\item Newtonian proper distance. 
	\item Post-Newtonian (PN) distances up to second-order (PN1, PN2).
\end{itemize}

We explain each of these definitions below.

\begin{figure}[htpb]
	\centering
	\includegraphics[width=0.9\columnwidth]{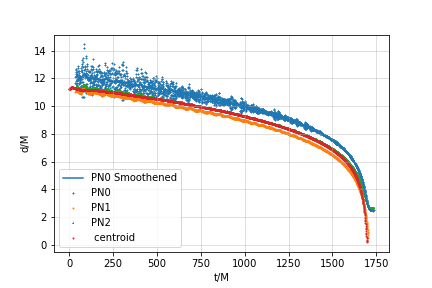}
	\caption{The various inter BH distance measures for the $q=0.4$ simulation. }
	\label{fig:dist_measures_q0p4}
\end{figure}

\section{Euclidean distance}

We use the coordinate locations of the geometric centroids ($(x_1, y_1, z_1)$ and $(x_2, y_2, z_2)$) of the individual horizons of the two black holes to define a simple Euclidean distance measure.
The distance is then defined by: 
\begin{align}
	d = \sqrt{(x_1 - x_2)^2 + (y_1 - y_2)^2 + (z_1 - z_2)^2} \,.
\end{align}

\subsection{Newtonian proper distance}

Kepler's third law for orbiting binaries can be used to calculate the physical distance of separation between the center of masses of the orbiting objects:
\begin{align}
	T^2 = \left(\dfrac{\pi^2}{2G(M)}\right) d^3 \,\,\,\, \textit{or} \,\,\,\, d = \left(\dfrac{GM}{\omega^2}\right)^{1/3} 
\end{align}
Where $M$ is the sum of the horizon masses of the black holes and $d$ is the distance of separation between them.

We compute the instantaneous angular frequency $\omega$ of the quasi-circular configuration by using the instantaneous frequency of the $\ell=2, m=2$ mode of the gravitational waves emitted from the system. We use the extracted gravitational radiation at $100 M$ from the center of the system for this purpose.

The gravitational waves emitted from the system provided by the WaveExtractCPM thorn in the Llama package in the Einstein toolkit infrastructure. 

\subsection{Post-Newtonian distances}

A version of the above Kepler's law can be derived for post-Newtonian corrected dynamics at a chosen order\cite{MaggioreI}. Then, using a method similar to the one described above for computing the Newtonian proper distance, we compute proper distances to Post Newtonian I and II order. The expressions are given by:
\begin{equation}
\omega^2 = \dfrac{G M}{d^3} \left(1 + (-3 + \nu)\gamma + \left(6 + \dfrac{41}{4}\nu + \nu^2\right) \gamma^2 \right) +\mathcal{O}(\dfrac{1}{c^5})
\end{equation}
Where $\gamma = \dfrac{2GM}{dc^2}$.

For smoothening the numerically computed distance, we use a moving window.

\subsection{Choosing the distance measure}

The plot for these distance measures for one of our simulations is shown in \ref{fig:dist_measures_q0p4} for illustration. 

It is observed that although the simple Euclidean distance measure seems less appropriate in comparison with the rest of the proper distance measures, it is the measure closest to the second-order Post-Newtonian distance measure. For any given simulation, the maximum cumulative (RMS) deviation between the simple Euclidean and other distance measures was found to be less than a few times the initial separation. Further, the cumulative deviation with the PN2 distance measure was found to be the least for the simple Euclidean distance measure. Therefore, throughout the analysis, we choose to work with this distance measure. 

The simple Euclidean distance measure also closely matches the Newtonian evolution of the separation between the two black holes. This can be verified through a simple fitting (see also \cite{etk_student_guide_2020}).

\clearpage

\cleardoublepage %
\flushbottom
\newpage
\cleardoublepage
\chapter{Waveform preparations}
\label{appendix:waveform_prep}

In this section, we discuss the process of extrapolation of gravitational waveforms from the numerical simulations run. \\

\section{Introduction}
Numerical relativity is crucial to the understanding of the dynamics of binary black holes and provides crucial inputs to the modelling of gravitational waveforms from, and dynamics of such systems. These waveforms informed by numerical relativity are in turn seminal to the construction of templates for the detection and parameter estimation of BBH mergers. Furthermore, tests of GR are carried out by comparing the observed signal in the gravitational wave detectors (like LIGO, Virgo, KAGRA) with these waveforms (from numerical relativity or templates i.e. waveform modes informed by NR). Thus, accuracy is not only important in the evolution of Einstein's field equations in numerical relativity codes but also in accurately computing and extracting the observable gravitational modes from evolved numerical systems.

The signals received in the gravitational wave detectors can be well approximated as being located at an infinite distance from the system responsible for these waves. However, in a numerical domain, it is not possible to have an infinite extent for evolving the spacetime using $3+1$ formalism. Thus, gravitational waveforms are conveniently extracted at large distances from the center of the numerical domain and are then extrapolated to infinity using one of the available methods (explained in the later sections).

The dominant information about the gravitational radiation is contained in the Weyl scalar $\Psi_4$. Consider a complex null tetrad ($n^{+}, m, \bar{m}, n^{-}$) where $n^{+}$ and $n^{-}$ denote the outgoing and ingoing null vectors. The Weyl scalar $\Psi_4$ is defined using the following contraction of the Weyl tensor $C^{\alpha \beta \gamma \delta}$:

\begin{equation}
\Psi_4 = C^{\alpha \beta \gamma \delta} n^{-}_{\alpha} \bar{m}_\beta n^{-}_\gamma \bar{m}_{\delta}  
\end{equation}

\begin{figure}
\centering
\includegraphics[width=\columnwidth]{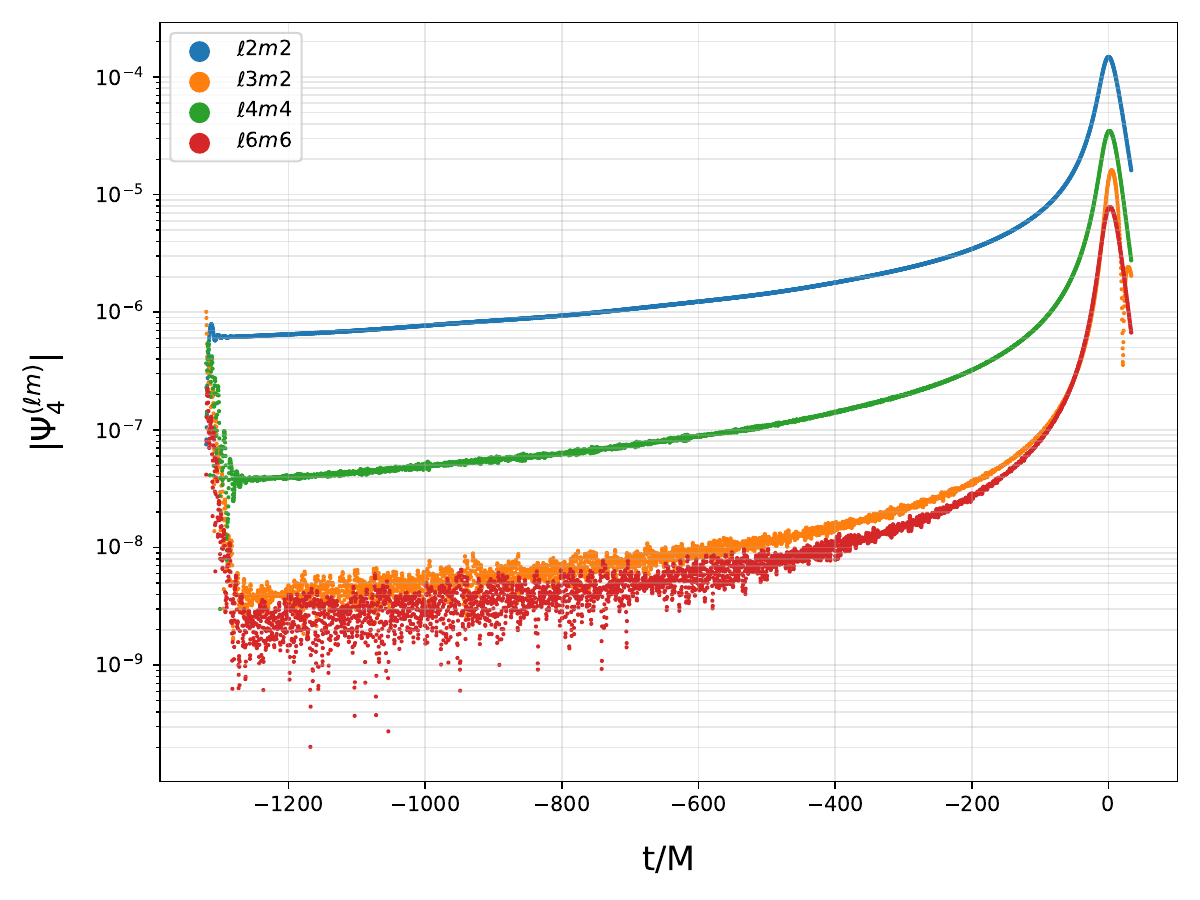}
\includegraphics[width=\columnwidth]{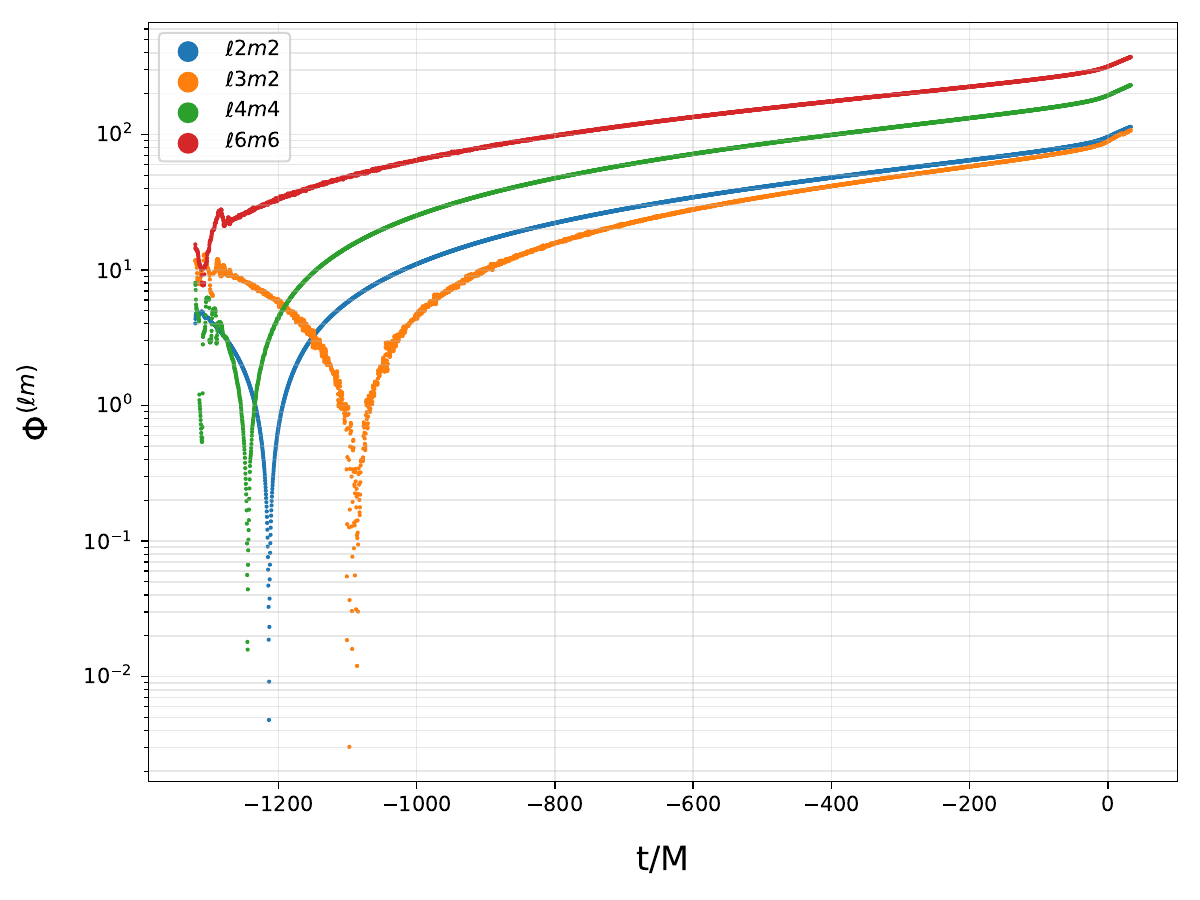}
\caption[The dominant waveform modes of $\Psi_4$ for a BBH system.]{The amplitude (top) and phase (bottom) evolution of the first four dominant modes of $\Psi_4$ an equal mass non-spinning BBH merger. The $\Psi_4$ was extracted at $r=500M$ coordinate sphere.}
\label{fig:modes_q1}
\end{figure}

In the Einstein toolkit, there are several thorns available for the computation of Weyl scalars. Some of these are:
\begin{itemize}
\item WeylScal4 \cite{yosef_zlochower_2022_6588641}
\item Psikedelia \cite{psik1}
\end{itemize}
The WeylScal4 thorn has been thoroughly tested and validated against the Psikadekia thorns \cite{ws4_t1_2011}. We use this for the computation of the Weyl scalars.
For convenience in use and analysis, one decomposes the above function in the Spin Weighted Spherical Harmonic (SWSH hereafter) bases $_{s}Y_{\ell m}$ instead of the angular coordinate basis ($\theta, \phi$), and works with the coefficients $C_{\ell m}$. Note that the above quantity is a rank two tensor and carries spin-weight $-2$ in the Newman-Penrose formalism.
\begin{equation}
C_{\ell m}(u) = \int_{R_{ext}} \Psi_4{(\theta, \phi), u} _{-2}Y_{\ell m} (\theta, \phi) dS
\end{equation}
Here $u$ is the retarded time, $R_{ext}$ is the radius of the extraction sphere, and $dS$ is the proper area element on the sphere.  \ref{fig:modes_q1} shows the first four dominant modes of $\Psi_4$.

The \colorbox{gray}{waveformtools.waveforms} module offers methods to expand data defined on a sphere into SWSH expansion and compute the coefficient. In particular the \\ \hl{waveformtools.waveforms.spherical\_array} class has a \hl{`to\_modes\_array'} methods that converts the spherical data into modes.

This is then corrected for the gauge effects due to the limitation of the numerical domain, CoM drift, etc. The gravitational wave news and strain (that is directly relevant in the detectors) is computed by a single and double time integration of $\Psi_4$ respectively.
\begin{align}
\mathcal{N}_{\ell m} &= \int_{R_{ext}} C_{\ell m} (u) du \\
h_{\ell m} &= \int_{R_{ext}}\int_{R_{ext}} C_{\ell m} (u) du 
\end{align}
Note that the integration is carried out over a fixed coordinate $r$ and therefore the integration is equivalent to an integration in time.

The total power of transverse gravitational radiation crossing the sphere at $R_{ext}$ is given by the News:
\begin{align}
\dfrac{dE}{dt} = \int_{R_{ext}}\int_{R_{ext}} \vert \mathcal{N} \vert^2 dS  
\end{align}

This integration cannot be directly carried out in the time domain as it introduces low-frequency drifts. This is computed in the frequency domain using the fixed frequency integration technique \cite{Reisswig_2011}. 

\begin{figure}
\centering
\includegraphics[width=\columnwidth]{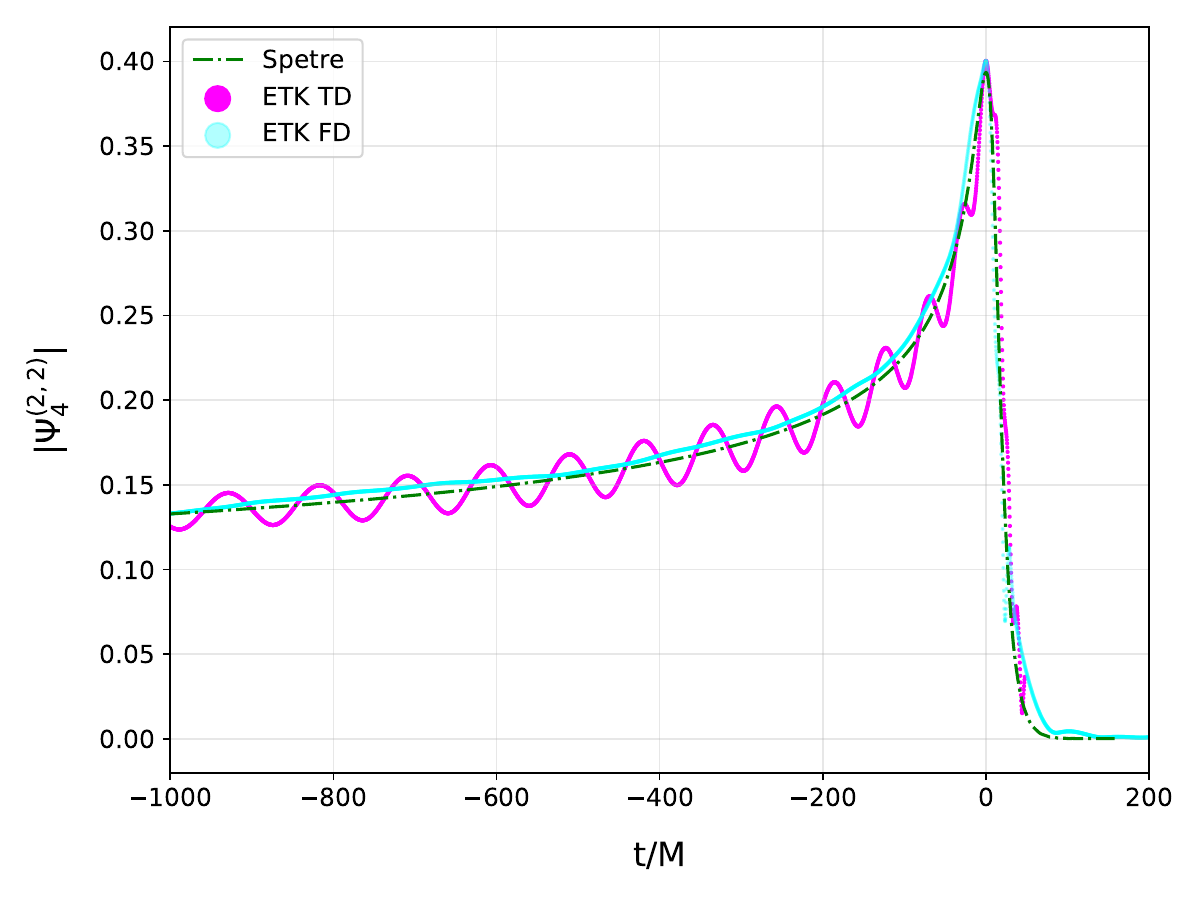}
\caption[The time integration of $\Psi_4$.]{The integration of $\Psi_4 ^{(22)}$ to obtain the strain $h^{(2 2)}$. The amplitudes of the respective waveforms are shown. Here, the curve in magenta is integrated in the time domain using a piece-wise finite difference integrator, whereas the curve in cyan is integrated using a fixed frequency integration in the frequency domain. For comparison, the strain obtained from SpECTRE using CCE is shown in green.}
\label{fig:integ}
\end{figure}

A routine for this is implemented in the \hl{waveformtools} module. The relevant function is\\ \hl{waveformtools.integrate.fixed\_frequency\_integrator}. There are also methods implemented for differentiating waveforms obtaining News, $\Psi_4$ from the strain, etc. These include finite difference and spectral methods and can be found in the module\\ \hl{waveformtools.differentiate.} The \hl{waveforms.modes\_array} class also contains methods to inter-convert strain, news and $\Psi_4$.  \ref{fig:integ} shows the strains obtained from various integration methods.

We use the Einstein Toolkit for the numerical evolution of the spacetime. In the ETK, the Weyl scalar is computed in the numerical domain using a quasi-Kinnersley tetrad. This tetrad adapts to and aligns with the Petrov type D behaviour of the Kerr spacetime at late times. The thorn \textsc{WeylScal4} is used to compute this Weyl scalar and the data is extracted at various coordinate spheres ($r= 100, ...0$). These coordinate spheres are placed at equal intervals in $1/r$, which aids in the extrapolation of the waveforms to infinity, which will be explained in later sections.

\section{Waveform extrapolation}

The largest available extraction radius is $500M$. This waveform is sufficiently accurate for most purposes. However, some gauge effects are still present. These include:
\begin{enumerate}
\item Finite radius effects. The ideal waveforms reach and reside at future null infinity $\mathcal{I}^{+}$. Due to the limited extent of the numerical domain, the waveforms extracted at finite radii do not contain the information from their evolution all the way to $\mathcal{I}^{+}$.
\item Gauge effects. The numerical gauge used to evolve the spacetime is dynamical and changes with simulation time. Thus, e.g., the time coordinate used to label the extracted gravitational waveforms can have small gauge effects. 
\item Center of mass drifts. Due to asymmetry in numerical evolution (e.g. reflection at boundaries, etc.), and physical effects, the center of mass of the system drifts during the course of the numerical simulation. However, the spheres on which the waveforms are extracted are at fixed coordinates. This introduces mode mixing in the extracted gravitational waveforms. 
\item BMS freedom. Even after computing/ extrapolating the waveforms at future null infinity, the waveforms possess the BMS symmetry. 
\end{enumerate}
Point (1) above is partly handled by extrapolation procedures which will now be described, or by using Cauchy characteristic evolution which has been currently developed \cite{SpECTRE_CCE_2021, Spectre_weyl2021}. 

\subsection{Perturbative extrapolation}

In order to extrapolate the gravitational waveform to infinity from the finite radius data, one uses approximate methods involving the Teukolsky equation \cite{Nakano2015}. In this procedure, the dynamics interior to the extraction sphere is considered as a perturbed Kerr spacetime. The formula (Equation 29 of \cite{Nakano2015}) for the extrapolation used is:
\begin{eqnarray}
\left. r\,\psi_4^{\ell m}\right|_{r=\infty}
&=&
\left(1-{\frac {2M}{{r}}}\right) \left( r\psi_{4\ell m}^{\rm NR}(t,r) 
- \frac{(\ell -1)(\ell +2)}{2\,r} \int dt [r\psi_{4\ell m}^{\rm NR}(t,r)]
\right. \cr && \qquad \left.
+ \frac{(\ell -1)(\ell +2)(\ell^2+\ell-4)}{8\,r^2} \int \int dt dt [r\psi_{4\ell m}^{\rm NR}(t,r)]
\right)
\cr &&
+ \frac{2\,i\,a}{(\ell+1)^2} \sqrt{\frac{(\ell+3)(\ell-1)(\ell+m+1)(\ell-m+1)}{(2\ell+1)(2\ell+3)}} \left( [r\partial_t{\psi}_{4\ell+1 m}^{\rm NR}(t,r)] 
\right. \cr && \qquad \left.
- \frac{\ell (\ell+3)}{r} [r\psi_{4\ell+1 m}^{\rm NR}(t,r)] \right)
\cr &&
- \frac{2\,i\,a}{\ell^2} \sqrt{\frac{(\ell+2)(\ell-2)(\ell+m)(\ell-m)}{(2\ell-1)(2\ell+1)}}
\left( [r\partial_t{\psi}_{4\ell-1 m}^{\rm NR}(t,r)]
\right. \cr && \qquad \left.
- \frac{(\ell-2)(\ell+1)}{r} [r\psi_{4\ell-1 m}^{\rm NR}(t,r)] \right)
\cr &&
+ \textrm{[higher}\,\,\textrm{order]} 
\,.
\label{eq:improved_formula}
\end{eqnarray}

Presently, we use this method for the extrapolation of waveforms. The improved second-order method corrects for second-order effects in mass and first-order effects in spin, along with the correction for the use of Spin Weighted Spherical Harmonics instead of Spin Weighted Spheroidal Harmonics.

\begin{figure}
\centering
\includegraphics[width=\columnwidth]{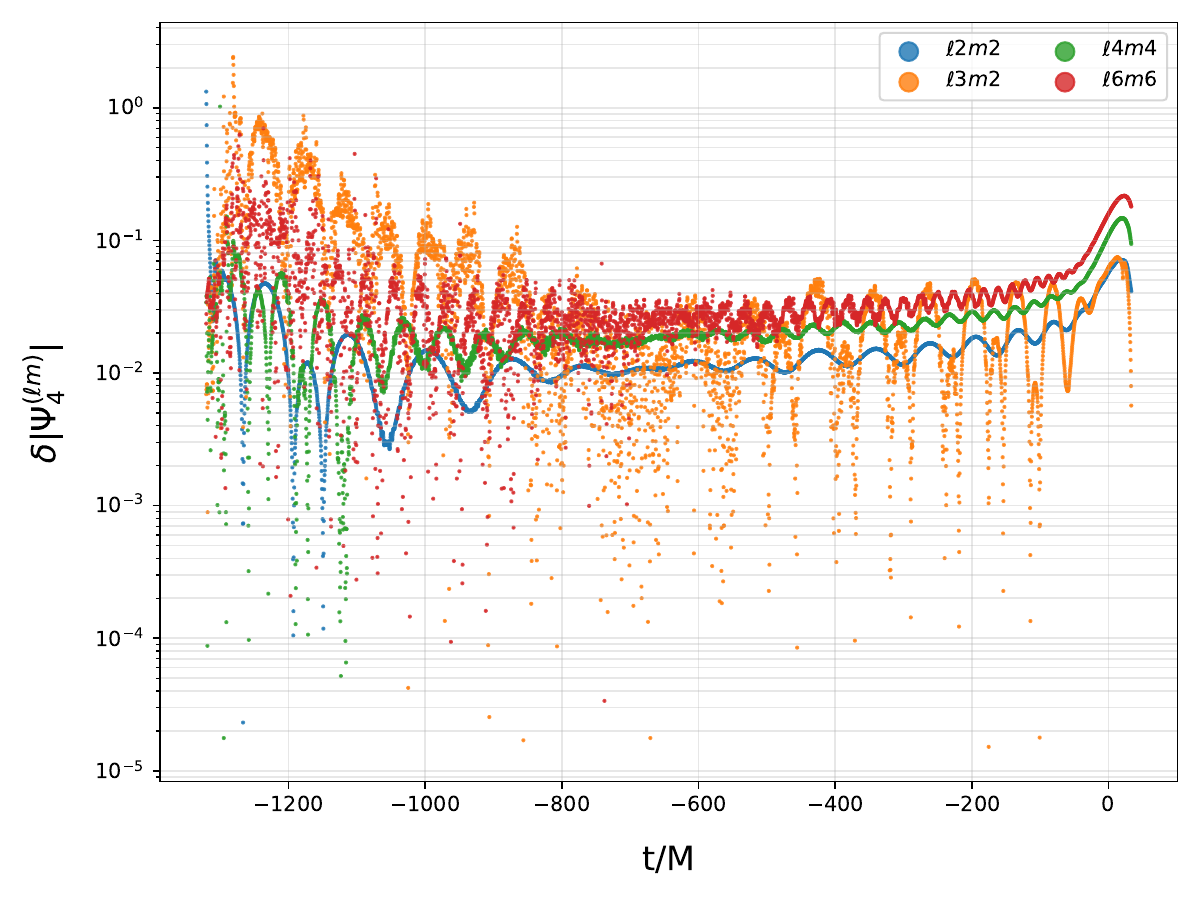}
\includegraphics[width=\columnwidth]{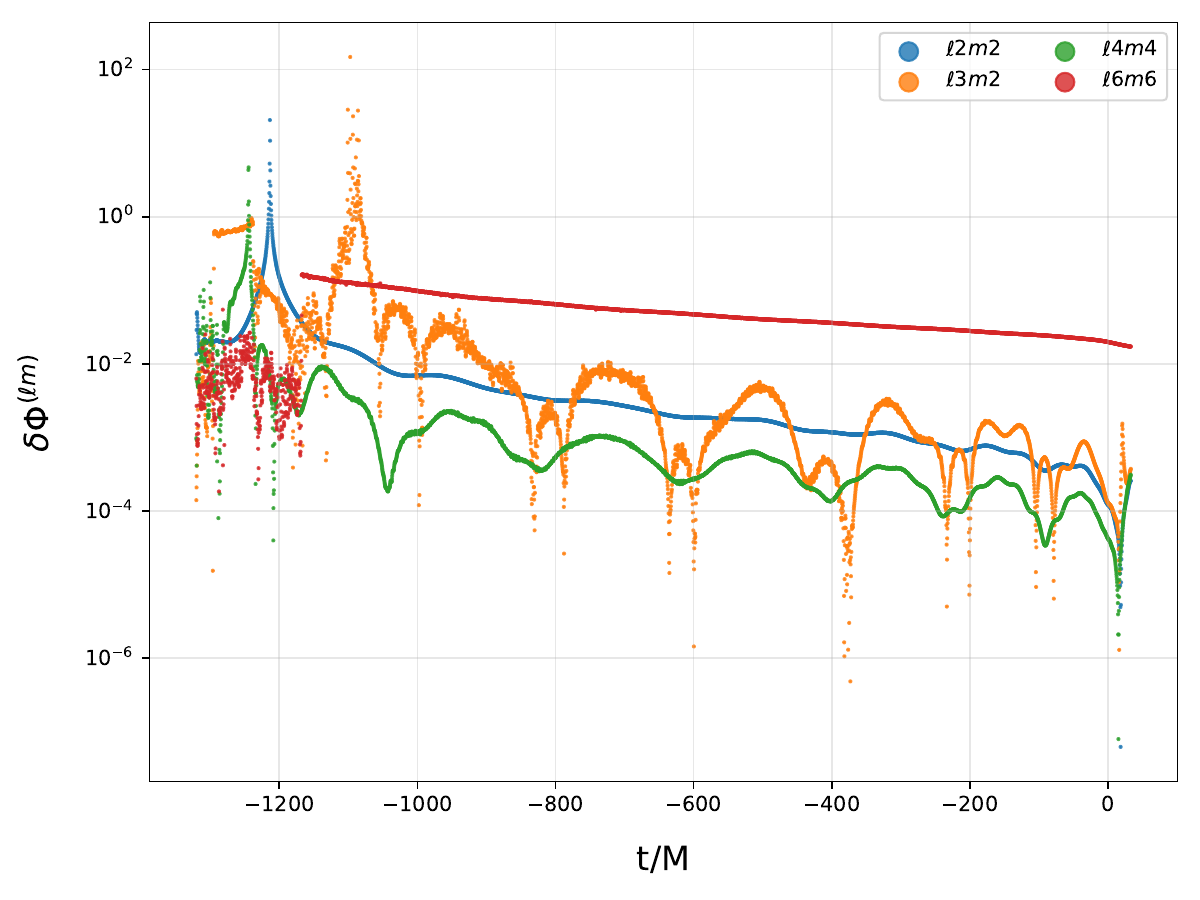}
\caption[The changes to the waveforms due to extrapolation to infinity.]{The fractional absolute difference in the values ($\vert \frac{(val_{500} - val_{\mathcal{I}^+})}{val_{500}}\vert$) of the amplitude (top) and phase (bottom) evolution of the first four dominant modes of $\Psi_4$ of an equal mass non-spinning BBH merger extracted at $r=500M$ and extrapolated to infinity.}
\label{fig:extrap}
\end{figure}

\begin{figure}[htpb]
\centering
\includegraphics[width=\columnwidth]{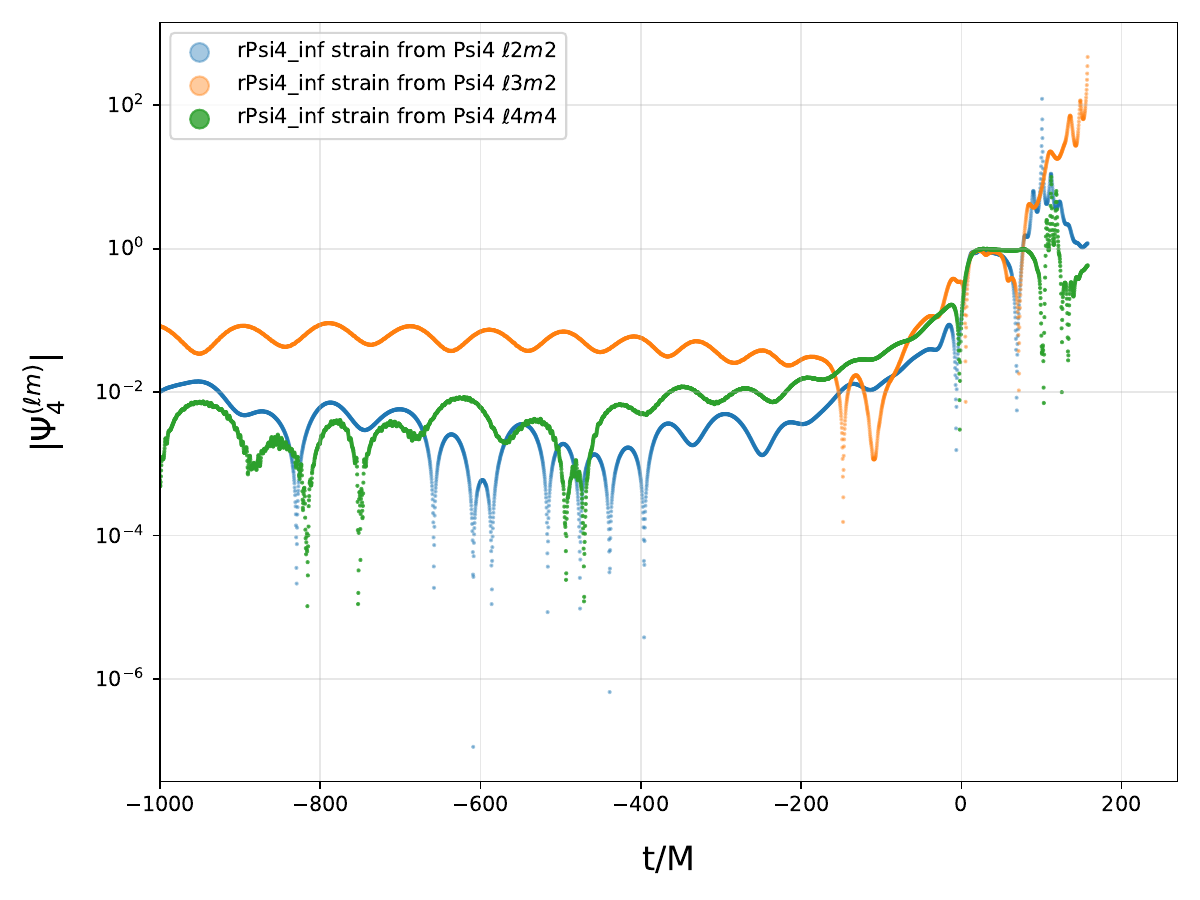}
\includegraphics[width=\columnwidth]{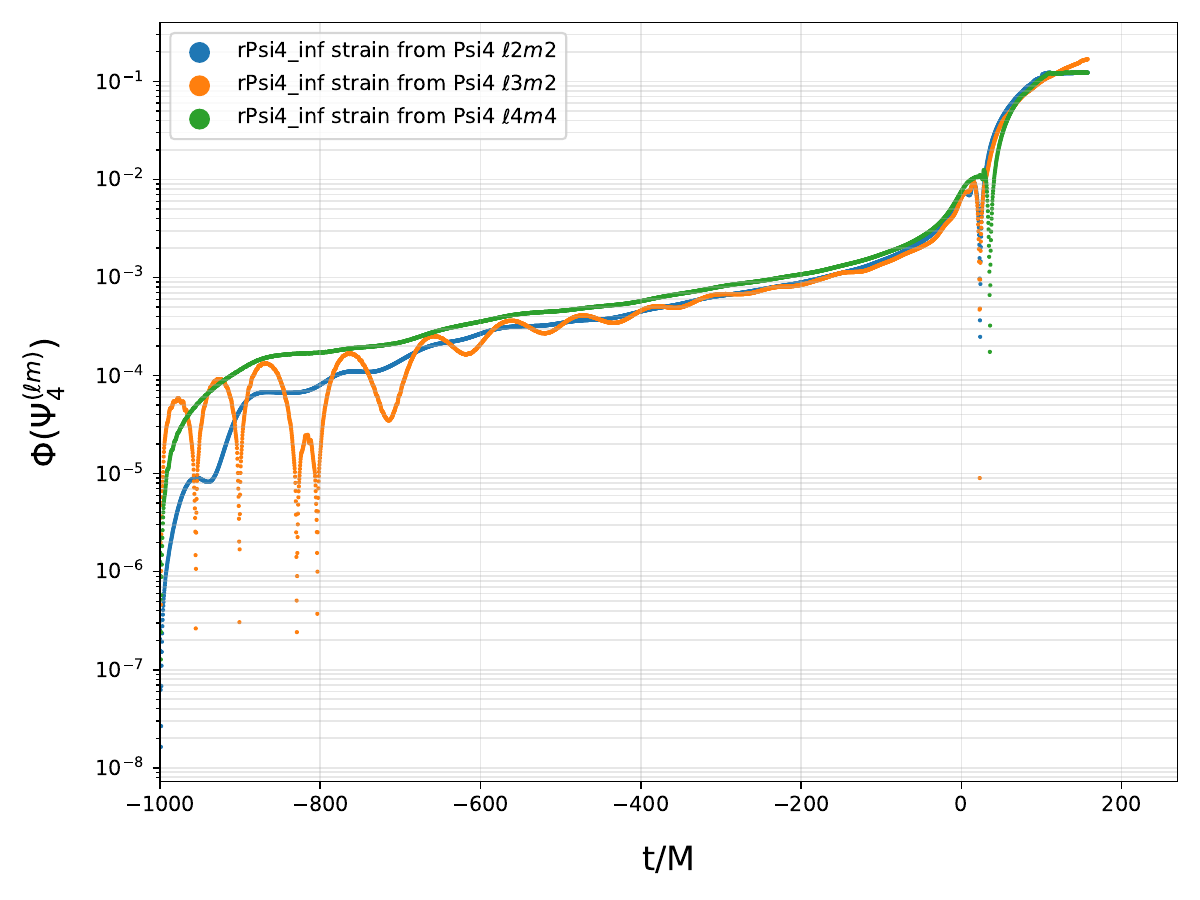}
\caption[Waveform comparison against SpECTRE CCE.]{The fractional absolute difference in the amplitudes (top) and phase (bottom) evolution of the first four dominant modes of the extrapolated strain and the strain from SpECTRE. The simulation corresponds to the same equal mass non-spinning BBH merger. The post-merger portion of the waveform from ETK (in dotted lines) is not usable as it is affected by the tapering function due to the smaller number of samples.}
\label{fig:extrap_comp}
\end{figure}

Methods for extrapolating waveforms are implemented in the \hl{extrapolate} sub-module. The \hl{waveforms.modes\_array} class contains methods to achieve the same. See  \ref{fig:extrap} and \ref{fig:extrap_comp} for a comparison between the extrapolated and non-extrapolated waveforms.

\subsection{Numerical methods}

Numerical methods can be used to extrapolate the waveforms to infinity. This is done by noting that the Weyl scalar $\Psi_4$ has the peeling behaviour of $1/r$ as $r\rightarrow \infty$. Thus, the contribution at $\mathcal{I}^{+}$ can be computed as the lowest order coefficient $\Psi_4 ^{\mathcal{I}^+}$ to the following fit to the polynomial expansion:
\begin{equation}
r \Psi_{4 (\ell m)} |_{r_{ext}} (u)= r \Psi_{4 (\ell m)} ^{\mathcal{I}^+} (u) + \sum_{k = 1}^{k_{max}} \dfrac{\Psi_{4 (\ell m)} ^{(k+1)}(u)}{r_{ext}^k}
\end{equation}
To carry this out, one needs the waveform data from several extraction radii. These are usually equally spaced in $1/r$ and the number of such available extraction spheres decides the order of the polynomial $k_{max}$ in $1/r$ to use for fitting. 

This method is being implemented in the \hl{extrapolate} sub-module and will be available shortly.

\section{A Note on \textsc{Llama/WaveExtractCPM} thorn}

\begin{figure}[htpb]
\centering
\includegraphics[width=\columnwidth]{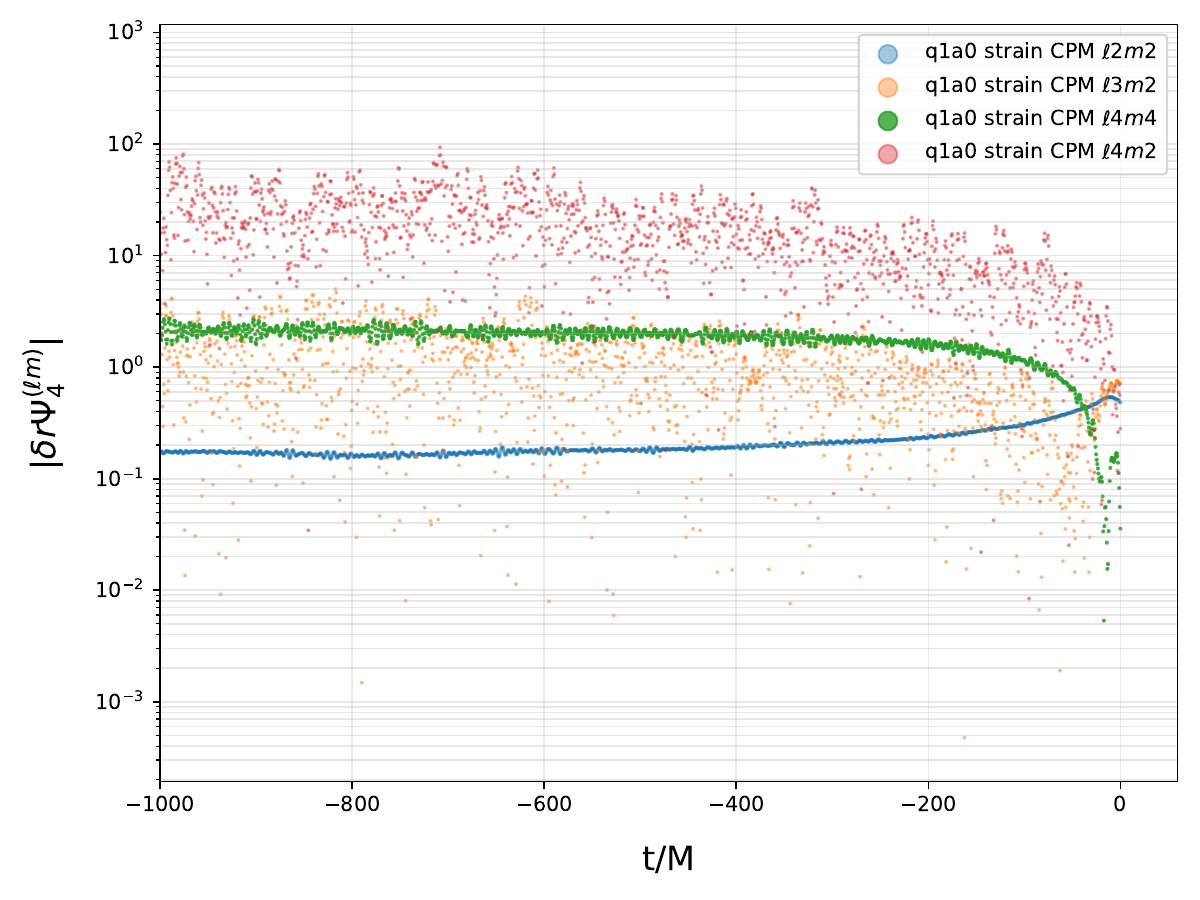}
\includegraphics[width=\columnwidth]{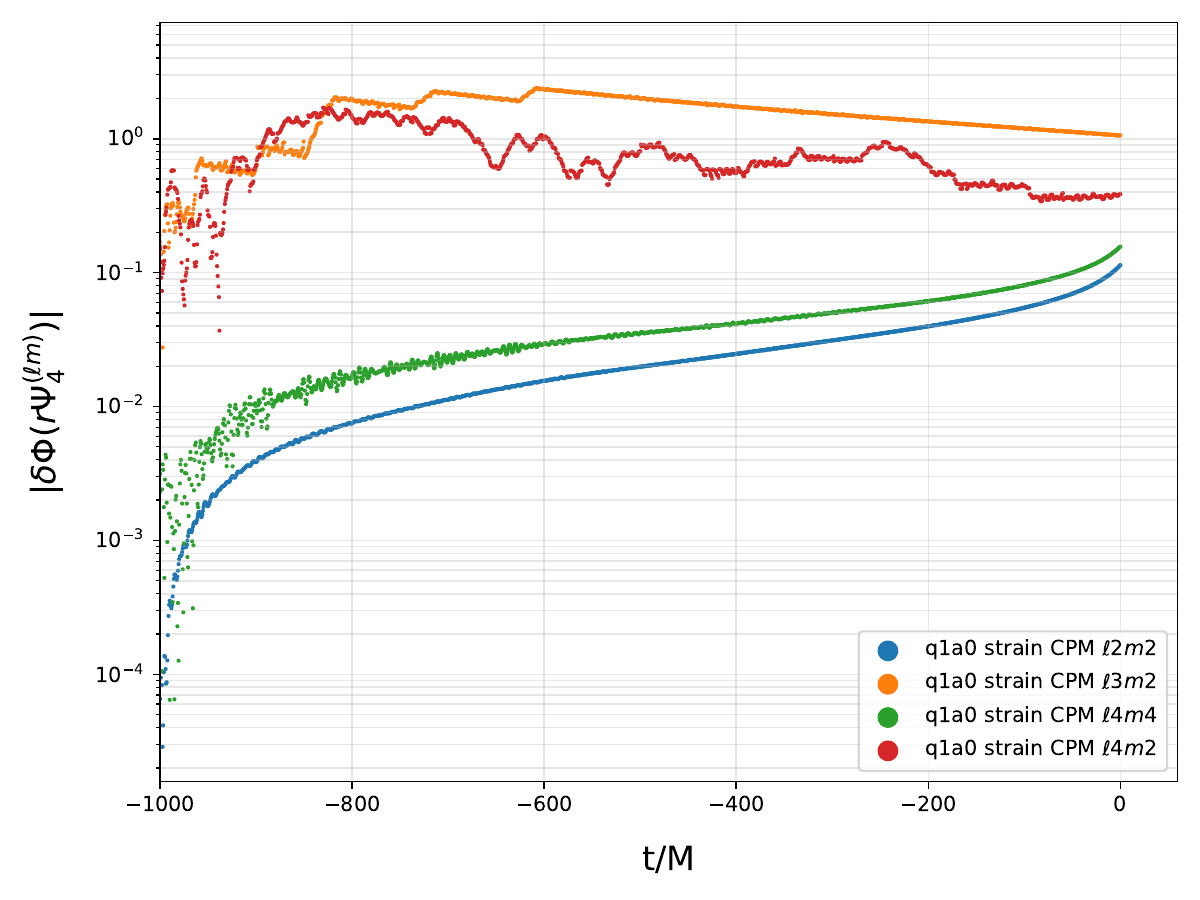}
\caption[Waveform comparison against WaveExtractCPM.]{The fractional absolute difference in the values ($\vert \frac{(val_{WCPM} - val_{\mathcal{I}^+})}{val_{WCPM}}\vert$) of the amplitude (top) and phase (bottom) evolutions of the first four dominant modes of the gravitational wave strain extracted using the thorn WaveExtractCPM and $\Psi_4$ of an equal mass non-spinning BBH merger. The $\Psi_4$ is extrapolated to infinity whereas the strain from WaveExtractCPM is extracted at $r=500M$.}
\label{fig:modes_cpm_vs_psi4int}
\end{figure}

The \textsc{WaveExtractCPM} thorn uses the Z-CPM formalism of the perturbations of the Schwarzschild spacetime to extract the gauge-fixed perturbations and compute the strain directly at the required extraction spheres. This is not a great choice to compute the gravitational waveform due to the approximations involved. Furthermore, the waveform obtained by this procedure does not match well with the other waveforms obtained from other methods. Please see  \ref{fig:modes_cpm_vs_psi4int} for further details.

\section{Centre of Mass corrections}

\begin{figure}[hbtp]
	\begin{tabular}{cc}
		\includegraphics[width=0.49\columnwidth]{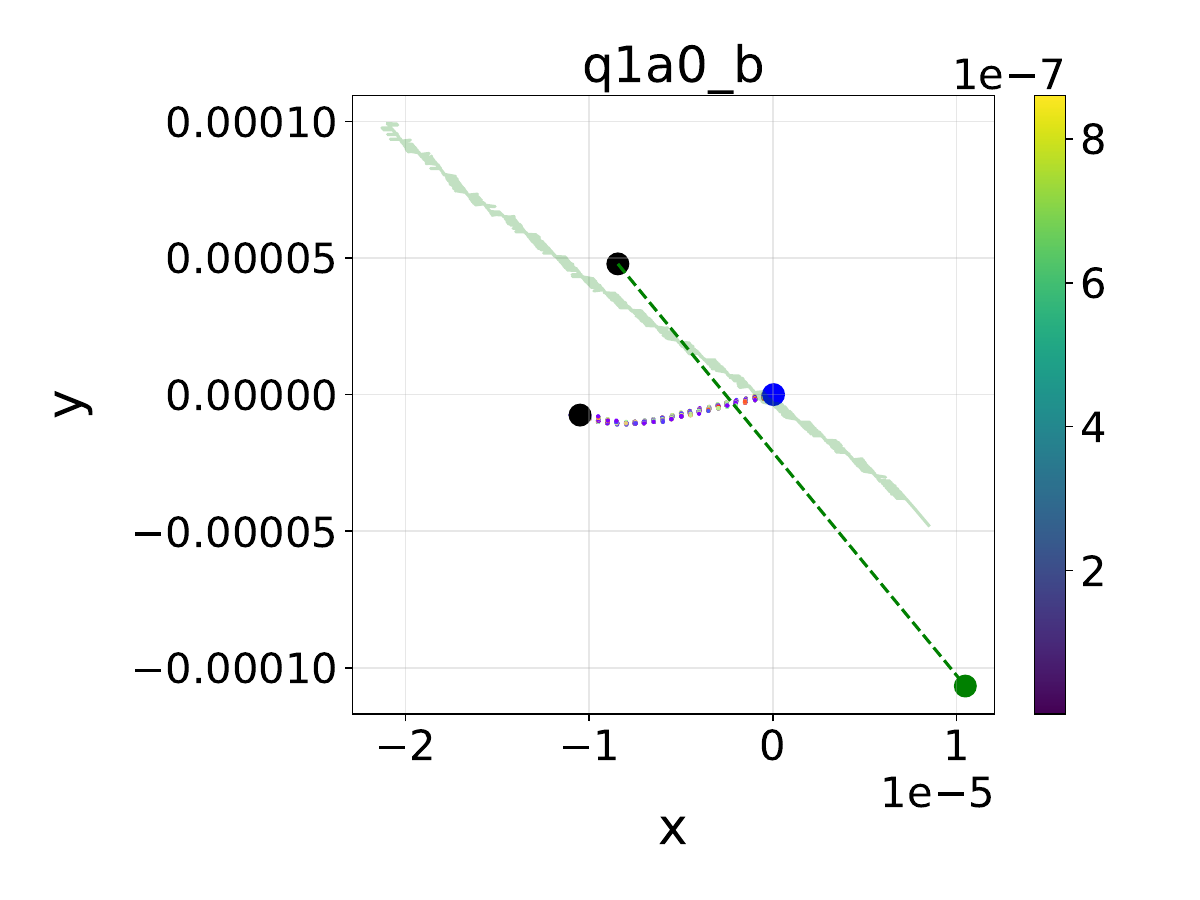} &
		\includegraphics[width=0.49\columnwidth]{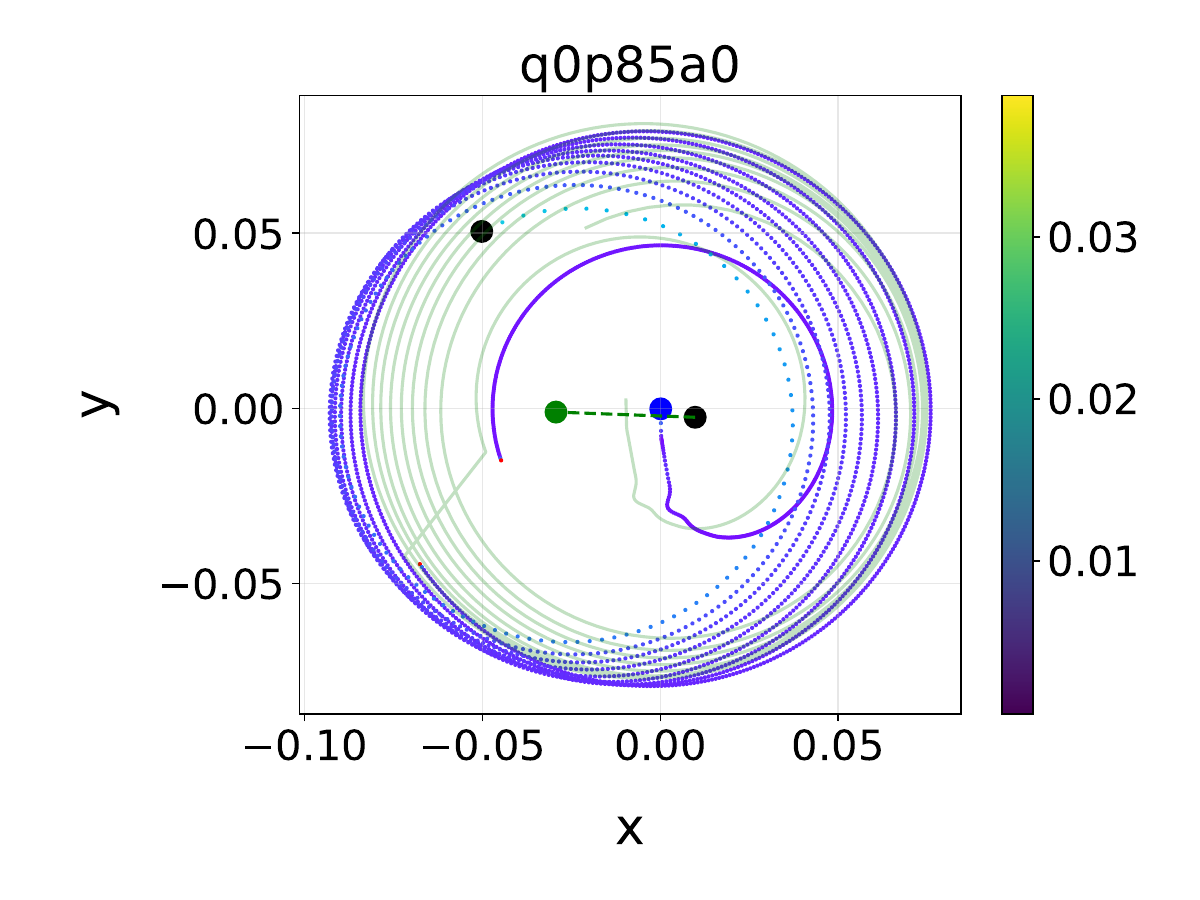} \\
		(i)   & (ii) \\[2pt]
	\end{tabular}
	\begin{tabular}{ccc}
		\includegraphics[width=0.49\columnwidth]{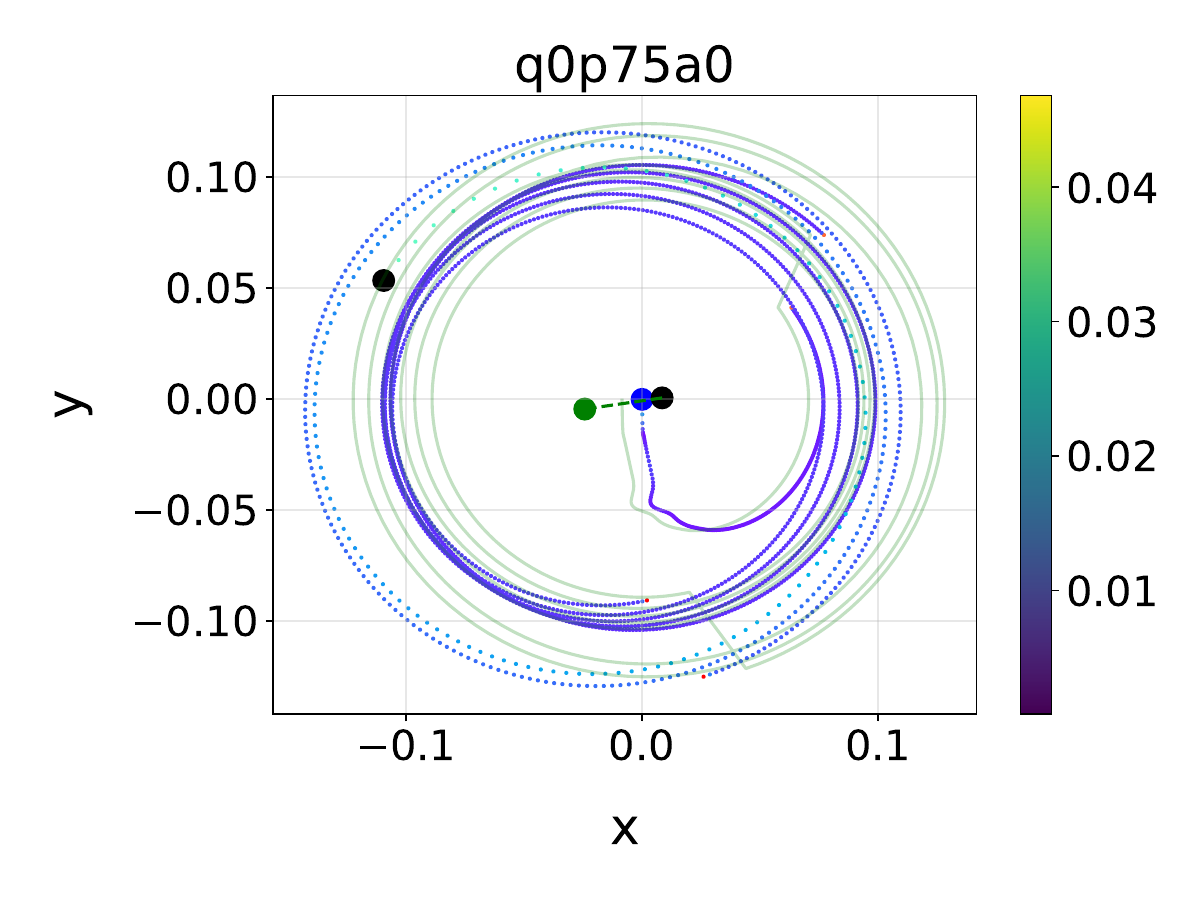} &
		\includegraphics[width=0.49\columnwidth]{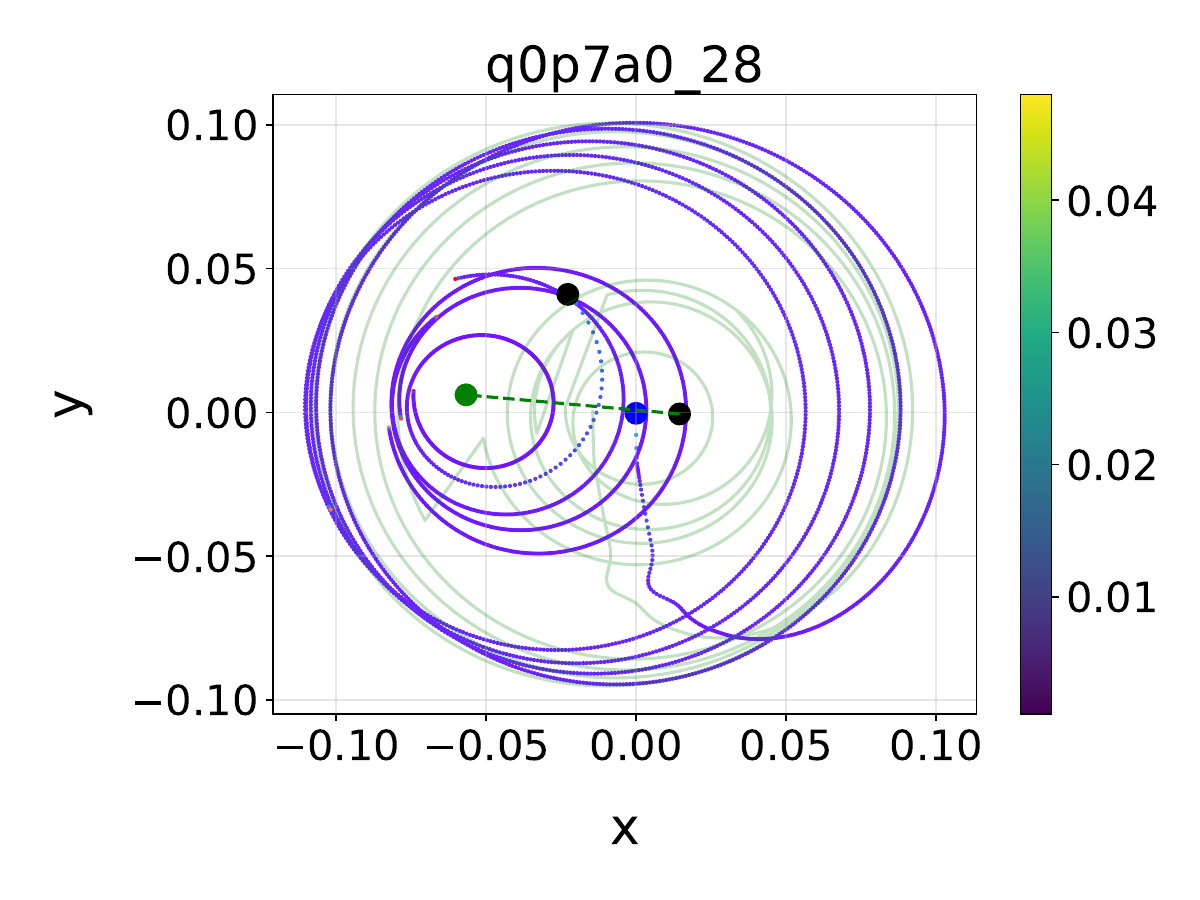} \\
		(iii)  & (iv) \\[2pt]
	\end{tabular}
	\caption[Drifts of the centre of mass of the BBH system - set I ($q<0.67$).]{The centre of mass drifts. Please see the next figure for details.}
\end{figure}

\begin{figure}[hbtp]
	\begin{tabular}{cc}
		\includegraphics[width=0.49\columnwidth]{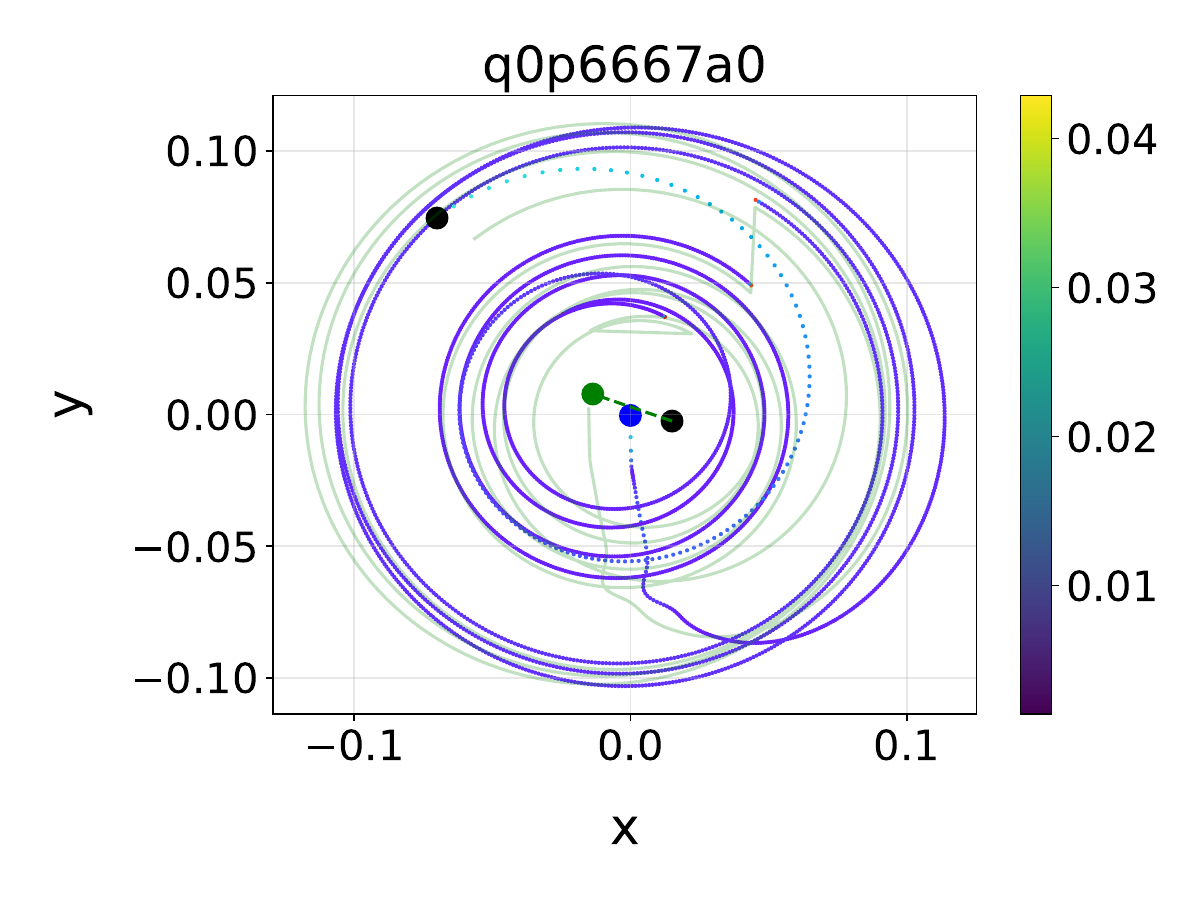} &
	    \includegraphics[width=0.49\columnwidth]{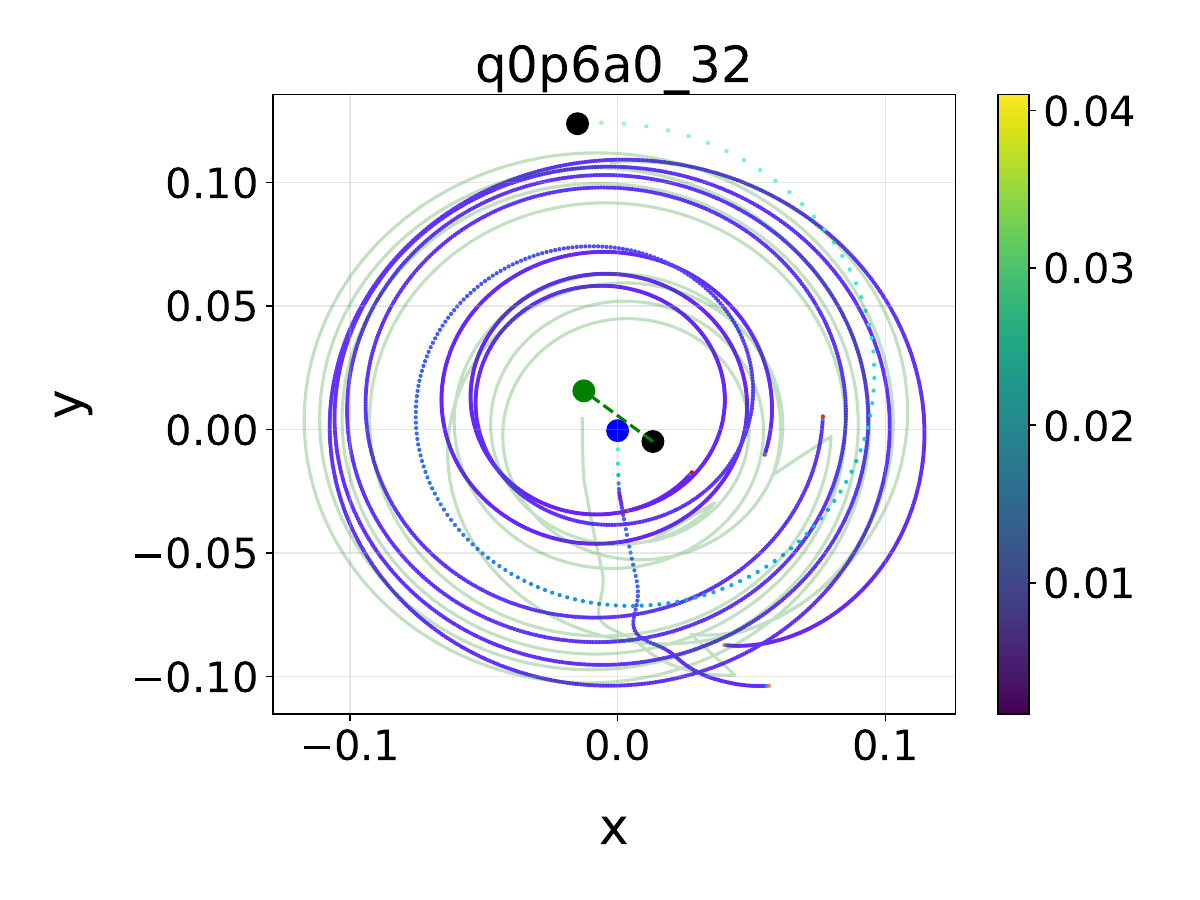} \\
		(v)   & (vi) \\[2pt]
	\end{tabular}
	\begin{tabular}{ccc}
		\includegraphics[width=0.49\columnwidth]{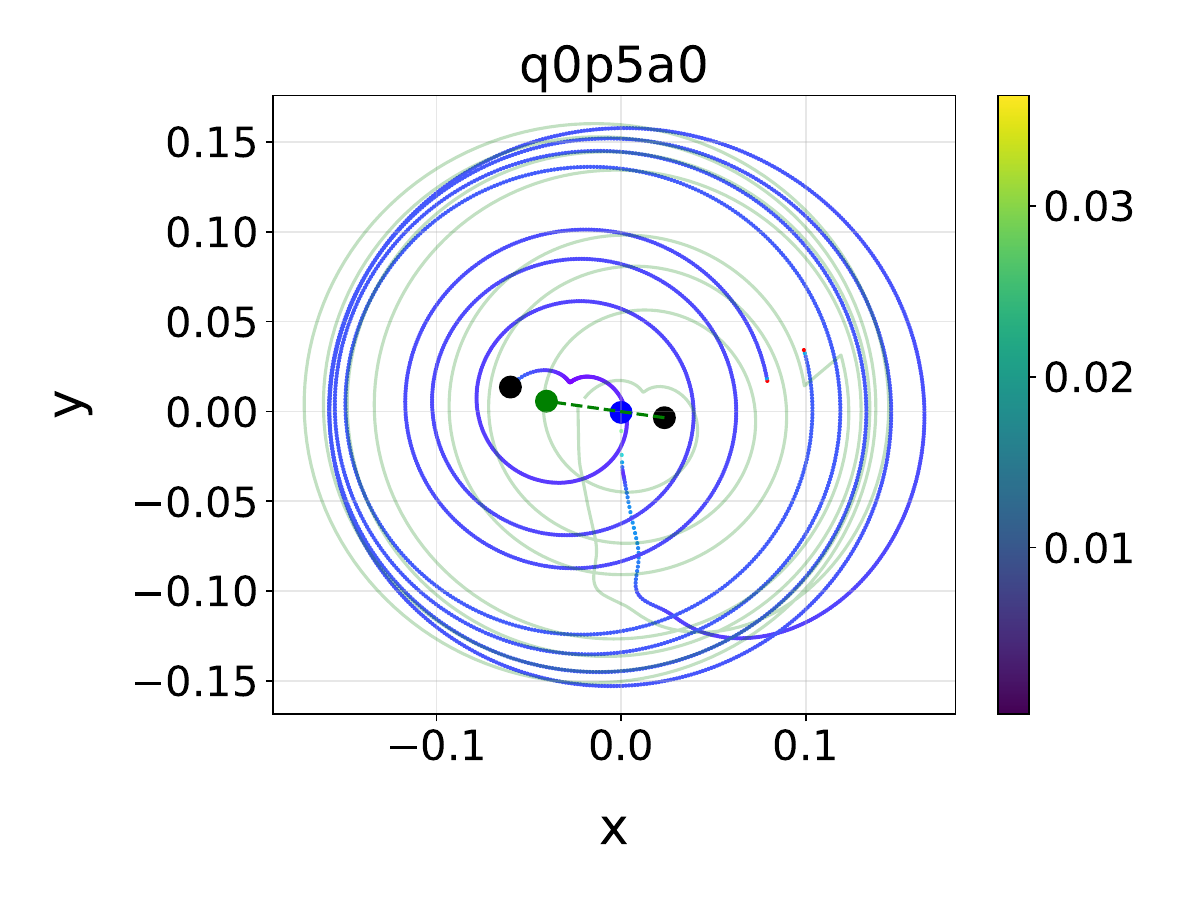} &
	     \includegraphics[width=0.49\columnwidth]{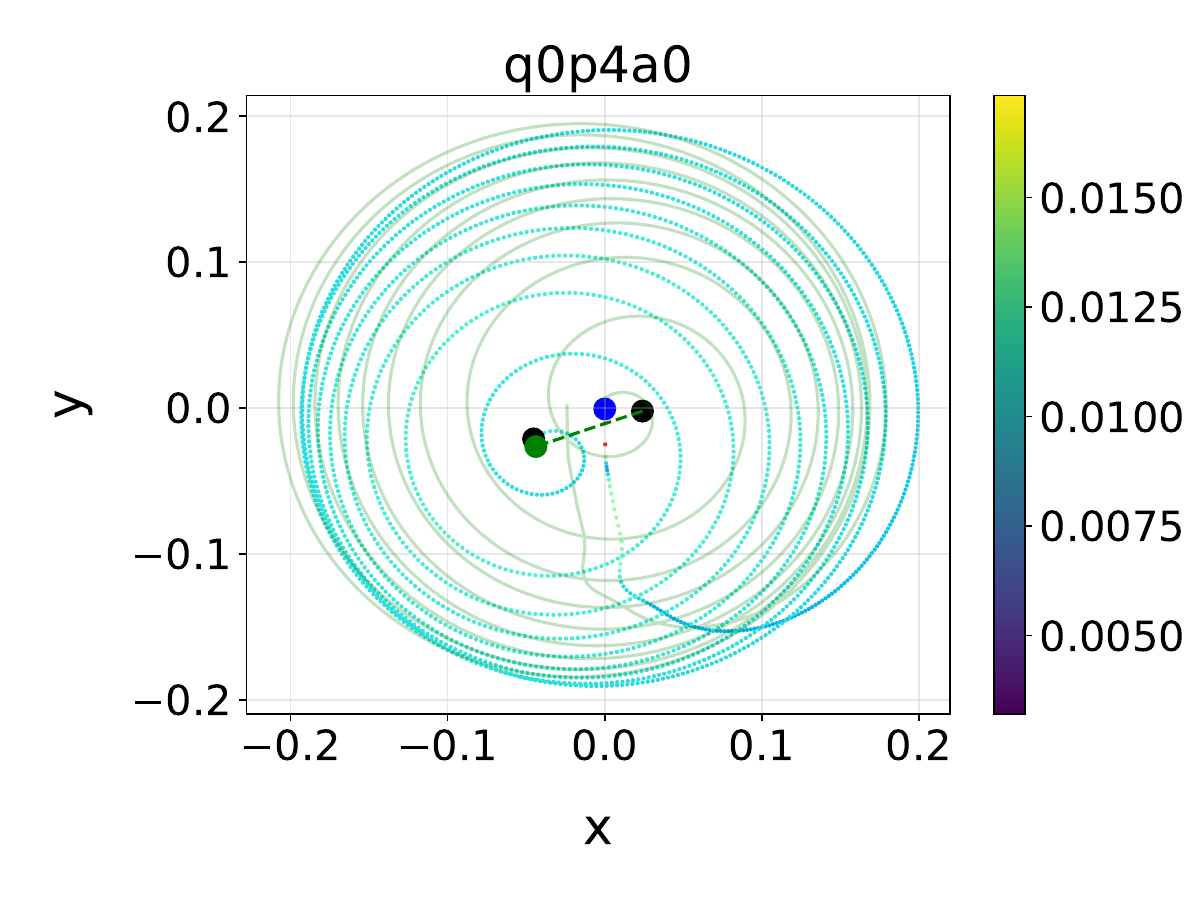} \\
		(vii)  & (viii) \\[2pt]
	\end{tabular}
\caption[Drifts of the centre of mass of the BBH system - set II ($q\geq0.67$).]{The centre of mass drifts. Each panel shows the center of mass motion in the due course of the inspiral for the a (non-spinning) simulation of a given mass-ratio (mentioned in its title) as dotted scatter points. The points have been coloured according to the instantaneous coordinate speed at that location. The CoM location can be seen to gradually drift with time. The initial location of the CoM is shown as a blue dot, and it terminates at a black dot. Also shown is the mean linear drift obtained by best fit as green dotted straight lines. The initial location of the best-fit curve is shown as a black dot and terminates at the green dot. The motion of the BHs in the CoM frame is shown as solid transparent curves in green. The CoM drift is absent for (i) q1a0\_b simulation as the system is the most symmetric of all. The drift increases with asymmetry between the masses  as the mass ratio decreases.} 
\label{fig:CoM_drift}
\end{figure}
\begin{figure}
		\includegraphics[width=\columnwidth]{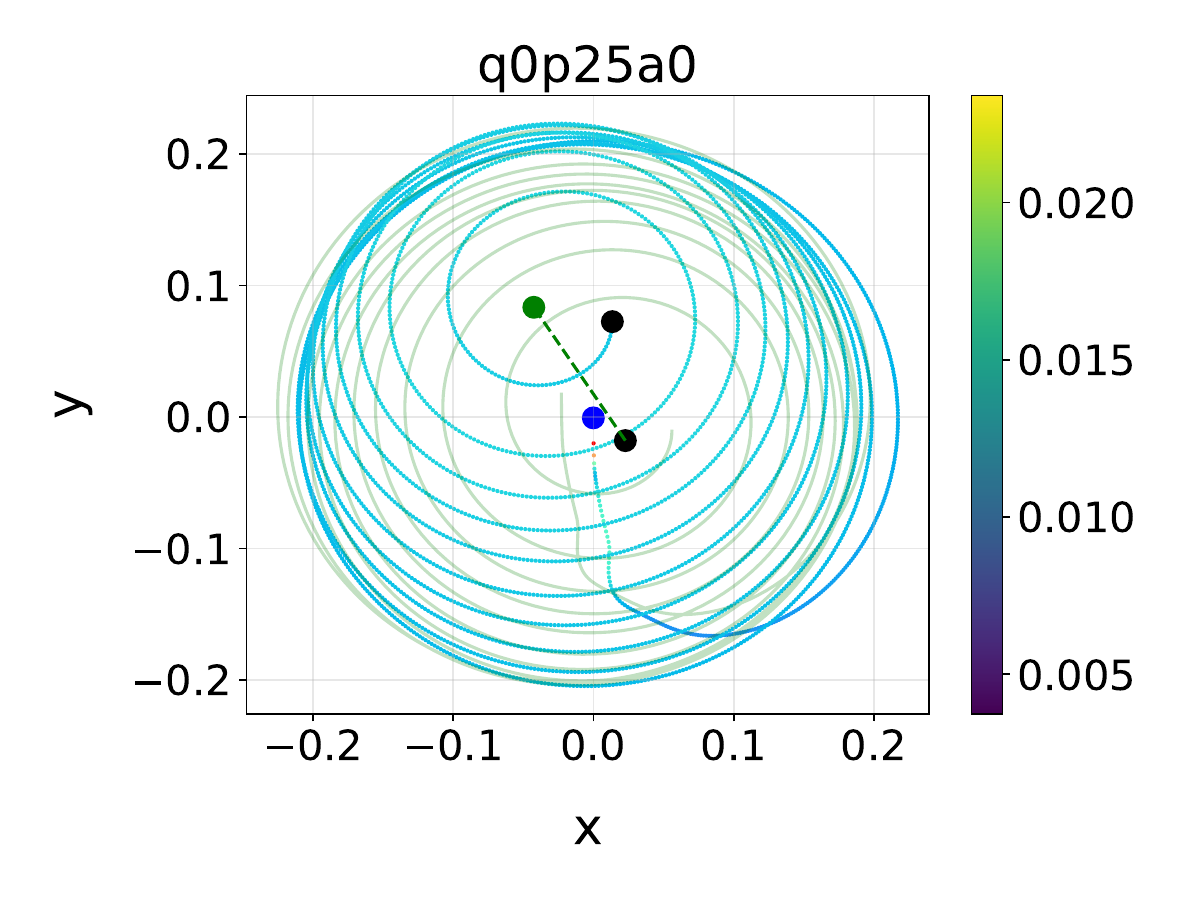}
		\caption[Drift of the centre of mass of the system $q=0.25$.]{The CoM drift for $q=0.25$}
\end{figure}

\begin{figure}[htpb]
	\centering
	\includegraphics[width=\columnwidth]{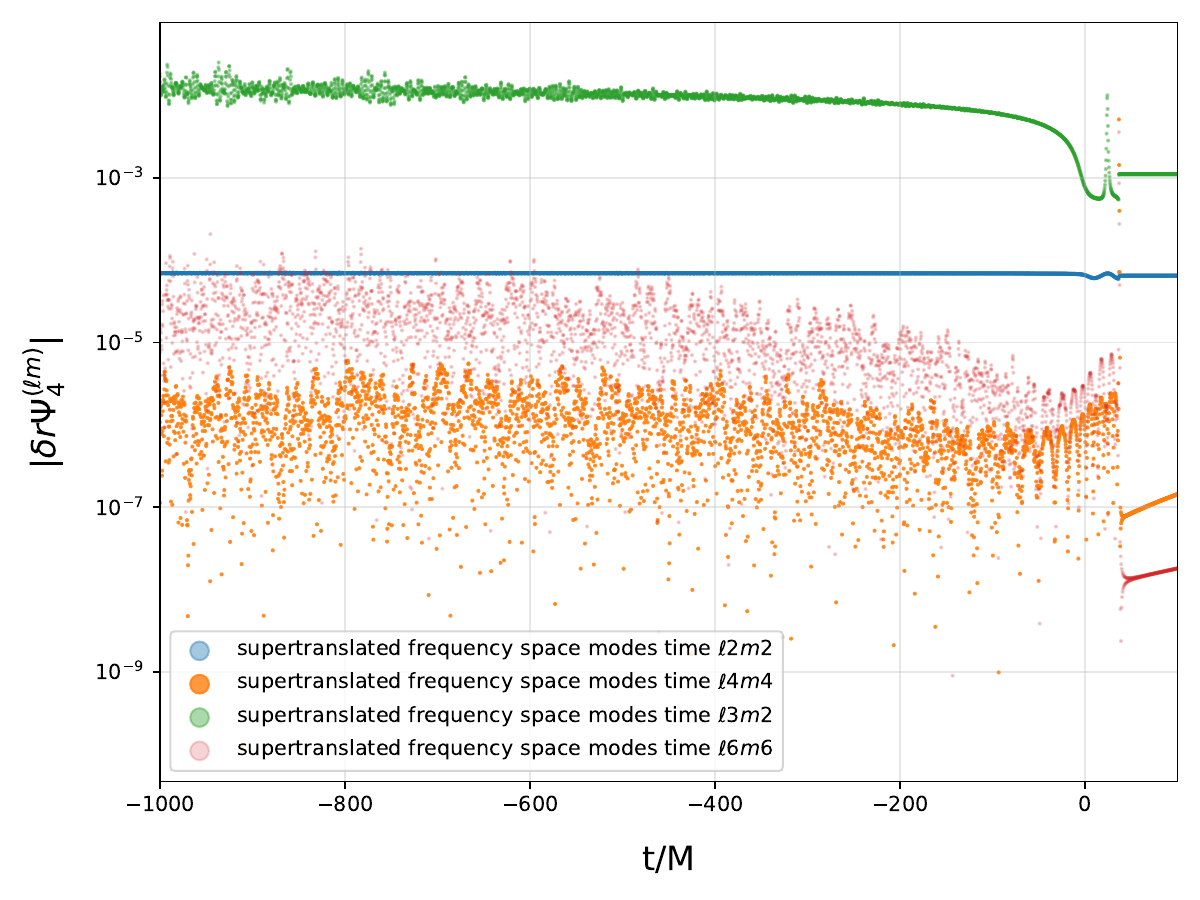}
	\includegraphics[width=\columnwidth]{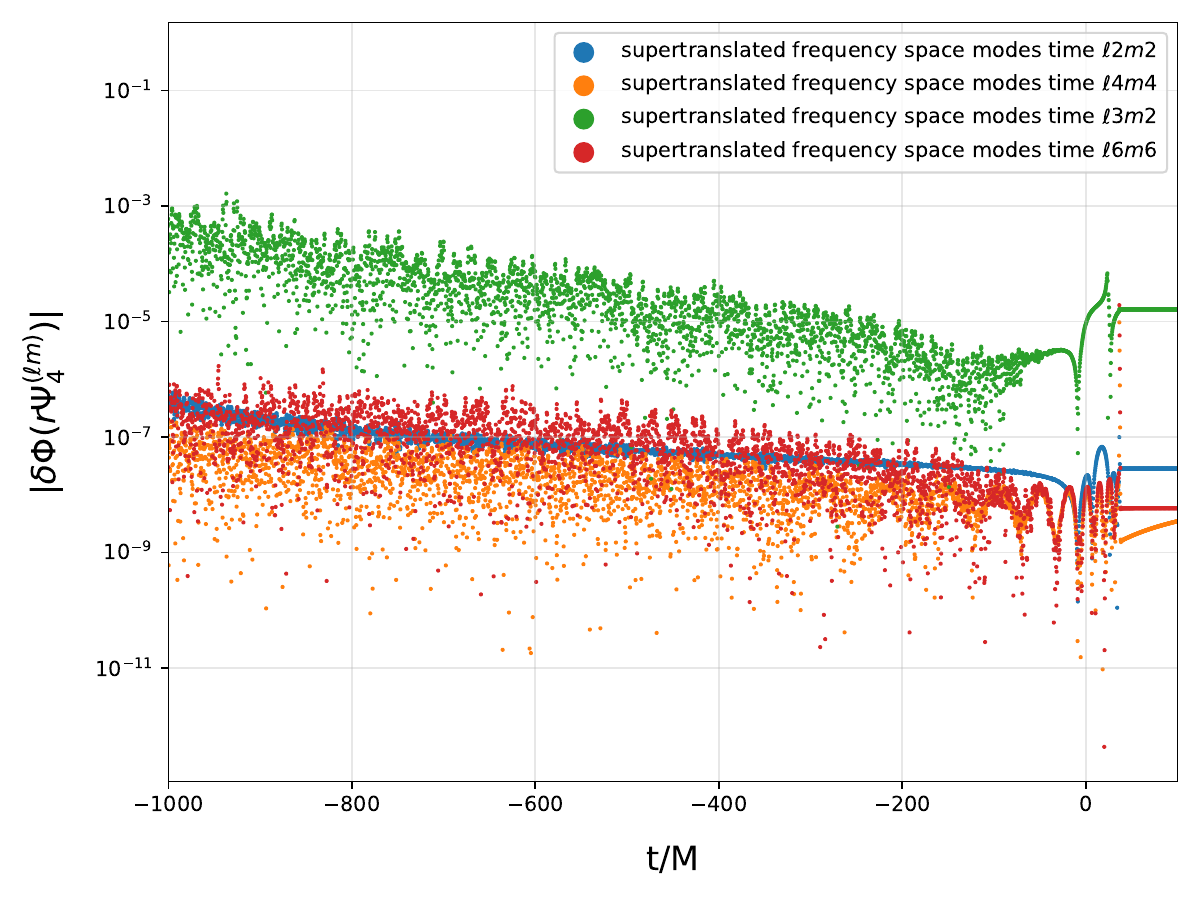}
	\caption[Changes in the waveforms due to centre of mass correction.]{The fractional absolute difference in the values ($\vert \frac{(val_{CoM} - val_{\mathcal{I}^+})}{val_{\mathcal{I}^+}}\vert$) of the amplitudes (top) and phase (bottom) evolution of the first four dominant modes of the gravitational waveforms $\Psi_4$ of the simulation $q=0.85$ extrapolated to infinity and also corrected for the CoM drifts shown in  \ref{fig:CoM_drift}. As can be seen, the CoM correction introduces corrections of the order of a per cent.}
	\label{fig:CoM_waveform}
\end{figure}

\begin{figure}[htpb]
	\centering
	\includegraphics[width=\columnwidth]{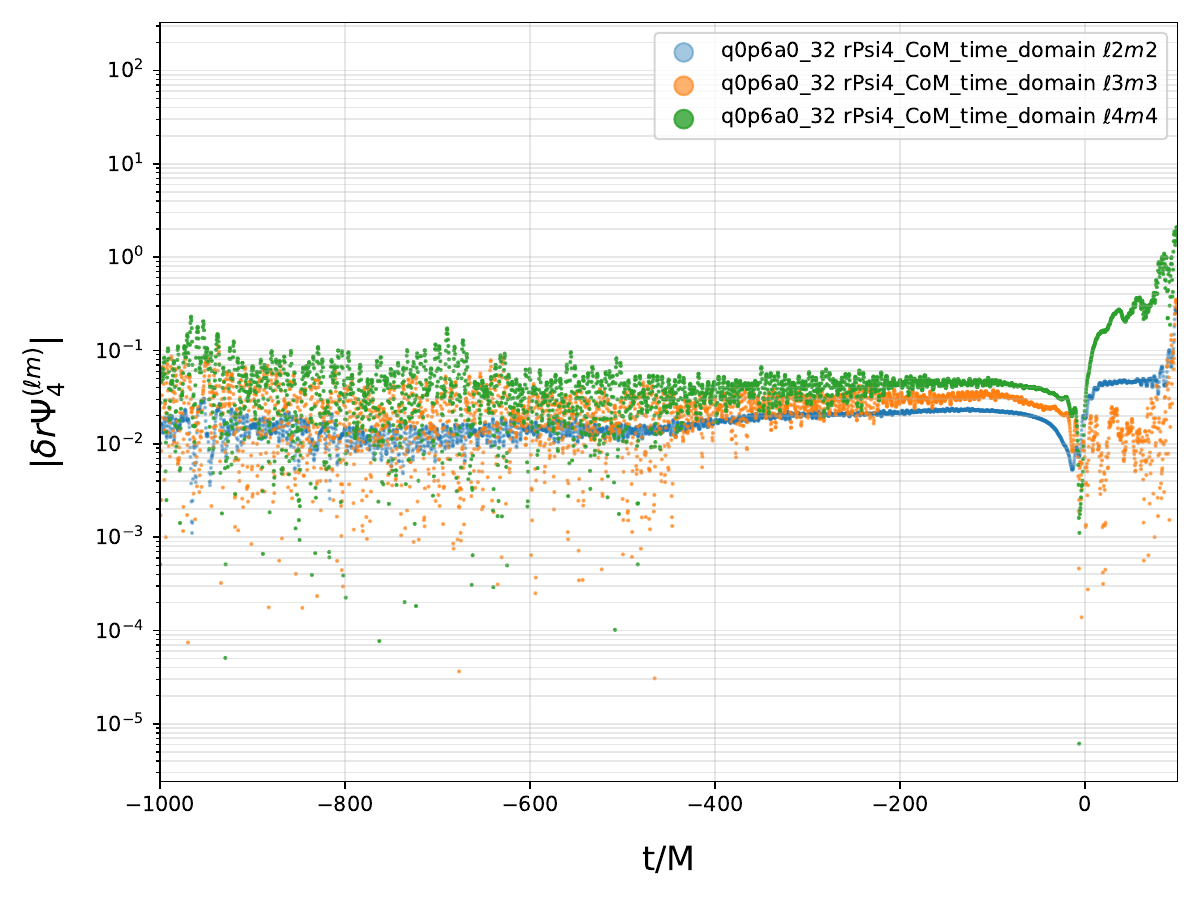}
	\includegraphics[width=\columnwidth]{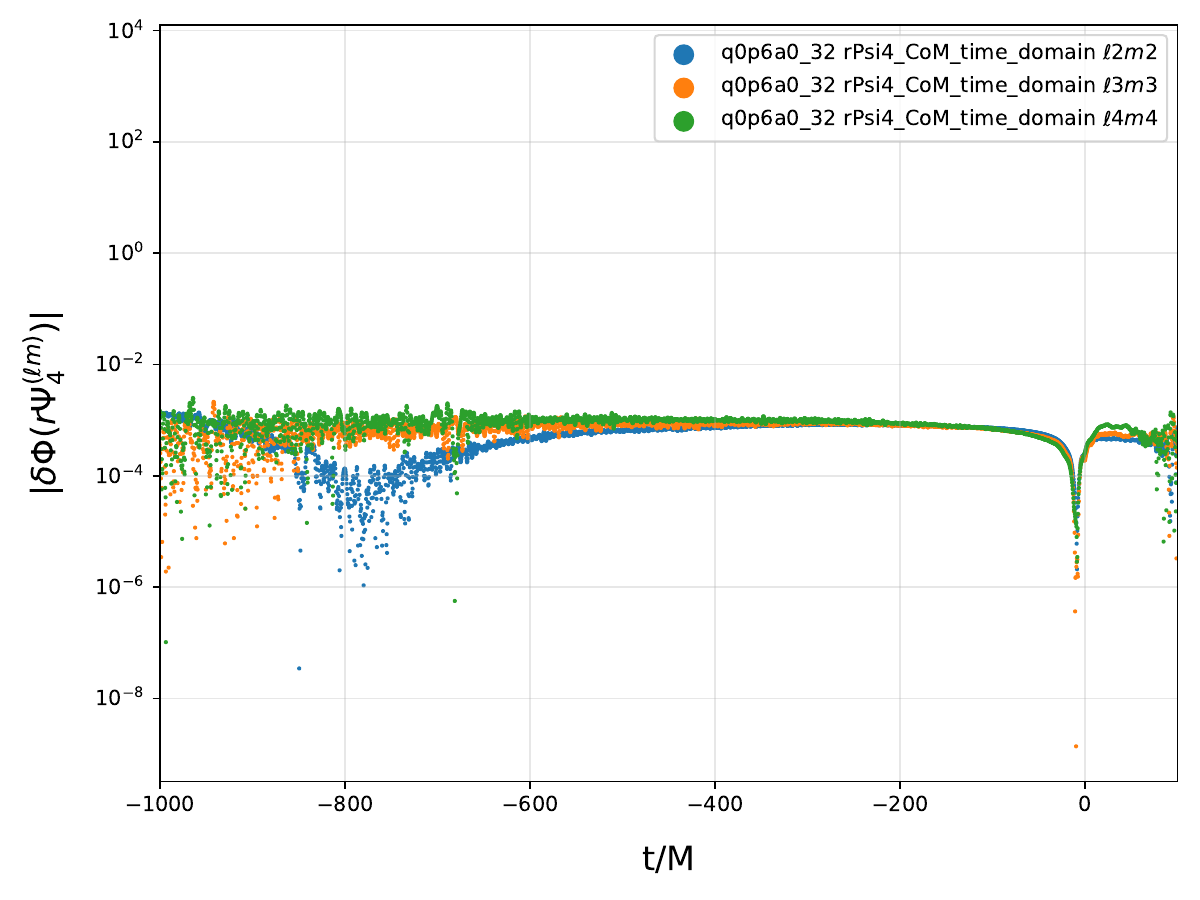}
	\caption[Waveform comparison against those of RIT catalogue.]{The fractional absolute difference in the values ($\vert \frac{(val_{CoM} - val_{\mathcal{I}^+})}{val_{\mathcal{I}^+}}\vert$) of the amplitudes (left) and phase (right) evolution of the first four dominant modes of the gravitational waveforms $\Psi_4$ of the simulation $q=0.25$ extrapolated to infinity and also corrected for the CoM drifts shown in \ref{fig:CoM_drift} with respect to the corresponding waveform from RIT catalogue. As can be seen, our waveforms agree well with those from the catalogue.}
	\label{fig:CoM_waveform_comp_vs_RIT}
\end{figure}

As mentioned earlier, the centre of mass (CoM) of a system drifts during the course of a simulation. In order to remove the effect of this from the gravitational waveform, one has to apply a boost and translation transformation on the waveform. This is carried out as follows:
\begin{enumerate}
\item Find the mean motion of the CoM. The average motion of the CoM is found by modelling its inspiral motion. The mean velocity and displacement are computed.
\item Apply a boost transformation followed by a translation on the waveform.
\end{enumerate}

We now describe all the steps of this procedure.

\subsection{The CoM mean motion}

The mean motion of the CoM is computed along the lines of \cite{CBW_waveform_correction_2019} and we describe the procedure here for the convenience of the reader. The CoM location is defined as:
\begin{equation}
	\label{eqn:comdef}
	\vec{x}_{\text{CoM.}} = \frac{M_{1}}{M} \vec{x}_1 + \frac{M_{2}}{M} \vec{x}_2,
\end{equation}
where $\vec{x}_j$ is the coordinate location of the BH$_{j}$. The mean CoM motion parameters are found by minimizing the quantity:
\begin{equation}
	\Theta(\vec{\alpha}, \vec{\beta}) = \int^{t_f}_{t_i} \vert \vec{x}_{CoM} - (\vec{\alpha} + \vec{\beta} t) \vert^2 dt
\end{equation}

The best fit values of $\vec{\alpha}$ and $\vec{\beta}$ can be expressed in terms of the temporal moments $E[t^n \vec{x}_{CoM}]$ of the CoM location for $n=0, 1$:
\begin{equation}
	E[t^n \vec{x}_{CoM}] = \dfrac{1}{t_f - t_i} \int_{t_i} ^{t_f} t^n \vec{x}_{CoM} dt 
\end{equation}

The solution for $\vec{\alpha}$ and $\vec{\beta}$ are then:
\begin{align}
	\vec{\alpha} &= \dfrac{4(t_f + t_i t_f + t_i ^2)E[\vec{x}]  - 6(t_i + t_f) E[t \vec{x}_{CoM}]}{(t_f - t_i)^2} \\
	\vec{\beta} &= \dfrac{12 E[t \vec{x}_{CoM}] - 6(t_f + t_i) E[\vec{x}_{CoM}]}{(t_f - t_i)^2}
\end{align}
The \hl{waveformtools.simulation} module contains the \hl{sim} class that can be used to compute the mean motion of the CoM. These functions read in the NR data of the simulations and compute and return the parameters $\vec{\alpha}$ and $\vec{\beta}$.

\subsection{Boost}

Then, the boost transformation is carried out on the waveform. To boost the waveform, we first transform the waveform to a coordinate basis ($\theta, \phi$) from a modal basis ($\ell, m$). This is done using the \hl{to\_spherical\_array} method of the \hl{modes\_array} class, returning an instance of the \hl{spherical\_array} class. We typically use a numerical grid of size ($121, 244$) in ($\theta, \phi$) directions, staggered about the north pole for the same. We then compute the boost conformal factor defined as:
\begin{equation}
	\Omega_{boost} = \gamma^3 (1 - \vec{\beta} \cdot \vec{r})^3 
\end{equation}
Here $\vec{r}$ is a radial unit vector that points in the direction of $(\theta, \phi)$, and $\gamma = (1 - \vert \vec{\beta} \vert^2)^{-1/2}$ is the Lorentz factor. The boost is applied on the waveform as:
\begin{equation}
	\Psi_4^{boosted} = \Omega_{boost} \Psi_4
\end{equation}

\subsection{Translation}

Now the boosted waveform is translated. The waveform at time $t$ has to be transformed by an amount $\vec{l} = \vec{\alpha} + \vec{\beta} t$. To apply the translation transformation, one notes that a super-translated waveform can be expressed in terms of the supertranslation parameter $\nu(\theta, \phi)$ and the original waveform as \cite{mode_mixing_supertranslation_2013,transf_asym_wavef_2016}:
\begin{align}
	\Psi_4 ^{st} (u) &= e^{-\left(\nu (\theta, \phi) \dfrac{\partial}{\partial u}\right)} ~\Psi_4 (u)
\end{align}
In the numerical implementation, only the first few terms of the above expansion are retained. We retain only the leading order correction term in the above expansion (although the code is capable of computing any number of terms).

At every instant of time $t$, the supertranslation parameter $\nu(t)$ is expressed on a spherical harmonic basis. Its modes are related to the displacement of the CoM $\vec{l}$ at time $t$ through the following expressions:
\begin{align}
	 \vec{l} &= (\delta x , \delta y, \delta z) \\
	  \nu^{(0, 0)} &= \sqrt{4\pi} t \\
	  \nu^{(1, \pm1)} &= \sqrt{\dfrac{2\pi}{3}} (\pm \delta x - i \delta y)\\
	  \nu^{(1, 0)} &= - \sqrt{\dfrac{4\pi}{3}} \delta z \\
\end{align}

Using the functions from \hl{waveformtools.BMS}, \hl{waveformtools.CoM}, and methods in \hl{waveformtools.waveforms} one can carry out all these computations and transformations. They can be accessed using \hl{supertranslate} method of the \hl{modes\_array} or \hl{spherical\_array} classes. The CoM corrected i.e. boosted and translated waveform on the spherical grid is then transformed back to ($\ell, m$) basis using the \hl{to\_modes\_array} method.

We compute the CoM drifts and apply CoM corrections to all the waveforms obtained from our numerical simulations set. The drift of the CoM of the system in the inspiral phase is shown in \ref{fig:CoM_drift}. Also shown here is the mean motion of the CoM of the system in green. The estimated mean motion is then used to apply the CoM corrections to the waveforms.

The CoM-corrected waveforms were computed. The comparison of the corrections for the $q=0.85$ case is shown in \ref{fig:CoM_waveform}. A comparison has also been made against the RIT catalogue \ref{fig:CoM_waveform_comp_vs_RIT}.

\section{Procedure for waveform preparation}

The gravitational waveforms from numerical simulations have to be corrected for the effects described above. The procedure followed to obtain `good' gravitational waveforms at future null infinity $\mathcal{I}^{+}$ are as under:
\begin{enumerate}
\item Load the $\Psi_4$ data, apply taper and add additional zeros.
\item Extrapolate the waveform to infinity using one of the available techniques.
\item Compute the mean CoM drift.
\item Apply the boost transformation first.
\begin{enumerate}
	\item Compute the conformal factor
	\item Resum $\Psi_4$ to angular coordinates.
	\item Apply the boost transformation.
\end{enumerate}
\item Apply the translation transformation.
\begin{enumerate}
	\item Compute the supertranslation parameter $\alpha(\theta, \phi, u)$ 
	\item Apply the supertranslation transformation.
\end{enumerate}
\item Decompose the waveform data into modes.
\item Integrate the waveform to obtain strain.
\end{enumerate}

\section{Conclusions}

The numerical relativity waveforms used in all the projects have been suitably extrapolated to infinity and integrated. They have also been corrected for the center of mass drifts and boosts. Methods for applying supertranslation transformation have also been implemented. The methods used for these corrections have been implemented using the Python programming language \cite{nrtools}. It was noted that each of these corrections brings about a per cent change in the amplitudes and phases of the waveforms.  

The gravitational waveforms from our simulations carried out using the ETK and those from the RIT and SpECTRE catalogue agree point-wise upto a percent level.
\cleardoublepage %
\flushbottom
\newpage
\cleardoublepage

\chapter{Installing the Einstein Toolkit on the Pegasus HPC}

\label{appendix:build_etk}

This document details the procedure for installing the Einstein Toolkit on the Pegasus HPC at IUCAA.

\section{Introduction}

The Pegasus HPC is a PBS/Torque system running on CentOS. This procedure should apply to similar systems.
\begin{bclogo}[logo=\bcinfo, couleurBarre=red, noborder=true, couleur=LightSalmon]{Note!}
\noindent\textbf{NOTE 1}: This procedure is designed for \textbf{ETK\_10\_2019.} Other versions may require improvisations.\\
\end{bclogo}
\begin{bclogo}[logo=\bcinfo, couleurBarre=red, noborder=true, couleur=LightSalmon]{Note!}
\noindent\textbf{NOTE 2}: This procedure is designed for Compiling and running the ETK on the Pegasus HPC at IUCAA, Pune, India. Although the strategy would remain the same for other clusters, the compiling of dependencies and configuration files would change from cluster to cluster. The changes will be indicated.
\end{bclogo}
\begin{bclogo}[logo=\bcinfo, couleurBarre=red, noborder=true, couleur=LightSalmon]{Note!}
\textbf{NOTE 3}: This assumes no root access.
\end{bclogo}

The Einstein toolkit \href{https://einsteintoolkit.org/}{https://einsteintoolkit.org/} \cite{EinsteinToolkit:web} contains documentation of the code, which is incomplete. The website also provides the procedure to install the software on a computer. This is present at this \href{https://github.com/nds-org/jupyter-et/blob/master/CactusTutorial.ipynb}{https://github.com/nds-org/jupyter-et/blob/master/CactusTutorial.ipynb}. We will present the details of this procedure.

This installation procedure is comprised of the following steps:
\begin{enumerate}
\item Setting up the dependency libraries.
\item Downloading the ETK
\item Setting up the configuration files.
\item Compiling the toolkit.
\item Post compilation modifications.
\item Running examples
\end{enumerate}

On a cluster, there are modifications that are required at each step in the list above.
These modifications depend (weakly) on the version of ETK being compiled and the libraries being used. They will be explained here.

\section{Setting up the dependency libraries}

\subsection{General libraries}
\noindent \textbf{General strategy}: Download and set up the dependency libraries mentioned in the 'tutorial for new users' link above. These are:
\vspace{0.75cm}
\begin{center}
\begin{tabular}{ c c c c }
	libjpeg-turbo-devel     &   gcc         &   git         &   lapack-devel \\
	make    &   subversion  &   gcc-c++     &   which       \\
	papi-devel  & hwloc-devel &   openmpi-devel   &   hdf5-openmpi-devel  \\
	openssl-devel & libtool-ltdl-devel  &  numactl-devel &  gcc-gfortran    \\
	hdf5-devel  & fftw-devel &    patch   &   gsl-devel    \\
\end{tabular}
\end{center}
\vspace{0.75cm}

The following libraries are Mandatory:
\vspace{0.75cm}
\begin{center}
\begin{tabular}{ c c c c }
	MPI compiler    &    HWLOC      &  GSL          &  LAPACK \\
	HDF5    &        PAPI       &     OPENSSL       & FFTW3 \\
\end{tabular}
\end{center}
\vspace{0.75cm}

The latest versions may be downloaded and installed.

Some of these libraries might already be centrally installed in the cluster. Thus two approaches may be considered here:
\begin{enumerate}
\item Locate individually the libraries already existing on the cluster out of this list and note down their paths i.e. the lib, lib64, bin and include directory paths. Download and compile the rest.
\item Download and compile the latest version of all libraries in a single directory exclusively for ETK.
\end{enumerate}

The latter may be preferable in some cases, while the former is better for those who need root access/ have compatibility issues with the cluster. Also, centrally compiled libraries may be optimized for use with the particular hardware. The latter may be preferred as all required libraries can be compiled in one location.

\begin{bclogo}[logo=\bcinfo, couleurBarre=red, noborder=true, couleur=LightSalmon]{Note!}
\textbf{Note4}: The less time coe ETK package also contains some of the major libraries like HDF5, MPI, HWLOC, etc. These will be built during compilation if they are not specified in the optionlist/ cannot be found in default locations. However, their compilation is not always successful as their versions may conflict with other packages on the cluster. It is therefore recommended to download and compile all libraries separately.
\end{bclogo}

The paths to these libraries are required for use in the configuration file for compiling the ETK (called the optionlists), which we will visit later. Please save the full paths for later use.

\subsection{MPI compiler}

The ETK can be set up in two configurations:
\begin{enumerate}
\item With MPI.
\item With MPI + OpenMP.
\end{enumerate}

The choices for the MPI compilers are: 1. Intel MPI, 2. MPICH, 3. OPENMPI.

Either of these can be used with OpenMP above.

\begin{bclogo}[logo=\bcinfo, couleurBarre=red, noborder=true, couleur=LightSalmon]{Note!}
\textbf{Note 5}:  The Intel compilers are usually the fastest, especially on a system whose hardware is supplied by Intel. For AMD, openmpi compilers may be preferred.

The Intel compilers for C, C++, and Fortran are also preferred and are usually the fastest. For AMD, the equivalent is the AOCC compilers.
\end{bclogo}

\subsubsection{Intel Parallel studio}

The Pegasus cluster has a centrally installed version of the Intel 2019.5 compiler. Alternatively, a student edition with a one year license can be directly obtained from the Intel website. The parallel studio contains all the basic compilers that ETK uses.

For downloading and compiling the libraries, prebuilt binaries for RHEL / CentOS 7 can be downloaded using a locator/ manager. One way is to use the user-yum third-party library \cite{useryum}. Another way is to manually download and compile all libraries using HPC-installed compilers.

\subsection{HDF5}

HDF5 is the library that provides the IO file system. Use Parallel HDF5 if the cluster is equipped with fast Disks like SSDs. Using parallel HDF5 on slow IO systems will make IO operations the bottleneck of runs.

The version of the OS can be checked using:
\begin{lstlisting}
cat /etc/centos-release
uname -a
\end{lstlisting}
commands.

\subsection{Checklist}

\begin{enumerate}
\item All packages downloaded.
\item Paths to include, lib, lib64, bin dirs noted.
\end{enumerate}

\section{Downloading the ETK}

This is the simplest of all the steps. Following the tutorial, one can use the curl and GetComponents to download the ETK components from the git and SVN repositories.

The GetComponents accepts thronlist as an argument. This thorn list contains the names of repositories and thorns that should be compiled and the links to the same.

If you need to add additional thorns/ functionalities to the ETK installation, you may stop the GetComponents execution after the thorn list downloads (the thorn list is also included in the components) and manually add the details of the required thorn.

It is recommended to include the following thorns:
\begin{enumerate}
\item Llama/WaveExtractCPM: Provides approximate strain computation based on perturbation theory.
\end{enumerate}

And run the GetComponents again by specifying the updated thorn list as an argument.

On the Pegasus cluster, the access to all the git/svn repositories is not stable and frequently results in failure of GetComponents routine. In this case, it is recommended to run the GetComponents from a local computer on the IUCAA LAN network and then transfer the files to the cluster account using \emph{rsync}/ \emph{scp}.

\begin{bclogo}[logo=\bcinfo, couleurBarre=red, noborder=true, couleur=LightSalmon]{Note!}
\textbf{Note 6}: Local LAN computer is recommended because the ETK installation is ~ 12 Gb.
\end{bclogo}

\subsection{Checklist}

\begin{enumerate}
\item Ensure all the optional thorns are downloaded (if any).
\item All components have been downloaded without any error messages.
\end{enumerate}

\section{Setting up the configuration files}

Please run the

\begin{lstlisting}
sim setup --silent
\end{lstlisting}
command and then proceed to the following steps.

Compiling the ETK is just like compiling any other source code. It requires a configuration file, called the option list, that contains information regarding:
\begin{enumerate}
\item The Compilers to be used
\item The setting for the compilers.
\item The optimization flags.
\item The proper linking of compilers with libraries.
\item Path to the libraries.
\item Other functionalities/modes/settings.
\end{enumerate}

While the proper setting of all but item 3 determines whether or not one can successfully compile the toolkit, item 3 is crucial for the proper performance of the toolkit.

Please keep the paths to the libraries noted in Sec. 2 handy.

The option lists for compiling the ETK\_10\_2019p and ETK\_05\_2020p can be requested from the author. Please keep in mind that this is only a template and the paths to the compilers will need to be changed. Please request for access to the libraries compiled, or if you want to use the libraries compiled by the author.

\subsection{Description of mdb files}

These are about the files in the corresponding sub-directories of the \emph{Cactus/simfactory/mdb} directory. Please find some examples of the mdb files below.

\subsubsection{Variables used}

\noindent\textbf{PREFIX}: will be used to denote the full directory path containing the variables.\\
\noindent\textbf{HOME}: Path to Home directory.\\
\noindent\textbf{SCRATCHDIR}: Path to scratch directory.\\

\subsubsection{Optionlist}

\textbf{Directory}: /optionlist

Description of the optionlists: An example optionlist is given below.
\newpage

\begin{lstlisting}
------------------------------------------------------------------------
##Optionlist
------------------------------------------------------------------------

# Example-pegasus.

# This option list is used internally by simfactory as a template during the
# sim setup and sim setup-silent commands
# Edit at your own risk

# Whenever this version string changes, the application is configured
# and rebuilt from scratch


VERSION = Pegasus-2019-07-18

#Compiler Options: Which C and Fortran compilers to use. Common options are intel and gcc #compilers. Intel compilers are : icc, icpc, ifort while the GNU compilers are gcc, g++ and gfortran.

#C Compilers
CPP = cpp
CC = icc
CXX = icpc

#FORTRAN compilers.
FPP = cpp
F77 = ifort
F90 = ifort



#Compiler flags: What flags to use with the compiler. Please google for a description :)
CPPFLAGS = -D_XOPEN_SOURCE -D_XOPEN_SOURCE_EXTENDED

#For C
CFLAGS = -g -std=gnu99 -march=native#c99 #gnu99
CXXFLAGS = -g -std=gnu++11 -march=native#c++11 #gnu++11

#For Fortran
FPPFLAGS = -g -traditional
F77FLAGS = -g -march=native
F90FLAGS = -g -march=native#-fcray-pointer -ffixed-line-length-none


#Carpet Flags
NDEBUG=1
CARPET_OPTIMIZE=1


#LD flags: Most compilers run in modes that use other libraries. LD flags are linker flags that help compilers identify the libraries they have to use when running.

LDFLAGS = -rdynamic -L$PREFIX/HDF5/hdf5-1.10.6p/lib -I$PREFIX/HDF5/hdf5-1.10.6p/include -I/mnt/csoft/compilers/intel-2019-5/compilers_and_libraries_2019.5.281/linux/mpi/intel64/include -L/mnt/csoft/compilers/intel-2019-5/compilers_and_libraries_2019.5.281/linux/mpi/intel64/lib/release -L/mnt/csoft/compilers/intel-2019-5/compilers_and_libraries_2019.5.281/linux/mpi/intel64/lib -Xlinker --enable-new-dtags -Xlinker -rpath -Xlinker /mnt/csoft/compilers/intel-2019-5/compilers_and_libraries_2019.5.281/linux/mpi/intel64/lib/release -Xlinker -rpath -Xlinker /mnt/csoft/compilers/intel-2019-5/compilers_and_libraries_2019.5.281/linux/mpi/intel64/lib -lmpifort -lmpi -lrt -lpthread -Wl,-z,now -Wl,-z,relro -Wl,-z,noexecstack -Xlinker --enable-new-dtags -ldl




#LIBDIRS: There are paths to the libraries corresponding to the linker options used in LDFLAGS above.

LIBDIRS= -L$PREFIX/MPI/intel/compilers_and_libraries/linux/mpi/intel64/lib -L$PREFIX/MPI/intel/compilers_and_libraries/linux/lib/intel64 -L$LD_LIBRARY_PATH -L$PREFIX/MPI/intel/compilers_and_libraries/linux/mpi/intel64/lib

#Libraries to link the compilers with.
LIBS =-lifcore -lhdf5 -lhdf5_hl


# Line directives. Yes, that's it.
C_LINE_DIRECTIVES = yes
F_LINE_DIRECTIVES = yes


#Debugging options. If you want to compile for just debugging purposes, this presents additional tools.
DEBUG = no
CPP_DEBUG_FLAGS = -DCARPET_DEBUG
C_DEBUG_FLAGS = -DCARPET_DEBUG
CXX_DEBUG_FLAGS =


#Optimization flags. Very important. -Ofast is the fastest but may introduce some inaccuracies.
OPTIMISE = yes
CPP_OPTIMISE_FLAGS =
C_OPTIMISE_FLAGS = -O3 -march=native -ipo -mtune=skylake
CXX_OPTIMISE_FLAGS = -O3 -march=native -ipo -mtune=skylake
F77_OPTIMISE_FLAGS = -O3 -march=native -ipo -mtune=skylake
F90_OPTIMISE_FLAGS = -O3 -march=native -ipo -mtune=skylake


#ToDo: Why cant we use -O3 here?

CPP_NO_OPTIMISE_FLAGS =
FPP_NO_OPTIMISE_FLAGS =
C_NO_OPTIMISE_FLAGS = -O0
CXX_NO_OPTIMISE_FLAGS = -O0
CUCC_NO_OPTIMISE_FLAGS =
F77_NO_OPTIMISE_FLAGS = -O0
F90_NO_OPTIMISE_FLAGS = -O0

#Profile flags.
#ToDo: find out.
PROFILE = no
CPP_PROFILE_FLAGS =
FPP_PROFILE_FLAGS =
C_PROFILE_FLAGS = -pg
CXX_PROFILE_FLAGS = -pg
F77_PROFILE_FLAGS = -pg
F90_PROFILE_FLAGS = -pg


#Warn Flags.
#ToDo: Fins all options and effects.
WARN = yes
CPP_WARN_FLAGS = #-Wall
C_WARN_FLAGS = #-Wall
CXX_WARN_FLAGS = #-Wall
F90_WARN_FLAGS = #-Wall

#OPENMP: If you want to use OPENMP with MPI, use this option.
OPENMP = no#yes
CPP_OPENMP_FLAGS = #-fopenmp
FPP_OPENMP_FLAGS = #-D_OPENMP
C_OPENMP_FLAGS = #-qopenmp
CXX_OPENMP_FLAGS = #-qopenmp
F90_OPENMP_FLAGS = #-qopenmp
F77_OPENMP_FLAGS = #-qopenmp


#Library paths:


#BLAS
BLAS_LIBS= -mkl
BLAS_DIR=/usr
BLAS_LIB_DIRS=/usr/lib64
BLAS_INC_DIRS=/usr/include

#MPI

MPI_DIR=/mnt/csoft/compilers/intel-2019-5/compilers_and_libraries_2019.5.281/linux/mpi/intel64/
MPI_INC_DIRS=/mnt/csoft/compilers/intel-2019-5/compilers_and_libraries_2019.5.281/linux/mpi/intel64/include
MPI_LIB_DIRS=/mnt/csoft/compilers/intel-2019-5/compilers_and_libraries_2019.5.281/linux/mpi/intel64/lib


#HWLOC
HWLOC_DIR=$PREFIX/soft1/root/usr
HLOC_INC_DIRS=$PREFIX/soft1/root/usr/include
HWLOC_LIB_DIRS=$PREFIX/soft1/root/usr/lib64
HWLOC_EXTRA_LIBS = numa

#PAPI_DIR = $PREFIX/soft1/root/usr

#LAPACK
LAPACK_LIBS=-mkl
LAPACK_DIR=/mnt/csoft/libraries/lapack-3.8.0
LAPACK_LIB_DIRS=/mnt/csoft/libraries/lapack-3.8.0/lib

#GSL
GSL_DIR=$PREFIX/other/gsl
GSL_INC_DIRS=$PREFIX/include
GSL_LIB_DIRS=$PREFIX/soft1/root/usr/lib64

#Boost
BOOST_DIR= BUILD

#PAPI
PAPI_DIR=$PREFIX/soft1/root/usr


#HDF5
HDF5_DIR=$PREFIX/HDF5/hdf5-1.10.6p
HDF5_LIB_DIRS=$PREFIX/HDF5/hdf5-1.10.6p/lib
HDF5_INC_DIRS=$PREFIX/HDF5/hdf5-1.10.6p/include

#OPENSSL
OPENSSL_DIR=$PREFIX/soft1/root/usr

#FFTW3
FFTW3_DIR=/mnt/csoft/libraries/fftw-3.3.8/bin
FFTW3_LIB_DIRS=/mnt/csoft/libraries/fftw-3.3.8/lib
FFTW3_INC_DIRS=/mnt/csoft/libraries/fftw-3.3.8/include
FFTW3_LIBS = -mkl

#Vectorization
VECTORISE          = yes
VECTORISE_ALIGNED_ARRAYS = no
VECTORISE_INLINE          = yes

#Pthreads
PTHREADS_DIR = NO_BUILD
---------------------------------------------------------------------------------------------------------------
\end{lstlisting}

\subsubsection{The Machine file}

\textbf{Directory}: \emph{/machines}

Contents of the machine file:

\begin{lstlisting}
------------------------------------------------------------------------
##Machine file
------------------------------------------------------------------------



[pegasus]

# This machine description file is used internally by simfactory as a template
# during the sim setup and sim setup-silent commands
# Edit at your own risk
# Machine description

nickname = pegasus
name = pegasus
location = IUCAA, Pune
description = The Pegasus cluster
status = production

# Access to this machine
hostname = <login-node>
aliaspattern = ^<compute-node-pattern>$ #Added proper aliaspattern on 17/07/2019

# Source tree management
sourcebasedir = $PREFIX/ETK_05_2020p
optionlist = pegasus.cfg
submitscript = pegasus.sub
runscript = pegasus.run
make      = make
basedir      = $SCRATCHLOC/simulations/ETK_05_2020p
spn         = 2
mpn         = 2
ppn      = 32
max-num-threads = 32
num-threads = 1
min-ppn      = 1
nodes = 24
submit = qsub -k oe @SCRIPTFILE@
interactivecmd = qsub -I -X -l nodes=@NODES@:ppn=@PPN@ -l walltime=@WALLTIME@ -l cput=@WALLTIME@
getstatus = qstat -x @JOB_ID@ |tail -n 1|awk {'print $5'}
stop = qdel @JOB_ID@
submitpattern = (\d+[.]hn1)
statuspattern = .+
queuedpattern = 'Q'
runningpattern = 'R'
holdingpattern = 'H'
finishedpattern = 'F'
scratchbasedir = $SCRATCHLOC

exechost = qstat -f @JOB_ID@
exechostpattern = exec_host = (\w+)/
stdout =$HOME/*.o@JOB_ID@ #ssh @EXECHOST@ cat /var/spool/PBS/spool/@JOB_ID@.hn1.OU
stderr =$HOME/*.e@JOB_ID@ #ssh @EXECHOST@ cat /var/spool/PBS/spool/@JOB_ID@.hn1.ER
stdout-follow = ssh @EXECHOST@ tail -n 100 -f$HOME/*.o@JOB_ID@ /$HOME/*.e@JOB_ID@ #/var/spool/PBS/spool/@JOB_ID@.hn1.OU /var/spool/PBS/spool/@JOB_ID@.hn1.ER



--------------------------------------------------------------------------------------------------------------------
\end{lstlisting}

\subsubsection{The submit script}

Contents of the submit scriptscript:
\begin{lstlisting}
------------------------------------------------------------------------
##Submit script
------------------------------------------------------------------------


#! /bin/bash
#PBS -l select=@NODES@:ncpus=@PPN@
#PBS -N @SHORT_SIMULATION_NAME@
#PBS -o @RUNDIR@/@SIMULATION_NAME@.out
#PBS -e @RUNDIR@/@SIMULATION_NAME@.err
cd @SOURCEDIR@
@SIMFACTORY@ run @SIMULATION_NAME@ --machine=@MACHINE@ --restart-id=@RESTART_ID@ @FROM_RESTART_COMMAND@
-----------------------------------------------------------------------------------
\end{lstlisting}

\subsubsection{The runscript}

\begin{lstlisting}
------------------------------------------------------------------------
##Ruscript
------------------------------------------------------------------------


#!/bin/sh

# This runscript is used internally by simfactory as a template during the
# sim setup and sim setup-silent commands
# Edit at your own risk

echo "Preparing:"
set -x # Output commands
set -e # Abort on errors
cd @RUNDIR@-active

echo "Checking:"
pwd
hostname
date
cat ${PBS_NODEFILE} > SIMFACTORY/NODES

echo "Environment:"
export CACTUS_NUM_PROCS=@NUM_PROCS@
export CACTUS_NUM_THREADS=1
export GMON_OUT_PREFIX=gmon.out
export OMP_NUM_THREADS=1
#Set the MPIDIR where mpirun recides.
MPIDIR="/mnt/csoft/compilers/intel-2019-5/compilers_and_libraries_2019.5.281/linux/mpi/intel64"
env | sort > SIMFACTORY/ENVIRONMENT

echo "Starting:"
export CACTUS_STARTTIME=$(date +%s)

if [ ${CACTUS_NUM_PROCS} = 1 ]; then
if [ @RUNDEBUG@ -eq 0 ]; then
@EXECUTABLE@ -L 3 @PARFILE@
else
gdb --args @EXECUTABLE@ -L 3 @PARFILE@
fi
else
$MPIDIR/bin/mpirun -np @NUM_PROCS@ @EXECUTABLE@ -L 3 @PARFILE@
fi

echo "Stopping:"
date
echo "Done."
--------------------------------------------------------------------------------------------------------------
\end{lstlisting}

\subsection{Checklist}
\begin{enumerate}
\item The sim setup command has been run.
\item The mdb files have been created.
\end{enumerate}

\section{Compiling the toolkit.}
\begin{enumerate}
\item Ensure that the appropriate compilers/libraries have been loaded.
\item Either run on the head node with max 32 threads or start an interactive job with one or two nodes and compile to the toolkit.
\item Please compile the toolkit now using the sim build command.
\end{enumerate}

\subsection{Checklist}
\begin{enumerate}
\item Ensure that the compilation is successful.
\end{enumerate}

\section{Cross Checks}

Check and modify (if necessary).
\begin{enumerate}
\item Ensure that the mdb files are intact. The Cactus/configs/RunScript and SubmitScripts have to be cross-checked too.
\item Ensure that the paths in the machine file are intact.
\end{enumerate}

\section{Run the examples!}
\begin{enumerate}
\item Run the HelloWorld! code on the head node.
\item Run the TOV code (NOT ON THE HEAD NODE)
\end{enumerate}

Some more examples can be found on the website \cite{EinsteinToolkit:web}.

For any questions/comments/ suggestions, please mail the author.

\subsection{Checklist}

Ensure the performance is up to the mark (physical units per hour i.e. me\_per\_hour).

Expected performance GW150914: On Pegasus ~ 25 units for 64 processors.
\vspace{5cm}
\begin{center}
{ \Large
	Enjoy simulating!}
\end{center}
\cleardoublepage
\cleardoublepage %
\flushbottom
\newpage
\cleardoublepage
\end{appendices}
\label{app:bibliography} %

\manualmark
\markboth{\spacedlowsmallcaps{\bibname}}{\spacedlowsmallcaps{\bibname}} 
\refstepcounter{dummy}

\addtocontents{toc}{\protect\vspace{\beforebibskip}} %
\addcontentsline{toc}{chapter}{\tocEntry{\bibname}}

\bibliography{references/references}

\end{document}